\documentclass[fleqn,usenatbib]{mnras}

\usepackage{newtxtext,newtxmath}

\usepackage[T1]{fontenc}

\DeclareRobustCommand{\VAN}[3]{#2}
\let\VANthebibliography\thebibliography
\def\thebibliography{\DeclareRobustCommand{\VAN}[3]{##3}\VANthebibliography}

\usepackage{graphicx}	
\usepackage{amsmath}	
\usepackage{cuted}
\usepackage{tasks}
\usepackage{esint}

\title[SPMHD for galaxy formation and cosmology]{Smoothed particle magnetohydrodynamics for simulations of galaxy and cosmic structure formation}

\author[O. A. Karapiperis et al.]{Orestis A. Karapiperis,$^{1,2}$\thanks{E-mail: karapiperis@strw.leidenuniv.nl}
Matthieu Schaller,$^{1,2}$
Nikyta Shchutskyi,$^{1,2}$
Federico A. Stasyszyn$^{3,4}$, and
\newauthor
Maarten Elion$^{1}$\\
$^{1}$Lorentz Institute for Theoretical Physics, Leiden University, PO Box 9506, 2300 RA Leiden, the Netherlands\\
$^{2}$Leiden Observatory, Leiden University, PO Box 9513, 2300 RA Leiden, the Netherlands\\
$^{3}$Instituto de Astronom\'ia te\'orica y experimental-Conicet, Laprida 854, X5000BGR C \'ordoba, Argentina\\
$^{4}$Observatorio Astron\'omico de C\'ordoba, Universidad Nacional de C\'ordoba, Laprida 854, X5000BGR C\'ordoba, Argentina}

\date{Accepted XXX. Received YYY; in original form ZZZ}

\pubyear{\the\year}

\begin{document}
\label{firstpage}
\pagerange{\pageref{firstpage}--\pageref{lastpage}}
\maketitle

\begin{abstract}
We introduce a novel formulation of cosmological smoothed particle magnetohydrodynamics (SPMHD), a discrete Lagrangian method which can be used to model magnetic field physics in a vast array of nonlinear astrophysical systems, and which we have implemented in the highly-parallel, entirely modular, and open-source simulation code SWIFT. Our numerical scheme is designed to offer optimal performance at a minimal computational cost, keep a low memory footprint, and most notably couple robustly to effective sub-resolution recipes of galaxy formation. This is achieved through expressing our evolution equations in a density-energy conservative form, and augmenting them with discontinuity-capturing terms tailored to high dynamic range simulations, which are further modulated by adaptive switches that drastically improve coupling to sub-grid models and limit spurious dissipation. We moreover present novel suggestions for the two major regularisation techniques used in modern SPMHD, namely a tensile instability correction and mixed hyperbolic/parabolic divergence-cleaning scheme, to ensure code stability in highly dynamical scenarios. We evaluate the performance of our method on a series of problems of increasing complexity, culminating in three astrophysical applications which have historically proven challenging for mesh-less methods: we study jet launching from a forming proto-stellar core, dynamo amplification in a massive galaxy cluster and magnetic field evolution in a Milky Way-like disk galaxy; the latter constitutes the first reported coupling of the EAGLE galaxy formation model to a magnetohydrodynamics solver. Keeping model hyperparameters fixed across our test suite to provide a transparent picture of our method's capabilities in production, we demonstrate sound performance and convergence with resolution on standard `laboratory' numerical experiments, as well as competitive capabilities in realistic applications.
\end{abstract}




\section{Introduction}


There is little remaining doubt that magnetic fields constitute a core component of our cosmos. This may have been the case from our universe's infancy onward, as posited by tentative indirect detections of non-zero extragalactic fields inferred from gamma-ray observations of powerful blazars~(e.g. \citealt{2010Sci...328...73N, 2010MNRAS.406L..70T, 2011MNRAS.414.3566T, 2011ApJ...733L..21D, 2011ApJ...727L...4D, 2011APh....35..135E, 2011A&A...529A.144T, 2012ApJ...747L..14V}; see however e.g.~\citealt{2012ApJ...752...22B} for an alternative interpretation of said observations). The existence of such magnetic fields which would be primordial in origin has, for instance, been proposed by~\citet{2020PhRvL.125r1302J} as a viable pathway through which to resolve the now longstanding `Hubble tension'~\citep[see e.g.][for an elaboration on other possible solutions]{2021CQGra..38o3001D}. 

Although the precise origin of magnetic fields in the universe remains largely unknown~\citep[see e.g.][for comprehensive reviews of the topic]{2008RPPh...71d6901K, 2012SSRv..166...37W, 2016RPPh...79g6901S}, a large body of evidence has unequivocally established their importance in a wide array of cosmic environments and astrophysical objects. Magnetic fields are believed to play a key role in the physics of proto-stellar cores, disks, and mass outflows~\citep{2023ASPC..534..317T}, as well as in the later stages of stellar evolution~\citep{2009ARA&A..47..333D}, up until the supernova phase~\citep{1970ApJ...161..541L, 1976Ap&SS..41..287B, 1976Ap&SS..42..401B}. They also play a crucial role in driving turbulent angular momentum transport in accretion disks through the magnetorotational instability~\citep[MRI;][]{1998RvMP...70....1B}, but also in launching powerful jets~\citep{1977MNRAS.179..433B, 1982MNRAS.199..883B, 1986PASJ...38..631S, 1996MNRAS.279..389L, 2003MNRAS.341.1360L, 2004ApJ...605..656V} from compact objects. The latter has also been studied~\citep[by e.g.][]{2017ApJ...834L..19G, 2017MNRAS.469.4879B} for the case of a relativistic jet powered by sources fed by tidal disruption events~\citep[such as the one reported in][]{2011Sci...333..203B}. Compact object mergers have been claimed to be events during which some of the strongest cosmic magnetic fields are generated~\citep{2006Sci...312..719P}, likely driving in turn highly energetic gamma-ray bursts~\citep{1992Natur.357..472U}. Diffuse interstellar molecular clouds are also believed to be magnetised, an attribute anticipated to have an impact on star formation occurring therein~\citep{2005LNP...664..137H, 2012ApJ...761..156F, 2015MNRAS.450.4035F}. Similarly, magnetic fields are also manifestly present in the gas phase of the interstellar medium, where they trigger fluid instabilities and partly dictate the topology of the baryonic matter distribution~\citep{1966ApJ...145..811P, 1990ARA&A..28..215D, 2011piim.book.....D}.

Another branch of astronomical research wherein magnetic fields physics and its repercussions are increasingly considered is galaxy and cosmic structure formation. Observations of both nearby~\citep[see e.g.][for an exhaustive account]{2013pss5.book..641B, 2015A&ARv..24....4B} and distant~\citep{2008Natur.454..302B, 2017NatAs...1..621M} galaxies, as well as of massive galaxy clusters~\citep[][for a recent review]{2019SSRv..215...16V}, point towards the presence of relatively strong ($\gtrsim 1-10 \; \mu \mathrm{G}$) and oftentimes spatially coherent magnetic fields on very large scales. Studying the mechanisms leading to such field strengths and topologies is naturally appealing in its own right~(refer to~\citealt{2023ARA&A..61..561B} and~\citealt{2018SSRv..214..122D} for reviews
on our theoretical understanding of magnetic fields in galaxies and clusters, respectively), but is also of interest as it has implications on other phenomena of interest such as cosmic ray transport~\citep{2011ARA&A..49..119K} and particle (re)acceleration~\citep[see again][and references therein]{2019SSRv..215...16V}. 

The framework of choice for theoretical studies of magnetic fields in astronomy is magnetohydrodynamics~\citep[see e.g.][for an introduction]{2016JPlPh..82c2001O}. Owing to the inherent non-linear nature of the equations that constitute it, as well as to the complexity and analytical intractability of astrophysical systems of interest, it is no surprise that a numerical treatment of MHD imposed itself as the natural approach to adopt. A considerable amount of effort has gone into devising and refining the corresponding computational tools, with progressively more complex discrete methods being implemented into increasingly sophisticated astrophysics codes; popular examples include 
\texttt{ZEUS}~\citep{1992ApJS...80..791S}, 
\texttt{FLASH}~\citep{2000ApJS..131..273F}, 
\texttt{HARM}~\citep{2003ApJ...589..444G}, 
\texttt{MAGMA}~\citep{2007MNRAS.379..915R}, 
\texttt{PLUTO}~\citep{2007ApJS..170..228M}, 
\texttt{ATHENA}~\citep{2008ApJS..178..137S}, 
\texttt{NIRVANA}~\citep{2008CoPhC.179..227Z}, 
\texttt{ENZO}~\citep{2014ApJS..211...19B}, 
\texttt{PHANTOM}~\citep{2018PASA...35...31P}, 
\texttt{ATHENA++}~\citep{2020ApJS..249....4S} and
\texttt{PENCIL}~\citep{2021JOSS....6.2807P}. The variety of available options is noteworthy, even when restricting oneself to the particular case of scientific software chiefly intended for galaxy evolution and non-linear structure formation studies; MHD solvers developed for such applications include the high-order Godunov scheme of~\citet{2006A&A...457..371F} that was incorporated into the~\texttt{RAMSES} adaptive mesh refinement (AMR) code~\citep{2002A&A...385..337T}, the mesh-less smoothed particle magnetohydrodynamics (SPMHD) method of~\citet{2009MNRAS.398.1678D} that was conceived as a constituent of \texttt{GADGET}~\citep{2001NewA....6...79S, 2005MNRAS.364.1105S}, the second-order finite-volume scheme of~\cite{2011MNRAS.418.1392P} implemented into the moving-mesh code \texttt{AREPO}~\citep{2010MNRAS.401..791S}, the `mesh-less finite mass'/`mesh-less finite volume' (MFM/MFV) approaches of~\citet{2016MNRAS.455...51H} that are a part of \texttt{GIZMO}~\citep{2015MNRAS.450...53H}, and `geometric density average' formulation of SPMHD conceived by~\citet{2020A&A...638A.140W} that was recently included in \texttt{GASOLINE2}~\citep{2017MNRAS.471.2357W} and \texttt{CHANGA}~\citep{2015ComAC...2....1M}. 

The major technical challenge any numerical MHD method has to address is maintaining the so-called `solenoidal condition' (SC), i.e. ensuring that any model magnetic field $\boldsymbol{B}$ satisfies the topological constraint $\nabla \cdot \boldsymbol{B} = 0$, as expected from Maxwell's equations. Although any initially divergence-less $\boldsymbol{B}$ will indefinitely fulfill $\nabla \cdot \boldsymbol{B} = 0$ in the continuum limit, this will no longer be the case for a discretised field: \citet{1980JCoPh..35..426B} discussed~\citep[see also][]{1985SSRv...42..153B, 1999JCoPh.149..270B} how for formally conservative MHD solvers, even small violations of the SC (owing for instance to the finite accuracy of discrete differential operators or time integration algorithms) can translate to unphysical forces proportional to $\nabla \cdot \boldsymbol{B}$ and directed along the magnetic field, which can self-amplify and corrupt numerical results entirely. 

A proposed solution to the aforementioned problem came from~\citet{1980JCoPh..35..426B}, who suggested favouring a non-conservative formulation of MHD, whereby a Helmholtz decomposition of the computed $\boldsymbol{B}$ is to be performed at each step to then retrieve a divergence-less projection perfectly perpendicular to the Lorentz force. A follow-up substitute recommendation was the `eight wave cleaning' approach of~\citet{1999JCoPh.154..284P}~\citep[see also][]{Powell1997} who proposed including additional source terms to the MHD equations of motion (EoM); at the cost of momentum non-conservation, these promote advection of $\nabla \cdot \boldsymbol{B}$ with (rather than its dispersion by) the fluid flow, preserving numerical divergence at the level of the truncation error and inhibiting the development of numerical instabilities. 

An elegant alternative was proposed by~\citet{1988ApJ...332..659E}, termed `constrained transport' (CT), where magnetic fields are defined as quantities averaged over the employed mesh's faces, while their evolution equation is considered in integral form; the assumed discrete counterpart of the latter is constructed in a way such that magnetic flux is conserved and $\nabla \cdot \boldsymbol{B} = 0$ to machine precision in the chosen discretisation. Initially tested on static Cartesian grids~\citep[see e.g.][]{1999JCoPh.149..270B}, CT was widely adopted soon thereafter in AMR codes~\citep[e.g.][]{2001JCoPh.174..614B,  
2005JCoPh.205..509G, 2008JCoPh.227.4123G, 2006JCoPh.218...44T, 2006A&A...457..371F, 2010ApJS..186..308C}, and has more recently been ported to unstructured moving-mesh codes as well~\citep{2014MNRAS.442...43M, 2016MNRAS.463..477M}; it is unclear for now whether, in the absence of a well defined grid on the faces of which the magnetic field could be represented, CT can be adapted to work with mesh-less MHD solvers.

\citet{2000JCoPh.161..605T} undertook an extensive comparative study of the divergence control capabilities of variants of the projection, eight wave and CT algorithms, to find the first's non-conservative nature leads to it predicting erroneous jump conditions at MHD shocks~\citep[later confirmed by e.g.][as well]{2010JCoPh.229.2117M, 2016MNRAS.455...51H}, while the latter two provide more robust results (albeit at an admittedly increased computational cost). This was later nuanced by~\citet{2004ApJ...602.1079B} finding that projection methods introduce unphysical small-scale features in tests involving interacting shocks, or~\citet{2013MNRAS.432..176P} defending that CT generates large advection errors in high Mach number flows, both configurations abounding in real astrophysical systems. 

Further alternatives for divergence control, compatible with a broader set of numerical methods, remained indispensable. An interesting suggestion came from~\citet{2002JCoPh.175..645D}, who proposed an incarnation of a `generalised Lagrange multiplier' formulation of MHD~\citep{2000JCoPh.161..484M}, which rather than aiming for a formally divergence-less magnetic field seeks instead to maintain $\nabla \cdot \boldsymbol{B}$ numerically small. This is achieved by coupling the MHD EoM to a scalar field obeying a mixed hyperbolic/parabolic evolution equation, which transports divergence at some maximal admissible speed away from where it is sourced in the form of waves, which are subsequently damped at a critical rate. This framework is method agnostic and can thus be implemented in any code; this, together with its empirically determined robustness and established capacity to dramatically reduce divergence errors, led to its widespread adoption by the research community~\citep[see e.g.][]{2006CQGra..23.6503A, 2007PASJ...59..905M, 2009ApJ...696...96W, 2010JCoPh.229.5896M, 2011MNRAS.418.1392P, WAAGAN20113331, 2012ApJS..198....7M, 2012JCoPh.231..718K, 2012JCoPh.231.7214T, 2013MNRAS.428...13S, 2016JCoPh.322..326T, 2016MNRAS.455...51H, 2020A&A...638A.140W, 2023MNRAS.518.4115G, 2026MNRAS.550g1121S}. It is fully compatible with and can be seamlessly coupled to other mitigation strategies~\citep[such as the eight-wave cleaning approach of][]{1999JCoPh.154..284P}, and can even be augmented with further algorithmic steps~\citep{2016MNRAS.462..576H}.

Although most of the aforementioned pioneering developments were carried out in the context of studies making use of Eulerian solvers, there is unquestionable value in incorporating MHD in Lagrangian and in particular particle-based codes as well, given the advantages these confer in the context of astrophysical fluid dynamics simulations. Highly desirable properties of such schemes include the fact that the discrete elements used to represent the fluid move with the bulk flow, and resolution is thus naturally coupled to mass; elements can intrinsically sample any non-trivial spatial geometry and the computational effort is automatically and adaptively targeted where the system under investigation is most dynamically active. Such methods are further inherently dissipation-less and allow for advection to be modeled perfectly, they are Galilean invariant and offer consistent performance irrespective of the presence of a (potentially supersonic) background flow, couple seamlessly to gravitational N-body solvers as well as additional physics modules, and can be derived such that they preserve a number of symmetries, conserving for instance mass, linear momentum, angular momentum, entropy or energy exactly. Compared to their Eulerian counterparts, particle methods are not subject to grid alignment effects or excessive numerical diffusion, and are also typically less demanding of time integration algorithms. SPMHD~\citep[see][for a recent review]{2023FrASS..1088219T} constitutes a prime example of a Lagrangian MHD scheme, which was conceived as an extension to smoothed particle hydrodynamics (SPH). 

SPH is a particle based method~\citep[first presented in][]{1977MNRAS.181..375G, 1977AJ.....82.1013L} that can be used to model the dynamics of a collisional fluid~\citep[see][for reviews]{2005RPPh...68.1703M, 2009NewAR..53...78R, 2010ARA&A..48..391S, 2012JCoPh.231..759P}. Its inherent simplicity, proven stability properties, low computational cost and ability to concurrently model spatial and temporal scales that vary by several orders of magnitude led to its early adoption as a tool for cosmological hydrodynamical simulations~\citep{1989ApJS...70..419H}. Original incarnations of the method suffered from shortcomings such as the development of spurious pressure forces at interfaces~\citep{2007MNRAS.380..963A} or the apparent inability to model subsonic turbulence~\citep{2012MNRAS.423.2558B}; these have since been largely overcome however, through improvements (among others) to discontinuity and shock capturing~\citep{2008JCoPh.22710040P, 2008MNRAS.387..427W, 2010MNRAS.408..669C, 2012MNRAS.422.3037R}, field interpolation fundamentals~\citep{2010MNRAS.405.1513R, 2012MNRAS.425.1068D} and the treatment of variable resolution terms~\citep{2013MNRAS.428.2840H}.

The development of SPMHD followed a qualitatively similar trajectory, it being perceived until recently as having limited applicability~\citep[see e.g.][]{2006A&A...457..371F, 2016arXiv160606972L}, prior to more than four decades of dedicated research compounding to a mature, robust and competitive formulation of the method. Extending SPH to magnetic field physics was already attempted by~\citet{1977MNRAS.181..375G}, who devised a non-conservative SPMHD scheme to study static magnetic polytropes; this was however shown to perform poorly on MHD shock problems~\citep{Morris1996-yn}. The first truly conservative SPMHD formulation was proposed by~\citet{1985MNRAS.216..883P} who expressed momentum and energy preserving discrete EoM in terms of the spatial gradient of a magnetic stress tensor; nevertheless, \citet{1985MNRAS.216..883P} also demonstrated that said EoM were `tensile unstable'~\citep[similarly to what is seen in particle simulations of solid media][]{1995JCoPh.116..123S, 2000JCoPh.159..290M}, as particles can clump together in the presence of negative stresses arising when the ratio of thermal to magnetic pressure $\beta$ drops below unity. Such unphysical attractive forces are attributable to the stress tensor-based acceleration equation including a force term parallel to $\boldsymbol{B}$, and proportional to an inevitably non-vanishing $\nabla \cdot \boldsymbol{B}$ `monopole' in the discrete limit. 

Notwithstanding a series of earlier proposed mitigation strategies~\citep[e.g.][]{1985MNRAS.216..883P, 1994MmSAI..65..991S, Morris1996-yn, 1996PASA...13...71B}, the solution that came to be widely accepted as being optimal to address the tensile instability problem was that of~\citet{2001ApJ...561...82B}, who suggested directly subtracting the spurious $\nabla \cdot \boldsymbol{B}$ force from the conservative acceleration equation; this coincidentally rendered the full EoM formally equivalent to the scheme of~\citet{1999JCoPh.154..284P}, forfeiting momentum preservation, but only to the extent that the SC is violated. Subsequent stability analyses demonstrated that subtraction of only $1/2$ the monopole term would in theory be sufficient~\citep{2004ApJS..153..447B, 2015JCoPh.302..359I}, which proved empirically true in some practical applications~(\citealt{2012MNRAS.420.3195B}, but see also~\citealt{2012JCoPh.231.7214T}). This was further refined by~\citet{2006ApJ...652.1306B} who proposed applying the correction adaptively on a per-particle basis, including the non-conservative corrective force only in the instability onset regime (i.e. $\beta < 1$), and disabling it when not needed to revert to a fully conservative system of evolution equations. It is variants thereof that are used in all contemporary SPMHD schemes~\citep[e.g.][]{2018PASA...35...31P, 2020A&A...638A.140W, 2023MNRAS.518.4115G}.

Further notable strides were made by~\citet{2004MNRAS.348..123P, 2004MNRAS.348..139P, 2005MNRAS.364..384P}, who proposed a more rigorous treatment of variable smoothing lengths (the spatial resolution scale associated with a given particle) in conservative SPMHD EoM. Even more importantly, they constructed discontinuity-capturing terms for improved handling of MHD shocks, by a scheme initially derived under the assumption of entropy conservation and which is therefore inherently incapable of treating discontinuous jumps in physical variables. \citet{2004MNRAS.348..123P} extended to MHD the formalism of~\citet{1997JCoPh.134..296C, 1997JCoPh.136..298M}, who had proposed artificial dissipation prescriptions for SPH in analogy with Riemann solvers, of a functional form that formally satisfies the second law of thermodynamics. \citet{2005MNRAS.364..384P} augmented this with a variable `switch', inspired by the work of~\citet{1997JCoPh.136...41M}, to apply artificial dissipation of the magnetic field only where necessary and drastically limit its excess dissipation in smooth flows. This was refined by~\citet{2013MNRAS.436.2810T}, who devised a novel switch that is globally less dissipative than that of~\citet{2005MNRAS.364..384P}, and better at capturing strong shocks in the $\beta >> 1$ regime, of foremost interest in the context of galaxy formation and cosmology~\citep[see however][]{2018MNRAS.476.2890B}. Further alternatives have been put forward~\citep{2018PASA...35...31P}, but it is unclear whether they yield improved performance in production simulations~\citep{2017arXiv170607721W}.

The question of the SC proved particularly challenging for SPMHD. Disappointing results of early attempts at incorporating the~\citet{2002JCoPh.175..645D} divergence-cleaning scheme into the method~(see~\citealt{2005MNRAS.364..384P}, but also~\citealt{2013MNRAS.428...13S}) pushed the community to instead consider expressing the magnetic field $\boldsymbol{B}$ using alternative fundamental variables, in a way such that $\nabla \cdot \boldsymbol{B} = 0$ by construction. Explored avenues included defining $\boldsymbol{B}$ through `Euler potentials' $(\alpha, \beta)$ as $\boldsymbol{B} = \nabla \alpha \times \nabla \beta$~\citep[see][]{1970AmJPh..38..494S}, or through a `vector potential' $\boldsymbol{A}$ as $\boldsymbol{B} = \nabla \times \boldsymbol{A}$~\citep[see][]{1976RvGSP..14..199S}. The former have restricted capabilities in terms of the magnetic field topologies they can represent, are incompatible with the linear diffusion operators one would use to model physical dissipative effects, and are incapable of capturing dynamo action~\citep{2010MNRAS.401..347B}; formulations of SPMHD in terms of the latter are either insufficiently tested or subject to severe numerical instabilities~\citep{2010MNRAS.401.1475P, 2015JCoPh.282..148S, 2023arXiv230615039T}. An effective treatment of divergence errors in SPMHD was only truly possible following the conception by~\citet{2012JCoPh.231.7214T} of a `constrained' formulation of~\citet{2002JCoPh.175..645D}'s mixed hyperbolic/parabolic divergence-cleaning algorithm. Formally requiring the scheme not to spuriously generate magnetic energy (through a careful choice of conjugate discrete differential operators in the relevant evolution equations, and by including additional energy conserving terms) yielded a divergence-cleaning prescription that is stable at discontinuities and free surfaces, providing excellent divergence control in realistic settings~\citep[see also e.g.][]{2012MNRAS.423L..45P}. The formalism was further improved by~\citet{2016JCoPh.322..326T} through a consistent treatment of particle-carried, variable cleaning speeds, which provided further guarantees of no runaway energy growth at no additional computational cost.

The aforementioned advances have jointly allowed SPMHD methods to now be able to successfully reproduce expected results on historically challenging problems, related to e.g. small-scale turbulent dynamos~\citep{2016MNRAS.461.1260T} or the MRI~\citep{2022A&A...659A..91W}. Said methods have in turn been employed to study a vast array of astrophysical systems, including proto-stars and proto-stellar disks~\citep[e.g.][]{2015MNRAS.452..278T, 2016MNRAS.457.1037W, 2017MNRAS.467.3324L}, tidally disrupted stars~\citep{2017MNRAS.469.4879B}, idealised disk galaxies~\citep[e.g.][]{2016MNRAS.461.4482D, 2019MNRAS.483.1008S, 2023A&A...673A..47W} and massive galaxy clusters~\citep[e.g.][]{2018MNRAS.476.2890B, 2022ApJ...933..131S, 2024ApJ...967..125S}. The method is also constantly being re-evaluated with suggestions for further improvements being continuously put forward, such as using polynomial fits to construct more accurate gradients~\citep{2003ApJ...595..564M, 2012ApJS..200....6M, 2012ApJS..200....7M}, replacing artificial dissipation prescriptions with Riemann solvers to improve shock capturing~\citep{2011MNRAS.418.1668I, 2013ASPC..474..239I}, using alternative discrete differential operators to eliminate low order field reconstruction errors~\citep{2020A&A...638A.140W}, and sub-cycling~\citep{2016JCoPh.322..326T, 2023A&A...673A..47W} or dynamically timestepping~\citep{2021MNRAS.502.2285D, 2026MNRAS.550g1121S} divergence cleaning to limit violation of the SC further.

The present paper introduces a novel formulation of SPMHD that was developed as part of the highly-parallel astrophysical and cosmological code SWIFT~\citep{2016pasc.conf....2S, 2024MNRAS.530.2378S}. Our Lagrangian MHD method was designed as an extension of the code's default module for emulating gas dynamics, the SPHENIX SPH scheme~\citep{2022MNRAS.511.2367B}. SPHENIX not only incorporates all latest proposed refinements to the SPH method, but is also conceived to couple optimally with the vast network of galaxy formation sub-grid recipes available in SWIFT~\citep[see e.g.][for a discussion of non-trivial implications of coupling hydrodynamics solvers to a galaxy formation model]{2018MNRAS.481..835O, 2021MNRAS.505.2316B}. These constitute effective numerical methods for capturing the impact of astrophysical processes that occur below the inevitably finite spatial and temporal resolution scales of a given numerical experiment; SWIFT ships with a broad range of such sub-grid prescriptions, targeted at modeling processes including radiative gas cooling~\citep{2009MNRAS.393...99W, 2020MNRAS.497.4857P}, dust grain evolution~\citep{2026MNRAS.545f2040T}, star formation~\citep{2008MNRAS.383.1210S, 2024MNRAS.532.3299N}, stellar feedback and chemical enrichment~\citep{2008MNRAS.387.1431D, 2012MNRAS.426..140D, 2009MNRAS.399..574W, 2023MNRAS.523.3709C} and active galactic nucleus feedback~\citep{2009MNRAS.398...53B, 2022MNRAS.515.4838N, 2022MNRAS.516..167B, 2022MNRAS.516.3750H, 2026MNRAS.547ag324H}, consideration of which has played a pivotal role in recent advances in numerical studies of galaxy formation~\citep[see e.g.][for reviews]{2017ARA&A..55...59N, 2020NatRP...2...42V, 2023ARA&A..61..473C}. Refined sub-grid models were notably crucial in allowing results from the recent FLAMINGO~\citep{2023MNRAS.526.4978S, 2023MNRAS.526.6103K} and COLIBRE~\citep{2026MNRAS.548ag375S, 2026MNRAS.548ag300C} simulation campaigns, both run with SWIFT (without magnetic fields), to be in excellent agreement with multimodal observational data~\citep[see e.g.][]{2024MNRAS.533.2656B, 2024MNRAS.534..251K, 2026MNRAS.548ag740C, 2026MNRAS.549ag947L, 2026arXiv260625995S}. 

Abiding by the philosophy~\citet{2022MNRAS.511.2367B} adhered to in developing SPHENIX, we conceived an MHD solver that does not necessarily offer the best possible performance per discrete resolution element or fastest convergence when compared to other state-of-the-art methods~\citep[e.g.][]{2011MNRAS.418.1392P, 2016MNRAS.455...51H, 2020ApJS..249....4S}. We rather propose a computationally inexpensive scheme with a minimal memory footprint that can freely and robustly be coupled with any (combination of) subgrid recipes implemented in SWIFT. Leveraging upon the discussed recent technical advances in SPMHD, we set forth a conservative formulation of the method which we couple to prescriptions for the tensile instability correction, artificial magnetic dissipation and divergence cleaning, that maximise performance and ability to accurately represent rich MHD phenomenology, while also allowing for sound results in realistic galaxy and structure formation applications. We validate the scheme's capabilities through running a large array of benchmarking problems of increasing complexity, keeping model hyperparameters fixed throughout to provide a transparent picture of its capabilities when configured as it would be used in production. We pay particular attention to motivating each test's relevance to our endeavour and discussing how our results compare to what is reported in other contemporary method presentation works, ensuring we provide detailed descriptions of how we set up our numerical experiments to facilitate their straightforward reproduction.

The remainder of this paper is structured as follows: in Section~\ref{sec:NumericalMethods} we first introduce the SWIFT simulation code, briefly describe the SPHENIX hydrodynamics solver, and finally detail the novel SPMHD method we developed on top of it.  In Section~\ref{sec:test_problems} we present results from numerical tests run with our method, starting from standard `laboratory' MHD experiments where the fluid is assumed to be a perfect conductor (the `ideal' MHD limit, where there is no physical dissipation of the magnetic field), followed by idealised set-ups concerned with imperfect conductors (the `non-ideal' MHD limit), and culminating in three astrophysical applications: we study, in turn, the formation and evolution of a proto-stellar core, a massive galaxy cluster, and an isolated Milky Way-like disk galaxy, the latter constituting the first reported coupling of an MHD solver with the full EAGLE galaxy formation model~\citep{2015MNRAS.446..521S, 2015MNRAS.450.1937C}. Finally, we provide concluding remarks in Section~\ref{sec:conclusions}.

\section{Numerical method}
\label{sec:NumericalMethods}

In this section we detail our novel implementation of SPMHD, after having briefly introduced the software framework as part of which it was developed, and the pure hydrodynamics solver on top of which our method was built and tested.

\subsection{The SWIFT simulation code}
\label{sec:SWIFT}

SWIFT~\citep{2024MNRAS.530.2378S} is a modern, highly parallel and fully open-source gravity, hydrodynamics, cosmology and galaxy formation code. It was designed with the core aim of exhibiting excellent scalability; this is achieved by adopting a task based parallelisation strategy, asynchronous communication using the Message Passing Interface (MPI) between compute nodes and graph-based domain decomposition algorithms, to take full advantage of present-day large-scale and multi-node computing infrastructure. This approach has been shown to lead to sound strong-~\citep{2016pasc.conf....2S} and weak-~\citep{2018arXiv180701341B} scaling behaviour on large calculations that are astrophysically relevant.

Another key attribute of SWIFT is its high flexibility, as users have great freedom in interchanging and combining the different available physics modules at compilation time. The code can be configured to account for gas physics through one of several smoothed particle hydrodynamics (SPH) solvers, from re-implementations of variants of the method widely used by the community~\citep{2002MNRAS.333..649S, 2017MNRAS.471.2357W, 2018PASA...35...31P} to a novel in-house scheme specifically designed with galaxy formation applications in mind~\citep{2022MNRAS.511.2367B}. Methods that extend the SPH formalism, to address some of its shortcomings, have also been incorporated into the codebase~\citep{2024arXiv240718587S}, while alternative approaches to modeling hydrodynamics inspired by~\citet{2010MNRAS.401..791S} \citep[moving mesh; see also][]{2016A&C....16..109V} or~\cite{doi:10.1142/S0218202599000117} \citep[arbitrary
Lagrangian-Eulerian renormalised mesh-free schemes; see also][]{2011MNRAS.414..129G, 2015MNRAS.450...53H} are in the works. SWIFT solves the equations of Newtonian Gravity through the fast multipole method~\citep{1987JCoPh..73..325G}, which can be complemented by a particle-mesh method~\citep{2003NewA....8..665B} to account for long-range gravitational interactions over a simulation box with periodic boundary conditions. Finally, SWIFT comes with two complementary ensembles of sub-resolution effective models for galaxy formation: the EAGLE model~\citep{2015MNRAS.446..521S}, aimed at studying massive galaxies in a cosmological context, and the GEAR model~\citep{2012A&A...538A..82R}, designed to model dwarf galaxies.

The SPMHD solver presented in this work was designed to be fully compatible with SWIFT's highly modular structure, and can in principle be combined with any of the SPH implementations available in the codebase, as well as with (any subset of) the aforementioned modules capturing additional physics. We moreover note that the code comes with extensive documentation, and a large ensemble of example problems which we complement with all tests detailed in the sections that follow.
The analysis of all data products we generate as part of the present study is performed using the \texttt{swiftsimio} software package~\citep{2020JOSS....5.2430B}. Our code is made publicly available~\footnote{\url{www.swiftsim.com}}. 

\subsection{The SPHENIX hydrodynamics scheme}
\label{sec:sphenix}

SPHENIX~\citep{2022MNRAS.511.2367B} is the default module for modeling gas dynamics with SWIFT, and is what the novel SPMHD solver presented in this work was built around and extensively tested with.

\subsubsection{Fundamentals}

SPHENIX makes use of the Smoothed Particle Hydrodynamics method, a numerical paradigm for solving the equations of hydrodynamics originally developed by~\cite{1977AJ.....82.1013L} and~\cite{1977MNRAS.181..375G}. SPH discretises the fluid under consideration on mass and is Lagrangian in nature: it uses point-like particles both as tracers of fluid flow, and as sampling points of physical quantities in space. Other than having resolution adaptively follow mass density, the method is prized for its stability, simplicity and low computational cost and couples naturally to gravity solvers. It thus naturally lends itself to astrophysical problems where a large range of spatial and temporal scales need to be modeled simultaneously. SPHENIX was moreover designed to optimally couple to sub-resolution galaxy formation models (see e.g.~\cite{2021MNRAS.505.2316B} which guided some of the design choices that were made in that respect). For a complete review of the SPH method, refer to e.g.~\cite{2012JCoPh.231..759P}.

The starting point of SPH is often taken to be the mass density estimator $\hat{\rho}$ evaluated at position $\boldsymbol{r}$, for a collection of particles $p_j$ labeled by Latin indices $j$, with $p_j$ at position $\boldsymbol{r}_j$ and of mass $m_j$
\begin{equation}
    \hat{\rho} ( \boldsymbol{r} ) =
    \sum_j m_j
    W ( |\boldsymbol{r} - \boldsymbol{r}_j|,
    h (\boldsymbol{r} ) )
\label{eq:density_estimator}
\end{equation}
where $W (r, h) \equiv W(| \boldsymbol{r} |, h)$ is the kernel function and $h$ the smoothing length. Note that equation~(\ref{eq:density_estimator}) can be used to estimate mass density at any point in space $\boldsymbol{r}$, and not just at the location of particles/interpolation points $\boldsymbol{r}_j$. Formally, the sum runs over all particles considered, but in practice only local neighbourhoods $\Omega (\boldsymbol{r})$ of $\boldsymbol{r}$ will need to be accounted for at a time, as in all practical applications the employed $W (r, h)$ have finite support.

$W(r, h)$ serves as a normalised, spatially isotropic weight function of finite support (we hereafter take $\Omega$ to be the extent of the kernel). Denoting the value of any attribute $Q (\boldsymbol{r})$ carried by particle $p_j$ as $Q_j$, one can use $W(r, h)$  to construct the summation interpolant
\begin{equation}
\label{eq:summation_interpolant}
    \hat{Q}(\boldsymbol{r}) = 
    \sum_j \frac{m_j}{\hat{\rho}_j} Q_j
    W ( |\boldsymbol{r} - \boldsymbol{r}_j|,
    h (\boldsymbol{r} ) )
\end{equation}
to obtain an estimate $\hat{Q} (\boldsymbol{r})$ of $Q (\boldsymbol{r})$. The interpolation weights can be defined through (we here follow the conventions of~\cite{2012MNRAS.425.1068D})
\begin{equation}
    W (r, h) = \frac{\kappa_\nu}{h^\nu} w (q)
\end{equation}
where $\nu$ is the number of spatial dimensions, $\kappa_\nu$ a constant prefactor and $w(q)$ an auxiliary function with argument $q \equiv r / (\gamma_K h)$. $w(q)$ is truncated at $q=1$, which corresponds to a radius $H \equiv \gamma_K h$ for $\gamma_K$ a kernel dependent constant; beyond that the function is set to zero. 

The most commonly adopted $W(r, h)$ in the SPH literature belong to one of two families of functions: the B-spline~\citep{1985A&A...149..135M} and Wendland~\citep{Wendland1995} kernels (see~\cite{2012MNRAS.425.1068D} for an extensive discussion of the impact of kernel choice on results on standard hydrodynamic tests). In the present work we make use of the $M_6$ (quintic) spline kernel, for which
\begin{equation}
    w (q) = 
    \begin{cases}
        (1-q)^5
        - 6 \left( \frac{2}{3} - q \right)^5
        + 15 \left( \frac{1}{3} - q \right)^5 ,
        & q \in [ 0, \frac{1}{3} ) \\    
        (1-q)^5
        - 6 \left( \frac{2}{3} - q \right)^5 ,
        & q \in [ \frac{1}{3}, \frac{2}{3} ) \\
        (1-q)^5 ,
        & q \in [ \frac{2}{3}, 1 ) \\
        0 ,
        & \text{otherwise}
   \end{cases}
\end{equation}
and $\kappa_\nu = \frac{3^7}{40 \pi}$ and $\gamma_K = 2.195775$ in $\nu = 3$ spatial dimensions. The results shown in~\cite{2022MNRAS.511.2367B} were obtained using a smaller kernel, but the authors found SPHENIX to perform equally well for any reasonable choice of $W(r, h)$. Use of a wider kernel is commonplace for SPMHD, both in testing the method~\citep[e.g.][]{2016MNRAS.455...51H, 2020A&A...638A.140W} as well as in production runs~\citep[e.g.][]{2024A&A...692A.232B}, as this reduces low order errors and allows for better estimates of spatial gradients~\citep[see e.g.][]{2010MNRAS.405.1513R, 2012MNRAS.425.1068D}, improving the accuracy and stability of MHD calculations.   

The smoothing length $h$ determines the rate at which $W(r, h)$ decays, in terms of the mean inter-particle distance within the kernel's support. In SPHENIX, the smoothing length is set by requiring that the number density
\begin{equation}
\label{eq:number_density}
    \hat{n} (\boldsymbol{r}) =
    \sum_j
    W ( |\boldsymbol{r} - \boldsymbol{r}_j|,
    h (\boldsymbol{r} ) )
\end{equation}
satisfies
\begin{equation}
\label{eq:constant_mass}
    \hat{n}(\boldsymbol{r}) \left( \frac{h (\boldsymbol{r})}{\eta_\text{res}} \right)^\nu = 1
    \text{,}
\end{equation}
where $\eta_\text{res}$ is a fixed parameter that sets the resolution scale. Equations~(\ref{eq:number_density}) and~(\ref{eq:constant_mass}) are solved simultaneously to a relative accuracy of $10^{-4}$, starting from an educated guess and then jointly using a Newton-Raphson and bisection root finding algorithm to obtain a given particle's $h$~\citep[refer to][for a complete discussion]{2024MNRAS.530.2378S}.

What remains to be specified is the thermodynamic variable considered by the scheme. SPHENIX makes use of the particle-carried specific internal energy $u_j$, which is related to the thermal pressure $P_j$ through the second law of thermodynamics
\begin{equation}
    \left. \frac{\partial u_j}{\partial \boldsymbol{q}_j} \right|_{A_j}
    = - \frac{P_j}{m_j} \frac{\partial V_j}{\partial \boldsymbol{q}_j}
    \text{,}
\end{equation}
with $V_j \equiv m_j / \hat{\rho}_j$ the particle volume, $A_j$ the entropy, $\boldsymbol{q}_j$ a placeholder for either of $\boldsymbol{r}_j$ or $h_j$, and $\partial / \partial \boldsymbol{q}_j |_{A_j}$ denoting differentiation with respect to $\boldsymbol{q}_j$ while $A_j$ is kept fixed. The gas is taken to obey, unless specified otherwise, the ideal gas equations of state
\begin{equation}
\label{eq:ideal_gas_law}
    P_j = (\gamma - 1) u_j \hat{\rho}_j = A_j \hat{\rho}_j^\gamma
\end{equation}
for $\gamma$ the adiabatic index.

\subsubsection{Equations of motion}
\label{sec:hydro_eom}

Conservative equations of motion (EoM) for SPH can be obtained from an action minimisation principle~\citep{1994MNRAS.270....1N, 2002MNRAS.333..649S, 2012JCoPh.231..759P}. For SPHENIX, \cite{2022MNRAS.511.2367B} make use of the formalism laid out in~\cite{2013MNRAS.428.2840H} to self-consistently derive evolution equations from the discrete particle Lagrangian
\begin{equation}
\label{eq:sph_lagrangian}
    \mathcal{L}_{\text{hydro}} = \sum_j m_j \left[
    \frac{1}{2} \boldsymbol{v}_j^2
    - u_j (\hat{\rho}_j, A_j)
    \right]
\end{equation}
where $\boldsymbol{v}_i$ is the particle velocity, using the Euler-Lagrange equations. They 
choose the density-energy flavour of SPH, which is advantageous for cosmological simulations, as it is free of the force errors alternative formulations of the method suffer from when coupled to dynamical timestepping and/or sub-grid models of galaxy formation~\citep[see][]{2018MNRAS.481..835O, 2021MNRAS.505.2316B}. These EoM have the advantage of exhibiting excellent conservation and stability properties and, using the shorthand notation $Q_{ij} \equiv Q_i - Q_j$ for any quantity $Q$, read
\begin{equation}
\label{eq:sph_force}
    \frac{\mathrm{d} \boldsymbol{v}_i}{\mathrm{d} t} = 
    - \sum_j m_j \left[ 
    \frac{f_{ij} P_i}{{\hat{\rho}}_i^2} \nabla_i W_{ij} (h_i) +
    \frac{f_{ji} P_j}{{\hat{\rho}}_j^2} \nabla_i W_{ij} (h_j)
    \right]
    \text{,}
\end{equation}
\begin{equation}
\label{eq:sph_energy}
    \frac{\mathrm{d} u_i}{\mathrm{d} t} =
    \sum_j m_j
    \frac{f_{ij} P_i}{{\hat{\rho}}_i^2}
    \boldsymbol{v}_{ij} \cdot \nabla_i W_{ij} (h_i)
\end{equation}
where $\nabla_i \equiv \partial / \partial \boldsymbol{r}_i$ a spatial gradient, $\mathrm{d} / \mathrm{d} t \equiv \partial / \partial t + \boldsymbol{v} \cdot \nabla$ the convective time derivative, $W_{ij} (h_i) \equiv W (|\boldsymbol{r}_i - \boldsymbol{r}_j|, h_i)$ and $f_{ij}$ the so called `grad-h' terms that account for the variability of smoothing lengths and read
\begin{equation}
    f_{ij} = 1 - \frac{1}{m_j}
    \left(
    \frac{h_i}{\nu \hat{n}_i} \frac{\partial {\hat{\rho}}_i}{\partial h_i}
    \right)
    \left( 1 +
    \frac{h_i}{\nu \hat{n}_i} \frac{\partial {\hat{n}}_i}{\partial h_i}
    \right)^{-1}
    \text{.}
\end{equation}
Equations~(\ref{eq:sph_force}) and~(\ref{eq:sph_energy}) can respectively be interpreted, through the lens of equation~(\ref{eq:summation_interpolant}), as discretised counterparts to the standard acceleration and energy equations of compressible hydrodynamics.

\subsubsection{Capturing discontinuities}
\label{sec:capturing_sph_discontinuities}

The conservative EoM presented in Section~\ref{sec:hydro_eom} on the assumption that entropy remains constant; they therefore need to be complemented with dissipative corrections, for the hydrodynamics scheme to be able to capture shocks~\citep{1992ARA&A..30..543M}. SPHENIX addresses this through a modified version of the artificial viscosity (AV) scheme of~\cite{2010MNRAS.408..669C}, whereby the terms
\begin{equation}
\label{eq:sph_artificial_viscosity_acceleration}
    \left. \frac{\mathrm{d} \boldsymbol{v}_i}{\mathrm{d} t} \right|_{\text{AV}} = 
    -
    \sum_j m_j \zeta_{ij} \left[
    f_{ij} \nabla_i W_{ij} (h_i) + f_{ji} \nabla_i W_{ij} (h_j) \right]
    \text{,}
\end{equation}
\begin{equation}
\label{eq:sph_artificial_viscosity_thermal_energy}
    \left. \frac{\mathrm{d} u_i}{\mathrm{d} t} \right|_{\text{AV}} = 
    \frac{1}{2} 
    \sum_j m_j \zeta_{ij} \boldsymbol{v}_{ij} \cdot \left[
    f_{ij} \nabla_i W_{ij} (h_i) + f_{ji} \nabla_i W_{ij} (h_j) \right]
\end{equation}
are added to equations~(\ref{eq:sph_force}) and~(\ref{eq:sph_energy}), respectively. Here, the scalar $\zeta_{ij}$ is given by
\begin{equation}
\label{eq:viscosity_zeta}
    \zeta_{ij} = - \alpha^{\text{AV}}_{ij} \mu_{ij} \frac{v^{\text{sig}}_{ij}}{\overline{\rho}_{ij}}
    \text{,}
\end{equation}
where we have made use of the shorthand notation $\overline{Q}_{ij} \equiv \frac{1}{2}(Q_i +Q_j)$ and
\begin{equation}
    \mu_{ij} = \text{min} 
    \left\{
    0, \boldsymbol{v}_{ij} \cdot \hat{\boldsymbol{r}}_{ij}
    \right\}
\end{equation}
for $\hat{\boldsymbol{r}}_{ij}$ the separation vector between particles $p_i$ and $p_j$ normalised to have unit length. Equation~(\ref{eq:viscosity_zeta}) makes use of the pair-wise signal velocity
\begin{equation}
\label{eq:sph_pairwise_sgnal_velocity}
    v^{\text{sig}}_{ij} = c_{s,i} + c_{s,j} - \beta^{\text{AV}} \mu_{ij}
    \text{,}
\end{equation}
where $c_{s,i}$ is the particle-carried sound speed, which for an ideal gas takes the value $c_s = \sqrt{\gamma P / \rho}$. The constant parameter $\beta^{\text{AV}}$ is set to $3$ by default. We moreover define 
\begin{equation}
    v^{\text{sig}}_i = \text{max}_{j \in \Omega_i} \{ v^{\text{sig}}_{ij} \}
    \text{.}
\end{equation}

The AV parameter $\alpha^{\text{AV}}$ is made particle dependent and evolved in time, in an attempt to limit AV to act only when and where necessary, reducing dissipation away from shocks. A source term for $\alpha^{\text{AV}}$ is computed as
\begin{equation}
\label{eq:sph_shock_indicator}
    S_i = H_i^2 \text{max} \left\{
    0, - \dot{\nabla} \cdot \boldsymbol{v}_i
    \right\}
\end{equation}
in converging flows (i.e. whenever $\nabla \cdot \boldsymbol{v}_i \leq 0 $, otherwise $S_i = 0$). $\dot{\nabla} \cdot \boldsymbol{v}_i$ is the change in $\nabla \cdot \boldsymbol{v}_i$ between consecutive time-steps, divided by the time-step length, and traces pre-shock regions well. $S_i$ is in turn used to compute the local value
\begin{equation}
    \alpha^{\text{AV}}_{\text{loc}, i} = 
    \alpha^{\text{AV}}_{\text{max}} \frac{S_i}{c_{s,i}^2 + S_i}
\end{equation}
where $\alpha^{\text{AV}}_{\text{max}} = 2$. A particle carried $\alpha^{\text{AV}}_i$ is instantaneously set equal to $\alpha^{\text{AV}}_{\text{loc}, i}$ if $\alpha^{\text{AV}}_{\text{loc}, i} > \alpha^{\text{AV}}_i$, and otherwise allowed to decay according to
\begin{equation}
    \frac{\mathrm{d} \alpha^{\text{AV}}_i}{\mathrm{d} t} =
    -\frac{\alpha^{\text{AV}}_{\text{loc}, i} - \alpha^{\text{AV}}_i}
    {\tau^{\text{AV}}_i}
\end{equation}
where we define the decay time scale $\tau^{\text{AV}}_i = l^{\text{AV}} H_i / c_{s,i}$ with $l^{\text{AV}} = 0.05$. Finally, SPHENIX makes use of the switch introduced by~\cite{1989PhDT.......206B}
\begin{equation}
\label{eq:sph_balsara}
    \xi_i = \frac{\Vert \nabla \cdot \boldsymbol{v}_i \Vert}
               {\Vert \nabla \cdot \boldsymbol{v}_i \Vert 
               +\Vert \nabla \times \boldsymbol{v}_i \Vert
               + 10^{-4} c_{s,i} / h_i}
\end{equation}
to modulate $\alpha^{\text{AV}}_i$, in order to suppresses unnecessary viscosity seeded by non-zero $\nabla \cdot \boldsymbol{v}_i$ in shear flows. We use $\Vert \boldsymbol{Q} \Vert$ here to denote the Euclidean norm of (either scalar or vector) quantity $\boldsymbol{Q}$. The pair-wise interaction AV parameter is taken to be
\begin{equation}
\label{eq:sph_pairwise_viscosity_alpha}
    \alpha^{\text{AV}}_{ij} = \frac{1}{4} 
    ( \alpha^{\text{AV}}_i + \alpha^{\text{AV}}_j) (\xi_i + \xi_j)
\end{equation}

\cite{1997JCoPh.134..296C} argued that shock capturing in SPH should involve dissipation terms in all energy components, beyond AV. To smooth over discontinuities in the internal energy, SPHENIX uses an artificial thermal diffusion (AD) scheme inspired by the work presented in~\cite{2008JCoPh.22710040P}, adding
\begin{equation}
\label{eq:sph_artificial_diffusion}
        \left. \frac{\mathrm{d} u_i}{\mathrm{d} t} \right|_{\text{AD}} =
        \sum_j m_j \alpha^{\text{AD}}_{ij} v^{\text{AD}}_{ij}
        u_{ij}
        \Biggl[
        \frac{f_{ij} F_{ij} (h_i)}{\hat{\rho}_i}
        + \frac{f_{ji} F_{ij} (h_j)}{\hat{\rho}_j}
        \Biggr]
\end{equation}
to equation~(\ref{eq:sph_energy}), where $F_{ij}$ is minus the norm of the kernel gradient defined as $F_{ij} (h_i) \equiv \hat{\boldsymbol{r}}_{ij} \cdot \nabla_i W_{ij} (h_i)$. The pair-wise interaction $\alpha^{\text{AD}}_{ij}$ is a pressure-weighted average of variable particle-carried, strictly non-negative diffusion coefficients
\begin{equation}
    \alpha^{\text{AD}}_{ij} =
    \frac{P_i \alpha^{\text{AD}}_i + P_j \alpha^{\text{AD}}_j}{P_i + P_j}
    \text{,}
\label{eq:alpha_AD}
\end{equation}
where the $\alpha^{\text{AD}}_i$ are evolved in time according to
\begin{equation}
\label{eq:time_evol_alpha_AD}
    \frac{\mathrm{d} \alpha^{\text{AD}}_i}{\mathrm{d} t} = 
    H_i \frac{\nabla_i^2 u_i}{\sqrt{u_i}}
    -\frac{\alpha^{\text{AD}}_i}
    {\tau^{\text{AV}}_i}
\end{equation}
and the Laplacian of the internal energy $\nabla_i^2 u_i$ is calculated using the discrete differential operator proposed by~\citealt{1985PASA....6..207B}~\citep[see also][]{2012JCoPh.231..759P} as
\begin{equation}
    \nabla_i^2 u_i = 
    2 \sum_j \frac{m_j}{\rho_j} u_{ij} 
    \frac{F_{ij} (h_i)}{\Vert \boldsymbol{r}_{ij} \Vert}
    \text{.}
\end{equation}
The first term on the RHS of equation~(\ref{eq:time_evol_alpha_AD}) acts as source seeding thermal diffusion at non-linear gradients in $u_i$, while the second term leads to a decay of $\alpha^{\text{AD}}_i$ over a timescale $\tau^{\text{AV}}_i = H_i / v^{\text{sig}}_i$. Diffusion coefficients are capped in viscous environments according to
\begin{equation}
    \alpha^{\text{AD}}_i \leftarrow 
    \text{min} 
    \left\{
        \alpha^{\text{AD}}_i ,
        \alpha^{\text{AD}}_{\text{max}}
        \left(
            1 - 
            \frac{
            \text{max}_{j \in \Omega_i} \{
            \alpha^{\text{AV}}_j \}
            }
            {\alpha^{\text{AV}}_\mathrm{max}}
        \right)
    \right\}
\end{equation}
where $\alpha^{\text{AD}}_{\text{max}} = 1$. $v^{\text{AD}}_{ij}$ is then set to be the average of two commonly employed diffusion speeds
\begin{equation}
\label{eq:sph_artificial_diffusion_signal_speed}
    v^{\text{AD}}_{ij}
    = \frac{1}{2} \left(
    \boldsymbol{v}_{ij} \cdot \hat{\boldsymbol{r}}_{ij} +
    \sqrt{\frac{P_{ij}}{\overline{\rho}_{ij}}}
    \right)
\end{equation}
These choices allow SPHENIX to correctly capture contact discontinuities, a challenge for density-energy SPH, while simultaneously avoiding excessive numerical radiative losses around point injections of energy, which is common in sub-grid galaxy formation models. For a more complete discussion, refer to~\cite{2022MNRAS.511.2367B}.

\subsection{A direct induction implementation of SPMHD in SWIFT}

The SPMHD solver we introduce in the present paper is largely inspired by that of~\cite{2018PASA...35...31P} (as are other recent implementations of the method, such as~\citealt{2018MNRAS.476.2890B, 2020A&A...638A.140W} and~\citealt{2023MNRAS.518.4115G}), who build upon the foundational work of~\cite{1985MNRAS.216..883P, 2004MNRAS.348..123P, 2004MNRAS.348..139P} and~\cite{2005MNRAS.364..384P} and augment their method with the major improvements presented in~\citet{2012JCoPh.231.7214T} and~\citet{2016JCoPh.322..326T}. We nevertheless deviate from~\citet{2018PASA...35...31P} with regard to several key elements of the algorithm (see particularly Sections~\ref{sec:tensile_instability_correction}, \ref{sec:capturing_spmhd_discontinuities}, \ref{sec:divergence_cleaning} and \ref{sec:time_inetgration}), opting instead for alternative recommendations made by ~\citet{2013MNRAS.436.2810T} and~\citet{2016MNRAS.455...51H} (see especially Appendix A of the latter) or novel suggestions for several (subcomponents of) regularisation strategies key to the method, so as to confer to our scheme optimal performance and stability capabilities in production rate galaxy and structure formation simulations, development and testing of SPMHD specifically targeting such applications remaining fairly limited to this day.  

All schemes detailed in the aforementioned works evolve the magnetic field $\boldsymbol{B}$ itself in time, falling under the `Direct Induction' class of solvers; this presents the most mature, robust and well-tested approach at incorporating magnetic field physics into SPH. Alternatives exist, such as methods that evolve the vector potential $\boldsymbol{A}$ defined through $\boldsymbol{B} = \nabla \times \boldsymbol{A}$ (see e.g. \cite{2015JCoPh.282..148S} but also \cite{2023arXiv230615039T}. These are nevertheless beyond the scope of the present discussion.

The development of our `Direct Induction' formulation of SPMHD was also guided by the design principles that shaped SPHENIX. We did not aim for maximal performance per resolution element, a metric by which moving-mesh~\citep[e.g.][]{2010MNRAS.401..791S} or mesh-less finite-volume~\citep[e.g.][]{2015MNRAS.450...53H} codes would yield improved results. Accuracy and precision remained key, but we concurrently sought for high computational and memory efficiency and good coupling to galaxy formation sub-grid models, which would render the method highly competitive when it comes to its intended scientific use case. We moreover chose not to employ the geometric force average formulation~\cite{2020A&A...638A.140W} advocated for, so as to remain consistent with SPHENIX but also have conservative MHD EoM that follow from an action minimisation principle.

\subsubsection{Discretised equations of ideal MHD}
\label{sec:discrete_ideal_mhd}

For a fully ionised fluid that can be regarded as a perfect electrical conductor, that is a fluid with vanishing magnetic diffusivity $\eta \rightarrow 0$, the mass continuity, acceleration, thermal energy and induction equations read
\begin{align}
    & \frac{\mathrm{d} \rho}{\mathrm{d} t} = 
    - \rho \nabla \cdot \boldsymbol{v} 
    \label{eq:mhd_mass_continuity} \\
    & \frac{\mathrm{d} \boldsymbol{v}}{\mathrm{d} t} = 
    - \frac{1}{\rho} \nabla \cdot \boldsymbol{\mathrm{S}}
    \label{eq:mhd_force} \\
    & \frac{\mathrm{d} u}{\mathrm{d} t} =
    -\frac{P}{\rho} \nabla \cdot \boldsymbol{v}
    \label{eq:mhd_thermal_energy} \\
    & \frac{\mathrm{d}}{\mathrm{d} t}
    \left( \frac{\boldsymbol{B}}{\rho} \right) =
    \left( \frac{\boldsymbol{B}}{\rho} \cdot \nabla \right) \boldsymbol{v}
    \label{eq:mhd_induction}
\end{align}
respectively, and together comprise the equations of Ideal MHD. Here we have defined the Maxwell stress tensor
\begin{equation}
    \mathrm{S}^{\mu \nu} = 
    \left(P + \frac{1}{2 \mu_0} B^2 \right) \delta^{\mu \nu} -
    \frac{1}{\mu_0} B^\mu B^\nu
\end{equation}
where Greek indices $\mu, \nu$ denote spatial components (as opposed to Latin indices $i,j$ used to label SPH particles) and where $\mu_0$ is the magnetic permeability of free space (taking the value $\mu_0 = 4 \pi \cdot 10^{-7} \; \mathrm{N} / \mathrm{A}^{2}$ in SI units) and $\delta^{ij}$ the Kronecker delta. Choosing the magnetic flux per unit mass $\boldsymbol{B} / \rho$ as the variable to evolve is presented as a natural choice in the literature, for a fluid represented by equal-mass particles~\citep{2005MNRAS.364..384P, 2016MNRAS.455...51H}; it has nevertheless been shown empirically that evolving $\boldsymbol{B}$ instead only leads to minor differences.

The magnetic field must also satisfy the solenoidal constraint
\begin{equation}
\label{eq:mhd_solenoidal_constraint}
    \nabla \cdot \boldsymbol{B} = 0
\end{equation}
at all times; even in the continuum limit, this can only be maintained by~(\ref{eq:mhd_mass_continuity})-(\ref{eq:mhd_induction}) if imposed as an initial condition. We note however, that the form chosen for~(\ref{eq:mhd_induction}) follows from~\cite{2000JCoPh.160..649J} and~\cite{2001JCoPh.172..392D} and remains consistent even in the presence of magnetic monopoles, where~(\ref{eq:mhd_solenoidal_constraint}) is violated (see~\cite{2004MNRAS.348..123P} for more information).

Discrete, conservative counterparts to~(\ref{eq:mhd_mass_continuity})-(\ref{eq:mhd_induction}) can be obtained through an action minimisation exercise, similarly to how the pure hydrodynamics EoM presented in Section~\ref{sec:hydro_eom} were derived, starting from the MHD Lagrangian~\citep{1985MNRAS.216..883P, 2004MNRAS.348..139P, 2012JCoPh.231..759P}
\begin{equation}
\label{eq:spmhd_lagrangian}
    \mathcal{L}_{\text{MHD}} = \sum_j m_j \left[
    \frac{1}{2} \boldsymbol{v}_j^2
    - u_j (\hat{\rho}_j, A_j)
    - \frac{1}{2 \mu_0} \frac{\boldsymbol{B}_j^2}{\hat{\rho}_j}
    \right]
\text{,}
\end{equation}
where the additional term compared to~(\ref{eq:sph_lagrangian}) corresponds to the magnetic energy per unit mass. Noting that both~(\ref{eq:mhd_mass_continuity}) and~(\ref{eq:mhd_thermal_energy}) are independent of $\boldsymbol{B}$ simplifies the problem considerably: in SPMHD, as in standard density-energy SPH, we take the mass density to follow from~(\ref{eq:summation_interpolant}) (an exact, time-independent solution to~(\ref{eq:mhd_mass_continuity}), see~\cite{2012JCoPh.231..759P}) and account for the time evolution of $u$ through~(\ref{eq:sph_energy}) (which is nothing but an SPH discretisation of~(\ref{eq:mhd_thermal_energy})).

We then follow what is common practice, that is to~\textit{assume} an SPH discretisation of~(\ref{eq:mhd_induction}) and use it as a constraint in the Euler-Lagrange equations applied to~(\ref{eq:spmhd_lagrangian}). The resulting discrete acceleration and induction equations read
\begin{equation}
\label{eq:spmhd_force}
    \frac{\mathrm{d} \boldsymbol{v}_i}{\mathrm{d} t} = 
    - \sum_j m_j \left[ 
    \frac{f_{ij}}{{\hat{\rho}}_i^2}
    \boldsymbol{\mathrm{S}}_i \cdot \nabla_i W_{ij} (h_i) +
    \frac{f_{ji}}{{\hat{\rho}}_j^2}
    \boldsymbol{\mathrm{S}}_j \cdot \nabla_i W_{ij} (h_j)
    \right]
\end{equation}
and
\begin{equation}
\label{eq:spmhd_induction}
    \frac{\mathrm{d}}{\mathrm{d} t}
    \left( \frac{\boldsymbol{B}_i}{\hat{\rho}_i} \right) =
    - \sum_j m_j 
    \frac{f_{ij} \boldsymbol{v}_{ij}}{\hat{\rho}_i^2}
    \boldsymbol{B}_i \cdot \nabla_i W_{ij} (h_i),
\end{equation}
respectively. The EoM thus derived preserve the symmetries of~(\ref{eq:spmhd_lagrangian}) and conserve both energy and linear momentum; they however, do not perfectly conserve angular momentum (in contrast to the pure hydrodynamics case) due to~(\ref{eq:spmhd_induction}) not being rotationally invariant. We note that~(\ref{eq:spmhd_force}) reduces to~(\ref{eq:sph_force}), with the Maxwell stress tensor reducing to an isotropic thermal pressure component as $\mathrm{S}^{\mu \nu} \rightarrow P \delta^{\mu \nu}$ in the limit $\boldsymbol{B} \rightarrow \boldsymbol{0}$. We moreover stress that the `grad-h' terms in~(\ref{eq:spmhd_induction}) do not appear self-consistently (as they do in~(\ref{eq:sph_force})) but their presence is rather an empirically motivated choice~\citep{2004MNRAS.348..139P} that leads to improved performance of the algorithm on standard tests.

\subsubsection{Non-ideal terms}

We extended the Ideal MHD formalism outlined in Section~\ref{sec:discrete_ideal_mhd} (relaxing the assumption that the fluid is a fully ionised, perfect conductor) to account for the simplest magnetic dissipation process, Ohmic diffusion, whereby the collision of charged particles leads to a reconfiguration of the magnetic field topology and an increase of the gas temperature through Joule heating. This effect has been included in several astrophysical MHD codes~\citep[see e.g.][]{2012ApJS..201...24M, 2018MNRAS.476.2476M, 2024MNRAS.527.8355Z} and has also been incorporated into SPMHD solvers~(see~\citealt{2013MNRAS.434.2593T}; also discussed in~\citealt{2014MNRAS.444.1104W}). 

We make use of the single-fluid approximation to describe the partially ionised plasma~\citep[see e.g.][]{2008MNRAS.385.2269P}, where a fluid mixture comprised of both ions and neutrals is represented by a single set of SPH particles. Ohmic diffusion takes the form of an additional term in the induction equation. Assuming a constant, spatially uniform magnetic diffusivity $\eta$~\citep[like][]{2011MNRAS.418.2234B, 2018MNRAS.476.2476M}, it reads
\begin{equation}
\label{eq:mhd_ohmic_induction}
    \left.
    \frac{\mathrm{d}}{\mathrm{d} t}
    \left( \frac{\boldsymbol{B}}{\rho} \right) \right|_\eta =
    \frac{\eta}{\rho} \nabla^2 \boldsymbol{B}
\end{equation}
We discretise~(\ref{eq:mhd_ohmic_induction}), using a symmetrised version of the SPH Laplacian operator suggested by~\cite{1985PASA....6..207B}, as
\begin{equation}
\label{eq:spmhd_ohmic_induction}
    \left.
    \frac{\mathrm{d}}{\mathrm{d} t}
    \left( \frac{\boldsymbol{B}_i}{\hat{\rho}_i} \right) \right|_\eta =
    \eta \sum_j \frac{m_j}{\hat{\rho}_i \hat{\rho}_j}
    \left[
    f_{ij} F_{ij} (h_i) + f_{ji} F_{ij} (h_j) \right]
    \frac{\boldsymbol{B}_{ij}}{\Vert \boldsymbol{r}_{ij} \Vert}
\end{equation}
The magnetic energy that is dissipated through~(\ref{eq:spmhd_ohmic_induction}) is converted into heat, an effect we model through an additional term in the thermal energy equation
\begin{equation}
\label{eq:spmhd_ohmic_thermal_energy}
    \left.
    \frac{\mathrm{d} u_i}{\mathrm{d} t} \right|_\eta =
    - \frac{\eta}{\mu_0} \sum_j \frac{m_j}{\hat{\rho}_i \hat{\rho}_j}
    \left[
    f_{ij} F_{ij} (h_i) + f_{ji} F_{ij} (h_j) \right]
    \frac{\boldsymbol{B}_{ij}^2}{\Vert \boldsymbol{r}_{ij} \Vert}
\end{equation}
as is customarily done when considering diffusive processes in SPH~(\citealt{1985PASA....6..207B}, see also~\citealt{2012JCoPh.231..759P}). This ensures energy conservation.

\subsubsection{Tensile instability correction}
\label{sec:tensile_instability_correction}

Self-consistently deriving the acceleration equation for SPMHD from an action minimisation principle comes at the cost of allowing for negative stresses along magnetic field lines. This is undesirable in SPH as it can lead to the infamous tensile instability~\citep[see e.g.][]{2000JCoPh.159..290M} where particles unphysically clump together.

The occurrence of the instability can be understood through consideration of the continuum counterpart to~(\ref{eq:spmhd_force}), by which the MHD acceleration can be expressed as the sum of an isotropic and anisotropic contribution, the first corresponding to a `pressure' term and the latter being further decomposable into a `tension' and `monopole' component as
\begin{equation}
\label{eq:mhd_force_components}
    \begin{split}
        \frac{\mathrm{d} \boldsymbol{v}}{\mathrm{d} t} 
        & = - \underbrace{\frac{1}{\rho} \nabla 
        \Bigl(
        P + \frac{1}{2 \mu_0} \boldsymbol{B}^2 
        \Bigr)}_{\text{isotropic}}
        + \underbrace{\frac{1}{\rho} \nabla \cdot
        \Bigl(
        \frac{1}{\mu_0} \boldsymbol{B} \boldsymbol{B}
        \Bigr)}_{\text{anisotropic}}
        \\
        & =
        - \underbrace{\frac{1}{\rho} \nabla 
        \Bigl(
        P + \frac{1}{2 \mu_0} \boldsymbol{B}^2 
        \Bigr)}_{\text{pressure}}
        + \underbrace{\frac{1}{\mu_0 \rho} (\boldsymbol{B} \cdot            \nabla) 
        \boldsymbol{B}}_{\text{tension}}
        + \underbrace{\frac{1}{\mu_0 \rho} (\nabla \cdot \boldsymbol{B})
        \boldsymbol{B}}_{\text{monopole}}
    \end{split}
\end{equation}
The monopole force term on the RHS of the second line of equation~(\ref{eq:mhd_force_components}) would vanish if~(\ref{eq:mhd_solenoidal_constraint}) were satisfied. As this cannot be guaranteed to necessarily hold in the finite-accuracy discrete limit, the term can become non-zero (note this is a direct consequence of our discretisation of the anisotropic MHD force), and subsequently seed the tensile instability whenever pressure and magnetic tension cannot counteract it. A stability analysis~\citep{1985MNRAS.216..883P} reveals that this can occur whenever magnetic pressure exceeds thermal pressure, leading to instability onset for
\begin{equation}
\label{eq:tensile_instability_criterion}
    \beta < 1
\end{equation}
where $\beta \equiv P_{\text{thermal}}/P_{\text{magnetic}} = 2 \mu_0 P / \boldsymbol{B}^2$ is the plasma beta. Approaches to address the issue include simply ignoring it if $\beta > 1$ is expected~\citep[i.e. for problems in the weak field regime, see e.g.][]{1999A&A...348..351D, 2001A&A...378..777D, 2002A&A...387..383D, 2005JCAP...01..009D}, subtracting from a given particle's stress the maximal negative stress in the whole simulation volume~\citep{1985MNRAS.216..883P}, including an `artificial stress' in the force calculation~\citep{1997JCoPh.136...41M, 2000JCoPh.159..290M, 2004MNRAS.348..123P} or using non-conservative formulations of the EoM~\citep{Morris1996-yn}.

The solution which has been favoured in the literature, and which we adopt here, is based on~\cite{2001ApJ...561...82B} who suggest explicitly subtracting the monopole force term (computed in the same discretisation as all other MHD forces) from the acceleration equation. The inclusion of this term leads to violation of both energy and linear momentum conservation. Different suggestions have therefore been made for prefactors $\lambda$ by which to pre-multiply it~\citep{2001ApJ...561...82B, 2004ApJS..153..447B, 2006ApJ...652.1306B, 2012MNRAS.420.3195B, 2012JCoPh.231.7214T, 2018PASA...35...31P, 2020A&A...638A.140W}, to limit its action exclusively to where necessary and reduce non-conservation errors elsewhere. We find the best compromise between stability and computational expense to be the suggestion of~\cite{2018PASA...35...31P}, who modulate the corrective force term applied to each particle $p_i$ on an adaptive basis by a $\lambda$ that explicitly depends on a local estimate $\beta_i^\mathrm{loc}$ of $\beta$,
\begin{equation}
\label{eq:tensile_instability_prefactor}
    \lambda (\beta_i^\mathrm{loc}) = 
    \begin{cases}
          1,  & \beta_i^\mathrm{loc} < 2\\
          (10 - \beta_i^\mathrm{loc}) / 8,  & 2 < \beta_i^\mathrm{loc} < 10 \\
          0, & \text{otherwise}
    \end{cases}
\end{equation}
This `switch' is introduced to ensure that, in the regime where it is required (which is identified through a discrete equivalent of the stability criterion~(\ref{eq:tensile_instability_criterion})), the full correction is applied, while it is conversely disabled in thermal pressure dominated environments where it is not needed; we then interpolate linearly between these two regimes. Given this, the stabilising term we add to~(\ref{eq:spmhd_force}) reads
\begin{equation}
\label{eq:spmhd_tensile_instability_correction}
    \begin{split}
        \left.
        \frac{\mathrm{d} \boldsymbol{v}_i}{\mathrm{d} t}
        \right|_\mathrm{TIC} =
        - \lambda (\beta_i^\mathrm{loc}) \boldsymbol{B}_i
        \sum_j m_j \Biggl[
        & \frac{f_{ij}}{{\hat{\rho}}_i^2}
        \boldsymbol{B}_i \cdot \nabla_i W_{ij} (h_i) \\
        & + \frac{f_{ji}}{{\hat{\rho}}_j^2}
        \boldsymbol{B}_j \cdot \nabla_i W_{ij} (h_j)
        \Biggr]
        \text{.}
    \end{split}
\end{equation}
Contemporary SPMHD solvers~\citep{2018PASA...35...31P, 2020A&A...638A.140W, 2023MNRAS.518.4115G} all take $\beta_i^\mathrm{loc}$ to be the ratio of thermal to magnetic pressure $\beta_i$ carried by particle $p_i$,
\begin{equation}
\label{eq:beta_loc_others}
    \beta_i^\mathrm{loc} = \beta_i
    \equiv 2 \mu_0 P_i / \boldsymbol{B}_i^2
    \text{,}
\end{equation}
but this does not necessarily have to be the case. The formal derivation of the condition for instability onset is based on a linear stability analysis of `mean field' SPMHD~\citep[see][]{1985MNRAS.216..883P, 1996PASA...13...97M, 2004ApJS..153..447B, 2015JCoPh.302..359I}, which considers the evolution of small perturbations induced on top of a uniform background. The $\beta$ appearing in equation~(\ref{eq:tensile_instability_criterion}), the value of which dictates when unphysical tensile forces take over, corresponds in fact to the pressure ratio of the unperturbed state; an alternative estimate to~(\ref{eq:beta_loc_others}) of the locally prevailing $\beta$ (setting the scale of the correction through~(\ref{eq:tensile_instability_prefactor})) could therefore be defended as being equally well-motivated. Something that should simultaneously be addressed is the fact that computing the corrective force~(\ref{eq:spmhd_tensile_instability_correction}) involves a symmetric SPH gradient operator, which is known to be sensitive `$E_0$' errors that converge sub-linearly with resolution and translate to relatively poor accuracy, particularly in settings where particles are highly disordered~\citep{2010MNRAS.405.1513R, 2012MNRAS.425.1068D}. We thus sought to identify a $\beta_i^\mathrm{loc}$ biased towards the higher end of the distribution of particle-carried $\beta$ in a given particle's local neighbourhood, to cap scale of the tensile instability correction in heterogeneous environments where its calculation is dominated by discretisation errors. After extensive experimentation we settled on
\begin{equation}
\label{eq:beta_loc_us}
    \beta_i^\mathrm{loc} = 
    \sqrt{
        \sum_j \beta_j^2 / \sum_j 1
    }
    \text{,}
\end{equation}
an \textit{unweighted} root mean square averaging over the attributes of all neighbours $p_j$ of any particle $p_i$. In the limit of infinite resolution, (\ref{eq:beta_loc_others}) and~(\ref{eq:beta_loc_us}) formally tend towards the same value. An extensive suite of numerical tests we performed (see Section~\ref{sec:test_problems}) demonstrated that our novel prescription for $\beta_i^\mathrm{loc}$, when passed as an argument to $\lambda$, robustly prevented unwarranted pairing of particles and limited non-conservation errors. Moreover, it largely improved code stability by effectively preventing the application of the tensile instability correction when its calculation was error dominated (and it amounted to a spurious repulsive force) in e.g. `full physics' galaxy formation experiments, where employed sub-grid models induced a high degree of particle disorder; a naive implementation of~(\ref{eq:beta_loc_others}) proved noticeably less robust in the same setting, being prone to the occasional violent acceleration or ejection of particles found in highly dynamical, irregular neighbourhoods.

Having described the individual constituents equation~(\ref{eq:spmhd_tensile_instability_correction}) is comprised of, we further add that it corresponds to a discrete counterpart to the third term on the RHS of the second line of equation~(\ref{eq:mhd_force_components}), pre-multiplied by $- \lambda$. The continuum limit equivalent of equation~(\ref{eq:spmhd_tensile_instability_correction}) thus reads 
\begin{equation}
    \left.
    \frac{\mathrm{d} \boldsymbol{v}}{\mathrm{d} t}
    \right|_\mathrm{TIC}
        =
        - \lambda (\beta )\frac{1}{\mu_0 \rho} (\nabla \cdot \boldsymbol{B})
        \boldsymbol{B}
    \text{.}
\end{equation}
We note that whenever $\lambda = 1$ our scheme becomes formally equivalent to the eight-wave cleaning approach of~\cite{1999JCoPh.154..284P}, which advects $\nabla \cdot \boldsymbol{B}$ with the fluid flow; this is the method adopted by other astrophysical MHD codes such as AREPO~\citep{2010MNRAS.401..791S} for divergence control~\citep[see e.g.][for an application to structure formation]{2018MNRAS.473.4077P}. We deem it necessary but insufficient in our case: the strength and detrimental effect of~(\ref{eq:spmhd_tensile_instability_correction}) will scale with the extent to which $\nabla \cdot \boldsymbol{B} \neq 0$ in our chosen discretisation. As this is something less trivial to control in SPH compared to grid codes~\citep[see again][]{2010MNRAS.405.1513R, 2012MNRAS.425.1068D}, we complement our method with additional corrective measures, which we describe next.

\subsubsection{Divergence cleaning}
\label{sec:divergence_cleaning}

Violation of the solenoidal constraint is undesirable, both in terms of the physical plausibility of the numerical solutions we obtain and the stability of our method; the detrimental side effects of otherwise necessary corrective measures such as the tensile instability correction are exacerbated by the degree to which $\nabla \cdot \boldsymbol{B} \neq 0$. 

Mitigation strategies commonly employed by finite-volume solvers~\citep[such as the `constrained transport' method of][]{1988ApJ...332..659E} that maintain the solenoidal constraint to machine precision cannot be ported to SPMHD, as they involve computing surface integrals which are ill-defined in the case of finite-mass methods. What has however proven a viable alternative are `divergence cleaning' approaches that aim to maintain the (relative) numerical value of $\nabla \cdot \boldsymbol{B}$ low: we here chose to implement the constrained hyperbolic/parabolic divergence-cleaning scheme proposed by~\cite{2016JCoPh.322..326T}, an elaboration on~\cite{2012JCoPh.231.7214T} which was the first truly effective and stable application to SPMHD of the method proposed by~\cite{2002JCoPh.175..645D}. This involves introducing a scalar $\psi$ which  couples to the magnetic field through an additional term to the induction equation
\begin{equation}
\label{eq:mhd_dedner_induction}
    \left.
    \frac{\mathrm{d}}{\mathrm{d} t}
    \left( \frac{\boldsymbol{B}}{\rho} \right) \right|_\psi =
    - \frac{1}{\rho} \nabla \psi
\end{equation}
and which is evolved in time according to
\begin{equation}
\label{eq:mhd_dedner_scalar_evolution}
    \frac{\mathrm{d}}{\mathrm{d} t}
    \left( \frac{\psi}{c_h} \right) =
    -c_h \nabla \cdot \boldsymbol{B}
    - \frac{1}{2} \nabla \cdot \boldsymbol{v} \frac{\psi}{c_h}
    - \frac{1}{\tau_p} \frac{\psi}{c_h}
\end{equation}
The first term on the RHS of~(\ref{eq:mhd_dedner_scalar_evolution}) is purely hyperbolic and acts to propagate divergence errors in a wave-like fashion at characteristic speed $c_h$. The second term was incorporated into the formalism by~\citet{2012JCoPh.231.7214T} to ensure divergence cleaning conserves energy in the presence of a time varying mass density field; it has nevertheless systematically been found to be subdominant, even if the fluid flow is solenoidal with $\Vert \nabla \cdot \boldsymbol{v} \Vert \gg 0$. The third term on the RHS of~(\ref{eq:mhd_dedner_scalar_evolution}) is purely parabolic and locally induces dampening of $\psi$ over a characteristic time-scale $\tau_p$. The joint action of these two terms therefore consists in dispersing divergence errors away from where they are sourced, and subsequently causing them to decay. This was made evident by~\cite{2012JCoPh.231.7214T} who showed that equations~(\ref{eq:mhd_dedner_induction}) and~(\ref{eq:mhd_dedner_scalar_evolution}) can be combined to yield a `generalised, dampened wave equation' for $\psi$, as well as for $\nabla \cdot \boldsymbol{B}$, for a given set of assumptions.

\cite{2012JCoPh.231.7214T} obtained a stable SPMHD formulation of \cite{2002JCoPh.175..645D}'s method by requiring that purely hyperbolic cleaning be energy conserving. This translates into requiring that the discrete gradient operators used to compute $\nabla \cdot \boldsymbol{B}$ and $\nabla \psi$ form a conjugate pair; it is also what prompted the inclusion of the term proportional to $\nabla \cdot \boldsymbol{v}$ in~(\ref{eq:mhd_dedner_scalar_evolution}), although this has empirically proven not to lead to any noticeable improvement. We follow their suggestion and use a symmetric gradient operator to compute $\nabla \psi$, adding to the induction equation
\begin{equation}
\label{eq:spmhd_dedner_induction}
    \left.
    \frac{\mathrm{d}}{\mathrm{d} t}
    \left( \frac{\boldsymbol{B}_i}{\hat{\rho}_i} \right) \right|_\psi =
    - \sum_j m_j \left[
    \frac{f_{ij} \psi_i}{\hat{\rho}_i^2} \nabla_i W_{ij} (h_i) +
    \frac{f_{ji} \psi_j}{\hat{\rho}_j^2} \nabla_i W_{ij} (h_j)
    \right]
\end{equation}
We then use a difference gradient operator to compute $\nabla \cdot \boldsymbol{B}$ in~(\ref{eq:mhd_dedner_scalar_evolution})
\begin{equation}
    \left( \nabla \cdot \boldsymbol{B} \right)_i =
    - \frac{1}{\hat{\rho}_i} \sum_j
    m_j f_{ij} \boldsymbol{B}_{ij} \cdot \nabla_i W_{ij} (h_i)
\end{equation}
and calculate the velocity divergence in the exact same way. These choices have as a consequence that the energy contribution of the parabolic term is negative definite: the full cleaning scheme is therefore either dissipative or at most conservative, and cannot lead to spurious growth of magnetic energy. The scheme has also been shown to be stable across large density contrasts and at free boundaries, the former being ubiquitous in the astrophysical systems we seek to simulate.

We investigated using a symmetric SPH discrete gradient operator for $\nabla \cdot \boldsymbol{B}$ instead (coupled to a difference operator for $\nabla \psi$), this being conceptually of interest as divergence errors would then be estimated in the discretisation in which they appear in the acceleration equation. Although stable, this approach proved to lead to excessive dissipation and unwarranted artefacts in the magnetic field in several standard MHD tests, as also reported by~\cite{2012JCoPh.231.7214T}; we did not pursue this further.

\cite{2016JCoPh.322..326T} showed that the formalism outlined above will remain conservative even in the presence of variable, particle-dependent cleaning speeds $c_h$ and decay timescales $\tau_p$, so long as the quantity evolved in time is $\psi/c_{h}$ rather than $\psi$~\citep[see also Appendix B in][]{2016MNRAS.455...51H}; hence the form of equation~(\ref{eq:mhd_dedner_scalar_evolution}). It is generally advised for $c_h$ to be set to the largest locally resolvable speed, prompting us to choose (see also the following section on time integration)
\begin{equation}
    c_{h,i} = \frac{1}{2} v^{\text{sig}}_i
\end{equation}
as suggested by~\cite{2016MNRAS.455...51H}. We find this to deliver better divergence control, compared to setting $c_h = c_\mathrm{ms}$ as advocated by e.g.~\cite{2018PASA...35...31P} and~\cite{2020A&A...638A.140W}, especially in problems involving a wide range of dynamical scales and sharp density jumps which, both of which are commonplace in structure formation simulations. It is in principle possible to choose an even larger $c_h$~\citep[see e.g.][]{2011MNRAS.418.1392P, 2016JCoPh.322..326T, 2021MNRAS.502.2285D}; this however necessitates introducing additional timestepping constraints, which would make calculations prohibitively expensive in terms of computational cost.

In a typical simulation run, the spatial extent of the divergence errors generated is expected to be of the order of the resolution scale. We therefore choose the characteristic damping time to be a multiple of the ratio of the smoothing length to the cleaning speed, setting the parabolic decay rate to 
\begin{equation}
\label{eq:parabolic_decay_rate}
    \frac{1}{\tau_{p,i}} = \frac{\sigma_p c_{h,i}}{h_i}
\end{equation}
where $\sigma_p$ is a free parameter. We find $\sigma_p=1$ to yield critical damping in three dimensions (see Section~\ref{sec:monopole_advection_test}), and make it our default choice. We tried all other alternatives for $\tau_p$ suggested in Appendix D of~\cite{2016MNRAS.455...51H}; none of them appeared to lead to a noticeable improvement on our tests.

For completeness, we note that formally our cleaning scheme is diffusive and we could in theory re-inject dissipated magnetic energy as heat. This effect is however dwarfed by other contributions to the thermal energy equation, and how much energy is lost locally is intractable due to the wave-like propagation of divergence errors~\citep[see][]{2012JCoPh.231.7214T, 2016JCoPh.322..326T}. We therefore take no measure to address this.

\subsubsection{Discontinuity capturing in SPMHD}
\label{sec:capturing_spmhd_discontinuities}

Devising dissipative terms to capture discontinuities in SPMHD (the equations of which were derived on the assumption of differentiability for all fields involved) can proceed similarly to SPH, following the principles laid out in~\cite{1997JCoPh.134..296C}. Such terms were first introduced by~\cite{2004MNRAS.348..123P} and \cite{2005MNRAS.364..384P}, and further improved by~\cite{2013MNRAS.436.2810T} and \cite{2018PASA...35...31P}.

The artificial viscosity prescription we employ in our scheme remains largely similar to that of SPHENIX, with terms~(\ref{eq:sph_artificial_viscosity_acceleration}) and~(\ref{eq:sph_artificial_viscosity_thermal_energy}) being added to the acceleration and thermal energy equations respectively. However, to more accurately capture MHD shocks, we now take the pair-wise signal velocity to be
\begin{equation}
\label{eq:spmhd_pairwise_sgnal_velocity}
    v^{\text{sig}}_{ij} = c_{\mathrm{ms},i} + c_{\mathrm{ms},j} - \beta^{\text{AV}} \mu_{ij}
\end{equation}
where
\begin{equation}
    c_{\mathrm{ms},i} = \sqrt{c_{s,i}^2 + v_{A,i}^2}
\end{equation}
is the fast magnetosonic wave speed and
\begin{equation}
    v_{A,i} = \sqrt{\frac{\boldsymbol{B}_i^2}{\mu_0 \rho}}
\end{equation}
the Alfvén speed.

Computation of the $\alpha^{\text{AV}}$ follows equations~(\ref{eq:sph_shock_indicator})-(\ref{eq:sph_pairwise_viscosity_alpha}), except for a few minor modifications. \cite{2009MNRAS.398.1678D} showed that including the~\cite{1989PhDT.......206B} switch as a multiplicative prefactor in the calculation of the $\alpha^{\text{AV}}_{ij}$ in equation~(\ref{eq:sph_pairwise_viscosity_alpha}) lead to poor treatment of MHD shocks, something corroborated by our own numerical experiments. While~\cite{2009MNRAS.398.1678D} chose simply not to use the switch, we deemed suppression of unnecessary viscosity in shear flows a desirable feature. We therefore chose to implement a variant of the alternative solution proposed by~\cite{2018PASA...35...31P}, a simplified version of the suggestion of~\cite{2010MNRAS.408..669C}. We compute the `Balsara-like' factor
\begin{equation}
    \tilde{\xi}_i = \frac{\Vert \nabla \cdot \boldsymbol{v}_i \Vert^2}
               {\Vert \nabla \cdot \boldsymbol{v}_i \Vert^2 
               +\Vert \nabla \times \boldsymbol{v}_i \Vert^2}
\end{equation}
which we use directly in the shock indicator, replacing~(\ref{eq:sph_shock_indicator}) by
\begin{equation}
\label{eq:spmhd_shock_indicator}
    S_i = \tilde{\xi}_i H_i^2 \text{max} \left\{
    0, - \dot{\nabla} \cdot \boldsymbol{v}_i
    \right\}
\end{equation}
We note that, unlike~\cite{2010MNRAS.408..669C} and~\cite{2018PASA...35...31P}, we do not rely on the gradient tensor of the velocity field to compute $\dot{\nabla} \cdot \boldsymbol{v}$ in~(\ref{eq:spmhd_shock_indicator}), but we rather make use of standard SPH gradient estimators~\citep[see][]{2022MNRAS.511.2367B}. These are admittedly less accurate, but favouring them reduces computational cost and minimises the memory footprint of individual particles. Having computed particle-carried AV parameters using our modified shock indicator, we finally replace equation~(\ref{eq:sph_pairwise_viscosity_alpha}) by
\begin{equation}
\label{eq:spmhd_pairwise_viscosity_alpha}
    \alpha^{\text{AV}}_{ij} = \frac{1}{2} 
    ( \alpha^{\text{AV}}_i + \alpha^{\text{AV}}_j)
\end{equation}
These changes drastically improved the scheme's ability to capture MHD shocks by reducing particle noise, all the while performing as well as SPHENIX on standard, pure hydrodynamics problems evaluating the suppression of unnecessary viscosity in shear flows~\citep[such as the vortex test of][]{1990IJNMF..11..621G}. We finally note that our SPMHD method makes use of the artificial thermal diffusion scheme described by equations~(\ref{eq:sph_artificial_diffusion})-(\ref{eq:sph_artificial_diffusion_signal_speed}), unchanged.

We complement the aforementioned dissipation terms with a prescription for artificial resistivity (AR), a method to smooth discontinuities in the magnetic field over the resolution scale. This comes as an additional term in the induction equation~\citep[compiling suggestions made by][]{2004MNRAS.348..123P, 2005MNRAS.364..384P, 2013MNRAS.436.2810T, 2016MNRAS.455...51H}
\begin{equation}
\label{eq:spmhd_AR_induction}
\begin{split}
    \left.
    \frac{\mathrm{d}}{\mathrm{d} t}
    \left( \frac{\boldsymbol{B}_i}{\hat{\rho}_i} \right) \right|_\text{AR} =
    \frac{1}{2} \sum_j m_j
    &
    \frac{
        \overline{
            \alpha^\mathrm{AR}
        }_{ij}
        \cdot
        \overline{
            v^\mathrm{AR}
        }_{ij}
    }{
        \overline{\rho}_{ij}
    } \\
    & \times
    \Biggl[ 
        f_{ij} F_{ij} (h_i) + f_{ji} F_{ij} (h_j)
    \Biggr]
    \boldsymbol{B}_{ij}
\end{split}
\end{equation}
where $v_i^\mathrm{AR}$ is a characteristic velocity scale (proper to AR and distinct from $v_{ij}^\mathrm{sig}$) that is modulated by the (potentially spatiotemporally adaptive) switch $\alpha_i^\mathrm{AR}$. The choice of $v_i^\mathrm{AR}$ is a non-trivial task in SPMHD: our inability to a priori determine which of the several possible hydromagnetic discontinuity types is to arise in any given scenario renders the construction of AR terms by analogy to Riemann solvers~\citep[as is done for standard SPH, see][]{1997JCoPh.136..298M} formally impossible. Contemporary works~\citep{2018PASA...35...31P, 2020A&A...638A.140W, 2023MNRAS.518.4115G} adopt an approach orthogonal to that and converge towards the optimal choice for $v_i^\mathrm{AR}$ being the pair-wise signal velocity
\begin{equation}
\label{eq:ar_signal_velocity_others}
    v^\text{AR}_{ij} = 
    \Vert \boldsymbol{v}_{ij} \times \hat{\boldsymbol{r}}_{ij} \Vert
    \text{,}
\end{equation}
which they subsequently scale by a constant $\alpha^{AR}_i$; this leads to AR contributions vanishing in the absence of relative motion and, although deceptively simple, has proven to limit excess dissipation and deliver satisfactory performance on challenging MHD problems~\citep[e.g.][]{2018MNRAS.481.2450W, 2022A&A...659A..91W, 2023A&A...673A..47W}. We therefore experimented extensively with~(\ref{eq:ar_signal_velocity_others}), to indeed find it to be effective in a variety of settings, but to also discover it induces surplus diffusion of the magnetic field in simulations of structure formation. With the concordance model of cosmology pointing toward spectral power in the matter distribution of the universe being distributed across a broad range of scales, cosmological simulations are bound to always involve coarsely represented objects no matter the resolution. The absence of a formal guarantee the velocity field in said structures will be converged can thus translate to potentially large velocity differences appearing in equation~(\ref{eq:ar_signal_velocity_others}), which in turn seed excess AR.

To address the aforementioned problem we choose to instead follow suggestions made by~\citet{2005MNRAS.364..384P, 2013MNRAS.436.2810T, 2016MNRAS.455...51H, 2018MNRAS.476.2890B} and set the AR signal velocity equal to a characteristic MHD wave speed. Having tested all three possible, we settle on the Alfvén speed with $v_i^\mathrm{AR} = v_{A,i}$ (similarly to~\citealt{2012JCoPh.231..759P} and~\citealt{2016MNRAS.455...51H}, and in contrast to~\citealt{2004MNRAS.348..123P} and~\citealt{2013MNRAS.436.2810T} who favour the fast magnetosonic wave speed). We then modulate it by
\begin{equation}
\label{eq:ar_swicth}
    \alpha_i^\mathrm{AR}
    =
    \mathrm{min}
    \left\{
        \alpha_\mathrm{max}^\mathrm{AR}
        ,
        \frac{
            h_i \Vert \left( \nabla \boldsymbol{B} \right)_i \Vert
        }{
            \Vert \boldsymbol{B}_i \Vert
        }
    \right\}
\end{equation}
where $\alpha_\mathrm{max}^\mathrm{AR}$ is a constant we set to $\alpha_\mathrm{max}^\mathrm{AR} = 0.1$~\citep[as do][]{2016MNRAS.455...51H}, and $\nabla \boldsymbol{B}$ is the gradient tensor of the magnetic field, the individual components of which we calculate using an anti-symmetric SPH derivative operator
\begin{equation}
    \left( \nabla^\mu B^\nu \right)_i =
    - \frac{1}{\hat{\rho}_i} \sum_j
    m_j f_{ij} B^\nu_{ij} 
    \left[
        \nabla_i W_{ij} (h_i)
    \right]^\mu
\end{equation}
and subsequently use to compute the tensor dual norm
\begin{equation}
    \Vert \left( \nabla \boldsymbol{B} \right)_i \Vert
    =
    \sqrt{
        \sum_{\mu, \nu}
        \left( \nabla^\mu B^\nu \right)_i^2
    }
    \text{.}
\end{equation}
The shock indicator defined in equation~(\ref{eq:ar_swicth}), originally introduced by~\citet{2013MNRAS.436.2810T}, is calculated using only the instantaneous values of particle-carried attributes and does not therefore require to be evolved in time independently (in contrast to $\alpha_i^\mathrm{AV}$ and $\alpha_i^\mathrm{AD}$, see Section~\ref{sec:capturing_sph_discontinuities}, or previous suggestions for $\alpha_i^\mathrm{AR}$ such as that of~\citealt{2005MNRAS.364..384P}, which~\citealt{2013MNRAS.436.2810T} found to under-perform in the weak field limit). The $\Vert \nabla \boldsymbol{B} \Vert$ term appearing in~\ref{eq:ar_swicth} `lights up' only in the presence of discontinuities in the magnetic field, while the $\Vert \boldsymbol{B} \Vert$ in the denominator renders $\alpha_i^\mathrm{AR}$ field strength independent, translating to robust discontinuity detection and consistent performance in both the weak and strong field limits. \citet{2013MNRAS.436.2810T} note that equation~(\ref{eq:ar_swicth}) is invariant to the rescaling $\boldsymbol{B} \rightarrow \lambda \boldsymbol{B}$ for $\lambda \in \mathbb{R}$, meaning our $\alpha_i^\mathrm{AR}$ will remain unchanged in dynamo amplification scenarios, which are commonplace and particularly relevant to galaxy and structure formation~\citep[see][]{2018SSRv..214..122D, 2023ARA&A..61..561B, 2025MNRAS.541.3427S}. The rationale behind us adopting the Alfvén speed as the rate of information propagation for our AR scheme is identical to that of~\citet{2016MNRAS.455...51H}; any plausible alternative, inevitably involving the sound speed, is subject to thermal energy fluctuations triggering excess AR~\citep[something that is observed, for instance, in the cosmological MHD simulations of][]{2018MNRAS.476.2890B}. The choices we outline here were empirically determined to compound to an optimal balance (by concurrent consideration of performance on all validation tests we present in Section~\ref{sec:test_problems}, for fixed code parameters) between appropriately capturing MHD shocks and limiting spurious dissipation.

Equation~(\ref{eq:spmhd_AR_induction}) is accompanied by an additional term to the thermal energy equation
\begin{equation}
\label{eq:spmhd_AR_thermal_energy}
\begin{split}
    \left.
    \frac{\mathrm{d} u_i}{\mathrm{d} t} \right|_\text{AR} =
    - \frac{1}{4 \mu_0} \sum_j m_j
    &
    \frac{
        \overline{
            \alpha^\mathrm{AR}
        }_{ij}
        \cdot
        \overline{
            v^\mathrm{AR}
        }_{ij}
    }{
        \overline{\rho}_{ij}
    } \\
    & \times
    \Biggl[ 
        f_{ij} F_{ij} (h_i) + f_{ji} F_{ij} (h_j)
    \Biggr]
    \boldsymbol{B}^2_{ij}
\end{split}
\end{equation}
the form of which guarantees conservation of the total energy, and a positive definite change in thermal energy and therefore in entropy~\citep{2004MNRAS.348..123P}, satisfying the second law of thermodynamics.

Up to the exact choice of SPH discretisation, equations~(\ref{eq:spmhd_AR_induction}) and~(\ref{eq:spmhd_AR_thermal_energy}) are very similar to equations~(\ref{eq:spmhd_ohmic_induction}) and~(\ref{eq:spmhd_ohmic_thermal_energy}) respectively. AR can therefore be interpreted as a non-ideal diffusion term with effective, particle-dependent magnetic diffusivity
\begin{equation}
\label{spmhd_AR_diffusivity}
    \eta^{\text{AR}}_{ij} \sim 
    \frac{1}{2} \alpha^{\text{AR}}_i v^{\text{AR}}_{ij}
    \Vert \boldsymbol{r}_{ij} \Vert
\end{equation}
We note that, unlike AV which acts only when $\boldsymbol{v}_{ij} \cdot\boldsymbol{r}_{ij} < 0$, AR can be applied to pairs of particles that are either approaching or receding, to capture magnetic shocks which can occur in both compression and rarefaction. The form of $\alpha^{\text{AR}}_i$ further implies that dissipation of the magnetic field scales to second order with resolution, by virtue of $\boldsymbol{r}_{ij} \sim h$ and $\alpha^{AR}_i \sim h$ in~(\ref{eq:ar_swicth}), ensuring good convergence behaviour of the method.

\subsubsection{Time integration}
\label{sec:time_inetgration}

Time integration of any physical field $Q(\boldsymbol{r},t)$ is performed using a generalised version of the Leapfrog integrator of~\cite{1967PhRv..159...98V}, which is time-reversible and accurate to second order. For these properties to hold when time derivatives $\dot{Q}$ depend on $Q$ explicitly (e.g. acceleration or induction depending on $\boldsymbol{v}$ and $\boldsymbol{B}$), the base algorithm needs to be modified to first compute a predicted, half-step estimate $\tilde{Q}$, subsequently used in  computing the full-step update to the variable~\citep[for a detailed discussion of this in the context of SPH see e.g.][for its application to SPMHD]{1989ApJS...70..419H, 2004ApJS..153..447B}; it is this strategy which confers excellent conservation properties to the scheme. We use superscripts $n \in \mathbb{Z}^+$ to label time-steps as $t^n$, $Q^n \equiv Q(t^n)$ to denote the value of $Q$ computed at $t^n$, and take $\delta t \equiv t^{n+1} - t^n$ to be the time-step length; the time integration algorithm we implement, formulated in its `kick-drift-kick' form, consists of the sequence of operations
\begin{align}
    & \boldsymbol{v}^{n + \frac{1}{2}}
    = \boldsymbol{v}^{n} + 
    \frac{1}{2} \delta t 
    \dot{\boldsymbol{v}} (\boldsymbol{r}^{n}, \boldsymbol{v}^{n}, \boldsymbol{B}^{n}) 
    \label{eq:kick_1}
    \\
    & \boldsymbol{r}^{n + 1}
    = \boldsymbol{r}^{n} + \delta t \boldsymbol{v}^{n + \frac{1}{2}}
    \label{eq:drift}
    \\
    & \tilde{\boldsymbol{v}}^{n + 1}
    = \boldsymbol{v}^n + 
    \delta t 
    \dot{\boldsymbol{v}} (\boldsymbol{r}^{n}, \boldsymbol{v}^{n}, \boldsymbol{B}^{n})
    \label{eq:predictor}
    \\
    & \boldsymbol{v}^{n + 1}
    = \boldsymbol{v}^{n + \frac{1}{2}} + 
    \frac{1}{2} \delta t 
    \dot{\boldsymbol{v}} (\boldsymbol{r}^{n+1}, 
    \tilde{\boldsymbol{v}}^{n+1}, \tilde{\boldsymbol{B}}^{n+1})
    \label{eq:kick_2}
\end{align}
where we have omitted the updates to $\boldsymbol{B}$ and $\psi$; these are identical to those for $\boldsymbol{v}$. Here, equation~(\ref{eq:kick_1}) is referred to as the first kick, equation~(\ref{eq:drift}) as the drift, equation~(\ref{eq:predictor}) as the predictor step and equation~(\ref{eq:kick_2}) as the second kick or corrector step.

SWIFT implements dynamical timestepping, where the $\delta t$ are made particle dependent and are allowed to vary~\citep[see e.g.][]{1997astro.ph.10043Q, 2005MNRAS.364.1105S}. The drift and predictor steps can be subdivided into smaller sequential updates, allowing for exact interpolation between kick steps and thus synchronous interaction between pairs of particles on different $\delta t$~\citep[see][for more information on individual timestepping in SWIFT]{2024MNRAS.530.2378S}. Individual $\delta t$ are set according to a Courant–Friedrichs–Lewy (CFL) criterion~\citep{1928MatAn.100...32C}
\begin{equation}
\label{eq:cfl_delta_t}
    \delta t_{\text{CFL},i} =
    C_{\text{CFL}} \frac{2 \gamma_K h_i}{v^{\text{sig}}_i}
\end{equation}
where $C_{\text{CFL}}$ is a constant which for all tests we present in this paper was set to $C_{\text{CFL}}=0.1$~\citep[consistent with what][used to test SPHENIX]{2022MNRAS.511.2367B}. Here, $v^{\text{sig}}_i / 2$ can be interpreted as the maximal speed of information propagation~\citep{1997JCoPh.136..298M}.

The CFL criterion is derived on the assumption that the ideal MHD approximation holds. To ensure code stability whenever we run with Ohmic diffusion, we need to consider an additional timestepping constraint, which introduces a new time-step size $\delta t_\eta$. We follow~\cite{2014MNRAS.444.1104W} and use
\begin{equation}
\label{eq:ohmic_diffusion_delta_t}
    \delta t_{\eta,i} =
    C_\eta \frac{h^2_i}{\eta}
\end{equation}
where we set $C_\eta = 1/8 \pi$ by default~\citep[identically to][up to a rescaling accounting for our different definition of $h$]{2018PASA...35...31P}, finding it to lead to satisfactory results on all non-ideal MHD tests we consider. We note that, unlike~\cite{2016MNRAS.457.1037W}, our SPMHD scheme does not come with an implementation of the super-timestepping algorithm of~\cite{1996CNES} to control $\delta t_\eta$ becoming prohibitively small, as we have not been confronted with such an issue throughout our experimentation with the method.

Expressions such as~(\ref{eq:cfl_delta_t}) and~(\ref{eq:ohmic_diffusion_delta_t}) are typically derived following analytical arguments, returning the minimal required $\delta t$ for stable discrete integration of the relevant PDEs, on the presupposition that fields and derivatives thereof entering them are continuous functions of time. This assumption is violated by several physics modules available in SWIFT we wish to ultimately couple our method to, such as the radiative cooling prescription of~\citealt{2009MNRAS.393...99W} or stellar feedback model of~\citealt{2012MNRAS.426..140D}, both of which involve discontinuous changes in thermal energy. We therefore propose two additional $\delta t$, consideration of which is intended to prevent destabilisation of time integration and runaway growth of either of the two primary fields evolved by our method; to restrict the per time-step fractional change in $\boldsymbol{B}$ and $\psi$ to respectively be at most $C_{\boldsymbol{B}}$ and $C_\psi$ (where $C_{\boldsymbol{B}}, C_\psi \in \mathbb{R}^+$), we define $\delta t_{\boldsymbol{B}}$ and $\delta t_\psi$ as
\begin{equation}
\label{eq:delta_t_B}
    \delta t_{\boldsymbol{B}, i} = 
    C_{\boldsymbol{B}}
    \dfrac{
        \boldsymbol{B}_i / \hat{\rho}_i
    }{
        \left( \boldsymbol{B}_i / \hat{\rho}_i \right)^\cdot
    }
\end{equation}
and
\begin{equation}
\label{eq:delta_t_psi}
    \delta t_{\psi, i} =
    \begin{cases}
        C_{\psi}
    \dfrac{
        \psi_i / c_{h,i}
    }{
        \left( \psi_i / c_{h,i} \right)^\cdot
    },
    \quad & \text{if} \quad
    \varepsilon_{\psi, i} / \varepsilon_{\boldsymbol{B}, i} > R_{\psi / \boldsymbol{B}}
    \\
    \infty, \quad & \text{otherwise}
    \end{cases}
    \text{.}
\end{equation}
The branching in~(\ref{eq:delta_t_psi}) ensures $\delta t_\psi$ is only taken into account when the dynamics of the divergence cleaning scalar is relevant to that of the magnetic field. This being the case is assessed by measuring the ratio of the specific energy in $\psi$ to specific energy in $\boldsymbol{B}$, $\varepsilon_{\psi, i} / \varepsilon_{\boldsymbol{B}, i}$, and verifying whether it exceeds a fixed threshold $R_{\psi / \boldsymbol{B}}$. Here, following~\citet{2012JCoPh.231.7214T}, we have that
\begin{equation}
    \varepsilon_{\boldsymbol{B}, i} = 
    \frac{1}{2 \mu_0} \frac{\boldsymbol{B}^2_i}{\hat{\rho}_i}
    \quad \text{and} \quad
    \varepsilon_{\psi, i} =
    \frac{1}{2 \mu_0 \hat{\rho}_i} \frac{\psi_i^2}{c_{h,i}^2}
    \text{.}
\end{equation}
We finally return a single MHD time-step size $\delta t_{\mathrm{MHD}}$ per particle, calculated as the minimum over~(\ref{eq:cfl_delta_t})-(\ref{eq:delta_t_psi})
\begin{equation}
    \delta t_{\mathrm{MHD}, i}
    =
    \mathrm{min}
    \left(
        \delta t_{\mathrm{CFL}, i},
        \delta t_{\eta, i},
        \delta t_{\boldsymbol{B}, i},
        \delta t_{\psi, i}
    \right)
\end{equation}
This is in turn compared with any additional time-step size computed by other physics modules of the code.

To ensure good conservation properties, especially in the presence of strong energy perturbations which are commonplace in sub-grid galaxy formation models, SWIFT complements the base time integration algorithm described above with a modified version of the time-step limiter of~\cite{2009ApJ...697L..99S}, as suggested by~\cite{2012MNRAS.419..465D}. The limiter requires the time-step size over which a given particle is evolved to lie within a factor $\Delta$ from that of any of its neighbours; we set $\Delta = 4$ by default.

\subsubsection{SPMHD in a cosmologically expanding frame}
\label{sec:cosmological_SPMHD}
The equations governing the dynamics of an astrophysical fluid in a cosmological context are usually expressed in terms of spatial coordinates $\boldsymbol{x}$ that are comoving with the expansion of the universe. Physical coordinates $\boldsymbol{r}$ are related to $\boldsymbol{x}$ through
\begin{equation}
    \boldsymbol{r} = a(t) \boldsymbol{x}
\end{equation}
where $a(t)$ is the scale-factor, related to the Hubble parameter $H$ through $H(t) \equiv \dot{a}(t) /a(t)$. We express the present-day value of the latter as $H_0 = H(t=t_\text{now}) = 100 h \, \text{km} \, \text{s}^{-1} \,\text{Mpc}^{-1}$ for $h$ the reduced Hubble constant. The time evolution of $a(t)$ is given by the Friedmann equation
\begin{equation}
    H(a) = H_0
    \left[ \Omega_m a^{-3} + \Omega_r a^{-4} + \Omega_k a^{-2} + \Omega_\Lambda
    \right]^{1/2}
\end{equation}
where $\Omega_m, \Omega_r, \Omega_k$ and $\Omega_\Lambda$ are respectively the present day dimensionless matter, radiation, curvature and dark energy densities.

One can get to the continuum fluid equations in an expanding frame through an action minimisation exercise~\citep[see e.g. Section II of][for a textbook example of how this is done]{1980lssu.book.....P}. In the case of pure hydrodynamics coupled to Newtonian gravity, the comoving EoM thus obtained are entirely equivalent to their non comoving counterpart, up to a reparameterisation of the physical variables. \cite{1996PhRvD..54.1291B} showed that for a given set of assumptions, this holds true for MHD as well. We are therefore prompted to derive cosmological EoM for our scheme, starting from their non-comoving counterpart presented in the previous sections, by simply making the coordinate change $(\boldsymbol{r}, t) \rightarrow (\boldsymbol{x}, t)$ and transforming any physical quantity $Q$ to its comoving twin $Q_c$ defined as $Q_c = a^n Q$ for some $n \in \mathbb{R} $. This is common practice in the development of simulation codes targeted at cosmological gas dynamics~\citep[see e.g.][]{2013MNRAS.432..176P}. Noting that there is no single possible comoving variable parametrisation (see e.g.~\citealt{2008ApJS..174....1L, 2014ApJS..211...19B, 2017MNRAS.472.4368R} for different suggestions, as well as~\citealt{2022MNRAS.515.3492B} and references therein), we define
\begin{equation}
\label{eq:comoving_thermo_variables}
    \rho_c = a^3 \rho \text{,}
    \quad P_c = a^{3 \gamma} P \text{,}
    \quad \text{and}
    \quad u_c = a^{3(\gamma-1)} u
\end{equation}
Plugging $\{\rho_c, P_c, u_c \}$ thus defined into the thermodynamic relations linking $\{\rho, P, u \}$, one finds that the functional interdependence of the former is identical to that of the latter. Thermodynamics remains, in this sense, cosmology independent; requiring said condition be satisfied served, in fact, as the guiding principle behind the choice of comoving variables we propose. We further introduce
\begin{equation}
\label{eq:comoving_B}
    \boldsymbol{B}_c = a^{3 \gamma / 2} \boldsymbol{B}
    \text{.}
\end{equation}
which, to our knowledge, is the first occurrence of such a definition of the comoving magnetic field~\citep[as opposed to the more commonplace $\boldsymbol{B}_c = a^2 \boldsymbol{B}$, see e.g.][]{2013MNRAS.432..176P, 2022MNRAS.515.3492B}. Our choice results in the Alfvén speed transforming identically to the sound speed when converting to the expanding reference frame, and can thus be seamlessly integrated into characteristic velocity calculations (entering for instance time-step, divergence-cleaning or discontinuity-capturing calculations, see Sections~\ref{sec:capturing_spmhd_discontinuities}, \ref{sec:divergence_cleaning}, \ref{sec:time_inetgration}) in a robust and infallible cosmology-agnostic fashion. We finally define the velocity variable
\begin{equation}
\label{eq:comoving_velocity}
    \boldsymbol{w} = a^2 \dot{\boldsymbol{x}}
\end{equation}
(which has no immediate physical interpretation, as opposed to e.g. the peculiar velocity $\boldsymbol{v}_\text{pec} = a \dot{\boldsymbol{x}}$ or Hubble flow $\boldsymbol{v}_\text{Hubble} = \dot{a} \boldsymbol{x}$). We then plug~(\ref{eq:comoving_thermo_variables}), (\ref{eq:comoving_B}) and~(\ref{eq:comoving_velocity}) into~(\ref{eq:mhd_force})-(\ref{eq:mhd_induction}) to obtain the continuum equations of cosmological ideal MHD
\begin{equation}
    \frac{\mathrm{d} \boldsymbol{w}}{\mathrm{d}t} 
    = - \frac{1}{a^{3(\gamma-1)}} 
    \frac{1}{\rho}_c \nabla_{\boldsymbol{x}} \cdot \boldsymbol{S}_c
\label{eq:comoving_momentum}
\end{equation}
\begin{equation}
    \frac{\mathrm{d}}{\mathrm{d}t}
    \left( \frac{\boldsymbol{B}_c}{\rho_c} \right)
    = \frac{1}{a^2} \left(
    \frac{\boldsymbol{B}_c}{\rho_c} \cdot \nabla_{\boldsymbol{x}} 
    \right)
    \boldsymbol{w}
    + \Gamma H \frac{\boldsymbol{B}_c}{\rho_c}
\label{eq:comoving_induction}
\end{equation}
\begin{equation}
    \frac{\mathrm{d} u_c}{\mathrm{d}t}
    = - \frac{1}{a^2}
    \frac{P_c}{\rho_c} \nabla_{\boldsymbol{x}} \cdot \boldsymbol{w}
\label{eq:comoving_energy}
\end{equation}
where partial time derivatives are now taken at constant $\boldsymbol{x}$, terms involving $\ddot{a} (t)$ are eliminated through a canonical transformation, composite comoving functions $f_c(\{Q_c\})$ have the same dependence on the set of comoving variables $\{Q_c\}$ as their physical counterpart $f(\{Q\})$ on $\{Q\}$, and where we have defined $\Gamma \equiv 3 \gamma / 2 - 2$. Note that other than multiplicative prefactors depending on $a(t)$, which are absorbed into the time integration operators following~\cite{1997astro.ph.10043Q}, and the source term in~(\ref{eq:comoving_induction}) which is typically subdominant, the functional form of equations~(\ref{eq:comoving_momentum})-(\ref{eq:comoving_energy}) is identical to that of~(\ref{eq:mhd_force})-(\ref{eq:mhd_induction}). 

The equations of divergence cleaning in a comoving frame are obtained by considering
\begin{equation}
\label{eq:comoving_dedner_variables}
    \psi_c = a^{3 \gamma / 2 +1} \psi 
    \quad \text{and} \quad
    c_{h,c} = a^{3 (\gamma - 1) /2} c_h
\end{equation}
which when plugged into~(\ref{eq:mhd_dedner_induction}) and~(\ref{eq:mhd_dedner_scalar_evolution}) result in a contribution to the induction equation
\begin{equation}
\label{eq:comoving_mhd_dedner_induction}
    \left.
    \frac{\mathrm{d}}{\mathrm{d} t}
    \left( \frac{\boldsymbol{B}_c}{\rho_c} \right) \right|_\psi =
    - \frac{1}{a^2}
    \frac{1}{\rho_c} \nabla_{\boldsymbol{x}} \psi_c
\end{equation}
and a prescription for the time evolution of $\psi_c$ that reads
\begin{equation}
\label{eq:comoving_mhd_dedner_scalar_evolution}
    \begin{split}
        \frac{\mathrm{d}}{\mathrm{d} t}
        \left( \frac{\psi_c}{c_{h,c}} \right) =
        - \frac{1}{a^2} \Biggl\{
        & 
        a^{5 - 3 \gamma} c_{h,c} \nabla_{\boldsymbol{x}} \cdot \boldsymbol{B}_c
        + \frac{1}{2} \nabla_{\boldsymbol{x}} \cdot \boldsymbol{w} \frac{\psi_c}{c_{h,c}} \\
        & + a^{(5 - 3 \gamma) / 2} \frac{1}{\tau_{p,c}} \frac{\psi_c}{c_{h,c}}
        \Biggr\} \\
        & + \left( \frac{5}{2} - \frac{\nu}{2} \right) H \frac{\psi_c}{c_{h,c}}
    \end{split}
\end{equation}
We note that whereas the expression for the comoving cleaning speed $c_{h,c}$ in~(\ref{eq:comoving_dedner_variables}) is entirely dictated by our picks in~(\ref{eq:comoving_thermo_variables}) and~(\ref{eq:comoving_B}) for the `fundamental' comoving variables ($c_{h,c}$ having de facto the same functional dependence on comoving attributes $\{ Q_c \}$ as $c_h$ on their non-comoving counterparts $\{ Q\}$), we are entirely free to decide how we define $\psi_c$. Our choice in~(\ref{eq:comoving_dedner_variables}) is a direct consequence of us requiring that the Dedner contribution~(\ref{eq:comoving_mhd_dedner_induction}) to be added to the comoving induction equation~(\ref{eq:comoving_induction}) preserve the structure of the non-cosmological PDEs \citep[up to a cosmological multiplicative prefactor that is again to be absorbed into the employed time integration operator, see][]{1997astro.ph.10043Q, 2024MNRAS.530.2378S}. Regarding the derived evolution equation for $\psi_c$, we note that in the particular case of a monoatomic gas with $\gamma = 5/3$ (a commonly assumed parameter choice in galaxy and structure formation production runs), equation (\ref{eq:comoving_mhd_dedner_scalar_evolution}) reduces to
\begin{equation}
\label{eq:comoving_mhd_dedner_scalar_evolution_gamma_5o3}
\begin{split}
    \frac{\mathrm{d}}{\mathrm{d} t}
    \left( \frac{\psi_c}{c_{h,c}} \right) =
    - \frac{1}{a^2} \Biggl\{
    & c_{h,c} \nabla_{\boldsymbol{x}} \cdot \boldsymbol{B}_c
    + \frac{1}{2} \nabla_{\boldsymbol{x}} \cdot \boldsymbol{w} \frac{\psi_c}{c_{h,c}} \\
    & + \frac{1}{\tau_{p,c}} \frac{\psi_c}{c_{h,c}}
    \Biggr\}
    + \left( \frac{5}{2} - \frac{\nu}{2} \right) H \frac{\psi_c}{c_{h,c}}
\end{split}
\end{equation}
which again, up to time integration details and a generally subdominant Hubble source term, preserves the equation structure of the base, non-comoving constrained hyperbolic/parabolic divergence-cleaning scheme. We repeat the exercise we discuss in the present section for all remaining non-ideal and corrective terms to obtain the full set of equations of SPMHD in an expanding frame, which we present in Appendix~\ref{appendix:cosmo_SPMHD}. 

\section{Test problems}
\label{sec:test_problems}

In this section, we present numerical predictions obtained with our scheme for a set of standard problems commonly considered in performance testing (cosmological) MHD codes. The ensemble of shown examples is chosen so that we can evaluate the method's ability to reproduce the full, varied phenomenology of MHD. We present these base problems in order of increasing physical and numerical complexity, considering first the method in isolation and subsequently coupling it to an incrementally more complex network of complementary code modules: we begin with `laboratory' tests of ideal and non-ideal MHD in Sections~\ref{sec:ideal_mhd_tests} and~\ref{sec:non_ideal_mhd_tests} respectively, continue with an example pairing MHD with gravity in Section~\ref{sec:mhd_and_gravity}, further account for cosmology in Section~\ref{sec:mhd_and_cosmology_and_gravity}, and finally evaluate our method's ability to operate in tandem with a full galaxy formation model in Section~\ref{sec:mhd_and_subgrid}. Unless otherwise specified, the simulated fluid is considered to obey the ideal gas law given by equation~(\ref{eq:ideal_gas_law}), with adiabatic index $\gamma = 5/3$, which is that for a monoatomic gas. For all `laboratory' examples (Sections~\ref{sec:ideal_mhd_tests} and~\ref{sec:non_ideal_mhd_tests}) we fix the vacuum permeability to $\mu_0 = 1$, while for all other entries in our test suite we set $\mu_0$ to its experimentally determined value of $\mu_0 \approx4 \pi \cdot 10^{-7} \; N/A^2$.

We show performance for fixed model parameters and with all corrective terms enabled across all tests (unless otherwise stated), in line with the approach adopted by e.g.~\citet{2016MNRAS.455...51H}, and most importantly by~\cite{2022MNRAS.511.2367B} when benchmarking SPHENIX. We thus seek to paint an unbiased, representative picture of the precision and accuracy of our solver as it would be used in practice for simulations targeting our science case of interest. Adjusting parameters for each simulation set-up could undoubtedly lead to improved results on an individual per-problem basis; we nevertheless choose to show results for what we consider to be the globally optimal, fixed code configuration instead (the one exception to this is Section~\ref{sec:monopole_advection_test}, where we vary the parabolic divergence cleaning parameter ($\sigma_p$ in equation~(\ref{eq:parabolic_decay_rate})) to motivate the value we select, as it modulates a divergence cleaning velocity that is chosen to be slightly different in our scheme compared to what is most commonly used elsewhere). Our code configuration can then be summarised as follows:
\begin{tasks}
    \task We use the quintic spline kernel~\citep[see][]{1985A&A...149..135M} with $\eta_\mathrm{res} = 1.595$ (corresponding to $\sim 180$ weighted neighbours).
    \task We compute ideal and non-ideal time-step sizes using prefactors set to $C_\mathrm{CFL}=0.1$ and $C_\eta = 1/8 \pi$ in equations~(\ref{eq:cfl_delta_t}) and~(\ref{eq:ohmic_diffusion_delta_t}) respectively. Dynamical timestepping is always enabled.
    \task We do not make use of the novel time-step conditions we introduce in Section~\ref{sec:time_inetgration}; we achieve this by setting $(C_{\boldsymbol{B}}, C_\psi, R_{\psi/\boldsymbol{B}}) = (\infty, \infty,\infty)$ in equations~(\ref{eq:delta_t_B}) and~(\ref{eq:delta_t_psi}).
    \task We configure our mixed hyperbolic/parabolic divergence-cleaning scheme (see Section~\ref{sec:divergence_cleaning}) with $\sigma_p = 1$ in equation~(\ref{eq:parabolic_decay_rate}).
    \task We configure our artificial resistivity scheme (see Section~\ref{sec:capturing_spmhd_discontinuities}) with $\alpha_\mathrm{max}^\mathrm{AR} = 0.1$ in equation~(\ref{eq:ar_swicth}).
    \task We set all parameters of the underlying hydrodynamics scheme to their defaults, as discussed in Section~\ref{sec:sphenix} and~\citet{2022MNRAS.511.2367B}, to preserve the successes of SPHENIX in non-magnetised scenarios.
\end{tasks}
Whenever a reference solution for a given physical attribute $Q$ (be it analytic or numerical) is readily available for any of the problems we consider, we quantitatively evaluate the extent to which our predictions deviate from it using a standard $\mathcal{L}_1$ metric defined as 
\begin{equation}
    \mathcal{L}_1 (Q) = 
    \frac{1}
    {\text{max}_{\boldsymbol{r} \in \mathbb{R}^\nu}
    \left\{ Q^\text{ref} (\boldsymbol{r})\right\}}
    \frac{1}{N}
    \sum_{i=1}^N
    \Vert Q_i^\text{SWIFT} - Q^\text{ref} (\boldsymbol{r}_i) \Vert
\end{equation}
where $Q_i^\text{SWIFT}$ is our scheme's prediction for the value of $Q$ at the position of particle $i$, and $Q^\text{ref} (\boldsymbol{r}_i)$ is its counterpart in the comparison solution. Following~\cite{2018PASA...35...31P}, we sum the absolute relative error for all particles in the simulation and scale the metric by the total particle number \textit{and} the maximal value of the reference solution over the simulation domain. This normalisation has the benefit of yielding a dimensionless quantity which makes resolution studies straightforward, without affecting convergence properties. Unless otherwise stated, 
the reference against which we compare our results when considering problems with no known analytic solution are the numerical predictions 
of~\citet{2008ApJS..178..137S}. These were obtained with the state-of-the-art, static grid code ATHENA; it is the point of comparison considered by e.g.~\cite{2016MNRAS.455...51H} as well in testing their mesh-less, finite-volume Lagrangian methods for MHD.

The extent to which the solenoidal constraint is violated is quantified through the dimensionless, particle-dependent error metric
\begin{equation}
    \varepsilon_{\nabla \cdot \boldsymbol{B}, i} \equiv 
    \frac{h_i \Vert \nabla \cdot \boldsymbol{B}_i \Vert}{\Vert \boldsymbol{B}_i \Vert}
\label{eq:dimensionless_divergene_error}
\end{equation}
For numerical results to be considered reliable and uncorrupted by divergence errors, it is widely accepted that the average of this quantity over the simulation domain should remain below $\sim 0.01-0.1$~\citep[as defended by e.g.][]{2018PASA...35...31P, 2020A&A...638A.140W}\footnote{Note that our SPH smoothing length $h$, which we compute according to the formalism laid out in~\citet{2012MNRAS.425.1068D}, will differ by a factor of $\sim 2- 3$ when measured against that of e.g.~\citet{2018PASA...35...31P}, who determine their $h$ slightly differently. Our calculated $\varepsilon_{\nabla \cdot \boldsymbol{B}}$, when compared to that returned by other codes for the same particle distribution, will be offset by a small amount; the metric is nevertheless intended as a rough diagnostic and its precise value is not that important.}. 

The initial condition and code parameter files, together with the post-processing routines, that were used to run the examples presented in the sections that follow are made publicly available as part of the SWIFT repository.

\subsection{Ideal MHD test problems}
\label{sec:ideal_mhd_tests}

We begin with a series of `laboratory' tests of ideal MHD, with our code configured in magnetohydrodynamics-only mode (i.e. with gravity, cosmology and other radiative physics modules disabled), so as to first evaluate our method in isolation. We lead with three simple problems (Sections~\ref{sec:monopole_advection_test}-\ref{sec:loop_advection_tests}) each independently testing one of the scheme's three major mitigation strategies against numerical instability (see Sections~\ref{sec:tensile_instability_correction}, \ref{sec:divergence_cleaning} and~\ref{sec:capturing_spmhd_discontinuities}); particularly, the example discussed in Section~\ref{sec:monopole_advection_test} was used to calibrate divergence cleaning~\citep[similarly to][]{2012JCoPh.231.7214T}. We subsequently follow up with a series of problems involving interacting, multidimensional MHD discontinuities (Sections~\ref{sec:MHD_shock_tubes}-\ref{sec:strong_magnetised_blast_wave}). Obtaining sound results on the first among these, the shock tube of~\citet{1988JCoPh..75..400B}, was what guided us in tuning the hyperparameters of our shock-capturing scheme, which were kept fixed for all experiments discussed in this paper; \citet{2022MNRAS.511.2367B} followed a similar approach when developing SPHENIX, using the shock tube of~\citet{1978JCoPh..27....1S} as the standard through which to calibrate their scheme. We finally conclude with two fluid mixing problems in Sections~\ref{sec:mhd_khi} and~\ref{sec:mhd_cloud_wind_int}.

\subsubsection{Advection of a magnetic monopole}
\label{sec:monopole_advection_test}

Our hyperbolic/parabolic cleaning scheme is designed to propagate divergence waves at the maximal locally resolvable speed, something shown to yield optimal balance between code stability and efficiently controlling $\nabla \cdot \boldsymbol{B}$ errors,  without having to introduce an additional timestepping constraint that would make computations prohibitively expensive~\citep{2016JCoPh.322..326T, 2016MNRAS.455...51H, 2020A&A...638A.140W}. There is therefore only one remaining free parameter that needs to be fixed, namely the parabolic cleaning multiplicative prefactor $\sigma_p$ in equation~(\ref{eq:parabolic_decay_rate}). 

We wish to identify an optimal value for $\sigma_p$ through which we can achieve critical damping for typical divergence waves generated by our cleaning scheme: we seek for them to lie in the transitory regime between being oscillatory and purely decaying, which maximises the gain from spreading divergence errors in space while simultaneously ensuring damping is not too weak. To this end, we reproduce the 3D monopole advection test of~\cite{2012JCoPh.231.7214T}, a generalisation of the 2D problem presented in~\cite{2002JCoPh.175..645D} to three dimensions, variants of which have routinely been used to study divergence cleaning in SPMHD~\citep[e.g.][]{2005MNRAS.364..384P}. It involves a localised, artificially induced magnetic monopole, introduced as an initial condition, that should subsequently be advected by the imposed uniform fluid flow for any~\cite{1999JCoPh.154..284P}-like MHD scheme, in the absence of dissipative and cleaning terms.

We initialise a cubic box of side length $L=2$ with periodic boundary conditions, within which we position $2 \times 32^3$ particles on the vertices of a body-centred cubic (BCC) lattice~\citep[see e.g.][]{1976itss.book.....K}. This ensures maximally symmetric initial conditions; using optimal particle packing is common practice in precision testing SP(M)H(D) codes. We assign constant masses $m$ and thermal energies $u$ to the particles, to obtain a spatially uniform gas density $\rho=1$ and thermal pressure $P = 6$. We impose a uniform fluid flow along the box diagonal with $\boldsymbol{v} = (1, 1, 1)$, and a constant magnetic field along the vertical $z$ direction, $\boldsymbol{B} = B_0 \hat{\boldsymbol{e}}_z$. $B_0$ here is a constant we set to $B_0 = 1/\sqrt{4 \pi}$. We therefore have that $\beta \gg 1$, i.e. we remain in the regime where we are completely unaffected by the tensile instability. We superimpose a monopole perturbation to the magnetic field along the horizontal $x$ direction
\begin{equation}
    \boldsymbol{B}_\text{mon} =
    B_0 \left[ 
    \left( \frac{\tilde{r}}{r_0} \right)^8
    - 2 \left( \frac{\tilde{r}}{r_0} \right)^4 + 1
    \right] \hat{\boldsymbol{e}}_x 
    \quad \text{for}  \quad \tilde{r} < r_0
\end{equation}
where $\tilde{r} = \Vert \boldsymbol{r} - \boldsymbol{r}_\text{mon} \Vert$ is the radial distance from the initial position of the centre of the monopole $\boldsymbol{r}_\text{mon}$, which w.l.o.g. we choose to be the centre of the box. $r_0$ is the monopole's radial size. As typically occurring divergence errors are expected to have a spatial extent comparable to the resolution scale, we set $r_0 = h$.

\begin{figure*}
 \includegraphics[width=\textwidth]{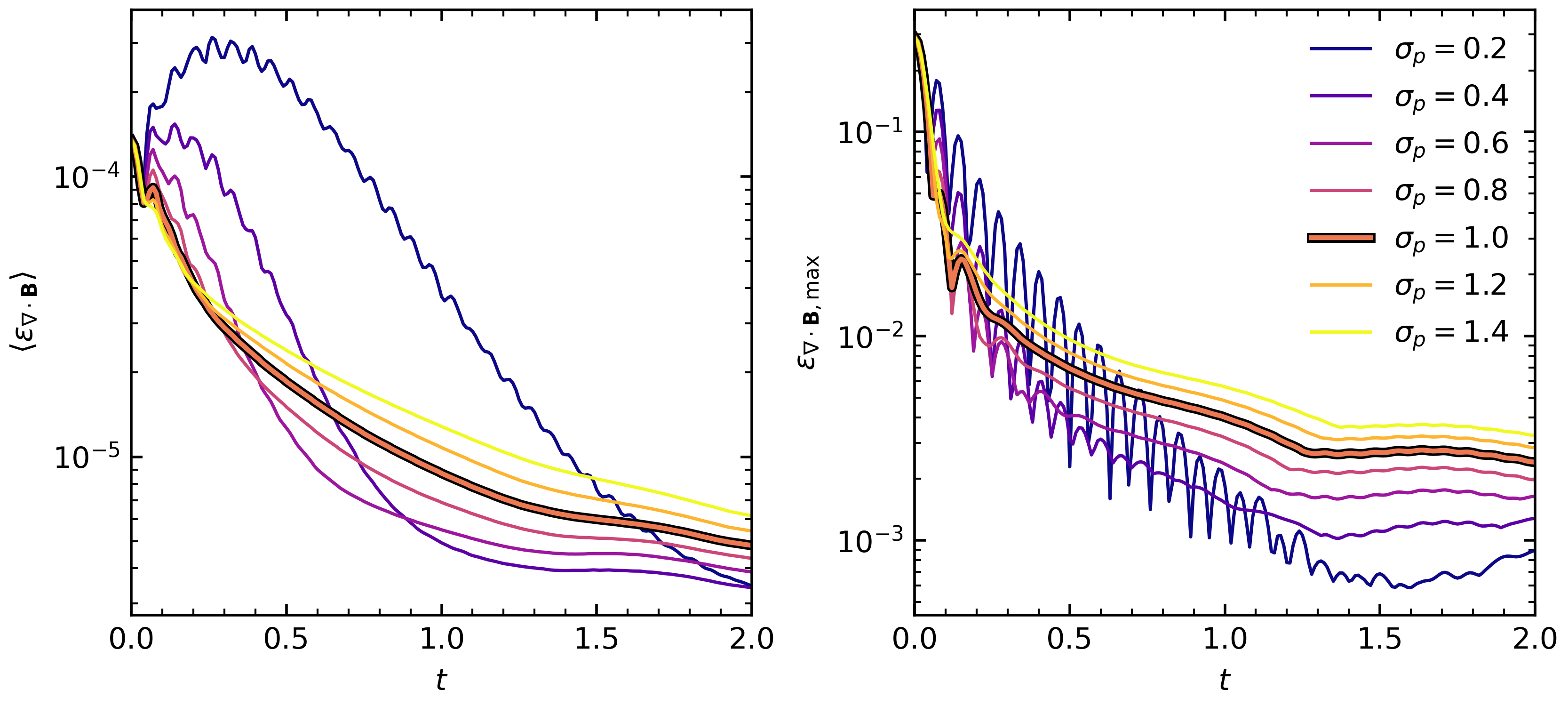} 
 \caption{Results of the three-dimensional magnetic monopole advection tests, for a set of simulations run with a range of parabolic divergence cleaning multiplicative prefactors $\sigma_p$ drawn from the interval $[0.2, 1.4]$. Left: time evolution of the box-averaged dimensionless divergence error. As $\sigma_p$ is increased, we observe a gradual transition from oscillatory to purely decaying behaviour, with the damping eventually becoming slower and slower. Right: time evolution of the maximal dimensionless divergence error. We again transition from wave-like to monotonically decaying solutions as larger and larger $\sigma_p$ are considered. By either of the two metrics shown here, our divergence-cleaning scheme effectively reduces divergence errors. Critical damping is achieved for $\sigma_p=1$; the corresponding lines are highlighted in both panels through use of a distinct line style.}
 \label{fig:monoploe_advection}
\end{figure*}

We evolve this set-up until $t=2$ in our arbitrary system of units (corresponding to one full box crossing of the monopole perturbation), having exceptionally switched off all corrective measures other than divergence cleaning, to independently and reliably assess the effect of varying $\sigma_p$ alone. We track the average and maximal divergence error in the system, for a set of simulations run with a range of $\sigma_p \in [0.2, 1.4]$, and present our results in Fig.~\ref{fig:monoploe_advection}.
For all runs presented here, the box-averaged dimensionless divergence error $\langle \varepsilon_{\nabla \cdot \boldsymbol{B}} \rangle$ decreases by more than 1.5 orders of magnitude over the time interval probed, demonstrating the effectiveness of our cleaning algorithm. In all runs, $\langle \varepsilon_{\nabla \cdot \boldsymbol{B}} \rangle$ converges asymptotically to some small value of $\mathcal{O}(10^{-6})$; at some fixed time, the decay rate drops abruptly, which~\cite{2012JCoPh.231.7214T} attribute to the cleaning scheme having eliminated small-scale modes and only being left to act on large-wavelength, slowly decaying errors. As $\sigma_p$ is incrementally increased, we note a shift from damped oscillatory solutions to monotonically decaying ones. This transition occurs around $\sigma_p = 1$, which we therefore make our default choice for this parameter. For larger $\sigma_p$, parabolic cleaning becomes increasingly less effective as the decay of divergence errors occurs over a longer characteristic time-scale. The observed time evolution of the maximal dimensionless divergence error leads to similar conclusions. Our scheme shows qualitatively similar behaviour to that of~\cite{2012JCoPh.231.7214T} on this test, and coincidentally performs optimally for the same value for $\sigma_p$ as them. We stress that this is a non-trivial result, given our different choice of cleaning speed and definition of $h$, which ultimately sets the damping scale for $\nabla \cdot \boldsymbol{B}$ waves per the parabolic decay rate given in equation~(\ref{eq:parabolic_decay_rate}).

\subsubsection{Travelling circularly polarised Alfvén wave}
\label{sec:circularly_polarised_alfven_wave}

Exact non-linear solutions to the equations of ideal MHD are incredibly valuable for quantitatively assessing the accuracy of a solver; they are nevertheless scarce. Moreover, if said solutions are differentiable and do not involve discontinuities, they can be used to study the convergence properties of a Lagrangian, finite-mass method ~\citep[convergence is a non-trivial problem in SPH, and is expected to be at best second order for orderly spatial distributions of particles away from shocks, see e.g.][]{1992ARA&A..30..543M}. The circularly polarised Alfvén wave test of \cite{2000JCoPh.161..605T} comes with both these attributes: it involves the propagation of an MHD wave in a periodic domain, something that can fully be described by a smooth, closed-form mathematical expression. We here reproduce the 3D travelling wave variant of the test, as first introduced in~\cite{2008JCoPh.227.4123G}.

We initialise a rectangular box of dimensions $L \times L/2 \times L/2$ for $L=3$ with periodic boundary conditions, within which we position particles on the vertices of a BCC lattice to which we assign constant masses and thermal energies to obtain a uniform density and pressure field, with $\rho=1$ and $P=0.1$. We seek to set an Alfvén wave of unit wavelength $\lambda=1$ propagating obliquely w.r.t. the orthogonal coordinate frame $\mathsf{S}$ defined by the sides of the simulation box; we choose the direction of propagation to be $\boldsymbol{r}_\text{prop} = (\text{cos}\alpha \text{cos} \beta, \text{cos}\alpha \text{sin} \beta, \sin \alpha)$, where the angles $\alpha$ and $\beta$ respectively determine the relative orientation of $\boldsymbol{r}_\text{prop}$ w.r.t. the $x$ and  $z$ directions. Following~\cite{2008JCoPh.227.4123G}, we set $\sin \alpha=2/3$ and $\text{sin} \beta=2/\sqrt{5}$, so that an integer number of wave periods fits along each direction in the box. In this context, expressing vector quantities $\boldsymbol{Q}$ can most easily be done by rotating to an orthogonal frame with its major axis aligned with the direction of wave propagation 
\begin{equation}
    \begin{pmatrix}
        Q_1 \\ Q_2 \\ Q_3
    \end{pmatrix}
    =
    \begin{pmatrix}
        \mathrm{cos} \alpha \mathrm{cos} \beta &
        \mathrm{cos} \alpha \mathrm{sin} \beta &
        \mathrm{sin} \alpha \\
        - \mathrm{sin} \beta &
        \mathrm{cos} \beta &
        0 \\
        \mathrm{sin} \alpha \mathrm{cos} \beta &
        - \mathrm{sin} \alpha \mathrm{sin} \beta &
        \mathrm{cos} \alpha
    \end{pmatrix}
    \begin{pmatrix}
        Q_x \\ Q_y \\ Q_z
    \end{pmatrix}
\end{equation}
where the indices $\{1, 2, 3\}$ denote components in the rotated frame $\mathsf{S}_\text{prop}$ and indices $\{x, y, z\}$ components in $\mathsf{S}$. Having defined the wavenumber $k = 2 \pi / \lambda$ we initialise velocities to $(v_1, v_2, v_3) =   (0, v_0 \text{sin} (k x_1), v_0 \text{sin} (k x_1))$ and magnetic fields to $(B_1, B_2, B_3) = (0, B_0 \text{sin} (k x_1), B_0 \text{sin} (k x_1))$, where $v_0=0.1$ and $B_0=0.1$. The solution to the problem (which coincides with the initial configuration at integer multiples of the wave period $T$ which for our choice of parameters is $T=1$) is invariant to the exact choice of these normalisation constants. Nevertheless, our choice translates into a $\beta <1$, putting us in the strong field regime where we are susceptible to the tensile instability: this allows us to moreover asses whether the implemented corrective measures are effective at countering it.

We evolve the system until $t=5$ (which corresponds to five wave periods) with all corrective measures enabled, for particle resolutions $N$ of $2 \times (32 \times 16 \times 16)$, $2 \times (64 \times 32 \times 32)$ and $2 \times (128 \times 64 \times 64)$. We present our results in Fig.~\ref{fig:circularly_polarised_alfven_wave}: 
\begin{figure*}
 \includegraphics[width=\textwidth]{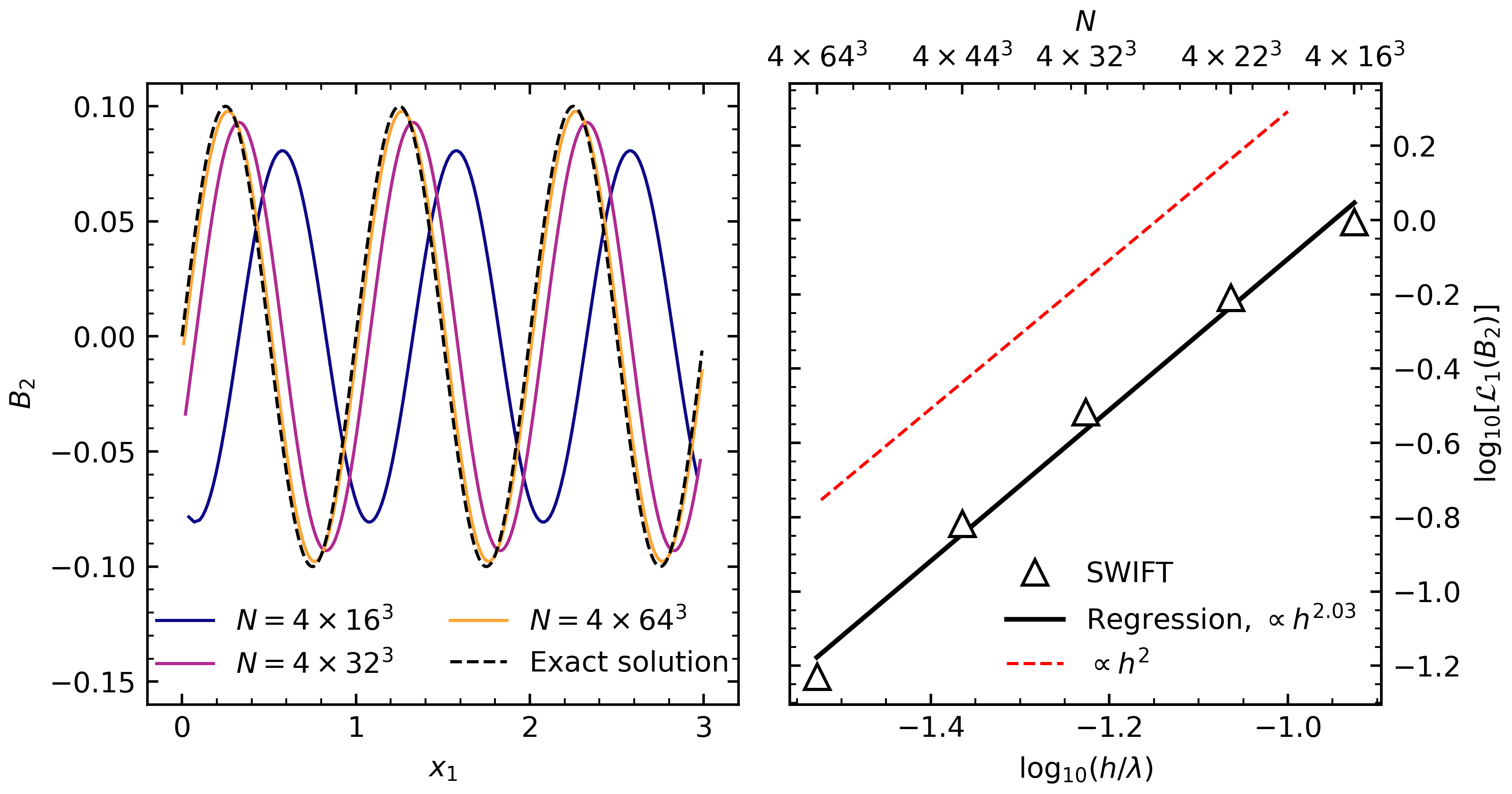} 
 \caption{Results of the Circularly polarised Alfvén wave test at t=5, which corresponds to five wave periods. Left: transverse magnetic field component $B_2$ plotted against the coordinate along the wave propagation direction $x_1$ for all particles in the simulation. We show results at three different resolution levels (solid, coloured lines), comparing them to the problem's known, exact solution (dashed black line). Our scheme's predictions converge to the ground truth, in terms of both phase and amplitude, with increasing resolution. Right: convergence rate of our full scheme as manifested through the relationship between $\mathcal{L}_1 (B_2)$ and resolution expressed in terms of the (single-valued, as gas density is uniform) smoothing length $h$ (we also show the corresponding total number of particles $N$ in the complementary abscissa axis at the top). We fit a power-law to our measurements (solid black line) to find convergence very close to second order (dashed red line).}
 \label{fig:circularly_polarised_alfven_wave}
\end{figure*}
The left panel plots the transverse magnetic field component $B_2$ against the coordinate along the wave propagation direction $x_1$, as a function of the number of resolution elements, making a comparison to the known closed-form solution. Our numerical predictions show an offset from expectation in both phase and amplitude, which is nevertheless reduced the more particles we consider, converging towards our point of comparison and eventually closely matching it for our highest resolution run. 

The right panel of Fig.~\ref{fig:circularly_polarised_alfven_wave} quantifies our scheme's convergence behaviour on this test by plotting the $\mathcal{L}_1$ error norm (computed for $B_2$ at $t=5$) against resolution, expressed through the (single-valued, as gas density is uniform) smoothing length $h$ scaled by $\lambda$. We fit a power law to our measurements by performing linear regression in logarithmic space, to find a dependence very close to $\mathcal{L}_1 \propto h^2$, which translates to second-order convergence. Notwithstanding dissipative and divergence-cleaning terms, our full scheme~\citep[like those of][]{2016MNRAS.455...51H, 2018PASA...35...31P, 2020A&A...638A.140W} shows here optimal convergence properties, comparable to those of state-of-the-art grid solvers~\citep[e.g.][]{2008ApJS..178..137S}. It moreover proves robust to particle clumping in the strong field regime, providing reliable predictions that remain uncorrupted by the tensile instability described by~\cite{1985MNRAS.216..883P}.

\subsubsection{Advection of a magnetic field loop}
\label{sec:loop_advection_tests}

We now proceed with a key tests of numerical MHD, introduced by~\cite{2005JCoPh.205..509G}, which considers the advection of a weak magnetic field loop by a spatially uniform, time-independent velocity field. The problem aims to evaluate the strength of the purely numerical, spurious diffusion - intrinsic to a given solver -  of a magnetic field chosen to be dynamically unimportant so that it would be expected to move with the fluid flow unchanged. This test has proved remarkably challenging for finite-volume schemes~\citep[][]{2008JCoPh.227.4123G}, particularly its 3D variant where the imposed velocity field is not aligned with the employed grid. Lagrangian methods such as SPMHD would be expected to fare better as they can (by construction) compute advection exactly~\citep{2007MNRAS.379..915R, 2012JCoPh.231..759P}; this problem nevertheless still serves as a valuable test of corrective measures that are inevitably dissipative, and more so in our case as contrary to~\cite{2018PASA...35...31P} and~\cite{2020A&A...638A.140W}, the artificial resistivity signal velocity we employ was not specifically chosen to minimise diffusion on this example.

We reproduce the 3D set-up used by~\cite{2018PASA...35...31P} and~\cite{2020A&A...638A.140W}, which is itself based on the version of the test discussed in~\cite{2008JCoPh.227.4123G}~\citep[whereas other notable entries in the MHD method description literature present a 2D variants of the test, e.g.][]{2011MNRAS.418.1392P, 2016MNRAS.455...51H}. We choose our resolution to broadly match that used in these works. We initialise a thin box of dimensions $L \times L \times L/4$ for $L=1$ with periodic boundary conditions, within which we position $2 \times (48 \times 48 \times 12)$ particles on the vertices of a BCC lattice. We assign constant masses and thermal energies to them to obtain a uniform density and pressure field, with $\rho = 1$ and $P = 1$. We initialise velocities to $\boldsymbol{v} = (2, 1, 0.1/\sqrt{5})$, meaning that as long as the flow remains constant, integer time corresponds to an integer number of box crossings in the $x$ and $y$ directions. We set the magnetic field to $\boldsymbol{B} = \boldsymbol{0}$ everywhere, except for a cylindrical region $\Omega$ of radius $R_0 = 0.3$ initially centred at the line $\boldsymbol{l}_\Omega = (l_{\Omega,y}, l_{\Omega,x}, z) = (L/2,L/2,z)$ ; there we initialise a magnetic field loop as
\begin{equation}
    \boldsymbol{B} =
    B_0 (- y / R, x / R, 0 )
\end{equation}
where $R$ is the cylindrical radial distance from $\boldsymbol{l}_\Omega$, defined as 
\begin{equation}
    R = \sqrt{{(x - l_{\Omega,x})}^2 + {(y - l_{\Omega,y})}^2}
\label{eq:cylindrical_radial_distance}
\end{equation}
and $B_0$ is a constant we set to $B_0 = 0.001$. We note that this profile produces a cylindrical sheet of infinite magnetic current $\boldsymbol{J} \equiv \nabla \times \boldsymbol{B}$ at $R=R_c$, and is invariant to translations in the vertical direction $z$.

We also explore a second, more demanding variant of the test where the mass density $\rho_\text{in}$ of material within $\Omega$ is set to double that of the medium that surrounds it, $\rho_\text{out}$. To this end, we again employ the set-up described above, except for the fact that we arrange particles within $\Omega$ on a lattice with inter-particle spacing smaller by a factor of $\sqrt[3]{2}$, to establish the sought-after density contrast of $\Delta \equiv \rho_\text{in}/\rho_\text{out} = 2$. This too represents a configuration that should remain in equilibrium and be passively advected by the bulk flow. We do not attempt to smooth over the perfectly sharp contact discontinuity at the loop boundary, a feature known to be particularly challenging to correctly capture for density-energy SPH solvers, so as to evaluate code performance on a `worst-case' possible configuration.

\begin{figure}
\includegraphics[width=.99\columnwidth]{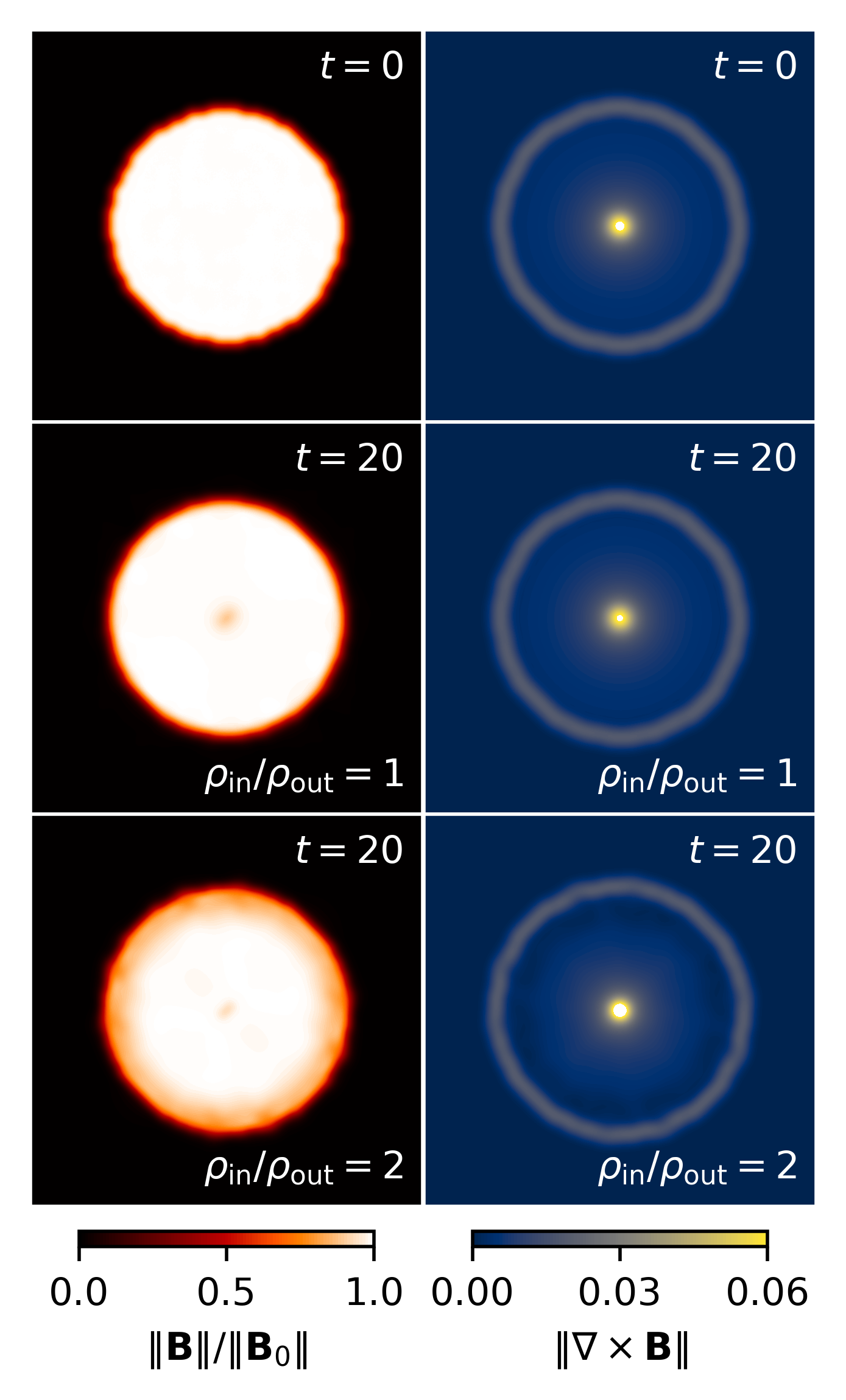} 
\caption{Results for the magnetic field loop advection test. We show projections in the $x-y$ plane of the norm of the magnetic field vector $\boldsymbol{B}$ measured in units of its initial value within the loop (left column), and of the norm of the magnetic current $\boldsymbol{J} = \nabla \times \boldsymbol{B}$ (right column). We compare our initial conditions (top row) to numerical predictions for the state of the system, after it has been evolved for 20 box crossings ($t=20$), for the two loop-to-ambient-medium density contrasts $\Delta \equiv \rho_\text{in} / \rho_\text{out}$ we consider (second and third row, $\Delta = 1$ and $\Delta = 2$ respectively). The profile in $\boldsymbol{B}$ is largely preserved in both cases, other than minor signs of excess diffusion at the centre and edge of the loop, discussed in more detail in the text. The cylindrical current sheet is also well-preserved.}
\label{fig:loop_advection}
\end{figure}

We present results for the two density contrasts we consider, $\Delta=1$ and $\Delta=2$, in Fig.~\ref{fig:loop_advection}.
We show maps of the magnetic field strength (measured in units of its initial value) and of the magnetic current, contrasting our initial conditions with numerical predictions for the state of the system after 20 box crossings (which corresponds to $t=20$ in our system of units). We see that, regardless of all dissipative discontinuity-capturing and divergence-cleaning measures being enabled and particularly of our artificial resistivity implementation not being specifically designed with this test case in mind~\citep[c.f. the AR prescription of][who favour the signal velocity given by equation~(\ref{eq:ar_signal_velocity_others}), which leads to AR terms formally vanishing in the absence of relative motion]{2018PASA...35...31P}, the evolution of the system has been largely marginal with the magnetic loop being well-preserved and minimally diffused, and with the cylindrical current sheet remaining close to intact. We moreover appear to be free of the noise present in the results of~\cite{2016MNRAS.455...51H} and~\cite{2020A&A...638A.140W} \footnote{It is worth mentioning that the emergence of such features can depend on the exact initial conditions that are made use of, and at this level of detail an uncontrolled comparison between different codes and runs is a non-trivial exercise}. We nevertheless wish to draw attention to two features that seem to be present, to a varying extent, in results obtained with most of the commonly employed astrophysical MHD solvers. First, we note the appearance of a `hole' at the centre of the current loop, the presence of which is owed to both the magnetic field and its gradient tensor varying over scales smaller than the spatial resolution limit there. Second, we note the smearing of the edge of the loop, particularly in the case of $\Delta=2$, as the abrupt discontinuity in fluid variables there cannot, by construction, be entirely captured by SPH. The components of the method sourcing magnetic diffusion and divergence cleaning will inevitably be somewhat inaccurate at these points in space; the effect of this is nevertheless thankfully small. 

\begin{figure}
 \includegraphics[]{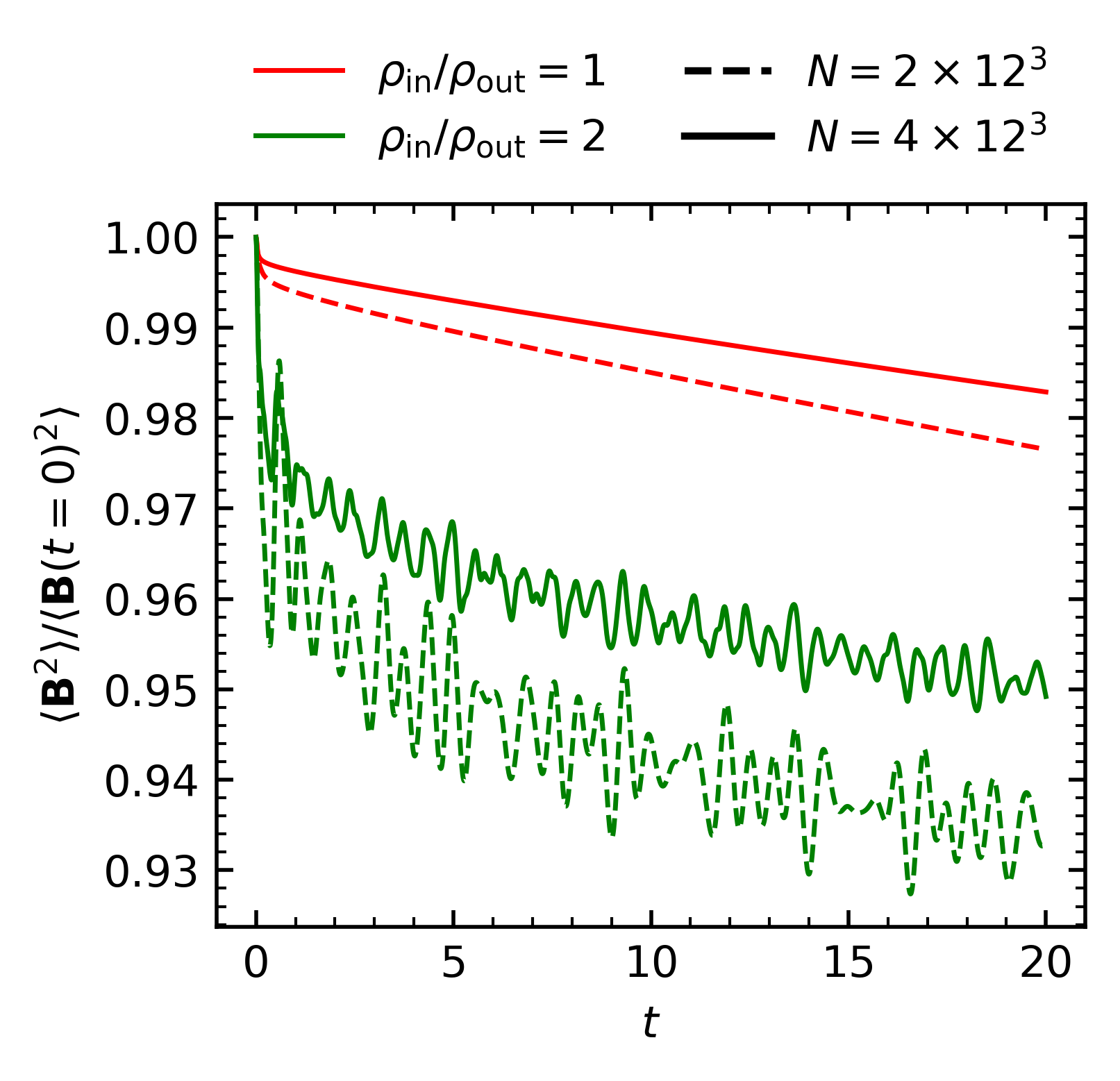} 
 \caption{Time evolution of the box-averaged magnetic energy $\langle \boldsymbol{B}^2 \rangle / 2$ (measured in units of its initial value) for the magnetic field loop advection test, for $\Delta = 1$ (red lines) and  $\Delta = 2$ (green lines), at low (dashed lines) and at our fiducial (solid lines) resolution. The effect of numerical diffusion (after settling of the initial conditions) is small, and the relative decrease in $\langle \boldsymbol{B}^2 \rangle / 2$ is of the order at most of just a few percent, even by $t=20$. Divergence cleaning leads to sub-percent level fluctuations for $\Delta = 2$ which, together with dissipative errors for both $\Delta = 1$ and $\Delta = 2$, are reduced as we increase resolution.}
 \label{fig:loop_advection_diffusion}
\end{figure}

To better quantify the strength of numerical diffusion produced by our code, we track the box-averaged magnetic energy $\langle \boldsymbol{B}^2 \rangle / 2$ in our simulations, measured in units of its initial value, and plot it as a function of time in Fig.~\ref{fig:loop_advection_diffusion}
for the two density contrasts we consider and for resolution levels. For the homogeneous $\Delta = 1$ case we observe a slow, monotonic decline in $\langle \boldsymbol{B}^2 \rangle / 2$ which is nevertheless preserved to better than $2\%$ by $t=20$ in our fiducial run. For $\Delta = 2$ we see that, after an initial, rapid but small change in $\langle \boldsymbol{B}^2 \rangle / 2$ which corresponds to the settling of the initial conditions through a snap rearrangement of the particle distribution, numerical diffusion proceeds at a similar rate, having in total only dissipated $\sim 5 \%$ of the magnetic energy by $t=20$. Much like~\cite{2020A&A...638A.140W} we here observe sub-percent fluctuations in $\langle \boldsymbol{B}^2 \rangle / 2$ owing to spurious $\nabla \cdot \boldsymbol{B}$ being measured by the discrete operator we use to estimate it, especially at the loop boundary, activating the divergence-cleaning scheme which leads to wave-like behaviour. The effect is nevertheless negligible and, much like the global diffusion of the field, reduced as we increase resolution. Numerical dissipation intrinsic to our method is comparable to that of other state-of-the-art MHD solvers~\citep{2011MNRAS.418.1392P, 2016MNRAS.455...51H, 2020A&A...638A.140W}, and we seem free from the unphysical magnetic field amplification~\cite{2016MNRAS.455...51H} observe using their implementation of SPMHD, likely because we employ a conservative divergence control scheme with variable cleaning speeds.
 
\subsubsection{MHD shock tubes}
\label{sec:MHD_shock_tubes}

The shock tube test introduced by~\citet{1988JCoPh..75..400B} is a standard coplanar Riemann problem routinely used to benchmark MHD solvers~\citep{1994JCoPh.115..485D, 1995ApJ...442..228R,1998ApJS..116..133B, 2004MNRAS.348..123P, 2004MNRAS.348..139P, 2007MNRAS.379..915R, 2009MNRAS.398.1678D, 2014CoPhC.185.1053V, 2016MNRAS.455...51H, 2018PASA...35...31P, 2020A&A...638A.140W}. It is an MHD analogue to the pure hydrodynamics Riemann problem of~\cite{1978JCoPh..27....1S} (although, unlike the hydrodynamics case, the MHD version of the example does not admit a closed-form solution). The ~\cite{1978JCoPh..27....1S} shock tube was used by~\cite{2022MNRAS.511.2367B} to adjust the hyperparameters that appear in the artificial viscosity and thermal diffusion prescriptions of SPHENIX; we here adopt a similar strategy to calibrate the single new free constant introduced when extending discontinuity capturing to MHD. Keeping the calibration of SPHENIX's shock-capturing scheme unchanged and employing the defaults suggested by~\citet{2022MNRAS.511.2367B}, we ran variations of the~\citet{1988JCoPh..75..400B} shock tube to settle on an optimal value for $\alpha_\mathrm{max}^\mathrm{AR}$, the global normalisation parameter entering our artificial resistivity implementation.

\begin{figure*}
 \includegraphics[width=\textwidth]{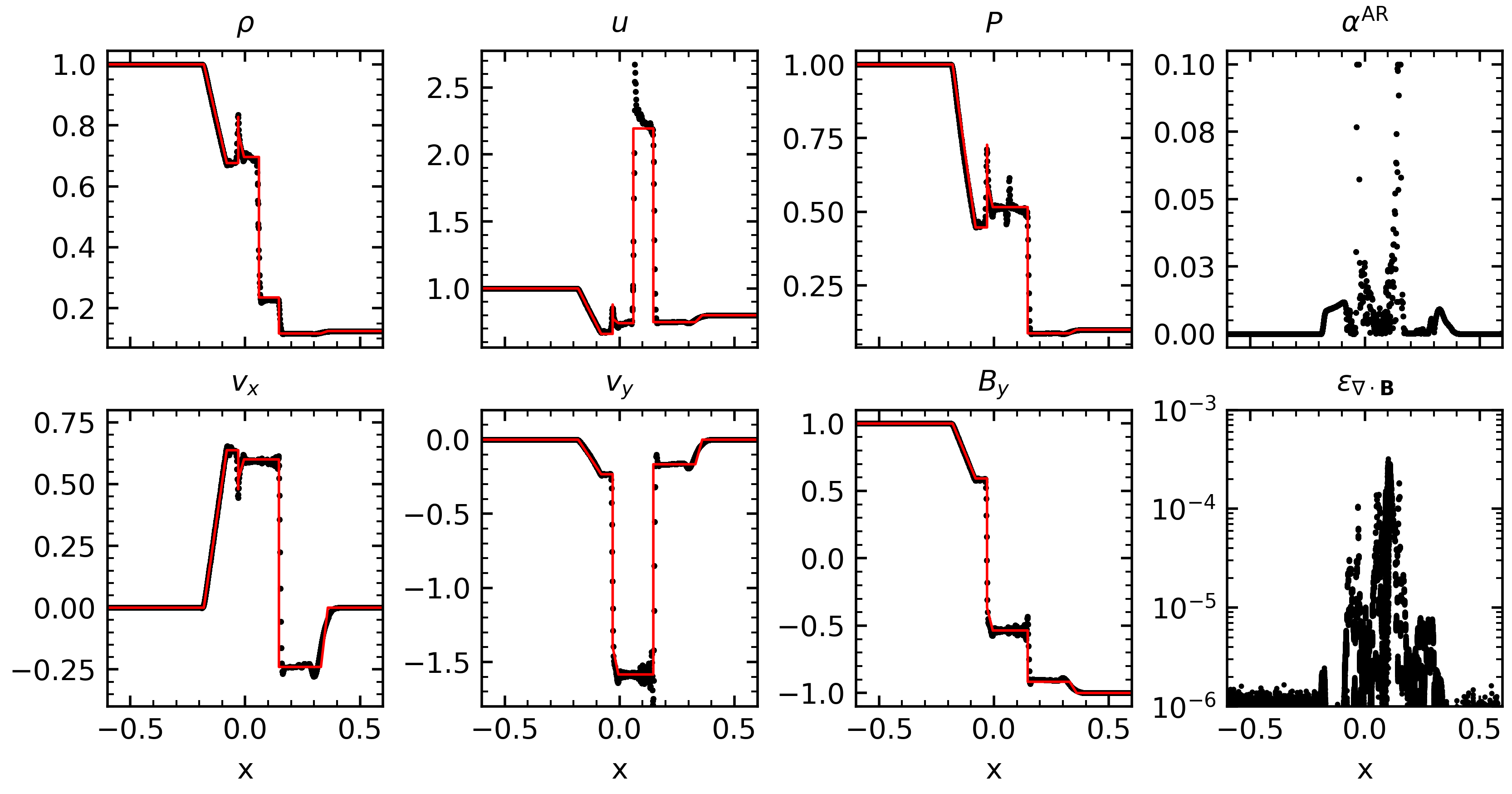} 
 \caption{Results for the Brio \& Wu shock tube test at t=0.1. The initial state left of the discontinuity is $S_L \equiv \{ \rho_L, P_L, \boldsymbol{v}_L, \boldsymbol{B}_L \} = \{ 1, 1, (0, 0, 0), (0, 1, 0) \}$, that to its right is $S_R \equiv \{ \rho_R, P_R, \boldsymbol{v}_R, \boldsymbol{B}_R \} = \{ 0.125, 0.1, (0, 0, 0), (0, -1, 0) \}$; we resolve these with $2 \times (264 \times 24 \times 24)$ and $2 \times (132 \times 12 \times 12)$ particles respectively. We show (in order, from left to right, top to bottom) profiles of the density, thermal energy, thermal pressure, artificial resistivity switch, longitudinal velocity component, transverse velocity component, transverse magnetic field component and dimensionless divergence error as functions of the position along the direction of shock propagation. We encounter, going from the pre- to post- shock region, a fast rarefaction wave and compound structure consisting of a slow rarefaction wave and a slow shock (moving to the left), and a contact discontinuity, slow shock and fast rarefaction wave (all three moving to the right). Our predictions agree well with the reference, with scatter around the ground truth kept at a minimum, and post-shock 'ringing' and overshoots in the velocity profiles being negligible. Artificial resistivity is only applied at instantaneous jumps in the magnetic field, and the dimensionless divergence error remains below the $0.1 \%$ level, even at shocks. There is non-negligible wall heating at the contact discontinuity, leading to a moderate `pressure blip' there (see text for details).}
 \label{fig:brio_wu_shock_tube}
\end{figure*}

\cite{1988JCoPh..75..400B}'s test is a one-dimensional initial value problem where the initial data takes the form of two distinct, uniform `left' and `right' states $S_L$ and $S_R$ that meet at a given point in space. This results in sharp discontinuities in physical variables being introduced there (in density, thermal pressure and thus thermal energy, and magnetic field); w.l.o.g. this point can be taken to be the origin of the chosen coordinate system. As the system is subsequently allowed to evolve, a rich set of diverse shocks develops due the fact that MHD admits three families of waves~\citep[Alfvén, and slow and fast magnetosonic waves, see e.g.][for a complete and concise derivation and discussion]{2016JPlPh..82c2001O} as opposed to pure hydrodynamics only allowing for sound waves; the problem is therefore more complex and demanding of the method, and serves as an excellent guide to gauge whether we can uniquely represent discontinuities and to evaluate the effectiveness of the implemented discontinuity-capturing terms. In order of appearance, going from the pre- to post- shock region, one is expected to encounter a fast rarefaction wave and compound shock structure, both moving to the left, and a contact discontinuity, slow shock and fast rarefaction wave, all three moving to the right. The compound structure comprises a slow rarefaction wave and slow shock, and it is debated whether this is physical in origin or a resultant of the reduced dimensionality of the problem~\citep{1996JCoPh.126...77B}. The left and right states are
\begin{equation}
    S_L
    \equiv 
    \{
    \rho_L, P_L, 
    \boldsymbol{v}_L, \boldsymbol{B}_L
    \}
    =
    \{
    1, 1, (0, 0, 0), (0, 1, 0),
    \}
\label{eq:brio_wu_left_state}
\end{equation}
and
\begin{equation}
    S_R
    \equiv 
    \{ 
    \rho_R, P_R, 
    \boldsymbol{v}_R, \boldsymbol{B}_R
    \}
    =
    \{ 
    0.125, 0.1, (0, 0, 0), (0, -1, 0),
    \}
\label{eq:brio_wu_right_state}
\end{equation}
where $\rho_i, P_i, \boldsymbol{v}_i$ and $\boldsymbol{B}_i$ correspond to state $S_i$'s mass density, thermal pressure, velocity and magnetic field respectively. The adiabatic index is exceptionally taken to be $\gamma=2$. Much like~\cite{2009MNRAS.398.1678D, 2018PASA...35...31P, 2020A&A...638A.140W} or~\cite{2022MNRAS.511.2367B}, we choose to simulate the Riemann problem in 3D, irrespective of it fundamentally being of lower dimensionality as per~(\ref{eq:brio_wu_left_state}) and~(\ref{eq:brio_wu_right_state}), to test our method in a configuration closer to the one meant for production runs. To this end we initialise an elongated rectangular box of dimensions $L \times L/11 \times L/11$ for $L=1$, taking the $x$-coordinate to be the direction of propagation of the shock features. Within it we position particles on the vertices of a BCC lattice (which ensures maximally symmetric lateral forces in the initial state), $2 \times (264 \times 24 \times 24)$ for $x < L/2$ and $2 \times (132 \times 12 \times 12)$ for $x > L/2$, corresponding to $S_L$ and $S_R$ respectively; this resolution is chosen to broadly match that of~\cite{2018PASA...35...31P} and~\cite{2020A&A...638A.140W}. We set the inter-particle separation in $S_L$ to double that in $S_R$ to achieve the sought-after density contrast of $8:1$ with equal-mass particles. Finally, we choose to employ periodic boundary conditions; this inevitably creates an additional Riemann problem across the periodic boundary in $x$. To ensure propagating features sourced there do not pollute the active region of interest, we extend the simulation box by $L/2$ along $x$ at both ends, making each of $S_L$ and $S_R$ twice as long. We then let the system evolve until $t=0.1$ and illustrate its state at that point in time in Fig.~\ref{fig:brio_wu_shock_tube}, showing the six primitive variables of interest (density, thermal energy, thermal pressure, the longitudinal and transverse components of the velocity vector, and the transverse component of the magnetic field) and two method diagnostics (artificial resistivity switch and dimensionless divergence error), plotting their values for all particles in black and contrasting them with reference solutions reproduced from~\cite{1998ApJS..116..133B}, shown in red. The origin of the primary axis is matched with the position of the discontinuity at $t=0$.

Numerical predictions obtained with SWIFT for the profiles of all six physical variables shown agree well with the reference, with all five expected features being reproduced by our method. The scale of the scatter or particle noise around the assumed ground truth (e.g. behind the slow shock at $x \sim 0.15$) is of the same order as what is seen with other state-of-the-art SPMHD codes~\citep[note also that post-shock oscillations are not specific to SPH but are observed with other methods as well, see e.g.~\cite{1995LNP...453..379A}][]{}. We appear to be free of severe post-shock `ringing' and significant overshoots in the velocity profiles. The profile of the artificial resistivity switch $\alpha^\text{AR}$ shows it to be non-zero only at the location of the contact discontinuity and right-moving slow shock, where instantaneous jumps in the transverse magnetic field component $B_y$ occur; $\alpha^\text{AR}$ is only minimally activated at linear gradients in $B_y$ and zero everywhere else, ensuring localised action of AR only where it is needed and avoiding unnecessary dissipation of the magnetic field. We also show a profile of the relative divergence error which peaks at shocks, but nevertheless remains below the $0.1 \%$ level across the whole simulation box, indicating excellent preservation of the solenoidal constraint. The one moderate but notable departure from~\cite{1998ApJS..116..133B}'s results is the wall heating occurring at $x \sim 0.08$ (notice the spike in the thermal energy profile at that point) which also translates to a non-negligible `pressure blip' there, comparable in scale to that reported by~\cite{2016MNRAS.455...51H} and notably less pronounced than the one observed by~\cite{2020A&A...638A.140W}. We have verified that this feature can be largely eliminated by employing an artificial thermal diffusion prescription better tailored to this problem, with a pair-wise signal velocity $v_{ij}^\text{AD}$ that depends only on pressure differences and not on relative approach velocities of particles (c.f. equation~(\ref{eq:sph_artificial_diffusion_signal_speed})), pre-multiplied by a constant rather than our own variable, particle-carried switch $\alpha^\text{AD}$; this is the approach favoured by~\cite{2018PASA...35...31P}. It has nevertheless been suggested that such a scheme would not perform optimally on simulations that include self-gravity~\citep{2008MNRAS.387..427W} and/or sub-grid prescriptions for energetic feedback~\citep{2022MNRAS.511.2367B}, both of which constitute core ingredients of all simulations targeting the primary scientific use cases of SWIFT. Marginal improvements on most profiles can also be obtained by employing constant viscosity switches $\alpha^\text{AV}$; this is, however, once again, not recommended for practical applications of the method as it would lead to over-dissipation of the velocity field in more realistic set-ups~\citep[see e.g.][]{2010MNRAS.408..669C}.

We now proceed with the shock tube test `1A' from~\cite{1995ApJ...442..228R}, also discussed in~\cite{1994JCoPh.111..354D}. This problem was used by~\cite{2000JCoPh.161..605T, 2002JCoPh.175..645D, 2010JCoPh.229.2117M} and~\cite{2016MNRAS.455...51H} to show that divergence control beyond that offered by the eight-wave cleaning approach of~\cite{1999JCoPh.154..284P}, through e.g. a divergence-cleaning scheme like the one we make use of here, is a necessity if one is to correctly capture jump conditions in MHD Riemann problems and avoid systematic offsets that persist even in converged numerical solutions. The test comprises two strong supersonic shocks, between which are enclosed a left-moving slow rarefaction wave, a contact discontinuity, and a right-moving slow magnetosonic shock. This has historically proved a particularly challenging example for SPMHD, with unphysical velocity oscillations and jumps in the magnetic field being especially non-trivial to eliminate~\citep{2005MNRAS.364..384P}, and intermediate states being almost entirely suppressed by the employed regularisation schemes~\citep{2009MNRAS.398.1678D}.

\begin{figure*}
 \includegraphics[width=\textwidth]{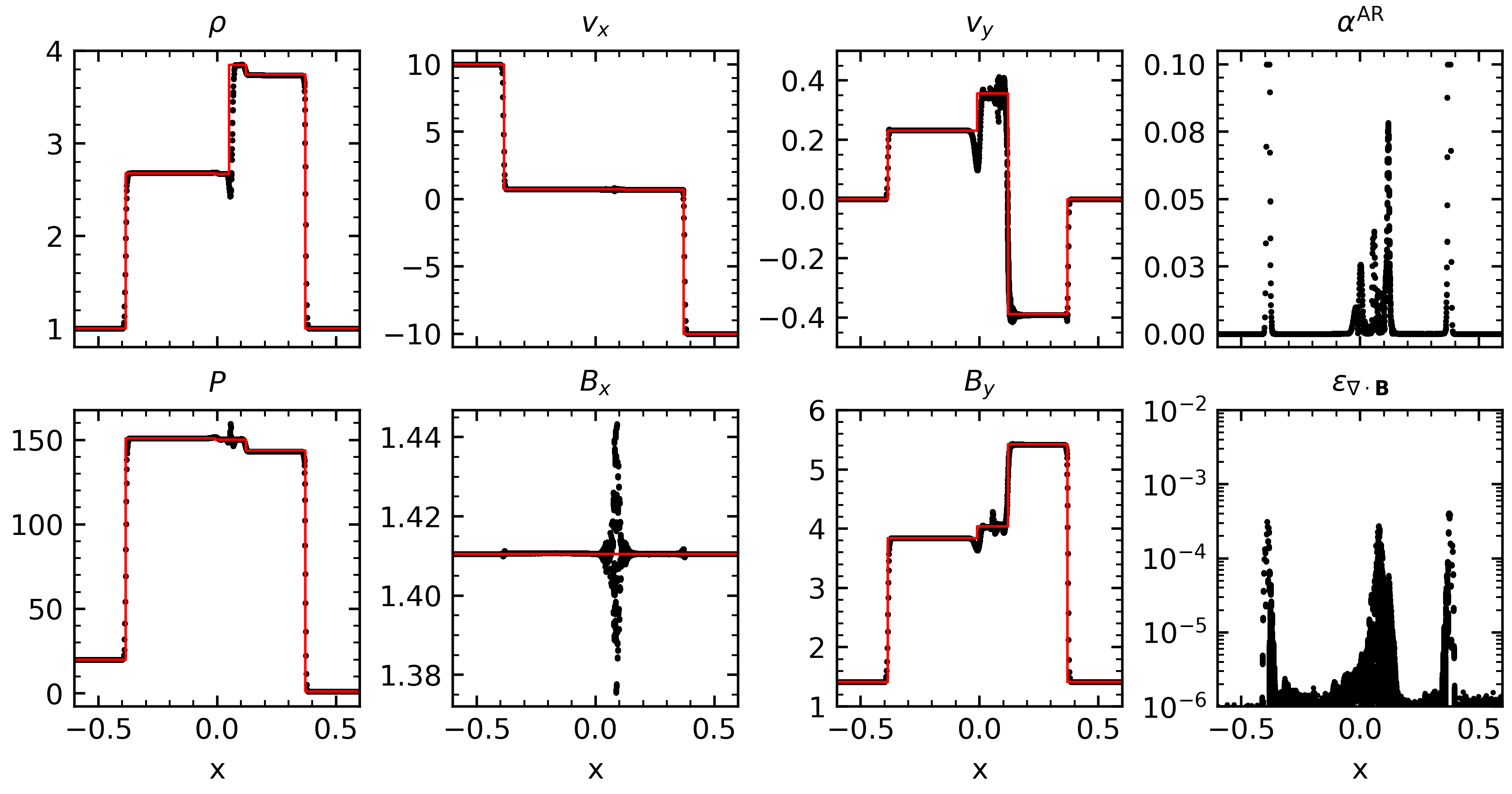} 
 \caption{Results for the Ryu \& Jones 1A shock tube test at t=0.08. The initial state left of the discontinuity is $S_L \equiv \{ \rho_L, P_L, \boldsymbol{v}_L, \boldsymbol{B}_L \} = \{ 1, 20, (10, 0, 0), (5/\sqrt{4 \pi}, 5/\sqrt{4 \pi}, 0) \}$, that to its right is $S_R \equiv \{ \rho_R, P_R, \boldsymbol{v}_R, \boldsymbol{B}_R \} = \{ 1, 1, (-10, 0, 0), (5/\sqrt{4 \pi}, 5/\sqrt{4 \pi}, 0) \}$. The region of interest is resolved by $2 \times (256 \times 24 \times 24)$ particles.  We show (in order, from left to right, top to bottom) profiles of the density, longitudinal and transverse velocity components, artificial resistivity switch, thermal pressure, longitudinal and transverse magnetic field components, and dimensionless divergence error as functions of the position along the direction of shock propagation. We encounter here two strong supersonic shocks which enclose a left-moving slow rarefaction wave, a contact discontinuity, and a right-moving slow magnetosonic shock. Our predictions agree well with the reference, with systematic offsets from and scatter around the ground truth kept at a minimum. Jump conditions for the intermediate states are correctly captured. The action of artificial resistivity is exclusively limited to where necessary, and the dimensionless divergence error only peaks at shocks, remaining nevertheless below the $0.1\%$ level everywhere. There are moderate over/under-shoots in the density, thermal pressure, velocity and magnetic field profiles around the contact discontinuity, due to our use of a conservative artificial thermal diffusion prescription; more on this in the text.}
 \label{fig:ryu_jones_1A_shock_tube}
\end{figure*}

We reproduce here the 3D version of the test presented in~\cite{2018PASA...35...31P}, with `left' and `right' states 
\begin{equation}
    S_L
    \equiv 
    \{
    \rho_L, P_L, 
    \boldsymbol{v}_L, \boldsymbol{B}_L
    \}
    =
    \{
    1, 20, (10, 0, 0), \frac{5}{\sqrt{4 \pi}}(1, 1, 0)
    \}
\end{equation}
and
\begin{equation}
    S_R
    \equiv 
    \{ 
    \rho_R, P_R, 
    \boldsymbol{v}_R, \boldsymbol{B}_R
    \}
    =
    \{ 
    1, 1, (-10, 0, 0),  \frac{5}{\sqrt{4 \pi}}(1, 1, 0)
    \}
\end{equation}
respectively. We set up a simulation box following a similar approach to that adopted for initialising the~\cite{1988JCoPh..75..400B} shock tube example discussed above: using once again $x$ to denote the position along the direction of shock propagation, we take $x=0$ to coincide with the point where $S_L$ and $S_R$ are taken to meet in the initial conditions. We populate the active region of interest (of dimensions $L \times 3L/32 \times 3L/32$ for $L=1$, taken to range from $x=-0.5$ to $x=0.5$) with $2 \times (256 \times 24 \times 24)$ particles positioned on the vertices of a BCC lattice, a number of resolution elements chosen to broadly match that considered by~\cite{2018PASA...35...31P}. We here too employ periodic boundary conditions, noting nevertheless that to make this work, simply extending the simulation box as before would be insufficient: the imposed velocity profile would quickly de-populate the region around the periodic boundary in $x$, which would be incompatible with SWIFT's neighbour-finding algorithm~\citep{1967PhRv..159...98V}. We thus extend $S_L$ and $S_R$ to $x=-3$ and $x=3$ respectively, \textit{and} interpolate the longitudinal velocity component $v_x$ linearly from $v_{R,x}$ to $v_{L,x}$ between $x=2.5$ and $x=-2.5$. This both allows the example to run and prevents the boundary Riemann problem from affecting the region of interest. We evolve the system until $t=0.08$ and present the results obtained in Fig.~\ref{fig:ryu_jones_1A_shock_tube}.

The predictions made with SWIFT (shown for all particles in the simulation, in black) agree well with the reference solution taken from~\cite{1998ApJS..116..133B} (shown in red), with all core features being closely reproduced. Systematic offsets that have historically proved hard to eliminate (e.g. in $v_y$ between the slow and fast shock fronts at $x \sim 0.1-0.4$, or in $B_x$ between the two strong shocks at $x \sim \pm 0.4$) are almost non-existent. What is also particularly noteworthy is our code's ability to correctly capture jump conditions for the intermediate states around $x \sim 0.0-0.1$. The particle noise in all quantities is minimal, except for that in $v_y$ behind the slow shock; it nevertheless remains at the sub-percent level when compared to the scale of the dominant component of the velocity field, $v_x$. We observe again a `pressure blip', which also brings about a moderate but noticeable undershoot in $\rho$, $v_y$ and $B_y$ behind the contact discontinuity at $x \sim 0$; this can again be cured by employing a different artificial thermal diffusion prescription, which we nevertheless do not opt for given the aforementioned reasons. The action of artificial resistivity is once more shown to be exclusively limited to where discontinuities in the magnetic field occur, and divergence errors peak at shocks but remain - a challenge in the present example - below the $0.1 \%$ level.

We finally turn to the `seven discontinuity' or `2A' shock tube from~\cite{1995ApJ...442..228R}. In contrast with the two Riemann problems presented above, this test does not consider a velocity and magnetic field that are coplanar; it rather examines fully three-dimensional field structures, notably a magnetic field the plane of which rotates across the initial discontinuity. It further comprises a contact discontinuity, either side of which propagate, in order, a slow shock, a rotational discontinuity, and a fast shock. This example has been discussed in~\cite{1994JCoPh.115..485D, 1995ApJ...442..228R, 1995ApJ...452..785R, 1998ApJS..116..133B, 2000JCoPh.161..605T, 2000ApJ...530..508L, 2008JCoPh.227.4123G} or~\cite{2009MNRAS.398.1678D}, but particularly relevant to our investigation is the study presented in~\cite{2005MNRAS.364..384P}; there, through the `seven discontinuity' test, the specifics of shock-capturing terms and correlation between divergence errors and erroneous jump conditions are investigated, and a case is made for wider kernels being used in SPMHD compared to SPH.

\begin{figure*}
 \includegraphics[width=\textwidth]{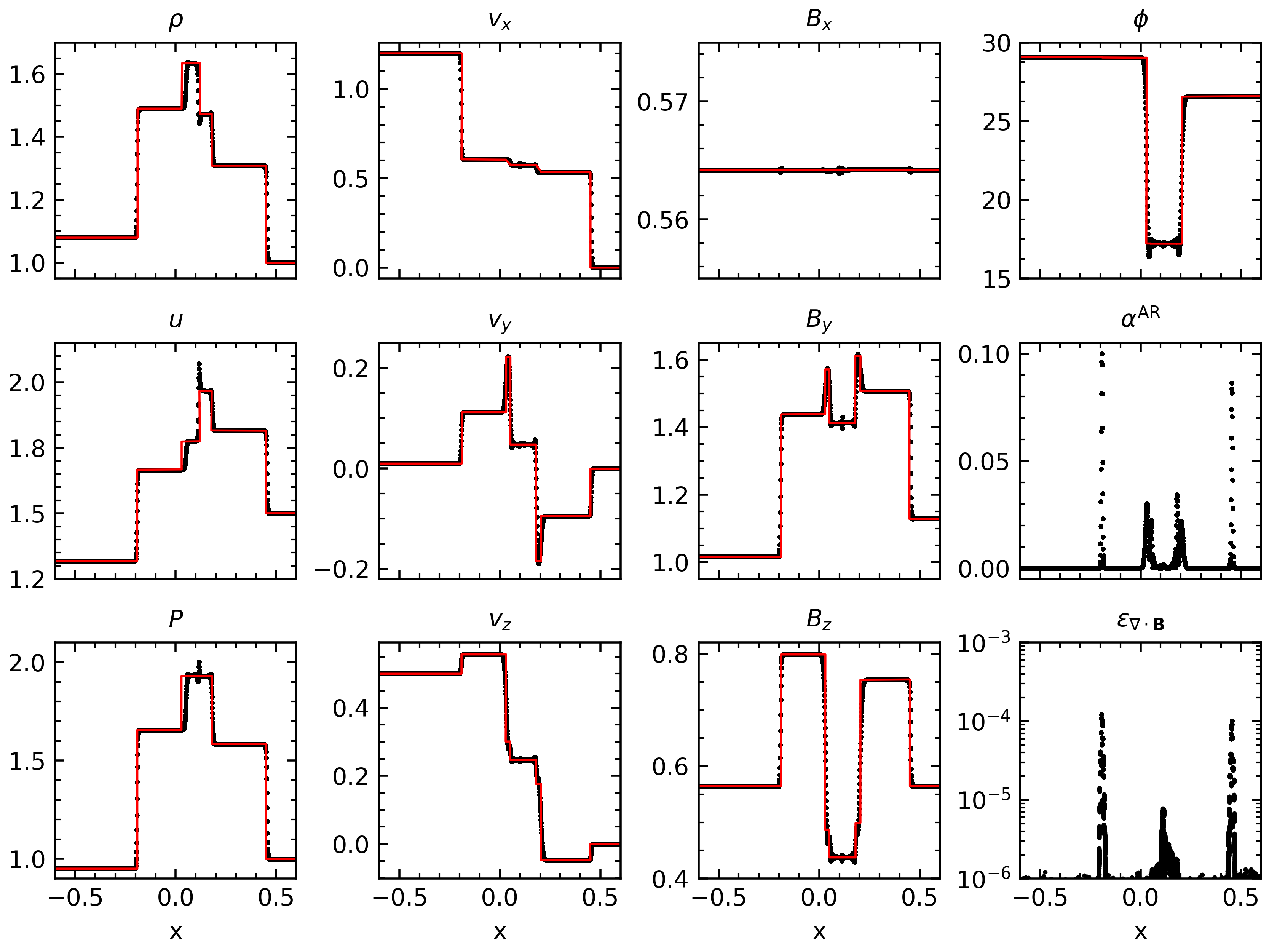} 
 \caption{Results for the Ryu \& Jones 2A shock tube test at t=0.08. The initial state left of the discontinuity is $S_L \equiv \{ \rho_L, P_L, \boldsymbol{v}_L, \boldsymbol{B}_L \} = \{ 1.08, 0.95, (1.2, 0.01, 0.5), (2/\sqrt{4 \pi}, 3.6/\sqrt{4 \pi}, 2/\sqrt{4 \pi}) \}$, that to its right is $S_R \equiv \{ \rho_R, P_R, \boldsymbol{v}_R, \boldsymbol{B}_R \} = \{ 1, 1, (0, 0, 0), (2/\sqrt{4 \pi}, 4/\sqrt{4 \pi}, 2/\sqrt{4 \pi}) \}$. We show (in order, from top to bottom, left to right) profiles of the density, thermal energy, thermal pressure, velocity field components, magnetic field components, rotation angle of the magnetic field $\phi \equiv \mathrm{tan}^{-1} (B_z / B_y )$, artificial resistivity switch, and dimensionless divergence error as functions of the position along the direction of shock propagation. We encounter here a contact discontinuity at $x \sim 0.1$. Propagating away from it we see, on either side of it, discontinuities from each MHD wave family: a slow shock, a rotational discontinuity and a fast shock. Our predictions agree well with the reference, with all jump conditions reproduced correctly and noise and post-shock ringing kept vanishingly small. We do not appear to be subject to over-dissipation, as even the sharpest changes in the transverse velocity and magnetic field components are captured. The action of artificial resistivity is, as intended, highly localised in space and the dimensionless divergence error is below the $0.01\%$ level everywhere. We observe moderate wall heating at the contact discontinuity, which nevertheless does not corrupt the solution.}
 \label{fig:ryu_jones_2A_shock_tube}
\end{figure*}

The `left' and `right' states of the Riemann problem are respectively
\begin{equation}
    \begin{split}
        S_L
        & \equiv 
        \{
        \rho_L, P_L, 
        \boldsymbol{v}_L, \boldsymbol{B}_L
        \} \\
        & =
        \{
        1.08, 0.95, (1.2, 0.01, 0.5), \frac{1}{\sqrt{4 \pi}}(2, 3.6, 2)
        \}
    \end{split}
\end{equation}
and
\begin{equation}
    S_R
    \equiv 
    \{
    \rho_R, P_R, 
    \boldsymbol{v}_R, \boldsymbol{B}_R
    \}
    =
    \{
    1, 1, (0, 0, 0), \frac{1}{\sqrt{4 \pi}}(2, 4, 2)
    \}
    \text{.}
\end{equation}
We set up the discontinuity in a three-dimensional box of dimensions $L \times L/40 \times L/40$ for $L=1$, at the centre of which $S_L$ and $S_R$ are taken to meet. We populate the box uniformly with $2 \times (480 \times 12 \times 12)$ particles positioned on the vertices of a BCC lattice, roughly matching the resolution~\cite{2018PASA...35...31P} used to run the same test. We extend both $S_L$ and $S_R$ by $3L/2$ to be able to run with periodic boundary conditions in all directions, without fear of the region of interest being corrupted by boundary effects. Finally, to establish the desired mild density contrast across the initial discontinuity, we appropriately rescale the $x$ coordinate of all particles in $S_L$. We let the system evolve until $t=0.2$ and show its state at that point in time in Fig.~\ref{fig:ryu_jones_2A_shock_tube}.

Numerical results obtained with SWIFT, plotted for all particles in the simulation box in black, agree well with the reference solution taken from~\cite{1995ApJ...442..228R}, which we overplot in red. All seven discontinuities are accurately reproduced by our method, and jump conditions are captured correctly, with divergence errors that remain below the $0.01\%$ level everywhere. Most notably, we retrieve the expected sharp jumps in the transverse velocity and magnetic field components at $x \sim 0.05$ and $x \sim 0.2$, while preserving the constant longitudinal magnetic field component near perfectly~\citep[c.f. the results of e.g.][]{2000JCoPh.161..605T, 2008JCoPh.227.4123G}. We also find spurious noise and post-shock oscillations to be vanishingly small, indicating that artificial viscosity and resistivity terms act, as intended, in a localised fashion \textit{while} not leading to over-dissipation~\citep[c.f. the results of e.g.][]{2005MNRAS.364..384P, 2009MNRAS.398.1678D}. We again observe limited wall heating and a small glitch in the pressure profile (and to a lesser extent in profiles of other variables) at the location of the contact discontinuity, which nevertheless does not appear to corrupt the solution. In light of this test's sensitivity to the number of neighbours used to compute discrete SPH interpolants~\citep[see][]{2005MNRAS.364..384P}, the results we obtain serve as validation of our choice of kernel and associated $\eta_\text{res}$ parameter.

\subsubsection{Orszag-Tang vortex}
\label{sec:oreszag_tang_vortex}

The planar vortex system first presented by~\cite{1979JFM....90..129O} is one of the most frequently reproduced examples used to validate MHD codes~\citep[see e.g.][]{1994SJSC...15..263Z, 1995ApJ...452..785R, 1998JCoPh.142..331D, 1998ApJ...509..244R, 1999JCoPh.150..561J, 2000JCoPh.161..605T, 2000ApJ...530..508L, 2005MNRAS.364..384P, 2006ApJ...652.1306B, 2006A&A...457..371F, 2007MNRAS.379..915R, 2008ApJS..178..137S, 2011MNRAS.418.1392P, 2016MNRAS.455...51H, 2020A&A...638A.140W}. It was originally introduced in the context of incompressible MHD turbulence studies, and subsequently extended to the case of compressible flows by~\cite{1989PhFlB...1.2153D} and~\cite{1991PhFlB...3...29P}, respectively. The problem starts with a fluid of uniform density within which two planar vortices are superimposed, one in the velocity and one in the magnetic field, which share a common singular point at the centre of the computational domain but have distinct modal structures. Although the initial set-up is smooth, the system gradually evolves into a complex configuration of propagating and interacting MHD modes and shocks, each moving at a different speed. Of particular interest is the emergence of a current sheet at the location of the aforementioned singular point: this is eventually destabilised by resistive tearing which leads to the formation of small-scale `magnetic islands'~\citep{1989PhFlB...1.2330P}. The system eventually transitions to a regime of decaying supersonic MHD turbulence, something which is particularly challenging to capture correctly; the quality of late-time solutions is non-trivial to assess as different solvers do not necessarily converge to the same solution~\citep[see e.g.][]{2011MNRAS.418.1392P, 2013MNRAS.428...13S}.

\begin{figure*}
 \includegraphics[width=\textwidth]{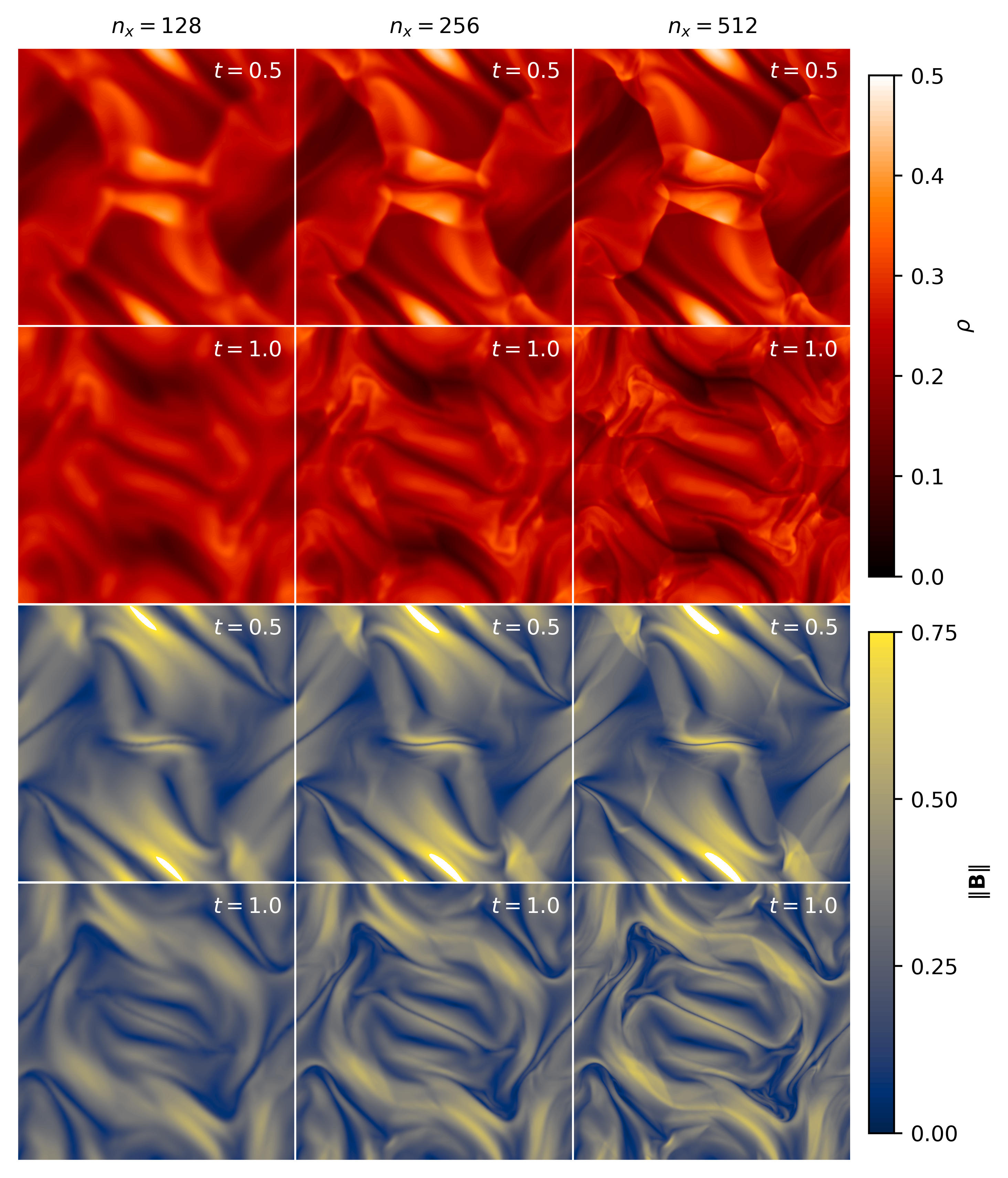} 
 \caption{Results for the Orszag-Tang vortex test at $t=0.5$ ($1^\textbf{st}$ and $3^\text{rd}$ rows) and $t=1$ ($2^\textbf{nd}$ and $4^\text{th}$ rows), run in 3D at a resolution of $n_x \sim 128$ ($1^\textbf{st}$ column), $n_x \sim 256$ ($2^\textbf{nd}$ column) and $n_x \sim 512$ ($3^\textbf{rd}$ column) particles along either of the two major sides of the computational domain. We show projections in the $x-y$ plane of the mass density ($1^\textbf{st}$ and $2^\textbf{nd}$ rows) and of the norm of the magnetic field vector $\Vert \boldsymbol{B} \Vert$ ($3^\textbf{rd}$ and $4^\textbf{th}$ row) measured in code units. The initially smooth superimposed velocity and magnetic field vortices evolve first into an intricate set of interacting discontinuities before transitioning to decaying supersonic turbulence. All major features are reproduced (e.g. system of four major interacting shocks, thin layer of trapped gas and current sheet in the middle of the simulation box at $t=0.5$) with sharp discontinuities being increasingly better resolved and small-scale features more truthfully reproduced as resolution is increased.}
 \label{fig:orszag_tang_vortex}
\end{figure*}

To set up the (originally 2D) problem in 3D, we reproduce the `thin box' configuration suggested by~\cite{2009MNRAS.398.1678D}, and use the test parameters proposed by~\cite{1995ApJ...452..785R} and~\cite{2000ApJ...530..508L}; these are the choices made by~\cite{2018PASA...35...31P} and~\cite{2020A&A...638A.140W} as well. Defining the integer parameter $s_{xy}$, to be used to specify the resolution along the $x$ and $y$ directions, we initialise a box of dimensions $L \times L \times 16L/s_{xy}$ for $L=1$ within which we position $2 \times (s_{xy} \times s_{xy} \times 16)$ particles on the vertices of a BCC lattice. We run the test at three different resolution levels, corresponding to $s_{xy} = 96, 192, 384$, which respectively translate to $n_x \sim 128, 256,512$ resolution elements along $x$ (and $y$). We employ periodic boundary conditions, and assign constant masses and thermal energies to particles to obtain a uniform density and pressure field with
\begin{equation}
    \rho = \frac{\gamma^2 B_0^2}{\mu_0 v_0^2}
    \quad
    \text{and}
    \quad
    P = \frac{\gamma B_0^2}{\mu_0}
    \text{.}
\end{equation}
Here, $v_0$ and $B_0$ are constant parameters that determine the characteristic velocity and magnetic field scales of the problem, which we set to $v_0=1$ and $B_0 = 1/ \sqrt{4 \pi}$. We initialise the velocity and magnetic vortices as
\begin{equation}
    \boldsymbol{v} = 
    v_0 ( - \mathrm{sin} (2 \pi y), \mathrm{sin} (2 \pi x), 0)
\end{equation}
and
\begin{equation}
    \boldsymbol{B} = 
    B_0 ( - \mathrm{sin} (2 \pi y), \mathrm{sin} (4 \pi x), 0)
\end{equation}
respectively, which gives an initial average Mach number of $\mathcal{M}(t=0) \equiv v_0/c_s(t=0) = 1$ and an initial plasma beta of $\beta(t=0)=10/3$. We evolve the system until $t=1$ and present results in Fig.~\ref{fig:orszag_tang_vortex}, showing projections in the $x-y$ plane of the density and magnetic field strength at $t=0.5$ and $t=1$ for the three resolution levels considered. 

All projections shown are highly symmetric, irrespective of the fact no particular care has been taken to impose exact point symmetry in the initial conditions~\citep[as opposed to e.g.][]{2013MNRAS.436.2810T, 2016JCoPh.322..326T}. By $t=0.5$, several bi-dimensional shocks and discontinuities have developed and started interacting. Most notably, all of the current sheet (visible, in the middle of the $\Vert \boldsymbol{B} \Vert$ projections, as a line of low magnetic field either side of which the field strength is at a maximum), the co-spatial thin layer of gas trapped within it (the filamentary feature in the middle of the $\rho$ projections), as well as the magnetic field reversal found in the middle of the $y$ boundaries of the computational domain are reproduced at all resolutions, with ever increasing sharpness and level of detail as $n_x$ in increased; the aforementioned attributes have historically been non-trivial to capture with SPMHD~\citep{2005MNRAS.364..384P, 2007MNRAS.379..915R}. We also draw attention to our scheme's ability to reproduce sharp features in the magnetic field which vary rapidly in space, even in regions of low mass density (i.e. that are sampled by only a handful of particles, such as near the midpoints of the $x$ boundaries). By $t=1$ the flow has locally broken down into turbulence and entered into a phase of chaotic evolution. All features appear again progressively better defined and resolved as resolution is increased, and are largely consistent with what has been reported in the literature. The one noteworthy departure from results obtained with grid codes here is the apparent absence of magnetic island formation in the centre of the computational domain; other contemporary SPMHD solvers are either also unable to capture the development of the tearing mode instability in this problem~\citep{2016MNRAS.455...51H, 2020A&A...638A.140W}, or do form magnetic beads that however migrate away from the box's midpoint and are of a size that is incompatible with what is expected~\citep[see e.g.][who use an artificial resistivity prescription different to ours, which we found to perform sub-optimally in more intricate, cosmic formation simulations, and therefore did not favour]{2017arXiv170607721W, 2018PASA...35...31P}. Similarly to~\cite{2018PASA...35...31P} and in contrast to previous formulations of SPMHD such as that of~\citet{2005MNRAS.364..384P}, we appear to be free of post-shock oscillations, owing to SPHENIX's improved discontinuity-capturing prescriptions. We have also verified, to corroborate the results of~\citet{2018PASA...35...31P} again, that running without artificial resistivity yields predictions that both (moderately) deviate from literature results and are considerably noisier; this confirms the necessity of incorporating magnetic discontinuity-capturing terms in SPMHD and reaffirms our chosen prescription's ability to do so as required.

\begin{figure*}
 \includegraphics[width=\textwidth]{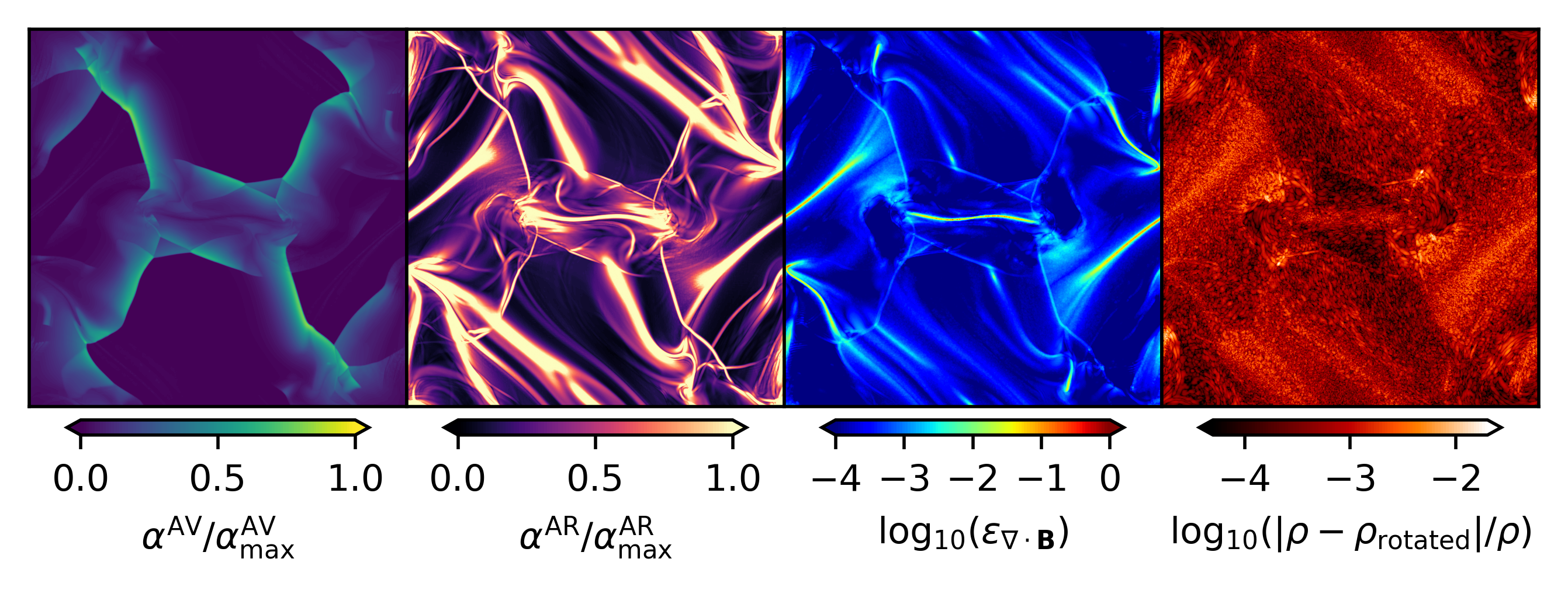} 
 \caption{Four method diagnostics in our highest resolution Orszag-Tang vortex run ($n_x \sim 512$ resolution elements along either of the two major sides of the computational domain). We show, from left to right, projections at $t=0.5$ of the artificial viscosity switch normalised to its maximum allowed value, the artificial resistivity switch normalised to its maximum allowed value, the dimensionless divergence error and the residual between the mass density field and itself rotated by 180 degrees about the $z$ axis. $\alpha^\text{AV}$ and $\alpha^\text{AR}$ only light up at the location of discontinuities and quickly decay back to zero as the flow reverts to being smooth; this is indicative of discontinuity-capturing terms acting in a highly localised fashion, minimally intervening only when necessary and greatly limiting excess diffusion of the fields being smoothed over the resolution scale. Divergence errors are kept well under control, with $\varepsilon_{\nabla \cdot \boldsymbol{B}} < {10}^{-3}$ everywhere, except at sharp discontinuities where it only moderately increases. The last panel illustrates our method's ability to preserve axial symmetry, even during the late-time non-linear phase, indicating that the corrective terms that violate it are numerically small and well behaved.}
 \label{fig:orszag_tang_vortex_corrections}
\end{figure*}

To better illustrate how our implemented corrective terms act, but also further motivate the need for allowing them to operate concurrently, we show in Fig.~\ref{fig:orszag_tang_vortex_corrections} maps at $t=0.5$ of four method diagnostic for our highest resolution run: the value of the artificial viscosity switch $\alpha^\text{AV}$, the value of the artificial resistivity switch $\alpha^\text{AR}$, the dimensionless divergence error $\varepsilon_{\nabla \cdot \boldsymbol{B}}$ and the residual between a projection of the density field and the same projection rotated by 180 degrees. The $\alpha^\text{AV}$ and $\alpha^\text{AR}$ maps point to highly localised action of the discontinuity-capturing terms (note in particular how our map of $\alpha^\text{AV}$ resembles the maps of $\nabla \cdot \boldsymbol{v}$, another commonly employed shock indicator, shown in e.g. Fig. 8 of~\citealt{1995ApJ...452..785R} or Fig. 3 of~\citealt{1998ApJ...509..244R}) as these are only seeded in the presence of shocks and swiftly decay to zero as the flow transitions back to being smooth, restricting their action exclusively to where necessary and limiting needless dissipation. The $\varepsilon_{\nabla \cdot \boldsymbol{B}}$ map illustrates the strong performance of our divergence-cleaning scheme on this test, globally maintaining errors below the $0.1\%$ level, allowing only for narrow, moderate peaks co-spatial with flow discontinuities. Finally, the low values seen in the map of density residuals upon rotation demonstrate ours code's ability to preserve point symmetry, even though the form of the induction equation we implement is not formally invariant under rotations; this is also indicative of good angular momentum conservation. As the corrective Powell acceleration term we use is antisymmetric and proportional to $\nabla \cdot\boldsymbol{B}$, results being uncorrupted here give further evidence the latter is kept entirely under control by our cleaning scheme. 

\begin{figure*}
 \includegraphics[width=\textwidth]{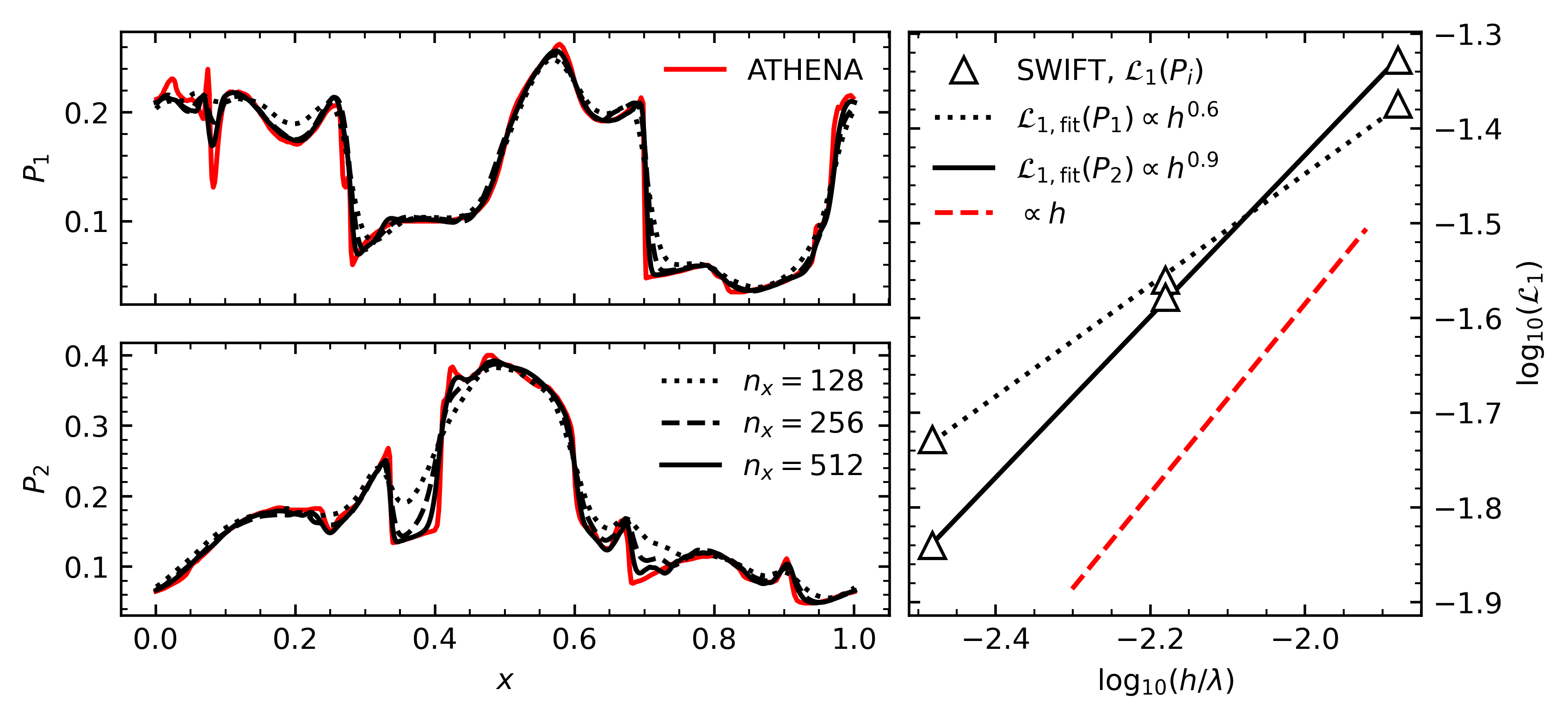} 
 \caption{Comparison of SWIFT and ATHENA~\citep[taken from][]{2008ApJS..178..137S} solutions for the Orszag-Tang vortex problem at $t=0.5$. Left: Horizontal slices of thermal pressure $P$ through $y=0.3125$ ($P_1$; top) and $y=0.427$ ($P_2$; bottom), retrieved through the 2D projections we show in Fig.~\ref{fig:orszag_tang_vortex}. We present SWIFT results in black, matching each time the line-style to the respective resolution level considered, as measured by the number $n_x$ of resolution elements used along the simulation box's major axis ($n_x \sim 128$ results are shown as dotted lines, $n_x \sim 256$ results as dot-dashed lines and $n_x \sim 512$ results as solid lines). The ATHENA `ground-truth' solutions are shown in red. Where the flow is smooth, SWIFT results quickly converge to those obtained with ATHENA and become indistinguishable. Convergence is slower at the location of discontinuities; these appear nevertheless increasingly sharper, better resolved and more accurately captured as resolution is increased. Right: $\mathcal{L}_1$ error norm computed on the entirety of each of the two profiles as a function of resolution, expressed in terms of box averaged smoothing length $h$; we normalise the latter to a characteristic scale of the problem, taken here to be the wavelength $\lambda$ of the velocity vortex perturbation in the initial conditions. We performed linear regression in logarithmic space on both sets of $\left\{ h (n_x) / \lambda, \mathcal{L}_1 ((n_x)| P_i) \right\}$ and show the resulting fits as black lines (dashed for $P_1$ and solid for $P_2$) and associated power laws in the legend. Convergence is slightly sublinear, as expected for a problem involving flow discontinuities; convergence for $P_1$ is moderately slower, as the profile contains sharper jumps which SPH is inherently incapable of representing exactly.}
 \label{fig:orszag_tang_vortex_convergence}
\end{figure*}

To quantify the quality of our solutions, as well as our method's convergence properties on intrinsically multidimensional flow problems, we retrieve from the 2D projections shown in Fig.~\ref{fig:orszag_tang_vortex} horizontal slices of the thermal pressure at $y=0.3125$ and $y=0.427$ from our simulation snapshots at $t=0.5$, which we compare with `ground-truth' results obtained from simulations run with ATHENA~\citep[as shown in][]{2008ApJS..178..137S}. Either or both of these two cuts are also shown in e.g.~\citet{1998ApJ...509..244R, 1999JCoPh.150..561J, 2000ApJ...530..508L, 2006ApJ...652.1306B, 2007MNRAS.379..915R, 2009MNRAS.398.1678D, 2013MNRAS.428...13S, 2018PASA...35...31P}. We show in Fig.~\ref{fig:orszag_tang_vortex_convergence} said slices for the three resolution levels considered, together with a convergence study; this takes the form of an $\mathcal{L}_1$ error norm computed on the profiles in their entirety as a function of initial box averaged smoothing length. SWIFT and ATHENA results are in excellent agreement, being nearly indistinguishable in the smooth parts of the flow. Increasing resolution brings us progressively closer to the reference in all flow features, most notably at the location of discontinuities, which are increasingly better defined and more accurately captured. Convergence of the $\mathcal{L}_1$ error with resolution is slightly sublinear~\citep[noting that performance of higher order Lagrangian methods, as measured through this metric, is not radically different, see e.g.][]{2016MNRAS.455...51H}, in line with the convergence properties of SPHENIX on problems involving hydrodynamic shocks, as reported by~\citet{2022MNRAS.511.2367B}. Convergence is slower for the slice through $y=0.31325$, which contains larger discontinuous jumps, something that SPH cannot perfectly represent by construction.

\subsubsection{Fast magnetic rotor}

The `rotor problem' was first presented in~\cite{1999JCoPh.149..270B} and is aimed at studying torsional Alfvén wave generation and propagation, as well as `magnetic braking', which are processes believed to play a key role in angular momentum redistribution in gravitationally collapsing molecular clouds during star formation~\citep{1980ApJ...237..877M}. The test can be thought of as an extension to two spatial dimensions of the 1D shear Alfvén wave example discussed in~\cite{1992ApJS...80..791S}, and has often been used to benchmark MHD solvers~\citep[e.g.][]{2000JCoPh.161..605T, 2000ApJ...530..508L, 2005MNRAS.364..384P, 2006ApJ...652.1306B, 2008ApJS..178..137S, 2009MNRAS.398.1678D, 2016JCoPh.322..326T, 2016MNRAS.455...51H, 2020A&A...638A.140W}. The problem starts with a fast-spinning dense disk of material (the rotor) embedded in a static diffuse medium that is threaded by a uniform magnetic field. With centrifugal forces within it being unbalanced, the rotor is expected to start expanding while strong torsional Alfvén waves are launched from its surface, resulting in angular momentum being transferred to the material surrounding it as its rotational velocity declines. The magnetic field being oriented perpendicularly to the axis of rotation causes pressure support to act preferentially along a specific direction, which results in the initially cylindrically symmetric rotor eventually becoming oblate in shape.

\begin{figure*}
 \includegraphics[width=\textwidth]{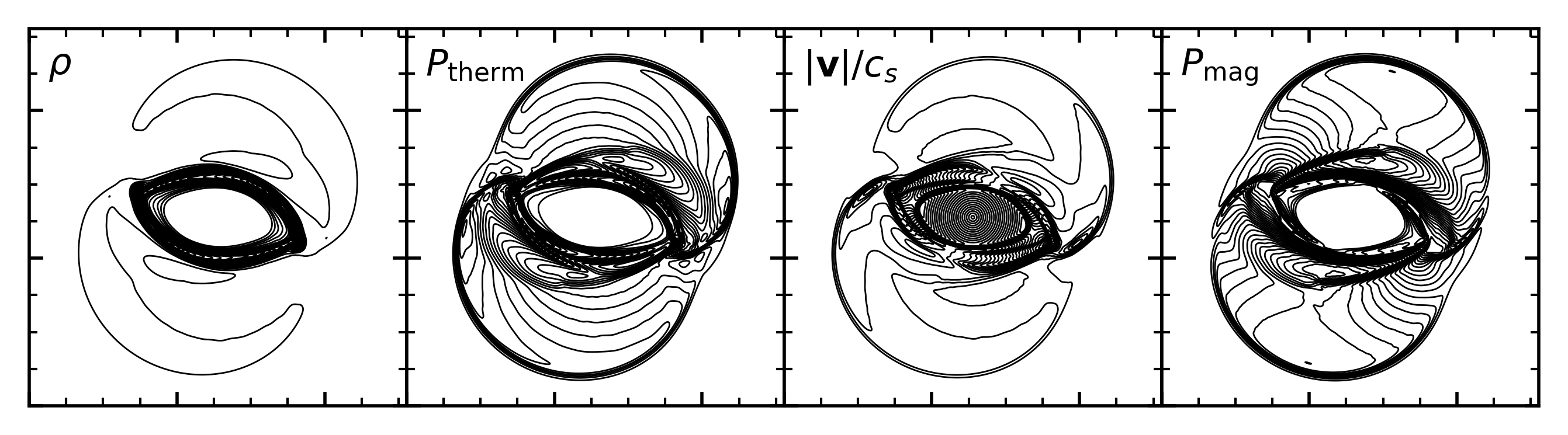} 
 \caption{Results for the fast magnetic rotor test at $t=0.15$, run at our fiducial resolution of $n_{x,y} \sim 128$. We show, going from left to right, contour plots of the density, thermal pressure, Mach number and magnetic pressure in a two-dimensional slice taken at half the depth of the simulated domain at $z=1/16$ (for a simulation box of unit length and width). For all four profiles, we use 30 contours evenly spaced between plotting limits chosen to match those of~\protect\cite{2000JCoPh.161..605T}, for ease of comparison: we show the ranges $0.483<\rho<12.95$, $0.0202 < P_\text{therm} < 2.008$, $0 < \Vert \boldsymbol{v} \Vert / c_s < 8.18$ and $0.0177 < P_\text{mag} < 2.642$. Torsional waves are launched from the surface of the rotor which, draped by the magnetic field lines and compressed vertically by magnetic pressure, has become oblate in shape. Within the central rarefaction, the uniform rotation profile and constant magnetic field strength are both preserved.}
 \label{fig:fast_rotor}
\end{figure*}

The variant of the example we reproduce here is the `first rotor problem' described in~\cite{2000JCoPh.161..605T}~(with the only difference being that we run in 3D rather than 2D). We initialise a thin box of dimensions $L \times L \times L/8$ for $L=1$ with periodic boundary conditions, within which we position $2 \times (208 \times 208 \times 26)$ particles on the vertices of a BCC lattice, to represent the low-density ambient medium. It is thus resolved by $n_{x,y}\sim 256$ particles along the $x$ and $y$ directions, a number chosen to match the fiducial resolution used by~\cite{2018PASA...35...31P} and~\cite{2020A&A...638A.140W} to run the same test. To embed a dense disk at the centre of this uniform medium, we replace the lattice of particles within a cylinder of radius $R_0=0.1$, centred around the line $\boldsymbol{l} = (l_x, l_y, z) = (L/2,L/2,z)$, with a more densely packed one with inter-particle spacing smaller by a factor of $\sqrt[3]{10}$. This allows us to establish a density contrast of $10:1$ between the rotor and its surroundings while using equal-mass particles. We assign constant masses so that the density inside the rotor is $\rho_\text{rotor}=10$ (and therefore $\rho_\text{out} = 1$ outside of it) and set appropriate thermal energies to obtain a uniform pressure field with $P=1$ everywhere. The rotor is set spinning at a uniform angular velocity $\omega = 20$ by imposing the velocity field
\begin{equation}
    \boldsymbol{v} =
    \begin{cases}
        \omega 
        [
        -(y - l_y), x-l_x, 0
        ]
        \quad & \text{if} \quad R <R_0 \\
        \boldsymbol{0} \quad & \text{otherwise}
    \end{cases}
\end{equation}
where $R$ is the cylindrical radial distance from $\boldsymbol{l}$ defined similarly to~(\ref{eq:cylindrical_radial_distance}), and the uniform field  permeating the box is taken to be $\boldsymbol{B} = (5/\sqrt{4 \pi}, 0, 0)$. The adiabatic index is exceptionally set to $\gamma=7/5$. We note that we do not smooth over the sharp discontinuities in density and velocity at the surface of the disk in the initial conditions~\citep[unlike e.g.][]{2000JCoPh.161..605T, 2000ApJ...530..508L, 2016MNRAS.455...51H}, to measure our code against a more challenging version of the test. We evolve the system until $t=0.15$ and present the results obtained in Fig.~\ref{fig:fast_rotor}, showing contour plots for the density, thermal pressure, Mach number $\mathcal{M} \equiv \Vert \boldsymbol{v} \Vert / c_s$ and magnetic pressure.

The magnetic pressure panel in Fig.~\ref{fig:fast_rotor} clearly shows torsional waves being seeded at the edge of the rotor and propagating through the ambient medium. We moreover observe that magnetic pressure, being maximal at the rotor's surface, compresses the structure vertically which leads to the expected elongated shape. All our profiles appear to maintain near perfect point symmetry, in line with predictions made with grid codes~\citep[e.g.][]{2008ApJS..178..137S}. Moreover, within the rarefaction around the centre of the computational domain, the constant magnetic field strength and uniform rotation (as seen in the Mach number profile) are both preserved, as anticipated. Discontinuities are sharp, and noise and spurious oscillations close to nonexistent: this is a notable improvement over early applications of SPMHD to this problem ~\citep{2005MNRAS.364..384P} and is in line with the observations of~\cite{2018PASA...35...31P}, bearing witness to the robustness and efficiency of the recently suggested improvements to discontinuity-capturing terms integrated in our method.

\begin{figure*}
 \includegraphics[width=\textwidth]{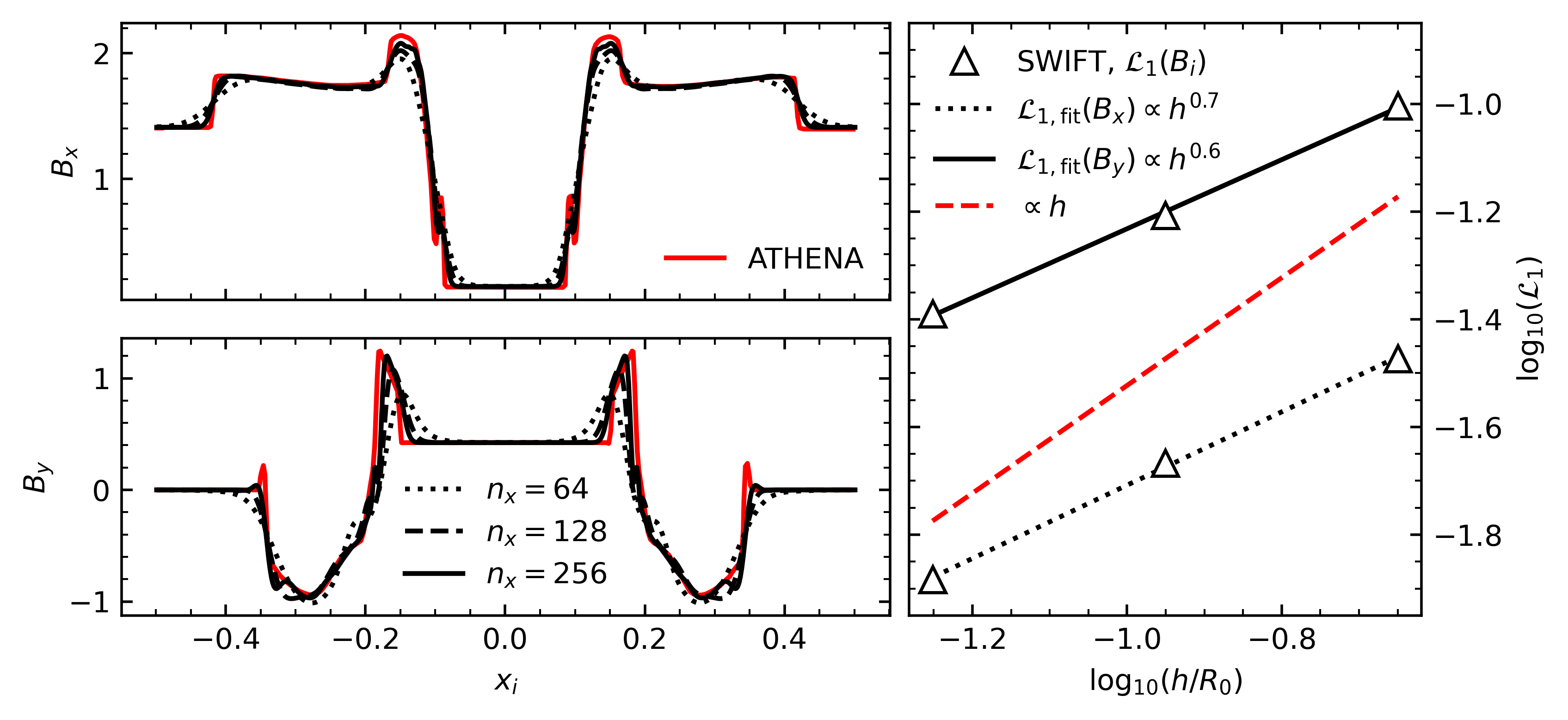} 
 \caption{Comparison of SWIFT and ATHENA~\citep[taken from][]{2008ApJS..178..137S} solutions for the fast magnetic rotor problem at $t=0.15$, similar to what is presented in Fig.~\ref{fig:orszag_tang_vortex_convergence} for the Orszag-Tang vortex. We show horizontal cuts of the $x$ component of the magnetic field through $x=0$ ($B_x$; top left) and $y$ component of the magnetic field through $y=0$ ($B_y$; bottom left), retrieved from 2D slices such as the ones depicted in Fig.~\ref{fig:fast_rotor}. We accompany these by a plot of an $\mathcal{L}_1$ error norm computed on the entirety of these two profiles as a function of resolution (right). These results corroborate the findings reported in Section~\ref{sec:oreszag_tang_vortex} with regards to our method's ability to correctly capture multidimensional non-linear flows, as well as its convergence properties.}
 \label{fig:fast_rotor_convergence}
\end{figure*}

Similarly to what we did for the Orszag-Tang vortex example, we here too evaluate our method's performance, by retrieving from two-dimensional maps such as those shown in Fig.~\ref{fig:fast_rotor}, one-dimensional slices of the $x$-component of $\boldsymbol{B}$ through $x=0$ and $y$-component of $\boldsymbol{B}$ through $y=0$. We contrast them with results obtained with ATHENA and presented in~\citep{2008JCoPh.227.4123G}. Having run two additional lower resolution (by a factor of $4$ and $16$ in terms of total particle number) counterparts to the simulation we detail above, we show in Fig.~\ref{fig:fast_rotor_convergence} slices extracted from SWIFT outputs at $t=0.15$ for the rotor test run at the three resolution levels corresponding to $n_{x,y} \sim 64, 128, 256$. We complement this with a convergence study, in the form of a panel showing $\mathcal{L}_1$ error norms computed on the entirety of both profiles as a function of resolution. Away from discontinuities, even for our lowest resolution run, the results we obtain appear indistinguishable from the reference we compare with. Shock features are initially smoother, but become increasingly sharp and tend toward the assumed ground truth as $n_{x,y}$ is increased. The convergence plot corroborates that our method converges slightly sub-linearly with smoothing length in problems involving sharp discontinuities.  

\subsubsection{Strong magnetised blast wave}
\label{sec:strong_magnetised_blast_wave}

The strong magnetised blast wave problem is an often-reproduced example which considers the expansion of a highly pressurised, hot parcel of gas of finite support embedded in a cool ambient medium with low plasma $\beta$. The high ambient magnetisation leads to strong tension forces which cause the emerging shock features to propagate anisotropically, preferentially along the direction of the initially uniform magnetic field lines. First presented in~\cite{1998ApJS..116..133B}, this problem and different variants thereof have been extensively studied in the numerical method benchmarking literature~\citep[e.g.][]{1999JCoPh.149..270B, 2000ApJ...530..508L, 2007MNRAS.379..915R, 2008JCoPh.227.4123G, 2009MNRAS.398.1678D, 2011MNRAS.418.1392P, 2012JCoPh.231.7214T, 2016MNRAS.455...51H, 2018PASA...35...31P, 2020A&A...638A.140W} to assess the ability of a given solver to correctly represent strong shocks, resolve sharp interfaces while preserving symmetry, and handle divergence errors which, if left uncontrolled, can give rise to pathological small-scale features and negative thermal pressures. In the case of finite-volume schemes, the MHD blast wave test can further be used to evaluate the severity of grid alignment effects, while in the case of SPMHD it has served as a guide to devise prescriptions for the tensile instability correction and divergence cleaning~\citep{2012JCoPh.231.7214T, 2016JCoPh.322..326T}. The problem is even more interesting in our case as it qualitatively recreates an often-encountered configuration in galaxy formation simulations, namely that of localised thermal energy deposition which is common in sub-grid modeling of energetic feedback from stars~\citep[e.g.][]{2012MNRAS.426..140D} and active galactic nuclei~\citep[e.g.][]{2009MNRAS.398...53B}. Our method robustly coupling to these would constitute one of the holy grails of the present endeavour.

\begin{figure*}
 \includegraphics[width=\textwidth]{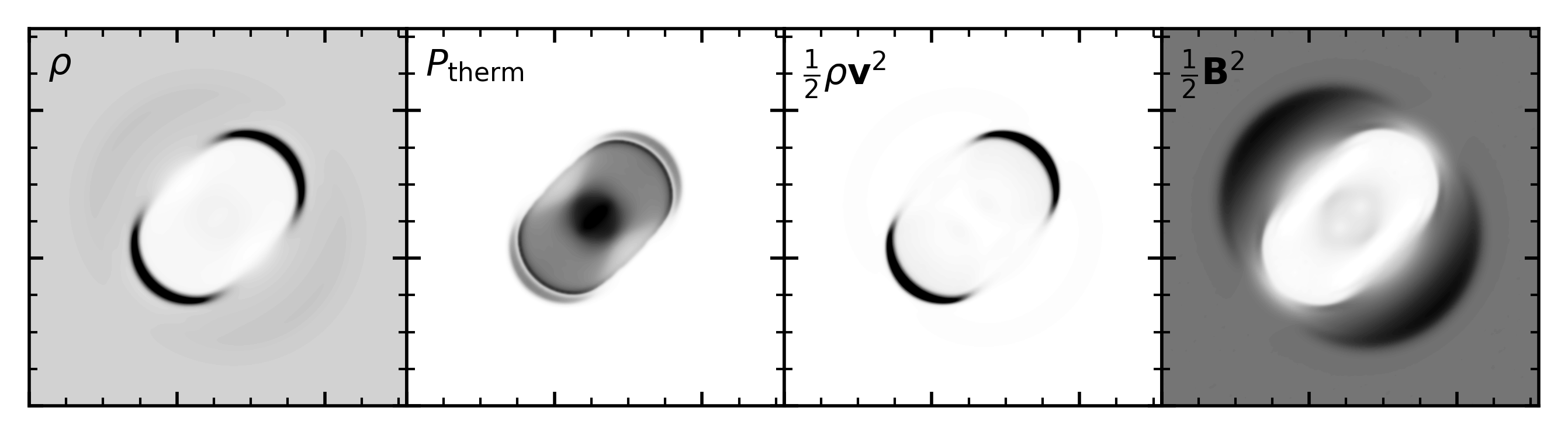} 
 \caption{Results for the 3D strong magnetised blast wave test, run in a cubic simulation domain at our fiducial resolution of $2 \times {200}^3$ particles. A spherical pulse of high thermal pressure is introduced at $t=0$ in the middle the box, which is permeated by a constant, uniform magnetic field pointing in the $x-z$ direction. We show, going from left to right, slices at $t=0.02$ through the $y=0$ plane of the mass density, thermal pressure, specific kinetic energy and specific magnetic energy. We have used the colour scheme and plotting limits of~\citet{2008ApJS..178..137S},~\citet{2018PASA...35...31P} and~\citet{2020A&A...638A.140W} to allow for immediate comparison with results presented therein: we show the ranges $0.19 < \rho < 2.98$, $1 < P_\text{therm} < 42.4$, $0 < \rho \boldsymbol{v}^2 /2 < 33.1$, and $25.2 < \boldsymbol{B}^2 / 2 < 64.9$. The blast wave propagates preferentially along the direction of the ambient magnetic field and consists of a nearly isotropic magnetically dominated fast magnetosonic wave that is only weakly compressive, and a highly directional slow magnetosonic wave and contact discontinuity which enclose $\sim 3$-fold and $\sim 4$-fold enhancements in mass density and specific kinetic energy respectively.}
 \label{fig:3d_blast_wave}
\end{figure*}

We first reproduce the three-dimensional set-up presented in~\cite{2008JCoPh.227.4123G}, which is also the version of the problem considered by~\cite{2018PASA...35...31P} and~\cite{2020A&A...638A.140W}. We initialise a cubic box of side length $L=1$ with periodic boundary conditions, within which we position $2 \times {200}^3$ particles on the vertices of a BCC lattice; this corresponds to $n_x \sim 256$ resolution elements along each of the computational domain's sides and matches the resolution employed by the aforementioned works to run this test. Exceptionally setting $\gamma = 7/5$, we assign constant masses to obtain a uniform density field with $\rho=1$. We moreover assign constant thermal energies to obtain a uniform pressure field with $P_\text{ambient} = 1$ everywhere, except for a spherical region of radius $R=L/8$ centred at $\boldsymbol{r}_\text{blast} = (L/2, L/2, L/2)$ within which we seed the blast wave by choosing constant thermal energies yielding $P_\text{blast} = 100$. We initially take particles to be still by setting velocities to $\boldsymbol{v} = \boldsymbol{0}$ and permeate the box by the constant, uniform magnetic field $\boldsymbol{B} = (10/\sqrt{2}, 0, 10/\sqrt{2})$. This results in an ambient medium with initial plasma beta $\beta_\text{ambient} = 2$ and a highly pressurised blast region with $\beta_\text{blast}$ initially. We evolve the system until $t=0.02$ and present slices through the $y=0$ plane of the resulting mass density, thermal pressure, specific kinetic energy and specific magnetic energy distributions in Fig.~\ref{fig:3d_blast_wave}; the color map and plotting ranges we employ are identical to those used by~\citet{2008ApJS..178..137S, 2018PASA...35...31P, 2020A&A...638A.140W} so that our solutions can be directly contrasted with results presented therein.

\begin{figure*}
 \includegraphics[width=\textwidth]{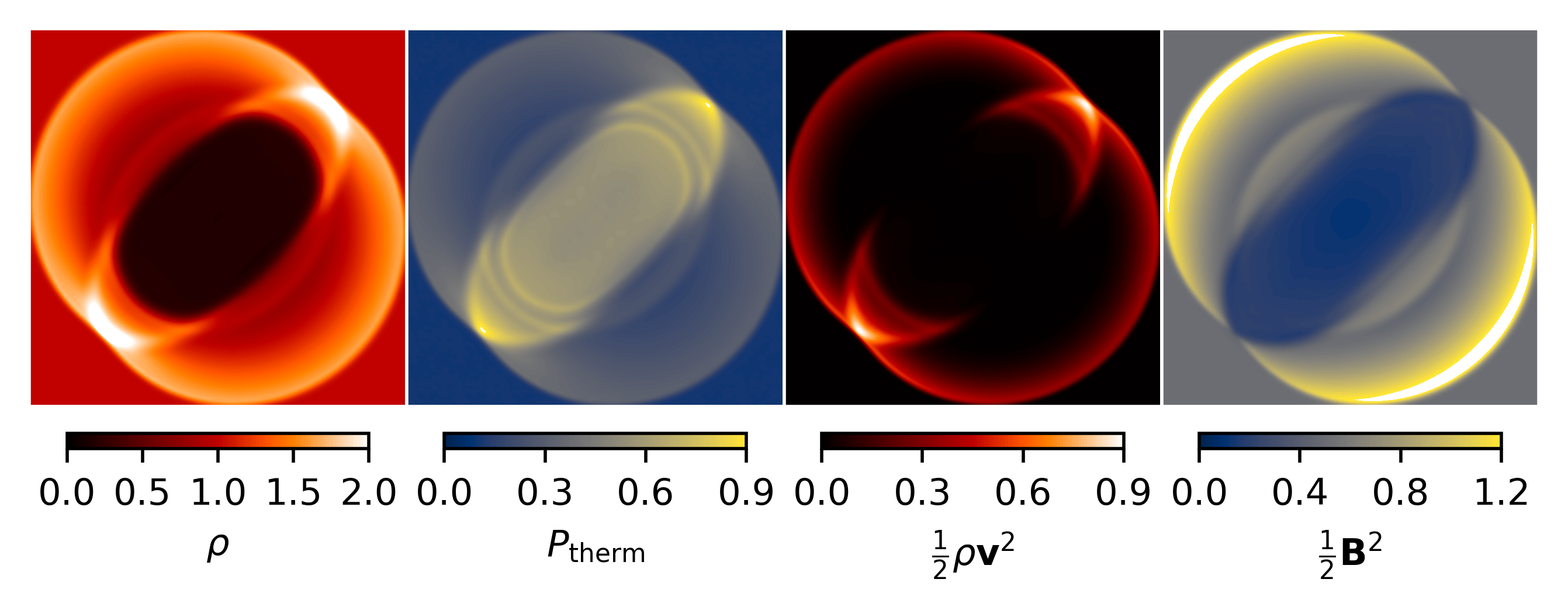} 
 \caption{Results for the 2D strong magnetised blast wave test, run in a thin 3D simulation box at our fiducial resolution of $2 \times (192 \times 192 \times 24)$ particles. A cylindrical pulse of high thermal pressure is introduced at $t=0$ in the middle the box, which is permeated by a constant, uniform magnetic field pointing in the $x-y$ direction. We show, going from left to right, slices at $t=0.2$ of the mass density, thermal pressure, specific kinetic energy and specific magnetic energy taken at half the depth of the simulation domain at $z=1/16$ (for a box of unit length and width). These can be directly compared to e.g. Fig. 4 in~\citet{2011MNRAS.418.1392P} or Fig. 17 in~\citet{2016MNRAS.455...51H}. The qualitative behaviour of the system is identical to that seen for the 3D case, shown in our Fig.~\ref{fig:3d_blast_wave}.}
 \label{fig:2d_blast_wave}
\end{figure*}

For ease of comparison with a wider selection of numerical methods, we also reproduce a variant of the problem which has been used to test MHD codes targeting similar science cases as SWIFT, such as AREPO~\citep{2011MNRAS.418.1392P} and GIZMO~\citep{2016MNRAS.455...51H}.Although said test configuration and associated realisations are two-dimensional, we here extend the set-up along the $z$ direction to produce a 3D `thin box' equivalent. We initialise an orthogonal parallelepiped domain of dimensions $L \times L \times L/8$ where $L=1$ which we populate with $2 \times \left( 192 \times 192 \times 24 \right)$ particles positioned on the vertices of a BCC lattice; this corresponds to $n_x \sim 256$ resolution elements along either of the box's major sides, matching the resolution used by~\citet{2016MNRAS.455...51H} and being half that used by~\citet{2011MNRAS.418.1392P}. We here assign constant particle masses to obtain a uniform density field with $\rho=1$. We again induce a blast wave perturbation by assigning thermal energies to obtain a uniform pressure field with $P_\text{ambient} = 0.1$ in the ambient medium and $P_\text{blast}= 10$ in the blast region, the latter being defined as the cylinder of radius $R=0.1$ centred around the line $\boldsymbol{l}_\text{blast} = (L/2,L/2,z)$. Particles here too are are initially at rest with velocities $\boldsymbol{v} = \boldsymbol{0}$, and the box is permeated by the constant, uniform magnetic field $\boldsymbol{B} = (1/\sqrt{2}, 1/\sqrt{2}, 0)$. The resulting plasma betas in the blast and ambient regions is identical to those in the 3D case. We evolve the system until $t=0.2$ and show mass density, thermal pressure, specific kinetic energy and specific magnetic energy projections at that point in time in Fig.~\ref{fig:2d_blast_wave}.

Both runs paint a qualitatively similar picture and produce results in excellent agreement with their respective counterparts from the literature. We observe an anisotropically expanding blast wave taking the shape of an oblate spheroidal aligned with the seed magnetic field. All expected individual MHD modes that constitute it~\citep[see e.g.][]{2008ApJS..178..137S} are retrieved, appearing sharp and well-resolved: the fast magnetosonic wave expanding close to isotropically, being magnetically dominated and only weakly compressive, followed by a propagating slow magnetosonic wave and contact discontinuity between which enhancements in density, thermal pressure and kinetic energy are trapped; these are co-spatial with the post-shock magnetic pressure minima in the direction along which the cavity preferentially expands, as strong magnetic tension forces perpendicular to it prevent such features from developing elsewhere. As with our shock tube results, we observe here a mild pressure blip at the contact surface, which does not however corrupt the solution. We deem noteworthy the perfectly preserved line symmetry about the explosion's two major axes, as well as the fact that our profiles seem to be free of noise and artifacts (as opposed to results shown in e.g.~\citealt{1998ApJS..116..133B} where poor divergence control led to small scale spurious fluctuations in the magnetic pressure, or~\citealt{2011MNRAS.418.1392P} and~\citealt{2016MNRAS.455...51H} who observed corrugations in the maxima of (some of) their density profiles). Given the above, and similarly to~\citet{2012JCoPh.231.7214T} and~\citet{2016JCoPh.322..326T}, we interpret encouraging results on the present test as supporting evidence that our implementations of both the tensile instability correction and constrained hyperbolic divergence cleaning work as desired.

\subsubsection{MHD Kelvin-Helmholtz instability}
\label{sec:mhd_khi}

Coexistence, interaction and mixing of distinct fluid phases is ubiquitous in astrophysical systems and of particular interest in the context of galaxy formation~\citep[see e.g.][]{2012MNRAS.424.2999S, 2012MNRAS.425.3024V}. 

An often studied mixing problem is the onset and evolution of the Kelvin-Helmholtz instability~\citep[KHI; see e.g.][]{1961hhs..book.....C}. A simple set-up where this can be observed is the interface between two fluid phases across which velocity shear is present. Any velocity perturbation perpendicular to the shear flow will grow exponentially, with shorter modes growing more quickly, giving rise at first to large vortices. The flow subsequently turns turbulent as further instabilities are seeded across the vortices' interfaces, until the two fluid phases ultimately become well-mixed.

The hydrodynamic version of the problem and the ability of SPH to correctly capture the associated phenomenology have been widely debated, especially for the particular case of the velocity shear being accompanied by a density discontinuity across the two fluid phases. An extensive body of research (most notably~\citealt{2007MNRAS.380..963A} but see also~\citealt{2003MNRAS.345..429O} or~\citealt{2010ARA&A..48..391S}) pointed out that systematic errors in the calculation of the pressure force in the presence of density gradients inherent to SPH suppressed phase mixing. These critiques were met with the suggestion of a series of mitigation strategies, including the introduction of an artificial thermal conduction scheme to smooth quantities across the fluid interface~\citep{2008JCoPh.22710040P, 2012A&A...546A..45V, 2013MNRAS.428.1968K, 2020ApJ...898...60R, 2020MNRAS.498.4230R}, redefining how the density estimate is computed~\citep{2010MNRAS.405.1513R, 2010MNRAS.406.2289H} or choosing the thermal pressure or energy (rather than the mass density) as the kernel-averaged attribute to be used in deriving conservative EoMs~\citep{2013MNRAS.428.2840H, 2013ApJ...768...44S}. \citet{2019MNRAS.488.5210T} demonstrated how an SPH implementation making use of the first among these options was able to converge to the KHI reference solution of~\citet{2016MNRAS.455.4274L} in both the linear and non-linear regimes. Artificial thermal conduction thus became the de facto method for properly capturing fluid mixing adopted in SPHENIX as well~\citep[see e.g.][on why smoothing over the pressure is not a fitting alternative when the hydrodynamics solver is coupled to sub-resolution galaxy formation models]{2021MNRAS.505.2316B}; \citet{2022MNRAS.511.2367B} showed that SPHENIX captures the KHI correctly across different resolution levels, and even yields accurate results for more demanding variants of the problem involving extreme density jumps, provided enough resolution elements are present to resolve the under-dense phase. 

We here seek to elaborate on the aforementioned discussion by investigating whether under our implementation of magnetic field physics on top of SPHENIX the artificial thermal conduction scheme still behaves as expected. We further consider whether the expected qualitative behaviour of the KHI upon inclusion of MHD effects is reproduced. This was studied early on through linear stability analyses (see~\citealt{1961hhs..book.....C} and~\citealt{1982JGR....87.7431M} for the case of an incompressible and compressible fluid respectively) which were followed soon thereafter by numerical studies~\citep[see e.g.][]{1995ApJ...452..785R, 1996ApJ...460..777F, 1996ApJ...456..708M, 1999JPlPh..61....1K}. The effect of a uniform magnetic field on the KHI depends on the relative orientation of the former with respect to the wave vector of the induced perturbation driving the instability. If the two are perpendicular, the magnetic field can only be compressed and the resultant behaviour is indistinguishable from that obtained in the hydrodynamic limit. If however the two are parallel, magnetic tension modes are excited which provide a stabilising effect inhibiting perturbation growth, potentially suppressing it entirely for Alfvén speeds that are too large compared to the velocity shear. The magnetic field itself is amplified at the interface, which subsequently halts further local excitation of the secondary KHI.

In the name of consistency and continuity, we reproduce the initial conditions used by~\citet{2022MNRAS.511.2367B} (which are themselves similar to those used by~\citealt{2008JCoPh.22710040P}) to test SPHENIX on the hydrodynamic version of the problem. We initialise a 2D box of dimensions $L \times L$ for $L = 1$ with periodic boundary conditions. We divide the simulation domain at $t=0$ into two: an outer region $\Omega_O = [0, 1]  \times ( [0, 0.25 L ] \cup [0.75 L , L] )$ (whose physical attributes are denoted by the subscript 'O'), and a central region $\Omega_C = [0, L] \times [0.25 L, 0.75 L]$ (corresponding to a horizontal band whose physical attributes are denoted by the subscript 'O'). We populate both with particles positioned on the vertices of regular square lattices, in $\Omega_O$ adding up to $n_{x,O} = 256$ along its longer side and in $\Omega_C$ to $n_{x,C} = \lfloor \sqrt{2} \cdot 256 \rfloor$ (where $\lfloor \cdot \rfloor$ is the floor function); we thus set up a mass density contrast of $\chi \equiv \rho_C / \rho_O = 2$. We assign constant masses and thermal energies to specifically set $\rho_O = 1$ and $\rho_C = 2$ and impose a uniform thermal pressure field as $P_O = P_C = 2.5$; this introduces a sharp contact discontinuity across the fluid phase interface which is intended to maximally stress test our method in this set-up. We initialise velocities to $\boldsymbol{v}_O = -0.5 \hat{\boldsymbol{e}}_x$ and $\boldsymbol{v}_C = 0.5 \hat{\boldsymbol{e}}_x$ to establish the relative shear $v = \Vert \boldsymbol{v}_O - \boldsymbol{v}_C \Vert$, and permeate the box by a constant magnetic field $\boldsymbol{B}_O = \boldsymbol{B}_C = B_0 \hat{\boldsymbol{e}}_x$ where $B_0$ is a constant we set to $B_0 = 0.1$ (in accordance with~\citealt{2016MNRAS.455...51H} and~\citealt{2020A&A...638A.140W}), so as to lie in the regime where MHD effects are evident but still not strong enough to entirely suppress the KHI. Finally, we seed the instability in a controlled manner by introducing the eigenmode velocity perturbation
\begin{equation}
    \delta \boldsymbol{v} =
    \delta v_0
    \mathrm{sin} \left( \frac{2 \pi}{\lambda} x \right)
    \left[
    \mathrm{exp} \left( - \frac{{(y - 1/4)}^2}{2\sigma^2} \right) + 
    \mathrm{exp} \left( - \frac{{(y - 3/4)}^2}{2\sigma^2} \right)
    \right] \hat{\boldsymbol{e}}_y
\label{eq:KHI_perturbation}
\end{equation}
where $\delta v_0$ sets its initial amplitude, $\lambda$ its wavelength and $\sigma$ the width of a thin transition layer; we set these parameters to $\omega_0=0.1$, $\lambda = 0.5$ and $\sigma = 0.05 / \sqrt{2}$~\citep[see e.g.][who make very similar choices]{2012MNRAS.422.3037R}. Defining the characteristic time-scale $\tau_{KH}$ as
\begin{equation}
    \tau_{KH} = 
    \frac{(1 + \chi) \lambda}{\sqrt{\chi} v}
\label{eq:KHI_timescale}
\end{equation}
(this simply being a scaled inverse of the characteristic growth rate given in~\citealt{1961hhs..book.....C}), we evolve the system until $t = 2 \tau_{KH}$.

\begin{figure}
    \includegraphics[width=.99\columnwidth]{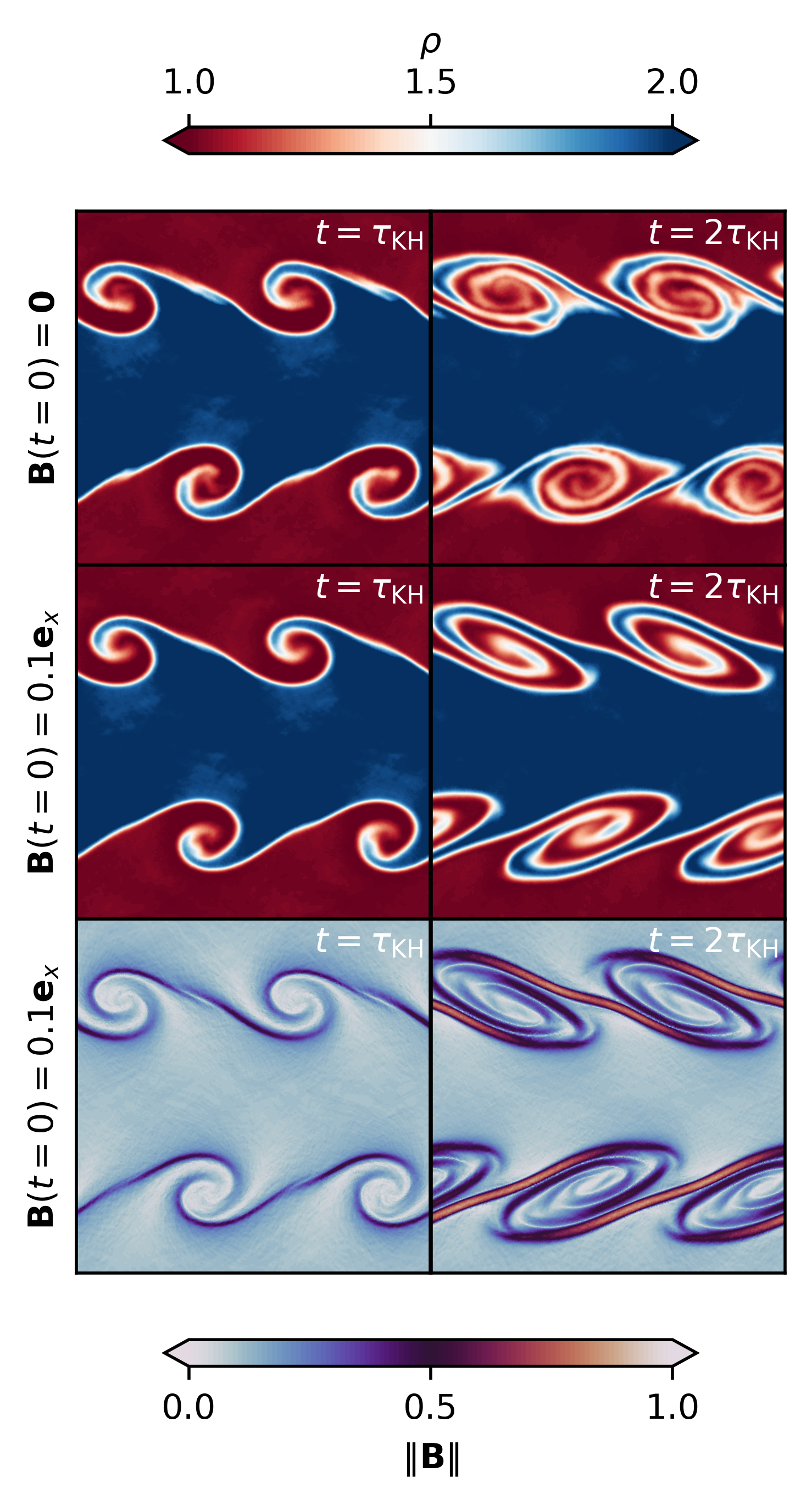}
    \caption{Results for the magnetic Kelvin-Helmholtz instability test at $t = \tau_{KH}$ (left column) and $t = 2 \tau_{KH}$ (right column), run at our fiducial resolution of $n_{x,O} = 256$ particles along the outer stream region's longer side. We contrast results for a control hydrodynamic run initialised with $\boldsymbol{B} (t = 0) = \boldsymbol{0}$ ($1^\text{st}$ row) with results for an MHD run initialised with $\boldsymbol{B} (t = 0) =  0.1 \hat{\boldsymbol{e}}_x$ ($2^\text{nd}$ and $3^\text{rd}$ row). We show projections in the $x-y$ plane of the mass density $\rho$ ($1^\text{st}$ and $2^\text{nd}$ row) and of the norm of the magnetic field vector $\boldsymbol{B}$ ($3^\text{rd}$ row). Growth of the KHI is well captured by our scheme in both runs, as swirls curling back onto themselves develop at the phase boundary and grow; as opposed to the pure hydrodynamic case, inclusion a non-zero magnetic field parallel to the initial shear flow suppresses the growing modes, giving rise to vortices that are more oblong in shape, and reduces phase mixing and excitation of smaller scale, secondary instabilities. The magnetic field itself is amplified at the interface as time progresses.}
    \label{fig:kelvin_helmholtz_instability}
\end{figure}

\begin{figure}
    \includegraphics[width=.99\columnwidth]{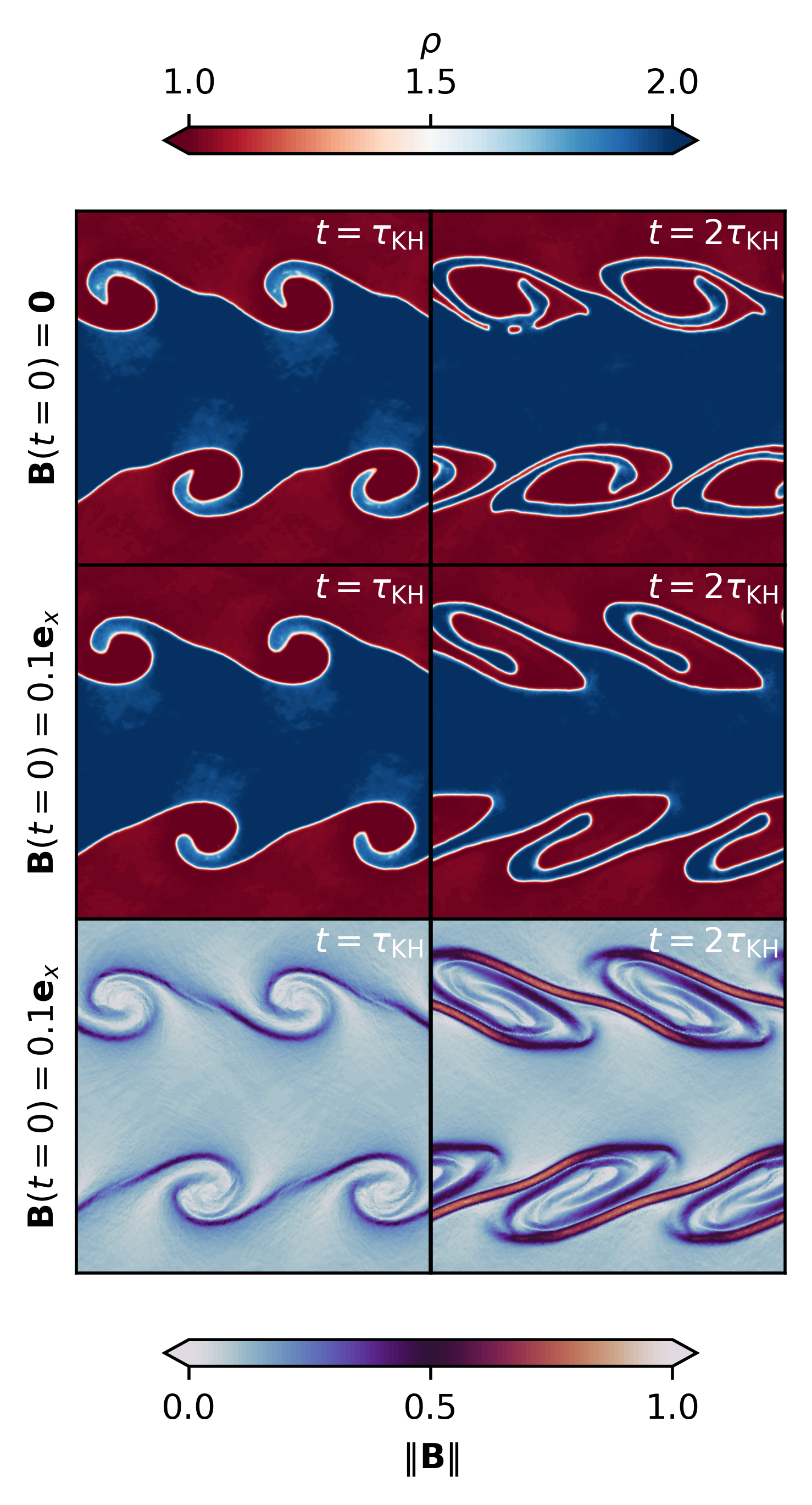}
    \caption{The same as Fig.~\ref{fig:kelvin_helmholtz_instability}, for a redo of the runs considered therein for which artificial thermal conduction has been disabled. The anticipated qualitative behaviour is in part still reproduced, although phase mixing is suppressed and the growth rate of the instability is reduced. We use this to demonstrate that absence of an appropriate treatment for discontinuities in thermal energy in SP(M)H(D) does give rise to the undesired behaviour reported in the literature; our artificial thermal conduction prescription nevertheless addresses this effectively and couples seamlessly to all other components of our method.}
    \label{fig:kelvin_helmholtz_instability_noAD}
\end{figure}

We first show results obtained with our full model in Fig.~\ref{fig:kelvin_helmholtz_instability}, contrasting a hydrodynamic control run (initialised with a uniform $\boldsymbol{B} (t = 0) = \boldsymbol{0}$) with the fiducial set-up described above at $t = \tau_{KH}$ and $t = 2 \tau_{KH}$; these match the timestamps of simulation outputs presented by both~\citet{2016MNRAS.455...51H} and~\citet{2020A&A...638A.140W} who tested their respective SPMHD implementations on the same problem, at the same resolution level. Mass density projections show a well captured KHI in both runs, as prominent vortices are generated at the interface which fold back onto themselves as time progresses. Mixing between phases is more prominent in the hydrodynamic run; when MHD effects are included, mass diffusion is suppressed, vortices are better delineated, more elongated in shape and have curled fewer times around their centres, and small scale modes are less present. Magnetic field strength projections indicate that the magnetic fields in the MHD run are progressively amplified at the surface boundary, as expected~\citep[compare e.g. to Fig. 7 in][]{1995ApJ...452..785R}. \citet{2016MNRAS.455...51H} commented that the performance of SPMHD on this problem is highly sensitive to the choice of $v^{AR}$, and that the particle neighbour number necessary to capture the instability is considerably large; we find our parameter choices to be sufficient to reproduce the desired results, and moreover argue that said parameter picks comfortably lie within the regime of what can reasonably be used for production runs~\citep[and are equally or less demanding than choices made in the literature for successful SPMHD simulation campaigns, see e.g.][]{2024A&A...692A.232B}. We also verify that our choice of AR prescription does indeed not suppress the anticipated magnetic field amplification at the interface. Finally, noting that~\citet{2020A&A...638A.140W} claimed that density-energy SPMHD was subject to surface tension effects that reduced mixing and killed growth of the KHI, we repeated the same numerical experiment but with our AD prescription disabled; results are shown in Fig.~\ref{fig:kelvin_helmholtz_instability_noAD}. The reported undesired behaviour is indeed reproduced, with fluid phases remaining distinct well into the non-linear evolution phase and vortices growing at a slower rate; both problems are nevertheless completely alleviated upon inclusion of SPHENIX's AD prescription, which couples seamlessly to the other components of our method, to produce results in good agreement with and of comparable quality to those of~\citet{2016MNRAS.455...51H} and~\citet{2020A&A...638A.140W}.

\subsubsection{MHD cloud-wind interaction}
\label{sec:mhd_cloud_wind_int}

Taking a step further in terms of complexity and (astro)physical pertinence, we consider a second fluid mixing study case in the shape of the `cloud-wind interaction' problem, most commonly referred to as the `blob' test. A toy model for multiphase fluid flows in the interstellar~\citep[e.g.][]{1977ApJ...218..148M}, circumgalactic~\citep[e.g.][and references therein]{2017ARA&A..55..389T} and intracluster media~\citep[e.g.][]{1990MNRAS.244P..26B}, the example considers an initially static patch of cold, dense gas (the `cloud') placed in a wind tunnel permeated by a hot, dilute ambient medium moving at a supersonic speed. 2D pure-hydrodynamic versions of the test~\citep[e.g.][]{1976ApJ...207..484W, 1982MNRAS.201..833N, 1993ApJ...407..588M, 1994ApJ...420..213K, 2002astro.ph..1398P, 2005MNRAS.364.1105S, 2006ApJS..164..477N}, as well as their 3D counterparts~\citep[e.g.][]{1992ApJ...390L..17S, 1995ApJ...454..172X, 2007MNRAS.380..963A, 2015MNRAS.450...53H, 2017MNRAS.471.2357W} showed that the cloud is expected to first be compressed along the direction of its relative motion with respect to the wind, as a bow shock forms in front of it; it is subsequently predicted to undergo fragmentation and dissolution through the joint action of Kelvin-Helmholtz (KH) and Rayleigh-Taylor (RT) instabilities seeded at its surface. This latter stage of the system's evolution has been claimed to be notoriously non-trivial for `traditional', density-energy SPH solvers to accurately capture~\citep[see ][]{2007MNRAS.380..963A, 2013MNRAS.428.2840H}; the core claim was that absence of interparticle entropy exchange and surface tension effects, much like in the case of the KHI, prevented the onset of mixing and maintained most of the cold gas in a single blob. Nevertheless, appropriately chosen dissipation terms, even in their earlier forms~\citep[such as those of][]{2005MNRAS.364.1105S}, already proved to largely correct for this. More modern flavours of SPH like SPHENIX now perform comparably well to other contemporary Eulerian and Lagrangian solvers on the problem, as demonstrated by~\citet{2022MNRAS.511.2367B}. Nevertheless, convergence (and the rate of convergence) of different methods to a unique solution remains an open problem; this was demonstrated by~\citet{2023MNRAS.523.1280B} through a series of controlled experiments undertaken with SWIFT, starting from a single set of well-posed initial conditions passed to seven different hydrodynamics solvers. Disagreement of predictions, as illustrated through consideration of a series of quantitative metrics, persisted at the highest investigated resolutions even among non-SPH methods.

The impact of MHD effects on the non-radiative, non-conducting and non-self-gravitating `blob' has been considered in both 2D~\citep[e.g.][]{1994ApJ...433..757M, 1996ApJ...473..365J} and 3D~\citep[e.g.][]{2000ApJ...543..775G, 2008ApJ...680..336S, 2008ApJ...677..993D, 2013ApJ...774..133L, 2015MNRAS.449....2M, 2020MNRAS.499.4261S}; strength, topology and orientation of the initial magnetic field were all found to distinctly affect the qualitative behaviour of the system and phenomenological departure from the hydrodynamical case. The minimally complex configuration whereby even a weak seed leads to a vastly different system response is that of a uniform magnetic field initially transverse to the wind velocity, and is what we shall consider here. Although the evolution at early times is not greatly altered by MHD effects, field amplification driven by the shear flow encircling the cloud leads to an increase in magnetic stress, which can locally become comparable in strength to the ram pressure it sees. Magnetic energy is accumulated upstream of the cloud in the form of a protective layer (something referred to as magnetic draping, as field lines embrace the cloud) which stabilises its surface by suppressing mixing instabilities and acting against cloud disruption and dissolution. The magnetic energy within the cloud is expected to eventually decrease as it is transported to its outer edges by turbulent phenomena, while a localised linear structure of highly magnetised material can be expected post-shock. The velocity field is expected to become smoother and manifest as a more laminar flow, while magnetic tension can boost the drag force experienced by the cloud. Inclusion of MHD effects has thus been presented as a plausible avenue through which one can reconcile hydrodynamics-only simulations of cloud-wind interactions that predict efficient mixing with observations of a low-redshift CGM that is clearly multiphase~\citep[e.g.][]{2014ApJ...792....8W}.

The version of the `blob' test we reproduce here is that described in~\citet{2023MNRAS.523.1280B}; they~\citep[similarly to][]{2022MNRAS.511.2367B} opt for a minimally complex initial configuration where a simple particle arrangement is used to set up a sharp `cloud-to-wind' contact discontinuity; this is in contrast to other contemporary works where this base state is perturbed using specific modes to controllably seed mixing instabilities~\citep[e.g.][]{2010MNRAS.405.1513R, 2014MNRAS.443.1173H}. We initialise a 3D box of dimensions $4L \times L \times L$ for $L=1$ with periodic boundary conditions which, given an integer $n$ setting the resolution level, we populate with $4n \times n \times n$ equal-mass particles positioned on the vertices of a regular cubic lattice; we here take $n=128$, which corresponds to the second to largest resolution level considered by~\citet{2023MNRAS.523.1280B} and is similar to that used for the SPH runs presented in~\citet{2007MNRAS.380..963A}.~\citet{2016MNRAS.455...51H}, the only other study that to our knowledge tests an SPMHD solver's performance on this problem, does not report the number of resolution elements that was employed. We carve out a sphere of radius $R_\text{cloud} = 0.1$ centred at $\boldsymbol{r}_\text{cloud} = (0.5, 0.5, 0.5)$ and populate it with equal-mass particles packed in a second, denser regular cubic lattice with $\chi^{- 1/3}$ times the inter-vertex separation of the former, where
\begin{equation}
    \chi \equiv \frac{\rho_\text{cloud}}{\rho_\text{wind}}
\end{equation}
is the `cloud-to-wind' mass density ratio, which we initialise to $\chi = 10$ by setting an appropriate particle mass such that $\rho_\text{cloud} = 10$ and $\rho_\text{wind} = 1$. Keeping cloud particles static, we initialise velocities of wind particles to $\boldsymbol{v}_\text{wind} = v_0 \hat{\boldsymbol{e}}_x$ with $v_0 = 1.5$. We set thermal energies so that the cold cloud and hot wind are initially in pressure equilibrium, and so that the resulting wind sound speed $c_{s, \text{wind}}$ yields a mach number $\mathcal{M}_\text{wind}$ of
\begin{equation}
    \mathcal{M}_\text{wind} \equiv \frac{v_0}{c_{s, \text{wind}}} = v_0
\end{equation}
\citet{2023MNRAS.523.1280B} argued for this choice of $\mathcal{M}$ on the grounds that it aligns with adopted conventions in recent research on the topic~\citep[][]{2018MNRAS.480L.111G, 2019MNRAS.482.5401S}. It is nevertheless worth mentioning that it is lower than the fiducial $\mathcal{M}_\text{fid}=2.7$ used by a large number of the aforementioned works. In contrast to results presented in these, we can therefore anticipate a milder evolution of the system.
We permeate the box by a constant, uniform magnetic field $\boldsymbol{B} = B_0 \hat{\boldsymbol{e}}_z$, where $B_0$ is a constant setting its relative dynamical importance. We chose it such that the resulting plasma beta matches a desired value; in the present work we examine $\beta = \infty, 250, 25$, corresponding to the hydrodynamic, weak field and strong field regimes respectively. Defining the characteristic `cloud-crushing' timescale~\citep[][]{2007MNRAS.380..963A, 2023MNRAS.523.1280B}
\begin{equation}
    \tau_\text{cc} = \frac{\sqrt{\chi} R_\text{cloud}}{v_\text{wind}}
\end{equation}
which corresponds to the time it takes for the incoming shock to traverse half the cloud (and is - up to a corrective additive factor of $+1$ in the numerator - identical to equation~(\ref{eq:KHI_timescale}) with $v_0$ as the shear and $R_\text{cloud}$ as the perturbation wavelength), we evolve the system until $t = 20 \tau_\text{cc}$ for the three $\beta$ considered.

\begin{figure*}
    \includegraphics[width=\textwidth]{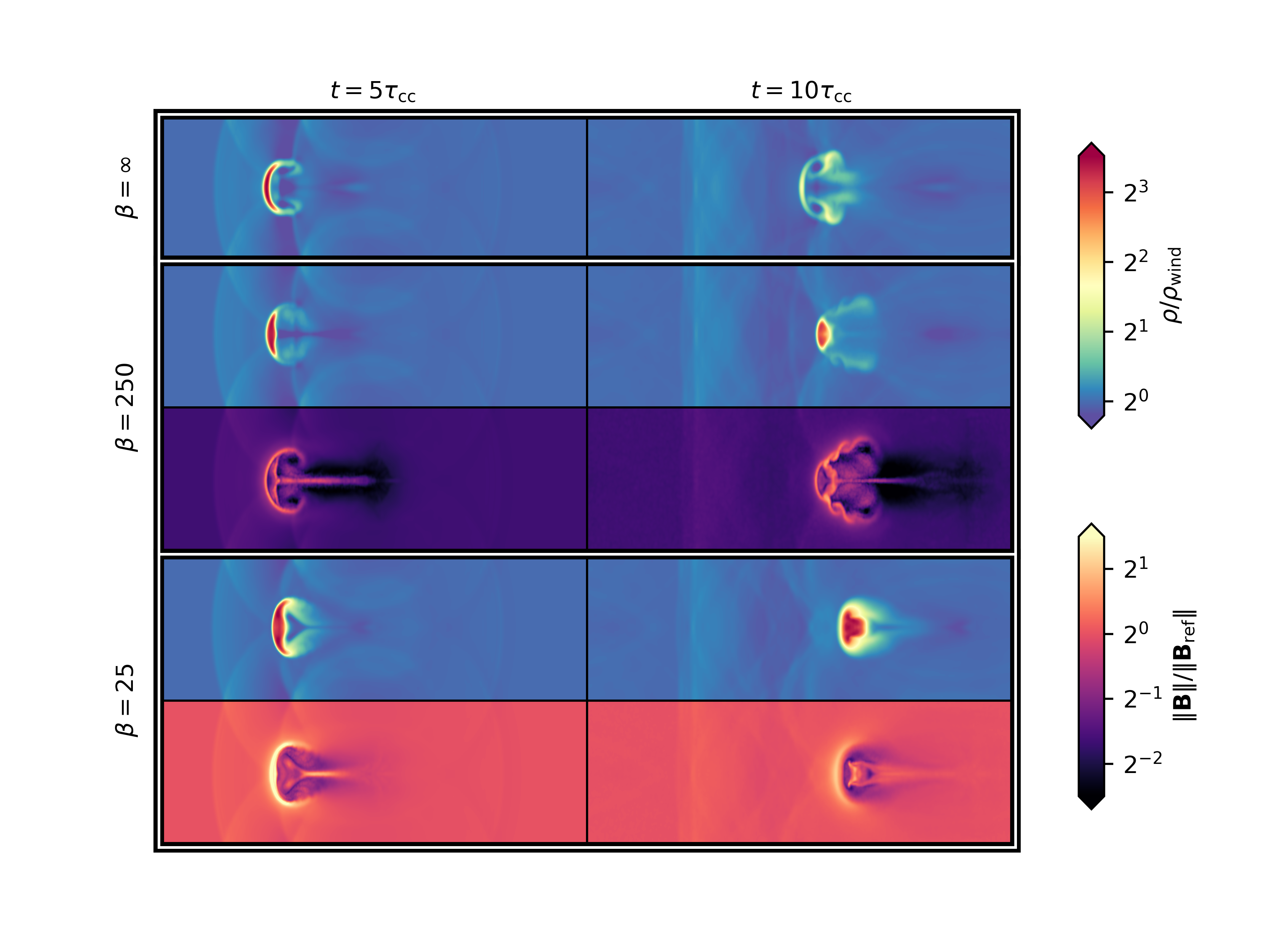}
    \caption{Results for the magnetic cloud-wind interaction test at $t=5\tau_{cc}$ (left column) and $t=10\tau_{cc}$ (right column), run at our fiducial resolution of $n = 128$ particles in the ambient medium along the simulation domain's short side (of length $L = 1$ in our code units). The assumed cloud-to-wind mass density ratio is $\chi=10$, and we have considered three different values for the strength of the uniform and constant seed magnetic field pointing in the $z$ direction, corresponding to a plasma beta of $\beta=\infty$ ($1^\text{st}$ row; the hydrodynamics case), $\beta=250$ ($2^\text{nd}$ and $3^\text{rd}$ row; the weak field case) and $\beta=250$ ($4^\text{th}$ and $5^\text{th}$ row; the strong field case). We show slices in the $x-y$ plane through $z=L/2$ of the mass density $\rho$ ($1^\text{st}$, $2^\text{nd}$ and $4^\text{th}$ row) normalised to the ambient medium's initial mass density $\rho_\text{wind}$, and slices of the magnetic field strength $\Vert \boldsymbol{B} \Vert$ ($3^\text{rd}$ and $5^\text{th}$ row) normalised to the strength of the $\beta=25$ seed field $\Vert \boldsymbol{B}_\text{ref} \Vert$. Differences are still minor at $t=5 \tau_\text{cc}$, as an upstream bow shock is established and the originally spherical cloud compressed into a crescent; its thickness nevertheless increases and the degree of phase mixing is reduced as $\beta$ is decreased. By $t=10 \tau_\text{cc}$ departures are much more pronounced; higher initial magnetisation leads to a higher survival rate for dense gas, a more laminar flow and a larger drag force acting on the cloud. The magnetic field is amplified at the surface boundary and forms a protective layer that drapes the cloud. The maximal field strength attained scales inversely with $\beta$, while magnetic energy is depleted from the cold core and linear magnetic structures form post-shock.}
    \label{fig:blob_projections}
\end{figure*}

We begin by showing mass density and magnetic field strength slices through the box midplane at $z=0.5L$ for our three runs at $t = 5 \tau_\text{cc}$ and $t = 10 \tau_\text{cc}$ in Fig.~\ref{fig:blob_projections}. Densities are normalised to the initial density of the ambient medium, while magnetic field strengths are normalised to the initial field strength of the $\beta = 25$ run. As far as the spatial distribution of mass is concerned, results from all three simulations do not vary vastly at early times; the initially spherical blob is `crushed' into a crescent shaped plate as the incident forward shock traverses it. In all cases, a bow shock is established upstream of the cloud which wraps around the periodic $y$ and $z$ directions (a feature confirmed not to affect results by~\citet{2023MNRAS.523.1280B}), while the cold gas layer's thickness and surface area in the examined cross section (moderately) increase as $\beta$ is lowered. There is nevertheless already unequivocal evidence of a progressively higher degree of mixing as $\beta$ is increased; having a weaker magnetic stabilising force facilitates the excitation and development of KH and RT modes which trigger dissolution, most evidently illustrated through the development of two trailing vortices on either vertical extremity of the cloud in the $\beta = \infty$ run (cold gas is depleted through these two points as this is where velocity is maximal and, by Bernoulli's principle~\citep[see e.g.][]{2016JPlPh..82c2001O}, pressure is minimal). By $t = 10 \tau_{cc}$, the hydrodynamics-only run shows a larger fraction of cold material that has been stripped, as well as a higher degree of mass diffusion. In the weakly magnetised case, a small dense cool core has survived which appears to nevertheless be continuously shredded through its edges by mixing instabilities that give rise to a downstream turbulent flow. The highly magnetised cloud, in turn, seems to have retained a considerable portion of its mass, behind which emerges a more ordered, laminar profile. Stronger magnetic fields seem to systematically lead to a stronger drag force acting on the cloud, as lower $\beta$ runs show it further to the right at late times when compared to the fiducial $\beta=\infty$ simulation. The observed trends mentioned above are consistent with AMR results~\citep[see e.g.][]{2008ApJ...680..336S} and are magnetohydrodynamical in nature; in the magnetised runs, the seed magnetic field is amplified at the cloud interface, forming a protective sheath, and reaches a maximal value that scales inversely with $\beta$. The magnetic energy within the cold gas is itself eventually suppressed, while a post shock filamentary magnetic structure is observed for both $\beta=250$ and $\beta=25$.

\begin{figure}
    \includegraphics[width=.99\columnwidth]{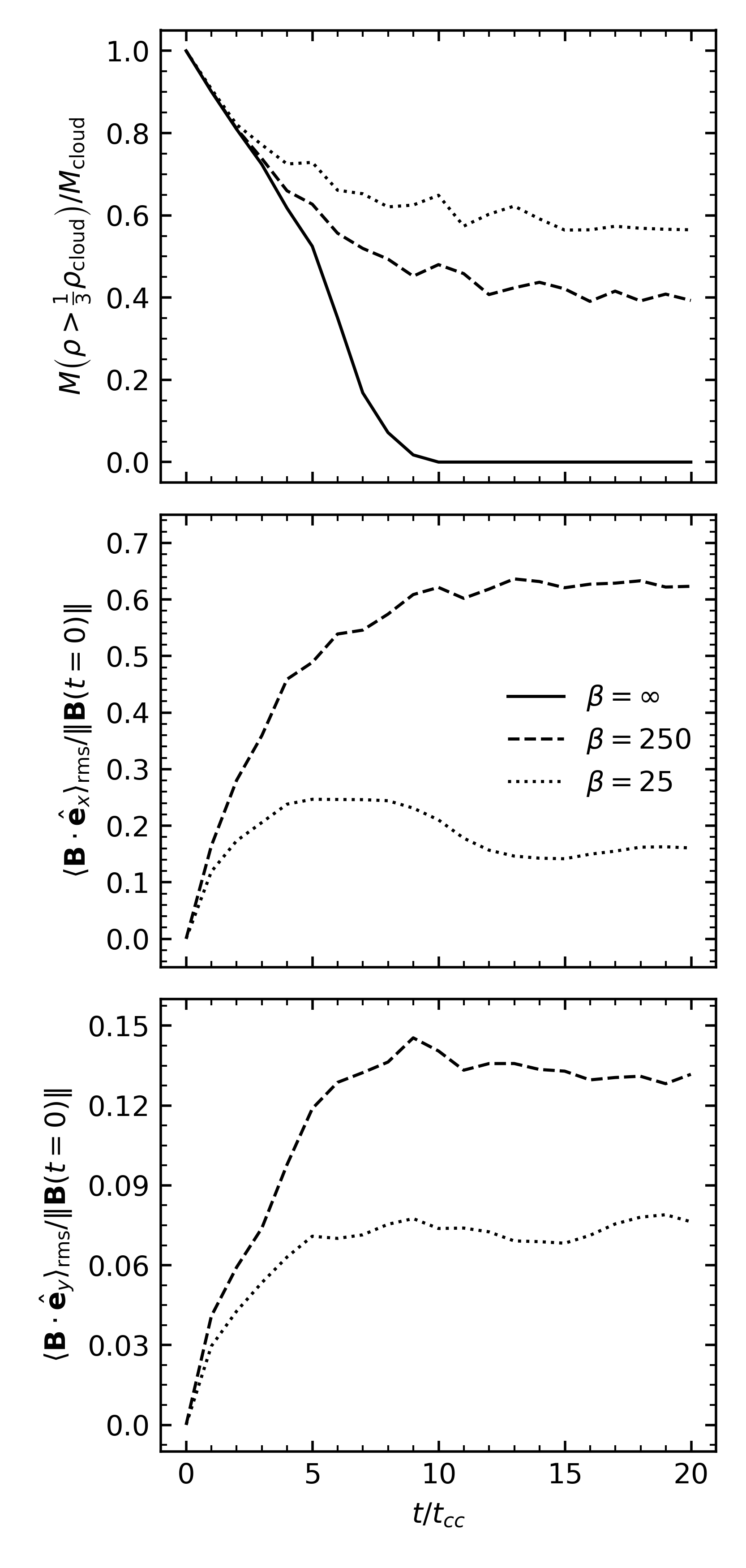}
    \caption{Time evolution of selected global simulation metrics for the cloud-wind interaction test. We differentiate between results from simulations run with a different initial plasma beta by line style (we plot $\beta = \infty$ results as solid lines, $\beta = 250$ results as dashed lines, and $\beta=25$ results as dotted lines). We show the surviving cloud mass (top panel) computed according to equation~(\ref{eq:cloud_wind_int_dense_gas}) and normalised to the cloud's original mass, as well as the quadratic means over all particles of the $x$ and $y$ components of the magnetic field normalised to its strength at $t=0$ (middle and bottom panels respectively). Differences become visible only after $t \sim \tau_\text{cc}$; larger initial $\beta$ leads to a larger fraction of dense gas surviving, while seed magnetic fields are proportionally more amplified the weaker they originally were (most notably in the case of the magnetic field component aligned with the wind velocity shown in the middle panel).}
    \label{fig:blob_metrics_evolution}
\end{figure}

To paint a fuller, more quantitative picture, we further consider the time evolution of three global metrics of interest. The first such quantity we track is the surviving cloud mass $M_\text{dense gas}$, which we define as (following the criterion proposed by~\citealt{2023MNRAS.523.1280B}, in accordance with conventions used in recent comparable works)
\begin{equation}
    M_\text{dense gas}
    =
    \sum_{p_i \in \text{ box}} 
    m_i  
    \left[ 
    \frac{\rho_i}{\rho_\text{cloud}}
     \in \left( \frac{1}{3}, \infty \right) 
     \right]
\label{eq:cloud_wind_int_dense_gas}
\end{equation}
where the Iverson bracket $[\cdot]$ returns $1$ if the statement within it is true and $0$ otherwise, to sum up all individual masses $m_i$ of particles $p_i$ within the simulation volume whose densities $\rho_i$ are larger than $1/3$ that of the cloud at $t=0$. We further consider the quadratic mean over all particles of the two components of the magnetic field orthogonal to $\boldsymbol{B} (t=0)$ (the evolution of the component parallel to it being solely dictated by compression engendered by the propagating shock, and thus being commonly omitted from such analyses), and plot everything in Fig.~\ref{fig:blob_metrics_evolution} as a function of time for each of our three simulations. We normalise $M_\text{dense gas}$ to the cloud's initial mass, and normalise the root mean square of $\boldsymbol{B}$ components for each run to the strength of the respective $\boldsymbol{B} (t=0)$. While up until $t \sim \tau_{cc}$ none of the examined metrics seem to discriminate between different $\beta$, the late time evolution is unquestionably different; stronger initial magnetic fields translate to a larger fraction of dense gas surviving. Weaker magnetic fields are proportionally more strongly amplified, with the longitudinal component (i.e. the component parallel to the wind velocity) showing more growth compared to the remaining transverse component (i.e. the component perpendicular to both the wind velocity and seed field), the former as opposed to the latter being stretched over the cloud phase boundary by the shear flow. These observations are in good agreement with the results of~\citet{2008ApJ...680..336S}.

\subsection{Non-ideal MHD test problems}
\label{sec:non_ideal_mhd_tests}

In this section, we present a series of test problems that were run to validate our treatment of non-ideal MHD terms. A distinct suite of numerical experiments is necessary for this purpose, as non-ideal terms involve discrete second-order differential operators acting on vector fields~\citep[for a discussion of the associated increment in complexity and interpretability when confronted with these see e.g.][]{2003PhRvE..67b6705E}; all method components tested thus far represented first derivatives at most. What follows not only allows us to confirm SWIFT performs non-ideal MHD calculations accurately, but also that these are correctly accounted for by the time-integration algorithm and therefore stable. Here, we reproduce three conceptually simple tests that were used in benchmarking codes targeting similar science cases as SWIFT (see e.g.~\citealt{2012ApJS..201...24M} and~\citealt{2024MNRAS.527.8355Z} for RAMSES and AREPO, respectively). For more involved and novel applications of SWIFT to non-ideal MHD problems, we redirect the reader to e.g.~\citet{2025MNRAS.541.1507K} who derived closed-form expressions describing the time evolution of diffusive Alfvén waves in a cosmologically expanding frame, and showed how these were in excellent agreement with numerical results obtained with SWIFT. Looking at more complex test problems, \citet{2025MNRAS.541.3427S} demonstrated that our method accurately reproduces a series of well-established kinematic dynamo benchmarks including the first Roberts flow, a large-scale slow dynamo illustrating the $\alpha$-effect, as well as the ABC flow, a fast dynamo that serves as an idealised single-scale model of small-scale magnetic field amplification. That work further presented an in-depth investigation of the relationship between resistivity and resolution in SPMHD.

\subsubsection{Diffusion of a magnetic pulse}
\label{sec:gaussian_pulse}

Similarly to~\citet{2012ApJS..201...24M, 2013MNRAS.434.2593T, 2018MNRAS.476.2476M} or~\citet{2024MNRAS.527.8355Z}, in order to first test the non-ideal terms independently, we consider the diffusive evolution of magnetic impulse perturbations on top of a uniform background for a fluid that is assumed to be static.

To emulate a formally static fluid we fix particle velocities to $\boldsymbol{v} = \boldsymbol{0}$ at all times. Doing so results in particles having constant mass density (it being a function of position alone in SPH) while the thermal pressure, although not strictly time invariant given the presence of Joule heating, can only indirectly affect the calculation through setting the time-step~\citep[this is in contrast to e.g.][who explicitly fix all these quantities]{2012ApJS..201...24M}. In a uniform fluid configuration, velocities being zero reduces the induction equation to the heat equation
\begin{equation}
    \frac{\mathrm{d} \boldsymbol{B}}{\mathrm{d} t}
    = \eta \nabla^2 \boldsymbol{B}
\label{eq:heat_equation}
\end{equation}
with the constant $\eta$ as diffusion coefficient. Any unidirectional Dirac pulse that is then induced in the magnetic field is constrained by the linearity of equation~(\ref{eq:heat_equation}) and the solenoidal condition to vary at most along the two coordinate directions perpendicular to it. Such a pulse can thus take either of two forms, which w.l.o.g. we can express as
\begin{equation}
    \boldsymbol{B}_\text{1D pulse} =
    \delta (x) \hat{\boldsymbol{e}}_z
\label{eq:1D_pulse_IC}
\end{equation}
and
\begin{equation}
    \boldsymbol{B}_\text{2D pulse} =
    \delta (x) \delta (y) \hat{\boldsymbol{e}}_z
\label{eq:2D_pulse_IC}
\end{equation}
corresponding to a 1D and 2D profile respectively, where $\delta ( \cdot )$ is the Dirac delta function. If introduced at a point $\boldsymbol{r}_c = (x_c, y_c, z_c)$ at time $t=0$, the pulses~(\ref{eq:1D_pulse_IC}) and~(\ref{eq:2D_pulse_IC}) will subsequently evolve according to the well known diffusive Gaussian solutions to equation~(\ref{eq:heat_equation})
\begin{equation}
    \boldsymbol{B}_\text{1D pulse} (\boldsymbol{r}, t; \boldsymbol{r}_c) = 
    \frac{1}{\sqrt{4 \pi \eta t}} 
    \mathrm{exp} \left(
    - \frac{{(x - x_c)}^2}{4 \eta t}
    \right)
    \hat{\boldsymbol{e}}_z
\label{eq:1D_pulse}
\end{equation}
and
\begin{equation}
    \boldsymbol{B}_\text{2D pulse} (\boldsymbol{r}, t; \boldsymbol{r}_c) = 
    \frac{1}{4 \pi \eta t} 
    \mathrm{exp} \left(
    - \frac{{(x - x_c)}^2 + {(y - y_c)}^2}{4 \eta t}
    \right)
    \hat{\boldsymbol{e}}_z
\label{eq:2D_pulse}
\end{equation}
respectively.

To simulate the diffusion of such pulses we begin by instantiating a cubic box of side length $L=1$, which we populate with $n^3$ particles positioned on the vertices of a regular cubic lattice. Boundary conditions are taken to be periodic. We assign constant particle masses and thermal energies to obtain a uniform density and pressure field with $\rho=1$ and $P=1$. We note that none of these choices affect our calculations and only mention them for completeness. Velocities and magnetic fields are first initialised to $\boldsymbol{v} = \boldsymbol{0}$ and $\boldsymbol{B} = \boldsymbol{0}$, while the magnetic diffusivity is set to $\eta = 1$. We take our runs to start at a finite time $t_0 = 10^{-3}$, so that we can introduce pulses in our initial conditions that can be expressed analytically according to equations~(\ref{eq:1D_pulse}) and~(\ref{eq:2D_pulse}); we place them in the centre of the simulation box $\boldsymbol{r}_\text{centre} = (L/2, L/2, L/2)$ by assigning individual particle magnetic fields according to $\boldsymbol{B}_\text{1D pulse}(\boldsymbol{r},t_0;\boldsymbol{r}_\text{centre})$ or $\boldsymbol{B}_\text{2D pulse}(\boldsymbol{r},t_0;\boldsymbol{r}_\text{centre})$. We evolve the system until $t=6 t_0$ for each of the two pulse profiles considered, repeating each experiment at the three resolution levels corresponding to $n=16,32,64$. We do not disable any of our corrective measures, so as to verify that neither our artificial resistivity nor divergence-cleaning prescription leads to excess magnetic diffusion.

\begin{figure*}
 \includegraphics[width=\textwidth]{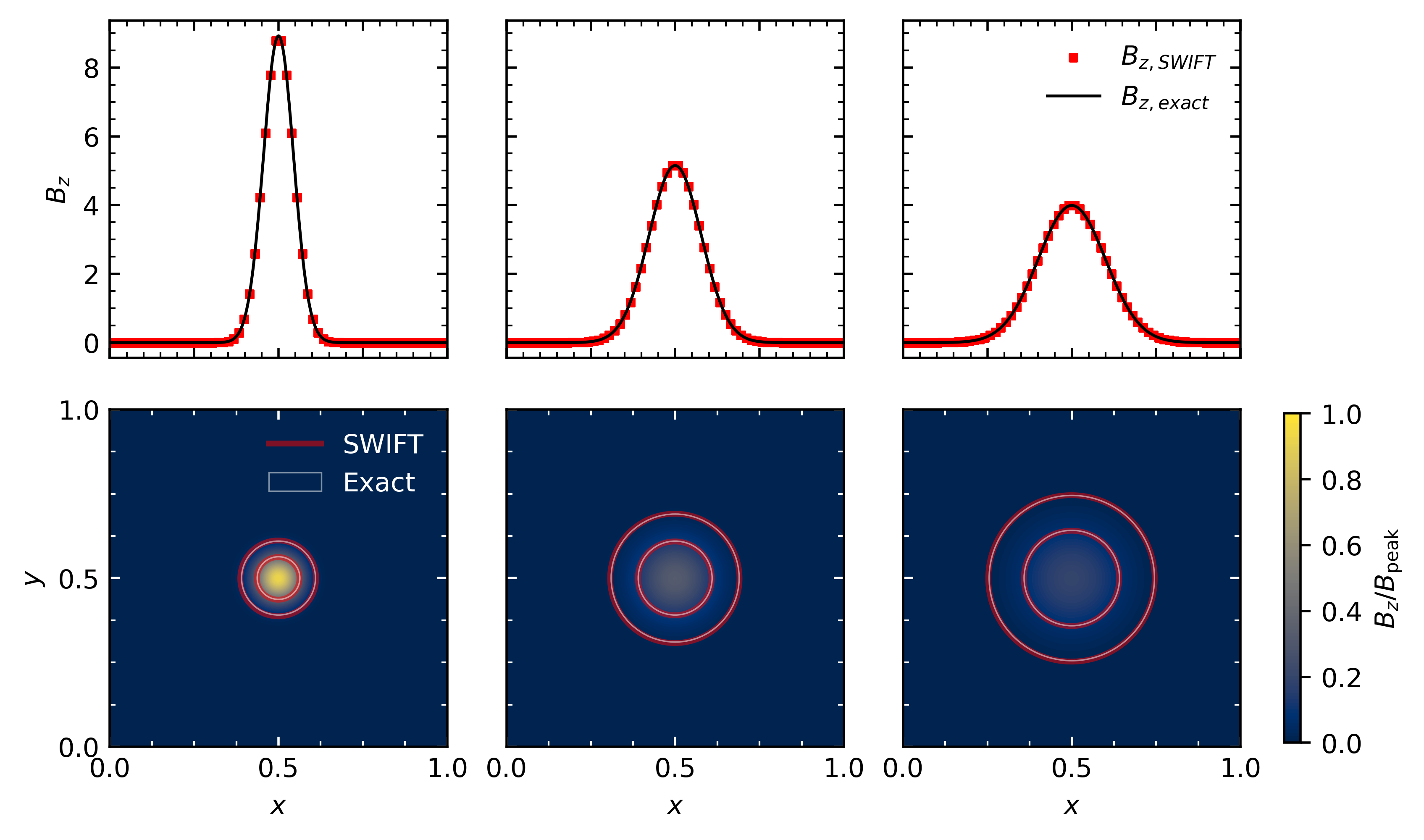} 
 \caption{Results for the magnetic pulse diffusion test, for a pulse along the $z$ direction with a profile that is 1D (top row) and 2D (bottom row), resolved in either case by ${64}^3$ particles in a cubic box of unit side length. The $1^\text{st}$, $2^\text{nd}$ and $3^\text{rd}$ columns show results at $t=2t_0$, $t=4t_0$ and $t=6t_0$ respectively, where $t_0={10}^{-3}$ is the physical time at the start of each simulation. For the 1D case, we overplot the $B_z$ of all particles in red with the expected solution in black. For the 2D case, we show projections in the $x-y$ plane of $B_z$ normalised to the pulse's initial height $B_\text{peak}$; we overplot measured and exact contours where the value of $B_z$ corresponds to that expected at a distance of $1$ and $2$ standard deviations from a Gaussian pulse's centre. Numerical and analytical predictions agree exceptionally well, with SWIFT results preserving the problems' original symmetry and showing no excess magnetic diffusion due to either artificial resistivity or divergence cleaning.}
 \label{fig:gaussian_pulse}
\end{figure*}

We present results for our highest resolution runs of the 1D and 2D pulses at three consecutive moments in time in Fig.~\ref{fig:gaussian_pulse}. We show readings of $B_z$, as a function of $x$ for \textit{all} particles in the 1D pulse run, and as projections in the $x-y$ plane for the 2D pulse run. For the former, we overplot SWIFT results with the associated analytic prediction. For the latter, taking the standard deviation of the model Gaussian distribution in $B_z$ at time $t$ to be $\sigma (t) = \sqrt{4 \eta t}$ as per equation~(\ref{eq:2D_pulse}), we show measured and exact contours where $B_z$ attains the value expected at a distance $\sigma(t)$ and $2 \sigma(t)$ from the pulse's centre. In either case, numerical and analytical results agree exceptionally well, and the symmetry of the problem is perfectly preserved. We further calculate for the final snapshot of all our runs an $\mathcal{L}_1$ norm on $B_z$ over all particles, which we plot as a function of the corresponding resolution expressed in terms of the smoothing length in Fig.~\ref{fig:gaussian_pulse}. The error appears to converge to greater than second order, and is smaller than $1\%$ and $0.1\%$ for the 1D and 2D pulses, respectively, even at our lowest resolution level where the feature of interest is resolved by a mere 16 particles. We attribute the systematic difference in $\mathcal{L}_1 (B_z)$ at fixed resolution between the two case studies investigated to the comparatively smaller spatial support of the 2D pulse; smaller support means a larger fraction of particles having negligible $B_z$ and contributing little to the error. SWIFT marginally outperforms both RAMSES~\citep{2012ApJS..201...24M} and AREPO~\citep{2018MNRAS.476.2476M} in terms of their quoted convergence behaviour on this problem. We nevertheless stress that an equally sound but qualitatively different strategy has been adopted to compute the error in each case, and that a method comparison based on such an artificial test is of limited value anyway. The only takeaway we aim to draw attention to is that the discrete second-order differential operator we use to model non-ideal MHD effects behaves as expected and reproduces known results.
\begin{figure}
 \includegraphics[width=.99\columnwidth]{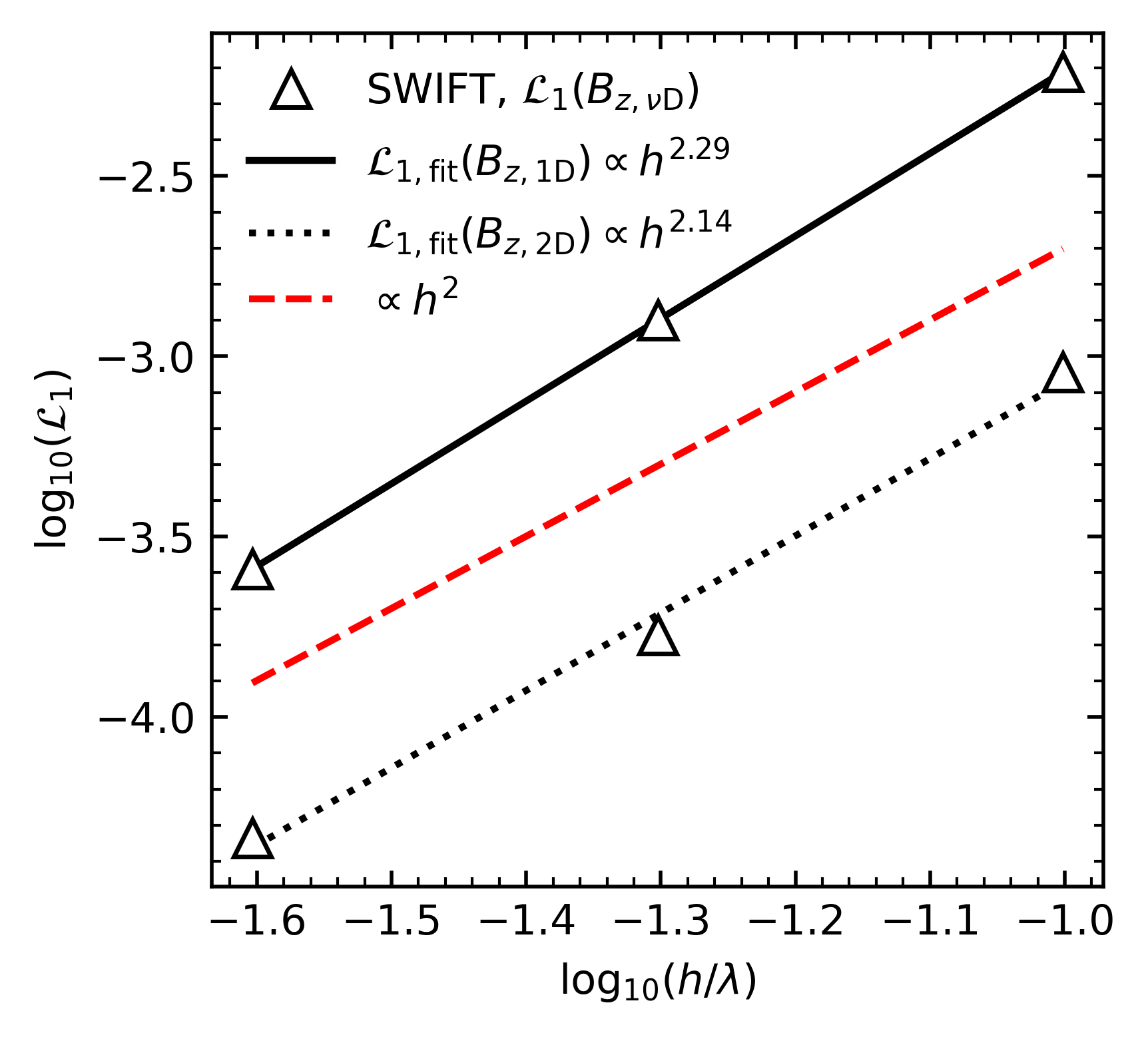} 
 \caption{Convergence study for both the 1D and 2D magnetic pulse diffusion tests. We show an $\mathcal{L}_1$ error on $B_z$ (triangle markers) computed over all particles in a given simulation at $t=6t_0$, for each of the three resolution levels considered for either variant of the problem. We plot them as a function of resolution expressed in terms of the box-averaged smoothing length $h$; we normalise the latter to a characteristic scale $\lambda$ of the set-up, taken here to be the box side length $L=1$. We performed linear regression in logarithmic space on $\{h(n), \mathcal{L}_1 (n | B_z)\}$ where $n$ is the number of resolution elements along the cubic simulation domain's side. We show the resulting fit through the points corresponding to the 1D pulse runs (solid black line), as well as the fit through the points corresponding to the 2D pulse runs (dotted black line), together with the associated scaling laws in the legend. We complement this with a line (red dashed) to guide the eye, corresponding to quadratic convergence. The error levels are particularly low for all six runs undertaken. Our non-ideal MHD terms, when tested independently, converge to beyond second order with resolution.}
 \label{fig:gaussian_pulse_convergence}
\end{figure}

\subsubsection{Diffusive Alfvén waves}
We next proceed with a couple of benchmarking exercises aimed at evaluating the accuracy and robustness of our non-ideal terms in the presence of a full magnetohydrodynamical flow. Following again~\citet{2012ApJS..201...24M, 2018MNRAS.476.2476M} and~\citet{2024MNRAS.527.8355Z}, we consider the case of both propagating and standing Alfvén waves in a homogeneous background medium subject to Ohmic diffusion. We more particularly focus on perturbative solutions to the equations of non-ideal MHD that correspond to incompressible single Fourier modes that are perpendicular to a constant background magnetic field. One such class of modes takes the form of torsional Alfvén waves that are circularly polarised, and constitute a subclass of the general family of solutions presented in~\citet{2007MNRAS.381..319L}.

We consider a uniform unperturbed state with constant mass density and thermal pressure~\citep[we use the same problem parameters as][]{2018MNRAS.476.2476M, 2024MNRAS.527.8355Z}
\begin{equation}
    \rho_0 = 1 \quad \text{and} \quad P_0 = 1
\end{equation}
The background velocity $\boldsymbol{v}_0$ and magnetic field $\boldsymbol{B}_0$ are fixed to
\begin{equation}
    \boldsymbol{v}_0 = \boldsymbol{0}
    \quad \text{and} \quad
    \boldsymbol{B}_0 = B_0 \hat{\boldsymbol{e}}_z
\end{equation}
where $B_0$ is a constant we set to $B_0=1$. 
On top of these, we respectively induce perturbations $\delta \boldsymbol{v}$ and $\delta \boldsymbol{B}$ propagating parallel to $\boldsymbol{B}_0$ which we express as
\begin{equation}
    \delta \boldsymbol{v}_{s,k} (\boldsymbol{r}, t)
    = \delta \boldsymbol{v}_0 e^{st + ikz}
    \quad \text{and} \quad
    \delta \boldsymbol{B}_{s,k} (\boldsymbol{r}, t)
    = \delta \boldsymbol{B}_0 e^{st + ikz}
\label{eq:diffusive_alfven_waves_general_dB_and_dv}
\end{equation}
where $\delta \boldsymbol{v}_0$ and $\delta \boldsymbol{B}_0$ are constant complex vectors, $s$ is the complex valued angular frequency and $k$ the real valued wave number. Assuming $\delta \boldsymbol{v}_0$ and $\delta \boldsymbol{B}_0$ are transverse to $\boldsymbol{B}_0$, meaning $\delta \boldsymbol{v}_0 \cdot \boldsymbol{B}_0 = 0$ and $\delta \boldsymbol{B}_0 \cdot \boldsymbol{B}_0 = 0$, travelling torsional Alfvén wave solutions that are planar and circularly polarised are obtained iff $\delta \boldsymbol{B}_0$ is constrained to be of the form~\citep[see section 4.1 of][]{2007MNRAS.381..319L}
\begin{equation}
    \delta \boldsymbol{B}_0^\text{travel.} =
    \delta B_0 \hat{\boldsymbol{e}}_x
    + i \delta B_0 \hat{\boldsymbol{e}}_y
\label{eq:diffusive_alfven_waves_specific_dB}
\end{equation}
where $\delta B_0$ is a real valued constant. Propagating~(\ref{eq:diffusive_alfven_waves_general_dB_and_dv}) and~(\ref{eq:diffusive_alfven_waves_specific_dB}) through the continuum equations of non-ideal MHD, one finds that the associated velocity perturbation reads
\begin{equation}
    \delta \boldsymbol{v}_0^\text{travel.} = 
    \frac{i k B_0}{\mu_0 \rho_0 s}
    \delta \boldsymbol{B}_0^\text{travel.}
\end{equation}
and that the angular frequency has to take either of two values given by the dispersion relation
\begin{equation}
    s_\pm = - \tilde{\eta} \pm
    \sqrt{\tilde{\eta}^2 - k^2 v_{A,0}^2}
\label{eq:diffusive_alfven_waves_dispersion_relation}
\end{equation}
where we have defined the proxy variable $\tilde{\eta} \equiv \eta k^2 / 2$ and the Alfvén speed of the unperturbed medium $v_{A,0} \equiv B_0 / \sqrt{\mu_0 \rho_0}$. Moreover, the same exercise shows that if (as is the case here) the mass density field is initially uniform, it remains constant in both space and time. 

We focus here on the weakly diffusive regime whereby
\begin{equation}
    \tilde{\eta} < k v_{A,0}
\label{eq:weakly_diffusive_regime_condition}
\end{equation}
and, by the dispersion relation~(\ref{eq:diffusive_alfven_waves_dispersion_relation}), the velocity and magnetic field perturbations take the form of the dampened travelling waves
\begin{equation}
    \delta \boldsymbol{v}_{s_\pm,k}^\text{travel.} (\boldsymbol{r}, t)
    = \delta \boldsymbol{v}_0^\text{travel.} e^{s_\pm t + ikz}
\label{eq:diffusive_alfven_waves_traveling_dv}
\end{equation}
and
\begin{equation}
    \delta \boldsymbol{B}_{s_\pm,k}^\text{travel.} (\boldsymbol{r}, t)
    = \delta \boldsymbol{B}_0^\text{travel.} e^{s_\pm t + ikz}
\label{eq:diffusive_alfven_waves_traveling_dB}
\end{equation}
The qualitative behaviour of such solutions is made evidently apparent by expressing the angular frequency $s$ in terms of its real and imaginary parts $s_r$ and $s_i$ as
\begin{equation}
    s_\pm \equiv s_r \pm i s_i \text{,}
    \quad \text{where} \quad
    s_r = -\tilde{\eta}
    \quad \text{and} \quad
    s_i = \sqrt{k^2 v_{A,0}^2 - \tilde{\eta}^2}
\end{equation}
The magnetic diffusivity $\eta$ being proportional to $\tilde{\eta}$, it is then clear that it being non-zero induces both a diffusive effect through $s_r$ as well as a dispersive effect through $s_i$ on both $\delta \boldsymbol{v}_{s_\pm,k}^\text{travel.}$ and $\delta \boldsymbol{B}_{s_\pm,k}^\text{travel.}$. Finally, it is also trivial to set up a standing wave by superimposing two such travelling waves with identical $s_r$ and opposite-valued $s_i$ as
\begin{equation}
    \delta \boldsymbol{v}_{s_\pm,k}^\text{stand.} (\boldsymbol{r}, t)
    = \frac{1}{2}
    \left[
    \delta \boldsymbol{v}_{s_+,k}^\text{travel.} (\boldsymbol{r}, t)
    + \delta \boldsymbol{v}_{s_-,k}^\text{travel.} (\boldsymbol{r}, t)
    \right]
\label{eq:diffusive_alfven_waves_standing_dv}
\end{equation}
and
\begin{equation}
    \delta \boldsymbol{B}_{s_\pm,k}^\text{stand.} (\boldsymbol{r}, t)
    = \frac{1}{2}
    \left[
    \delta \boldsymbol{B}_{s_+,k}^\text{travel.} (\boldsymbol{r}, t)
    + \delta \boldsymbol{B}_{s_-,k}^\text{travel.} (\boldsymbol{r}, t)
    \right]
\label{eq:diffusive_alfven_waves_standing_dB}
\end{equation}
\citet{2007MNRAS.381..319L} and \citet{2012ApJS..201...24M} showed that the solutions we detail here maintain a homogeneous thermal pressure field, which is, however, not time invariant as spatially uniform Joule heating transforms magnetic into thermal energy. They derived closed-form expressions for the pressure as a function of time, for both the case of travelling and standing Alfvén waves, which respectively read
\begin{equation}
    \begin{split}
        P^\text{travel.} (t)
        = 
        P_0 
        & + (\gamma - 1) \tilde{\eta} {\delta B_0^2}
            \frac{e^{2 s_r t} - 1}{s_r}
    \end{split}
\label{eq:diffusive_alfven_waves_traveling_P}
\end{equation}
and
\begin{equation}
    \begin{split}
        P^\text{stand.} (t)
        = 
        P_0 
        & + \frac{1}{2}(\gamma - 1) \tilde{\eta} {\delta B_0^2}
        \Biggl[
            \frac{e^{2 s_r t} - 1}{s_r} \\
             & + e^{2 s_r t}
            \left(
                \frac{s_r \text{cos}(2 s_i t) + s_i \text{sin}(2 s_i t)}
                     {{\Vert s \Vert}^2}
            \right)
            - \frac{s_r}{\Vert s \Vert}
        \Biggr]
        \text{.}
    \end{split}
\label{eq:diffusive_alfven_waves_standing_P}
\end{equation}
Both $P^\text{travel.} (t)$ and $P^\text{stand.} (t)$ can trivially be shown to be monotonically increasing functions of time that tend towards some finite bound as $t \rightarrow \infty$.

To simulate both of the aforementioned types of diffusive Alfvén waves with SWIFT, we initialise cubic boxes of side length $L=1$ with periodic boundary conditions, within which we place $2n^3$ particles (for $n$ an integer), which we position on the vertices of a BCC lattice. First, constant individual masses, thermal energies, velocities and magnetic fields are assigned to all particles to reproduce the homogeneous, unperturbed base state described by the set of physical variables $(\rho_0, P_0, \boldsymbol{v}_0, \boldsymbol{B}_0)$. We subsequently perturb velocities and magnetic fields to seed waves with amplitude $\delta B_0 = 1$ and wave number $k = 2\pi$ (corresponding to a wavelength equal to the box size), as prescribed by the real part of equations~(\ref{eq:diffusive_alfven_waves_traveling_dv}) and~(\ref{eq:diffusive_alfven_waves_traveling_dB}) in the case of travelling waves (where we have taken w.l.o.g. $s = s_+$), and equations~(\ref{eq:diffusive_alfven_waves_standing_dv}) and~(\ref{eq:diffusive_alfven_waves_standing_dB}) in the case of standing waves. We fix the magnetic diffusivity to $\eta=0.02$ in all our experiments, which ensures that condition~(\ref{eq:weakly_diffusive_regime_condition}) is always satisfied given our other parameter choices; this consequently means that all expected solutions are of the forms described above. For both types of waves, we run simulations at the three resolution levels, corresponding to $n=24, 36, 48$, with all corrective measures enabled. In each case, we evolve the system until $t=4\sim4T$ where $T \equiv 2 \pi / s_i$ is the wave period, allowing ample time for diffusive terms to act and for their impact to be appreciable.

\begin{figure*}
 \includegraphics[width=\textwidth]{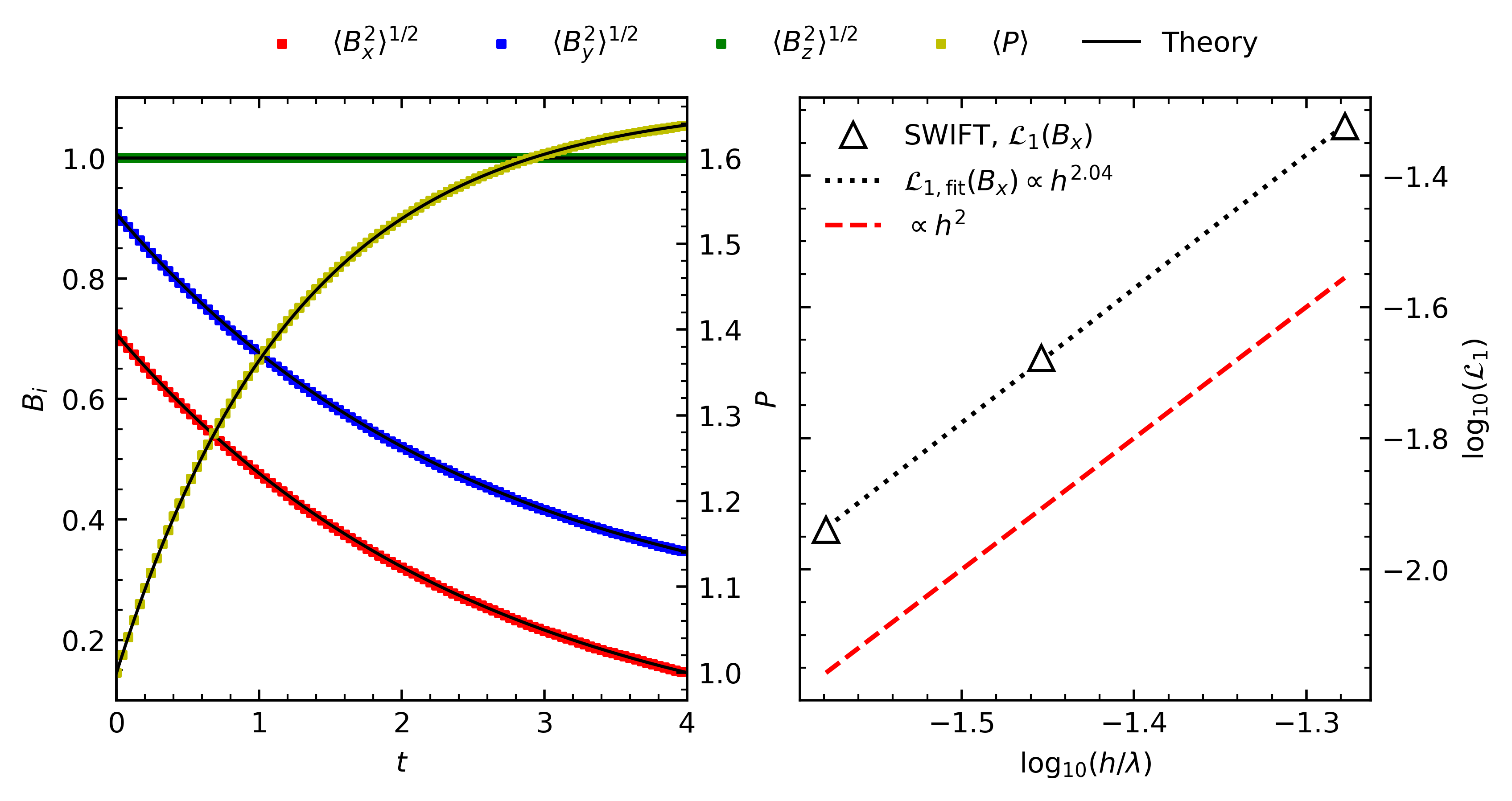} 
 \caption{Results for the travelling, circularly polarised, diffusive Alfvén wave test run with magnetic diffusivity set to $\eta=0.1$ and all our corrective measures enabled. Left: Time evolution of the system across $\sim4$ wave periods, corresponding to t=4 in our arbitrary system of units. We show time series of the root mean square of the three components of the magnetic field $\langle B_i^2 \rangle^{1/2}$, as well as of the average thermal pressure $\langle P \rangle$ (coloured markers, as per the legend). We extract these from consecutive simulation outputs from our highest resolution run ($n = 48$ resolution elements along each side of our cubic domain), averaging over all particles in each individual snapshot, and compare them to analytical expectations (solid black lines). We translate the time series for the transverse component $B_y$ vertically by 0.2, to distinguish it from $B_x$. Agreement between simulations and theory is excellent, with non-ideal MHD effects sourcing the diffusion of the transverse components of the magnetic field and a monotonic increase in thermal pressure through Joule heating. The longitudinal magnetic field, as expected, remains unaffected, and our corrective measures do not seem to source any excess diffusion. Right: $\mathcal{L}_1$ error (triangle markers) computed on the entirety of the spatial profile of $B_x$ at $t=4$ using all particles in the associated snapshots, plotted as a function of resolution expressed in terms of the box averaged smoothing length $h$; we normalise the latter to a characteristic scale of the problem, taken here to be the wavelength $\lambda=1$ of the Alfvén perturbation. We performed linear regression in logarithmic space on $\{ h(n), \mathcal{L}_1 (n |B_x)\}$ and show the resulting fit (dotted black line) and associated power law in the legend. We also include a line corresponding to quadratic scaling with $h$ (red dashed line) to guide the eye. Even in the presence of a full magnetohydrodynamical flow, our method with non-ideal MHD terms enabled displays convergence marginally beyond second order.} 
 \label{fig:traveling_diffusive_alfven_wave}
\end{figure*}

\begin{figure*}
 \includegraphics[width=\textwidth]{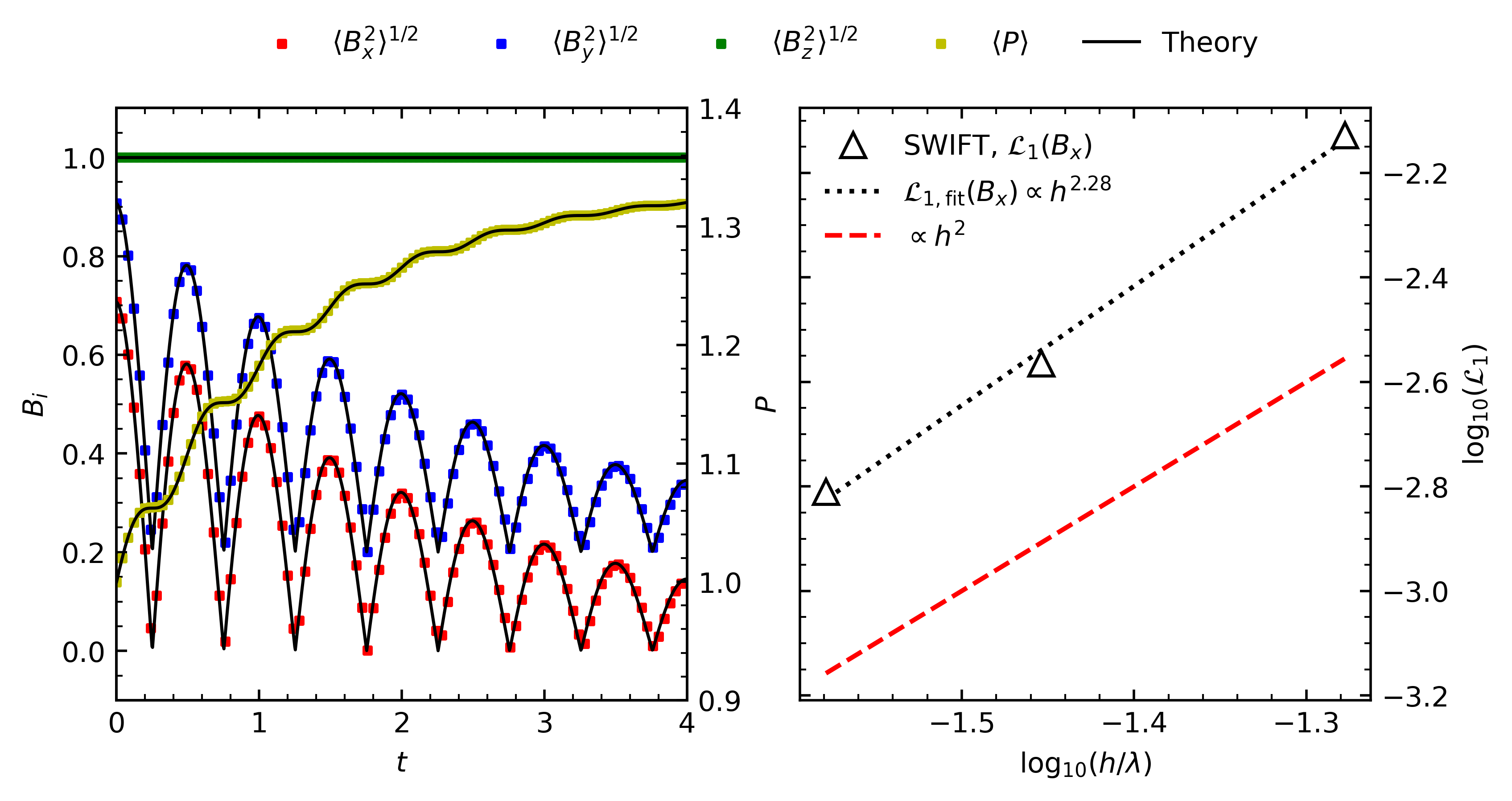} 
 \caption{The same as Fig.~\ref{fig:traveling_diffusive_alfven_wave}, but for the standing, circularly polarised, diffusive Alfvén wave test run with magnetic diffusivity set to $\eta=0.1$ and all our corrective measures enabled. The results are consistent with and corroborate the conclusion drawn from the travelling Alfvén wave test.}
 \label{fig:standing_diffusive_alfven_wave}
\end{figure*}

We present results for the travelling and standing diffusive Alfvén wave tests in Fig.~\ref{fig:traveling_diffusive_alfven_wave} and Fig.~\ref{fig:standing_diffusive_alfven_wave},respectively; although the exact magnetic field configuration and evolution history differ, conclusions drawn from either test are very similar. We calculate in both cases the root mean square of each magnetic field component ${\langle B_i^2 \rangle}^{1/2}$ and average thermal pressure $\langle P \rangle$ from the consecutive snapshots of our highest resolution runs, and plot them (in colour) as a function of time together with the corresponding exact solutions (in black). We accompany these with convergence studies showing an $\mathcal{L}_1$ error on $B_x$ computed over all particles at $t=4$ and plotted as a function of resolution; we choose not to repeat this for $B_y$ and $B_z$, as the former behaves qualitatively similarly to $B_x$ while the latter is of limited physical interest. We argue that our numerical results match the reference solutions, both at the level of fixed time spatial profiles, as well as at the level of time series of global simulation attributes. Through either lens we observe SWIFT predictions to be in excellent agreement with theory, with both transverse magnetic field components being diffused at the expected rates. The longitudinal field components remain constant as expected, whereas the thermal pressures increase monotonically in accordance with equations~(\ref{eq:diffusive_alfven_waves_traveling_P}) and~(\ref{eq:diffusive_alfven_waves_standing_P}). In good agreement with what was shown in Section~\ref{sec:gaussian_pulse}, our non-ideal MHD implementation displays (slightly) beyond second order convergence with resolution on this set of tests~\citep[this is again comparable to what is reported in][for their respective methods]{2012ApJS..201...24M, 2018MNRAS.476.2476M, 2024MNRAS.527.8355Z}. Coupling our non-ideal terms (tested independently in Section~\ref{sec:gaussian_pulse}) to a full magnetohydrodynamical flow has not brought about a degradation in accuracy, and our corrective measures seem once again not to have engendered any spurious diffusion; magnetic dissipation and Joule heating are here fully compatible with those expected for the $\eta$ chosen for our runs. We finally note that the method performance and convergence behaviour discussed in the present section are in perfect agreement with those seen for both the simpler, ideal MHD counterpart to the problem presented in Section~\ref{sec:circularly_polarised_alfven_wave}, as well as for the more involved cosmological, non-ideal MHD wave test suite explored in~\citet{2025MNRAS.541.1507K}.

\subsubsection{Oblique C-shock}
The final problem we evaluate our implementation of Ohmic diffusion against is that of propagating magnetic C-Shocks, a highly illustrative example of the impact of non-ideal MHD processes that is often reproduced in the numerical scheme benchmarking literature. Other than being of particular interest to interstellar medium (ISM) physics~\citep[see e.g.][]{1993ARA&A..31..373D}, the problem has a semi-analytic solution which allows for quantitative method performance evaluation.

The ideal MHD `jump-type' (J-type) shocks discussed in Section~\ref{sec:MHD_shock_tubes} comprised a rich set of concurrent discontinuous changes in hydrodynamic variables. For a broad range of physical parameters, non-ideal MHD effects can modify such solutions of initial value problems to be perfectly smooth, taking the form of `continuous-type' (C-type) shocks. \citet{1980ApJ...241.1021D} argued that MHD shock waves in a partially ionised plasma may be preceded by a `magnetic precursor', a propagating ion-electron disturbance which can travel ahead of the shock front, if the associated magnetosonic speed exceeds the shock speed. Given a strong enough ambient magnetic field, this could translate to compression and heating of the fluid upstream of the original discontinuity, `quenching' it and yielding continuous profiles in all physical attributes. This has been most widely discussed and verified in the context of ambipolar diffusion processes~\citep[][]{1991MNRAS.251..119W, 1995ApJ...442..726M, 2008MNRAS.391.1659D, 2009ApJS..181..413C, 2011ApJ...736..144B, 2014MNRAS.444.1104W}. Nevertheless, \citet{2012ApJS..201...24M} also ported the problem in the form of a shock tube example to the case of a non-isothermal, non-radiative fluid subject to Ohmic diffusion; it is this version of the test, reproduced in~\citet{2024MNRAS.527.8355Z} as well, that we consider here. The reference against which we compare our simulation results is a set of six coupled ordinary differential equations~\citep[see][]{2012ApJS..201...24M} which can trivially be solved numerically. They correspond to a reduced form of the equations of non-ideal MHD, which is obtained by assuming that all physical variables' profiles are time-invariant (i.e. in a steady state) and depend only on the coordinate along the shock propagation direction (taken w.l.o.g. to be $x$).

Shock tube simulations reproducing C-type discontinuities are typically run in the frame of the shock, having been initialised as a step function in un-conserved physical variables (in a similar fashion to what was done for the ideal MHD examples of Section~\ref{sec:MHD_shock_tubes}). Such discontinuous initial data are constructed by joining together, in the middle of the simulation domain, the homogeneous `left' and `right' states $S_L$ and $S_R$. For the present example these are taken to be
\begin{equation}
    \begin{split}
        S_L
        & \equiv 
        \{
        \rho_L, P_L, 
        \boldsymbol{v}_L, \boldsymbol{B}_L
        \} \\
        & =
        \{
        0.4, 0.4, (3, 0, 0), \frac{\sqrt{2}}{2}(1, 1, 0)
        \}
    \end{split}
\end{equation}
and
\begin{equation}
    \begin{split}
        S_R
        & \equiv 
        \{
        \rho_R, P_R, 
        \boldsymbol{v}_R, \boldsymbol{B}_R
        \} \\
        & =
        \{
        0.71, 1.19, (1.69, 0.43, 0), (\frac{\sqrt{2}}{2}, 1.44, 0)
        \}
    \end{split}
\end{equation}
which, respectively, correspond to the upstream and downstream state of the fluid a large distance from the C-type transition. Here, $\rho_i, P_i, \boldsymbol{v}_i$ and $\boldsymbol{B}_i$ correspond in order to state $S_i$'s mass density, thermal pressure, velocity and magnetic field. The strength of the shock scales inversely with the angle $\theta_s$ between $\boldsymbol{v}_L$ and $\boldsymbol{B}_L$~\citep{1991MNRAS.251..119W}; we take the two vectors to be oblique to one another with $\theta_s=\pi/4$, so that we can probe the physical regime of interest. 

\begin{figure*}
 \includegraphics[width=\textwidth]{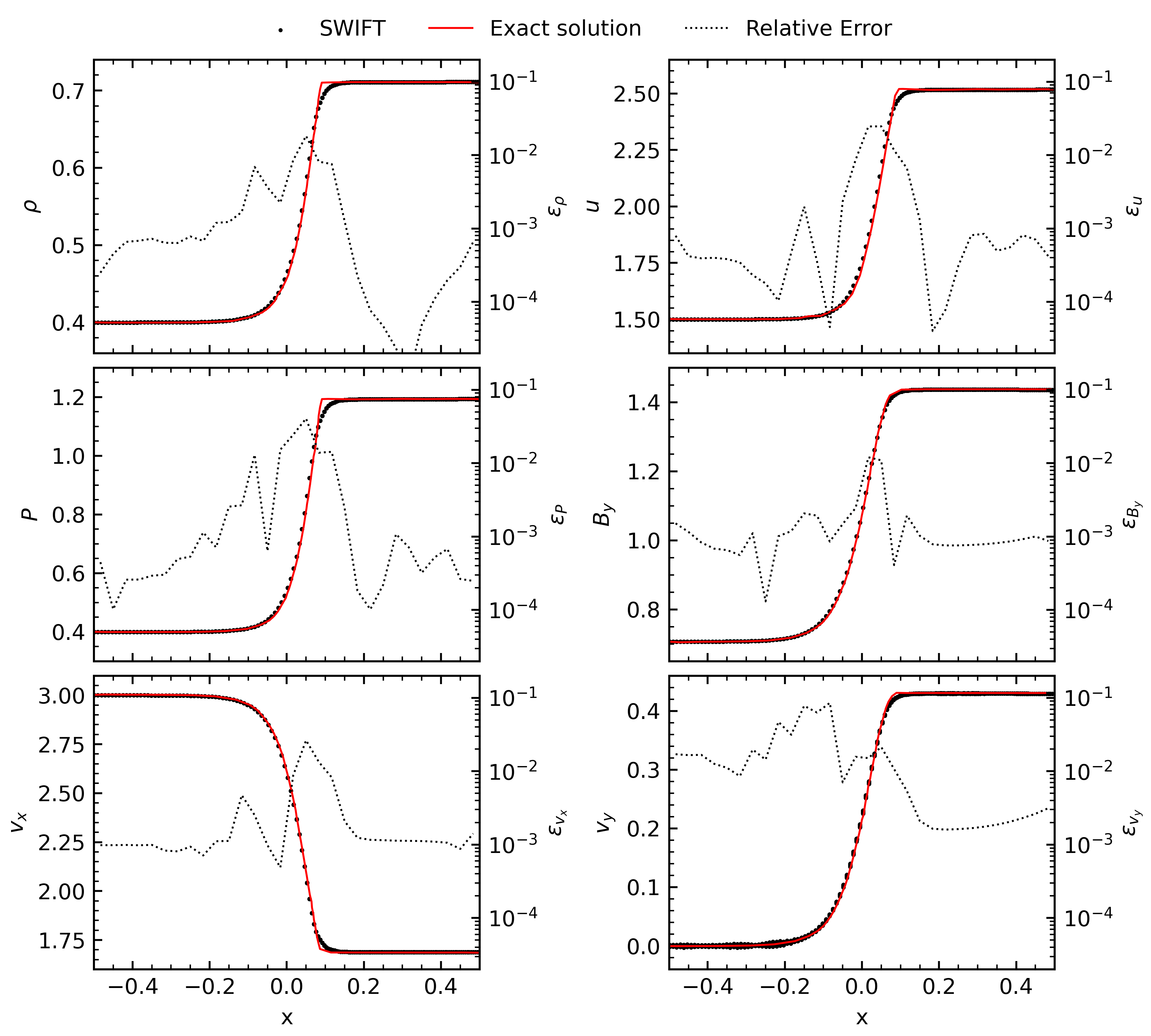} 
 \caption{Results for the non-isothermal, non-radiative, oblique C-Shock test at $t=1$. The initial discontinuity is established by joining in the middle of the simulation domain the left state $S_L \equiv  \{ \rho_L, P_L, \boldsymbol{v}_L, \boldsymbol{B}_L \} = \{0.4, 0.4, (3, 0, 0), (\sqrt{2}/2, \sqrt{2}/2, 0)\}$ with the right state $S_R \equiv \{\rho_R, P_R, \boldsymbol{v}_R, \boldsymbol{B}_R \} = \{0.71, 1.19, (1.69, 0.43, 0), (\sqrt{2}/2, 1.44, 0) \}$. The region shown is resolved by $\sim 2\times (48 \times 12 \times 12)$ particles. We display, going from left to right, top to bottom, profiles of the mass density $\rho$, thermal energy $u$, thermal pressure $P$, transverse magnetic field component $B_y$, longitudinal velocity component $v_x$ and transverse velocity component $v_y$ as functions of the position $x$ along the direction of shock propagation. We plot the attributes of all particles in our simulation (black markers, ordinate axis on the LHS), the semi-analytical solution reproduced from~\citet{2012ApJS..201...24M} (solid red lines, ordinate axis on the LHS), and the relative error between the two (dotted black lines, ordinate axis being on the RHS; note the different plotting range when compared to the LHS ordinate axis). The system has settled into a steady state whereby, owing to the effect of Ohmic diffusion, all quantities transition smoothly from their pre- to post- shock values across a C-type jump. Agreement between numerical and reference solutions is excellent, with the error remaining globally below the $1\%$ level. It only increases to a few per cent where quantities vary most rapidly, as particle attributes are smoothed over the resolution scale by discontinuity-capturing terms.}
 \label{fig:c_shock}
\end{figure*}

We instantiate the shock region $\Omega_\text{shock}$ as a rectangular domain of dimensions $L \times L/4 \times L/4$ for $L=1$, matching the origin of our coordinate system to $\Omega_\text{shock}$'s geometric centre. The position of the initial discontinuity, where $S_L$ and $S_R$ are taken to meet, coincides consequently with the midpoint along the domain's major axis. We populate $\Omega_\text{shock}$ with $2 \times (48 \times 12 \times 12)$ equal mass particles placed on the vertices of a BCC lattice, which corresponds to $\sim 60$ elements sampling the $[-L/2, L/2]$ interval along the direction of shock propagation; this broadly matches the resolution at which~\citet{2024MNRAS.527.8355Z} ran the same problem. The mild density contrast is set up by appropriately rescaling the $x$ coordinate of all particles in $S_L$. We then assign all remaining individual particle properties to obtain the desired uniform fields either side of the jump. While \citet{2012ApJS..201...24M} employ inflow-outflow boundary conditions and trivially handle the horizontal bulk fluid motion, SWIFT does not yet support this option. We can obtain a qualitatively identical set-up using periodic boundary conditions, provided we take precautionary measures to give the C-shock enough time to settle before it is corrupted by the propagating Riemann problem inevitably seeded at the periodic boundary in $x$. We achieve this by extending the simulation box at both ends, making each of $S_L$ and $S_R$ eight times as long, which allows for an ample time window during which the region of interest is not polluted by boundary effects. We fix the magnetic diffusivity to $\eta = 0.1$, and evolve the system until $t=1$. We display its state at that point in time in Fig.~\ref{fig:c_shock}.

We show measured profiles of the mass density, specific thermal energy, thermal pressure and selected velocity and magnetic field components in black (plotting the attributes of all particles in the simulation) and contrast them with the reference solution in red. Similarly to~\citet{2024MNRAS.527.8355Z}, we translate the latter along $x$ by a small amount; the magnitude of this shift is chosen to be the one minimising the offset between the numerical and exact solution for the density. This simulation-to-theory mismatch in shock location occurs both because our strategy for initialising the density contrast displaces the discontinuity slightly leftwards of $x=0$ in the initial conditions, as well as because the position of the C-shock can marginally drift during the course of the simulation; the latter is not specific to SWIFT and is also observed for both structured grid and moving-mesh codes like ATHENA++ and AREPO~\citep{2024MNRAS.527.8355Z}. We complement the figure with relative error profiles $\varepsilon_Q (x)$ for each quantity $Q$ we probe, which we compute as
\begin{equation}
    \varepsilon_Q (x) = \
    \Big\Vert
    \frac{Q_\text{ref}(x) - Q_\text{SWIFT}(x)}
    {Q_\text{ref}(x) + \delta \text{max}_x \{ Q_\text{ref}(x) \}}
    \Big\Vert
\label{eq:cshock_error_profile}
\end{equation}
$Q_\text{ref}(x)$ and $Q_\text{SWIFT}(x)$ correspond to reference and measured profiles respectively, and $\delta$ is a small $\mathcal{O} (10^{-2})$ constant introduced as part of an additive regularisation factor in the denominator of equation~\ref{eq:cshock_error_profile}, aimed at stabilising the calculation when the exact solution goes to zero. Our results demonstrate that the code has converged to a standing shock solution, with upstream compression of the mass density and magnetic field, accompanied by Joule heating and bulk deceleration in the post-shock region. The transition from left to right state is continuous for all quantities and in excellent agreement with theory; our numerical predictions systematically deviate from the semi-analytic solutions only where these vary most rapidly, a consequence of our discontinuity-capturing terms acting concurrently to smooth all fields over the resolution scale. The calculated errors remain safely below the $1\%$ level in general, only increasing to a few per cent at the location of the C-type jump~\citep[the magnitude and spatial distribution of our $\varepsilon_Q (x)$ are comparable to those reported in][]{2012ApJS..201...24M}. The one variable that is (marginally) less well captured is $v_y$~\citep[this is the case for][as well]{2024MNRAS.527.8355Z}; the associated error and noise are nevertheless negligible when viewed in light of the global velocity scale of the problem, largely dominated by $v_x$.

\subsection{Coupling MHD to gravity}
\label{sec:mhd_and_gravity}

We proceed by considering the coupling of our SPMHD method to SWIFT's gravity solver. The short-range gravitational potential and associated force fields acting on a distribution of point-like particles are calculated using a fourth-order Fast Multipole Method~\citep{1987JCoPh..73..325G, 1999JCoPh.155..468C, 2000ApJ...536L..39D, 2002JCoPh.179...27D, 2014ComAC...1....1D}, whose computational cost has the advantage of scaling linearly with the number $N$ of resolution elements employed; an exact, `brute-force' calculation would involve $\mathcal{O}(N^2)$ operations, and other popular alternatives like~\citet{1986Natur.324..446B}'s algorithm would typically only bring this down to $\mathcal{O}(N \mathrm{log}N)$. Long-range gravity acting over periodic boundaries is accounted for through a standard particle-mesh method~\citep{1988csup.book.....H}, operating in Fourier space according to the potential splitting formalism of~\citet{2003NewA....8..665B}. 

The mass density distribution $\rho_i (\boldsymbol{r})$ corresponding to an individual particle of mass $m_i$ located at $\boldsymbol{r}_i$ is modeled as the convolution of the Dirac delta distribution $m_i \delta (\boldsymbol{r} - \boldsymbol{r}_i)$ with a radially symmetric kernel $W(|\boldsymbol{r}|, H)$ of finite spatial support $H$. In SWIFT, $H$ is set to $H = 3 \varepsilon_\text{soft.}$, where $\varepsilon_\text{soft.}$ is a free model parameter referred to as the softening length whose value is chosen based on the resolution of any given simulation. $W(|\boldsymbol{r}|, H)$ is taken to be the Wendland C2 kernel~\citep{Wendland1995}, such that the resulting $\rho_i (\boldsymbol{r})$ corresponds within $|\boldsymbol{r} - \boldsymbol{r}_i| < H$ to an easy-to-compute, finite-valued, softened gravitational potential $\varphi_i (\boldsymbol{r})$ whose form prevents spurious two-body relaxation between converging particles. Beyond the softening scale, the expression for $\varphi_i (\boldsymbol{r})$ reverts to that corresponding to a Newtonian potential sourced by a point mass. SWIFT implements the adaptive gravitational softening prescription of~\citet{2007MNRAS.374.1347P}, a conservative formalism for self-gravity that allows for an $\varepsilon_\text{soft.}$ which varies with density; this removes the need for fine-tuning the parameter and improves the accuracy of results on non-uniform particle distributions, which are commonplace in astrophysical and cosmological simulations. In what follows, and in contrast to what is most commonly done in the literature, we nevertheless present results obtained with a spatially uniform, time-invariant $\varepsilon_\text{soft.}$, to test our scheme in a configuration that is as close as possible to that anticipated for production runs targeting our method's primary intended scientific use cases. Indeed, major galaxy and structure formation simulation campaigns carried out with SWIFT~\citep[e.g.][]{2023MNRAS.526.4978S, 2026MNRAS.548ag375S} opt for constant softening as it couples optimally with the effective sub-resolution models they make use of.

\subsubsection{Gravitational collapse of a magnetised molecular gas cloud: Problem statement}
\label{sec:magnetised_cloud}

The ability of a multi-purpose astrophysical code to reliably couple magnetic field physics to gravity calculations is commonly benchmarked through idealised simulations of the early stages of individual star formation~\citep[e.g.][]{2005A&A...435..385Z, 2006A&A...457..371F, 2011MNRAS.418.1392P, 2016MNRAS.455...51H}, where the gravitational collapse of a magnetised molecular cloud is followed down to the emergence of the `first Larson core'~\citep{1969MNRAS.145..271L}. The evolution of such cores is undeniably linked to MHD effects and has been the subject of extensive study~\citep[see e.g.][for a recent review]{2023ASPC..534..317T}; over the past two decades tremendous advances have been made, both from the observational side through e.g. the use of (sub)millimeter dust polarised continuum emission to map proto-stellar magnetic fields~\citep[e.g.][]{2018A&A...616A..56A, 2019ApJ...879...25K, 2021ApJ...920...71A}, as well as on the modeling front where simulations incorporating progressively more relevant physical processes have reproduced an ever increasing number of reported trends~\citep[e.g.][]{2021MNRAS.502.4911X, 2021MNRAS.507.2354W}. 

Non-ideal effects are believed to be dynamically relevant in the environments where star formation occurs, as the gas found there is forecast to be only weakly ionised~\citep{1956MNRAS.116..503M, 1957PhT....10Q..40C, 1986MNRAS.218..663N, 1990MNRAS.243..103U}. Nevertheless, a number of controlled numerical experiments~\citep[such as the ones of][]{1998ApJ...502L.163T, 2000ApJ...528L..41T, 2002ApJ...575..306T, 2005MNRAS.362..382M, 2006ApJ...641..949B, 2011MNRAS.417L..61B, 2011MNRAS.417.1054S, 2012MNRAS.422..347S, 2015MNRAS.451..288L, 2017MNRAS.467.3324L} have shown that part of the observationally established phenomenology~\citep[see e.g.][]{1996ARA&A..34..111B, 1999ApJ...513L..57A, 2002A&A...393..927B, 2004A&A...426..503W, 2010ApJ...715.1344C, 2011ApJ...742....1D, 2013ApJ...768..159H} can already be recreated using ideal MHD alone. The one such experiment we consider here is the gravitational collapse problem detailed in~\citet{2008A&A...477....9H} which, up to minor differences in the choices made for some of the physical parameters we detail below, is the gravity + MHD benchmarking test run in the context of studies similar to ours~\citep{2011MNRAS.418.1392P, 2016MNRAS.455...51H, 2020A&A...638A.140W, 2023MNRAS.518.4115G}; instead of a full quantitative description of the numerical results obtained, what is typically reported is a given code's ability to qualitatively match anticipated physical behaviour. 

\citet{2008A&A...477....9H} start by setting up a gas cloud as a sphere of constant mass density, in solid body rotation, and threaded by a uniform seed magnetic field parallel to the rotation axis. It is embedded in a diffuse homogeneous medium and allowed to contract under its own gravity. Over the course of one free-fall time, the gas at the centre of the cloud is compressed by $\sim 8$ orders of magnitude to yield a dense hydrostatic pre-stellar core, while differential motions lead to significant amplification of the axial and azimuthal components of the magnetic field $B_z$ and $B_\phi$. In the companion paper~\citet{2008A&A...477...25H} it is demonstrated that even weak fields are able to stabilise the core against fragmentation, which is clearly observed in twin pure hydrodynamical runs. If the initial magnetisation is below a given threshold, the collapse occurs quasi-spherically, and the emergence of the pre-stellar core is accompanied by the formation of a stable, centrifugally supported disk. Stronger seed fields, on the other hand, lead to anisotropic mass inflow and strong magnetic braking, which translate to efficient angular momentum redistribution that only allows for the development of an infalling magnetised pseudo-disk. \citet{2008A&A...477....9H} further show that, in either case, collimated bipolar jets are eventually launched from the core; these co-occur with a slowly expanding `magnetic tower' in weak seed scenarios, a structure associated with $B_\phi$ which emerges as field lines are tightly wound and stretched. These mass outflows are believed to be driven by a joint action of centrifugal~\citep{1982MNRAS.199..883B, 1983ApJ...274..677P, 1986ApJ...301..571P, 1992ApJ...394..117P} and toroidal magnetic pressure forces~\citep{1986PASJ...38..631S, 1996MNRAS.279..389L, 2003MNRAS.341.1360L, 2004ApJ...605..307K}, and have been particularly challenging for earlier implementations of SPMHD to capture. \citet{2012JCoPh.231.7214T} established that poor control of divergence errors, leading to large non-conservative tensile instability correction terms, could eject the core from the disk and disrupt the outflow. \citet{2017arXiv170607721W} showed that an inadequate choice of AR prescription could engender the formation of unphysical bubbles and overly dense proto-stellar disks, or cause the migration of the core (although it should be noted they included sink particles in their simulations, a potential source of additional complications). Finally, both~\citet{2016MNRAS.455...51H} and~\citet{2020A&A...638A.140W} reported that their respective `traditional' SPMHD schemes were overly diffusive, with AR quenching magnetic field amplification and preventing jet launching. We are thus confronted with a problem that stresses all of our scheme's components concurrently. It moreover serves as an assessment of our code's conservation properties, as maintaining the mirror symmetry about the disk midplane, as well as the axial symmetry about the rotation axis, requires near-perfect preservation of linear and angular momentum.

\subsubsection{Gravitational collapse of a magnetised molecular gas cloud: Initialisation}

We initialise a uniform spherical cloud of radius $R_\mathrm{cloud} = 1.5 \cdot 10^{-2} \; \text{pc}$ and total mass $M_\mathrm{cloud} = 1 \; \mathrm{M}_\odot$ (the corresponding constant mass density there is $\rho_\mathrm{cloud} \sim 5 \cdot 10^{-18} \; \mathrm{g} \; \mathrm{cm}^{-3}$) in the centre of a cubic box of side length $L=10 R_\mathrm{cloud}$ with periodic boundary conditions. Defining the integer $n_\mathrm{cloud}$, we populate the cloud region with $n_\mathrm{cloud}^3$ equal-mass particles positioned on the vertices of a BCC lattice, whose unit cell has an edge length we term $l_\mathrm{cloud}$. The resolution of a given simulation is entirely set by $n_\mathrm{cloud}$, which also fully determines the individual particle mass $m_i$ and inter-particle separation $l_\mathrm{cloud}$ through
\begin{equation}
    m_i(n_\mathrm{cloud}) = \frac{M_0}{n_\mathrm{cloud}^3}
    \quad \text{and} \quad
    l_\mathrm{cloud}(n_\mathrm{cloud}) = \frac{2^{2/3}}{n_\mathrm{cloud}}
\end{equation}
We fill the remainder of the simulation volume with a homogeneous diffuse atmosphere of mass density $\rho_\mathrm{atm.} = \lambda \rho_\mathrm{cloud}$, where $\lambda$ is a constant we set to $\lambda = 1/360$. We instantiate said atmosphere by distributing particles of mass $m_i$ on the vertices of a second BCC lattice, whose unit cell has an edge length of $l_\mathrm{atm.} = \sqrt[3]{1 / \lambda} l_\mathrm{cloud}$, to achieve the desired density contrast of $\lambda$. 

The collapsing molecular cloud is expected to initially be optically thin and behave quasi-isothermally, subsequently becoming opaque to radiation and undergoing adiabatic self-heating as it contracts to higher densities. To mimic this behaviour, \citet{2008A&A...477....9H} suggest not to track gas thermodynamics through solving a full energy equation, but to rather use the barotropic equation of state~\citep[EoS; see also e.g.][]{2004MNRAS.347.1001H, 2005A&A...435..385Z}
\begin{equation}
    P (\rho) =
    c_{s,0}^2 \rho
    \sqrt{1 + \left( \frac{\rho}{\rho_c} \right)^{4/3}}
\label{eq:barotropic_eos}
\end{equation}
where $c_{s,0} = 0.2 \; \mathrm{km} \; \mathrm{s}^{-1}$ is a characteristic isothermal sound speed and $\rho_c = 10^{-14} \; \mathrm{g} \; \mathrm{cm}^{-3}$ a characteristic critical density scale; these parameter choices set the cloud's initial temperature to $T\sim10 \; \mathrm{K}$, the specific value depending on its exact chemical composition. Equation~(\ref{eq:barotropic_eos}) transitions from the isothermal EoS $P \propto\rho$ at low densities to the polytropic EoS $P \propto \rho^{5/3}$ as the $\rho_c$ density threshold is exceeded, the latter corresponding to the EoS obeyed by a monoatomic gas with three translational degrees of freedom undergoing isentropic changes. Our choice of EoS incidentally also prevents an infinite density singularity from forming, allowing us not to have to introduce sink particles~\citep[in contrast to e.g.][]{1995MNRAS.277..362B, 2007MNRAS.377...77P, 2010ApJ...713..269F, 2011MNRAS.412..171B, 2012MNRAS.422..347S, 2012MNRAS.423L..45P, 2015MNRAS.451..288L, 2017MNRAS.467.3324L} since the maximal densities reached in our simulations ($\rho_\mathrm{max} \sim {10}^{-12}-{10}^{-11}  \; \mathrm{g} \; \mathrm{cm}^{-3}$, in line with other results from the literature) do not give rise to prohibitively small time-step sizes. The constant gravitational softening is set so that a particle's density cannot exceed an $\mathcal{O}(10)$ multiple $C$ of $\rho_\mathrm{max}$, i.e. we require $\rho_i \sim m_i / \varepsilon_\mathrm{soft.}^3 \leq C\rho_\mathrm{max}$; we nevertheless verified that reducing $\varepsilon_\mathrm{soft.}$ further, within reasonable bounds, does not qualitatively impact results. We use the definition of an isothermal sound speed to write~\citep[similarly to e.g.][]{2012A&A...543A.128J}
\begin{equation}
    c_s (\rho) \equiv \frac{P(\rho)}{\rho} = 
    c_{s,0}
    \sqrt[4]{1 + \left( \frac{\rho}{\rho_c} \right)^{4/3}}
\label{eq:barotropic_sound_speed}
\end{equation}
which we take to be the characteristic hydrodynamic speed entering our signal velocity calculation, to be used in discontinuity capturing and divergence cleaning. 

Having set individual particles' mass densities as described above, the following equations~(\ref{eq:barotropic_eos}) and~(\ref{eq:barotropic_sound_speed}) uniquely determine all remaining thermodynamic variables entering our scheme. Our initialisation procedure inevitably introduces sharp discontinuities in all of them at the cloud boundary; we nevertheless employ no mitigation strategy against this~\citep[as opposed to e.g.][]{2011MNRAS.418.1392P}, making the problem maximally challenging for our method. Although the cloud surface will not be in pressure equilibrium initially, the ratio of thermal to gravitational specific energies being
\begin{equation}
    \alpha 
    \equiv
    \frac{\epsilon_\mathrm{therm.}}{\epsilon_\mathrm{grav.}}
    \sim
    \left.    
    \frac{3}{2} c_{s,0}^2
    \middle/
    \frac{3}{5} \frac{G M_\mathrm{cloud}}{R_\mathrm{cloud}}
    \right.
    \approx 0.35
\end{equation}
where $G$ is the gravitational constant, means the expansion seeded by this spurious pressure differential will immediately be overcome by gravitational forces. The collapse can thus be safely considered to be unaffected by it.

The atmosphere is set to be static, while the cloud is put in solid body rotation with constant angular acceleration $\boldsymbol{\omega} = \omega_\mathrm{cloud} \hat{\boldsymbol{e}}_z$, where $\omega_\mathrm{cloud} \equiv 2 \pi / T_\mathrm{cloud}$ is determined in terms of the rotation period $T_\mathrm{cloud}$ we set to $T_\mathrm{cloud} = 4.7 \cdot10^5 \; \mathrm{yr}$. This parameter choice yields a ratio of rotational to gravitational specific energies of
\begin{equation}
    \beta 
    \equiv
    \frac{\epsilon_\mathrm{rot.}}{\epsilon_\mathrm{grav.}}
    \sim
    \left.
    \frac{1}{5} M_\mathrm{cloud} R_\mathrm{cloud}^2 \omega_\mathrm{cloud}^2
    \middle/
    \frac{3}{5} \frac{G M_\mathrm{cloud}}{R_\mathrm{cloud}}
    \right.
    \approx 0.045
\end{equation}
which establishes rotation as being dynamically subdominant. Moreover, it situates the test in a physically plausible regime, our $(\alpha, \beta)$ pair being broadly consistent with observations and the theoretical studies these have motivated~\citep[see e.g.][for a review]{2007ARA&A..45..565M}. We further instantiate throughout the entirety of the simulation volume the uniform seed magnetic field $\boldsymbol{B}_0 = B_0 \hat{\boldsymbol{e}}_z$, noting $\boldsymbol{B}_0 \parallel \boldsymbol{\omega}$, whose strength $B_0$ we determine through the variable $\mu$ defined as
\begin{equation}
    \mu \equiv
    \left.
    \left( \frac{M_\mathrm{cloud}}{\Phi_B} \right)
    \middle/
    \left( \frac{M_\mathrm{cloud}}{\Phi_B} \right)_\text{crit.}
    \right.
\end{equation}
where $M_\mathrm{cloud}/\Phi_B$ is the ratio of mass to magnetic flux threading the cloud, and $\left(M_\mathrm{cloud}/\Phi_B \right)_\mathrm{crit.}$ the critical value of this ratio beyond which magnetic support prevents gravitational collapse ($\mu=1$ therefore corresponds to a marginally stable cloud). The magnetic flux $\Phi_B$ is calculated over the surface $S$ of the cloud with surface element $\mathrm{d} \boldsymbol{S}$ as
\begin{equation}
    \Phi_B \equiv 
    \oiint_S \boldsymbol{B}_0 \cdot \mathrm{d} \boldsymbol{S} = 
    \pi R_\mathrm{cloud}^2 B_0
\end{equation}
meaning $B_0 \propto 1/\mu$, and the critical mass-to-flux ratio is computed using an expression derived through virial arguments by~\citet{1968dms..book.....S}, adapted for our choice of magnetic field units to
\begin{equation}
    \left( \frac{M_\mathrm{cloud}}{\Phi_B} \right)_\mathrm{crit.}
    =
    \frac{2 c_1}{3} \sqrt{\frac{5}{\pi \mu_0 G}}
\end{equation}
where $c_1$ is a constant numerically determined by~\citet{1976ApJ...210..326M}  to be $c_1=0.53$. We begin with simulations targeting the fiducial value $\mu=10$, corresponding to $B_0 = 60.7 \; \mathrm{\mu G}$, and subsequently proceed by investigating the super-critical cases $\mu = 20, 5, 2$ as well as the sub-critical cloud $\mu=0.7$, to probe a representative range of plausible initial magnetisations~\citep[see e.g.][]{1999ApJ...520..706C, 2008ApJ...680..457T}.

One final consideration we take into account is the resolution criterion proposed by~\citet{1997MNRAS.288.1060B}~\citep[but see also][]{1993MNRAS.265..271N, 1998MNRAS.296..442W, 2003MNRAS.339..577B}, to ensure that results are physical in nature and free from spurious effects such as numerically induced stabilisation and/or fragmentation of the gas. The claim is that in simulations of collapsing isothermal spheres, the minimal resolvable mass $M_\text{min res.}$ must at any time be smaller than the local Jeans mass $M_J$. Taking the former to be twice the mass contained within the support of a single particle's kernel (i.e. $M_\text{min res.} = 2 N_\mathrm{neigh.} m_i$ for $N_\mathrm{neigh.}$ a particle's typical number of neighbours, where $N_\mathrm{neigh.} \sim 180$ for the quintic spline we use), and calculating the latter through the conservative estimate~\citep[see e.g.][for various definitions]{1902RSPTA.199....1J, 1982FCPh....8....1T, 2006MNRAS.373.1039N, 2008gady.book.....B}
\begin{equation}
    M_J (\rho) =
    \left(
        \frac{5 \epsilon_\mathrm{therm} (\rho)}{3G}
    \right)^{3/2}
    \left(
        \frac{4}{3} \pi \rho
    \right)^{-1/2}
\end{equation}
the stability criterion of~\citet{1997MNRAS.288.1060B} would require us to resolve the cloud region with at least
\begin{equation}
    N_\mathrm{min} \sim
    2 N_\mathrm{neigh.} \frac{M_\mathrm{cloud}}{M_J (\rho_c)} \approx 4.7 \cdot 10^4 \; \text{particles}
\label{eq:cloud_resolution_requirement}
\end{equation}
We have chosen~\citep[similarly to e.g.][]{2011MNRAS.417L..61B} to evaluate $M_J(\rho)$ at $\rho_c$ in equation~(\ref{eq:cloud_resolution_requirement}), as this is roughly the maximal density scale at which the cloud still behaves isothermally and the above stability analysis is valid. As it subsequently enters the adiabatic evolution stage, the local Jeans mass starts increasing with density and is always guaranteed to be adequately resolved. Our resolution requirement can equivalently be expressed as $n_\mathrm{cloud} \gtrsim \sqrt[3]{N_{min}} \approx 36$, which prompts us to run simulations at resolution levels corresponding to $n_\mathrm{cloud}=32, 64, 128, 256$. Expressing time in units of the gravitational free-fall timescale (see e.g.~\citealt{2010gfe..book.....M}, and note our definition differs from the unconventional one used by~\citealt{2016MNRAS.455...51H, 2020A&A...638A.140W, 2023MNRAS.518.4115G})
\begin{equation}
    t_\mathrm{ff} = \sqrt{\frac{3 \pi}{32 G \rho_0}} 
    \approx 3 \cdot {10}^{4} \; \mathrm{yr}
\end{equation}
we evolve the system until $t \sim 1.1-1.4 t_\mathrm{ff}$, providing enough time for a hydrostatic core to form at the centre of the computational domain; the exact time at which this occurs in a given simulation depends on both the specific initial conditions and the resolution employed.

\subsubsection{Gravitational collapse of a magnetised molecular gas cloud: Results}

\begin{figure*}
 \includegraphics[width=\textwidth]{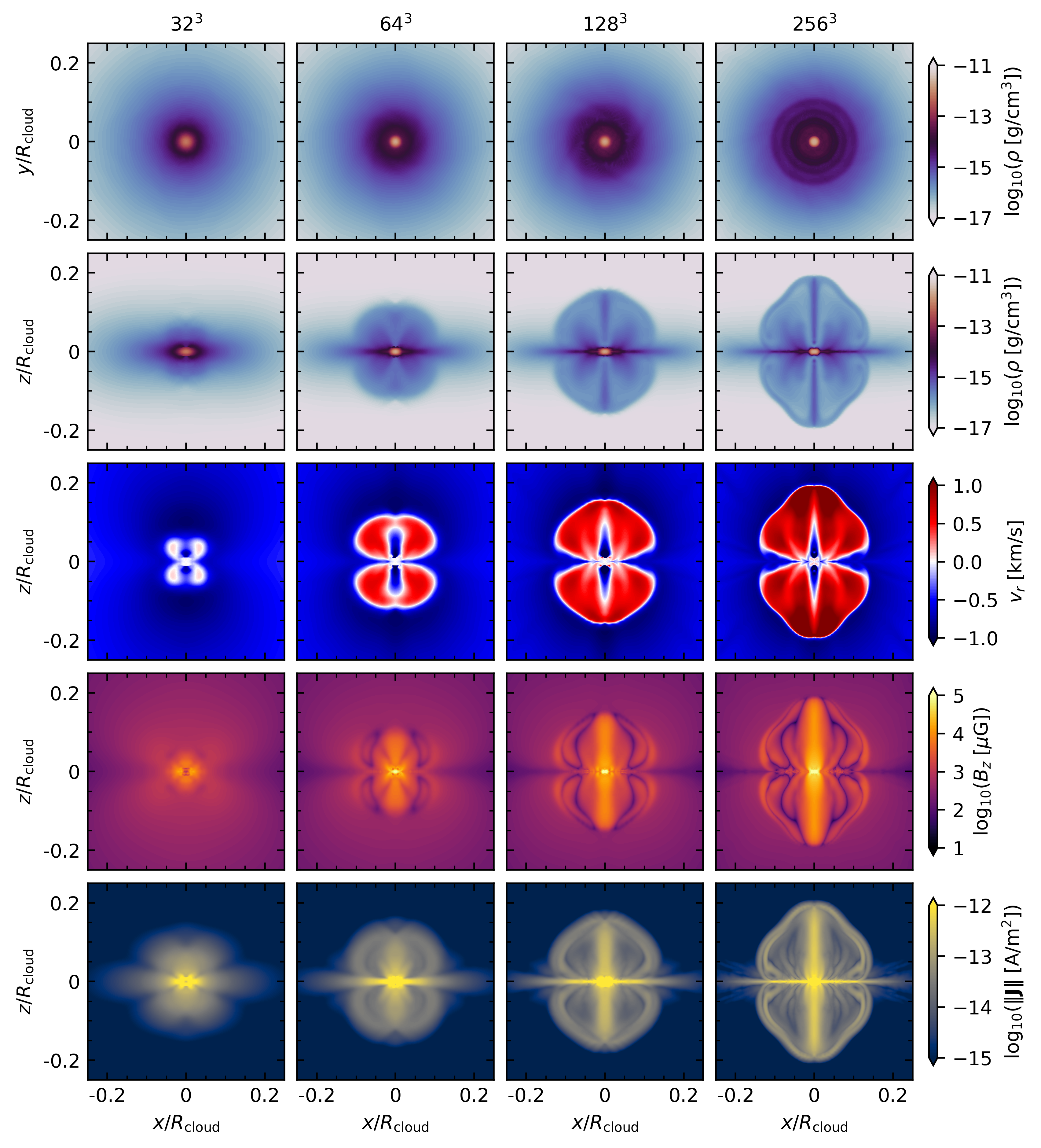} 
 \caption{Gravitational collapse of a $1 \mathrm{M}_\odot$ molecular cloud. Instantiated as a uniform sphere of constant mass density in a homogeneous diffuse atmosphere $360$ times less dense, the cloud is initially set in rigid body rotation about the $z$ axis with a period of $T_\mathrm{cloud} = 470 \; \mathrm{Myr}$, and permeated by a uniform magnetic field $\boldsymbol{B} = B_0 \hat{\boldsymbol{e}}_z$ whose strength $B_0$ is chosen to give a mass-to-magnetic-flux ratio of $\mu=10$. We allow the collapse to proceed under the effect of self-gravity while modeling gas thermodynamics through a barotropic equation of state, and portray the state of the system at $t/t_\mathrm{ff} = 1.17$ (where $t_\mathrm{ff} \sim 30 \; \mathrm{Myr}$ is the gravitational free-fall time). Each column shows results from a re-simulation of the same problem at one of four different resolution levels: from left to right, these respectively correspond to ${32}^3,{64}^3,{128}^3$ and ${256}^3$ equal mass particles resolving the cloud in the initial conditions. We present `face-on' maps of the mass density $\rho$ in the equatorial plane (first row), followed by `edge-on' maps of $\rho$, of the radial/outflow velocity component $v_r$, of the axial magnetic field $B_z$ and of the norm of the magnetic current vector $\boldsymbol{J}$ in the $y=0$ midplane (second to fifth row, respectively). The cloud has collapsed to form a nearly spherical, dense pre-stellar core which is surrounded by an accretion (pseudo)disk. Magnetic fields are amplified by compression and differential motions, being stretched and tightly wound about the $z$ axis. This results in the launching of magnetically driven, bipolar collimated jets from the core, which entrain larger scale diffuse outflows. Hints of all these features are already present in the ${32}^3$ simulation (where the local Jeans mass is formally not entirely resolved), and are clearly in place from the second resolution level onward. Higher resolution simulations produce profiles with more and finer detail.}
 \label{fig:magnetised_cloud_convergence}
\end{figure*}

We first present a convergence study for simulations initialised with $\mu=10$ in Fig.~\ref{fig:magnetised_cloud_convergence}, displaying results after $t=1.17 t_\mathrm{ff}$ for each of the four resolution levels considered. We show `face-on' slices (through the $z=0$ plane) of the mass density $\rho$, followed by `edge-on' slices (through the $y=0$ plane) of $\rho$, of the radial component $v_r$ of the velocity field $\boldsymbol{v}$ (calculated as $v_r = \Vert \boldsymbol{v} \cdot \hat{\boldsymbol{e}}_r \Vert$ for $\hat{\boldsymbol{e}}_r$ the unit radial vector of a spherical coordinate system whose origin is at the computational domain's centre), of the axial magnetic field $B_z$ and of the magnitude of the magnetic current $\Vert \boldsymbol{J} \Vert = \Vert \nabla \times \boldsymbol{B} \Vert / \mu_0$. All `face-on' maps show perfect axial symmetry about the rotation axis $\hat{\boldsymbol{e}}_z$, demonstrating our method excels at preserving angular momentum in this highly dynamical example, even though our assumed form of the induction equation does not formally guarantee this would be the case. In turn, all `edge-on' images display clear mirror symmetry about the $z=0$ plane, indicating linear momentum is conserved as well. The only symmetry-breaking term in this regard being the tensile instability correction, which scales linearly with $\nabla \cdot \boldsymbol{B}$, we interpret these results as suggestive of our divergence-cleaning prescription operating efficiently, even for a considerably heterogeneous particle distribution proper to an astrophysically relevant problem.

All expected qualitative physical features are largely in place from our second resolution level onward, with hints already present in the lowest resolution simulation as well; we note that the latter uses a number of particles marginally below what our estimate of $N_\mathrm{min}$ would recommend. A compact core has formed in all runs, of radial size $\sim 0.025 R_\mathrm{cloud}$ and reaching densities of $\sim {10}^{-11}  \; \mathrm{g} \; \mathrm{cm}^{-3}$ (i.e. seven orders of magnitude larger than the cloud's initial density). It is surrounded by a thin gas disk that extends to $\sim 0.1 R_\mathrm{cloud}$; the non-null $v_r$ there suggests that this is not a centrifugally supported structure, but rather a contracting pseudo-disk providing continuous mass accretion onto the core as a consequence of magnetic braking. Bipolar collimated jets are launched from the core owing to MHD effects, extending to $z \sim \pm 0.2 R_\mathrm{cloud}$ at time of plotting and moving at a few $\mathrm{km} \; \mathrm{s}^{-1}$ in our highest resolution simulation; as noted by~\citet{2012MNRAS.423L..45P}, this is in good agreement with the velocities of outflows emanating from observed pre-stellar cores~\citep[see e.g.][]{2011ApJ...742....1D}. The thin jets are enveloped in a wider outflow structure, whose footprint in the midplane extends over the entire pseudo-disk, and which propagates vertically to establish two symmetric bow shocks in the ambient medium. An elongated structure of a strong axial magnetic field is established as a product of the joint action of compression and a shearing motion-powered dynamo. It reaches values of $B_z \sim {10}^4 - {10}^5 \; \mathrm{\mu G}$ in the core (which amounts to an amplification by $\sim 4$ orders of magnitude when compared to the seed value, and is in good quantitative agreement with the findings of e.g.~\citealt{2008A&A...477....9H, 2011MNRAS.418.1392P}), and is co-spatial with a slender tower of high magnetic current, corresponding to a tightly wound up toroidal magnetic field; note that $B_\phi = 0$ in our initial conditions, so what we witness is a conversion of axial to toroidal magnetic field, which is subsequently amplified by gas motions that twist and stretch it. This is a non-trivial feat for an SPMHD solver, earlier incarnations of the method having proven unable to represent such intricate field geometries if they were to preserve the solenoidal constraint~\citep[see e.g.][]{2007MNRAS.377...77P, 2008MNRAS.385.1820P}. The topology of the magnetic current generated in our runs is in excellent qualitative agreement with that shown in e.g.~\citet{2012MNRAS.423L..45P}, and is what controls the degree of collimation of the jets.

To our knowledge at time of writing, this constitutes the first report of a `traditional' SPMHD method (i.e. one that does not rely on the `geometric force averaging' of e.g.~\citet{2017MNRAS.471.2357W} and is consistent with a Lagrangian derivation) that is able to jointly reproduce, using only a modest ${64}^3$ particles, all of the expected qualitative features of the $\mu=10$ cloud collapse problem. Increasing resolution beyond ${64}^3$ yields late-time profiles with a finer level of detail, with outflows having propagated slightly further as the timing of jet launching is weakly resolution dependent, but the qualitative picture remains largely consistent and unchanged. This represents a clear improvement over what was reported by e.g.~\citet{2016MNRAS.455...51H} or~\citet{2020A&A...638A.140W}, whose respective `traditional' SPMHD schemes displayed considerably slower convergence on this example, being either entirely unable to produce bipolar outflows or requiring $\mathcal{O}(10^2)$ times more particles to do so. Leaving results of low-resolution simulations aside (these being un-converged for any method), our scheme appears to fare comparably well with the best performing `beyond SPMHD' models detailed in the aforementioned works. The definition of `well' being partly subjective in this context, we can but encourage the community to pursue a methodical study of the formation of pre-stellar cores using different astrophysical codes, employing common initial conditions, code and physical parameters, as well as simulation post-processing routines, to establish a reliable and controlled comparison between methods. Such enterprises have have for instance proven to be very informative in the context of pure hydrodynamics~\citep[e.g.][]{2007MNRAS.380..963A, 2023MNRAS.523.1280B}, supporting advances in both the understanding of the highly non-linear problem under consideration and of the numerical technique used to simulate it.

\begin{figure*}
 \includegraphics[width=\textwidth]{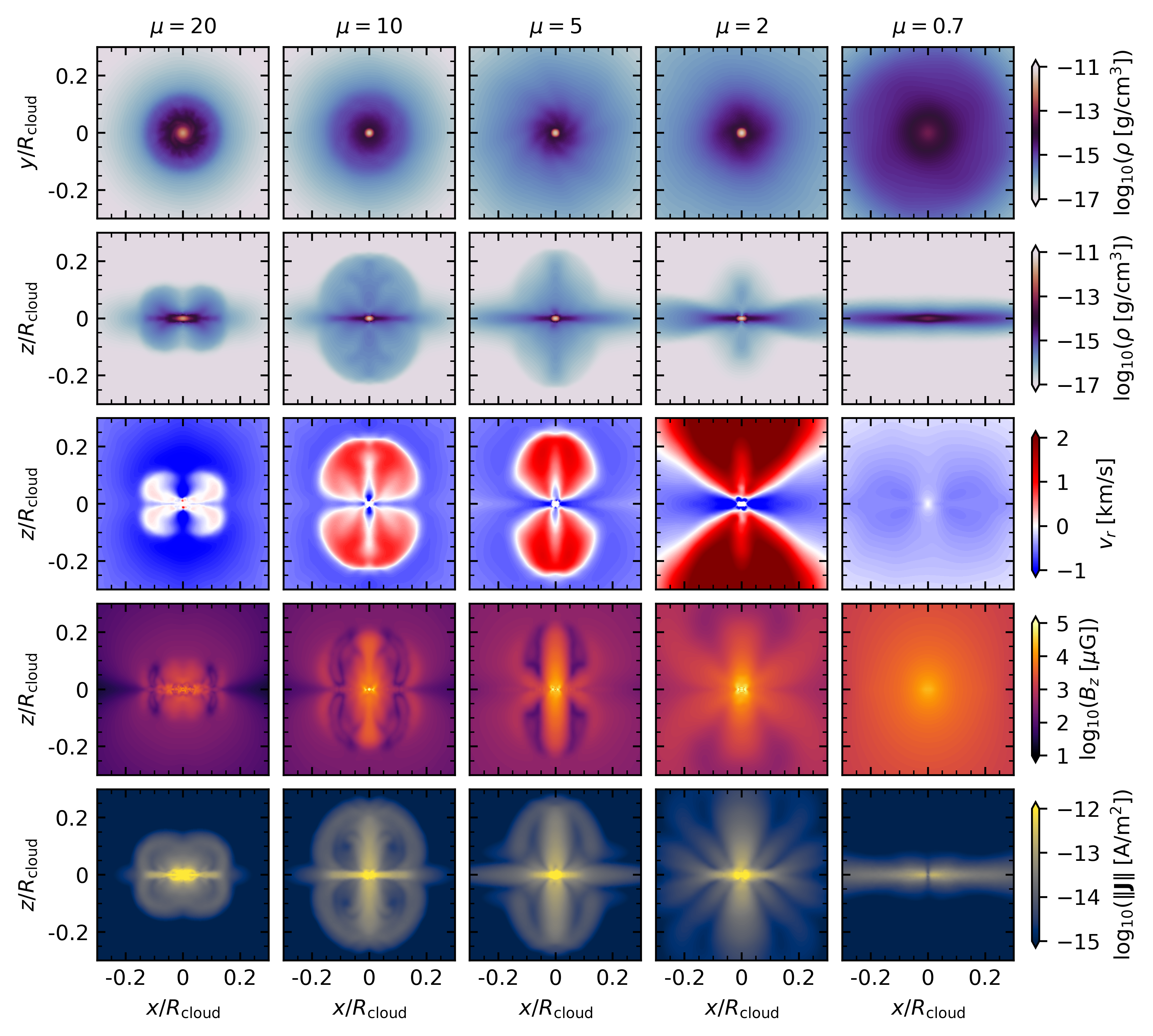} 
 \caption{Similar to Fig.~\ref{fig:magnetised_cloud_convergence} but for re-simulations of the gravitational collapse problem pictured therein at fixed resolution (${64}^3$ particles in the molecular cloud region in the initial conditions), for which the strength of the uniform seed magnetic field was varied. Individual columns show results at $t > t_\mathrm{ff}$ for an initial magnetisation corresponding to one of five cloud mass-to-magnetic-flux ratios $\mu$: from left to right, these are respectively $\mu=20,10,5,2$ and $0.7$. Our code is conservative in both the weakly magnetised and magnetically dominated cases, as evidenced by all projections being highly symmetric. In the supercritical regime ($\mu > 1$), the cloud collapses to form a dense pre-stellar core in the centre of the domain; in the sub-critical regime ($\mu < 1$) magnetic support prevents this from happening. For $\mu=20$, a rotationally supported pre-stellar disk forms in tandem with the magnetic tower structure described by~\citet{2008A&A...477....9H}, which drives a diffuse outflow propagating through the ambient medium. For $\mu=10,5,2$, we observe the emergence of bipolar collimated jets launched from the core, entrained larger scale outflows, accreting pseudo-disks and largely amplified axial and toroidal magnetic fields.}
 \label{fig:magnetised_cloud_mu}
\end{figure*}

We complement our study with an examination of the impact of initial magnetisation on the cloud's late time evolution. We fix $n_\mathrm{cloud}=64$ and run simulations with $\mu=20, 10, 5, 2, 0.7$, showing the corresponding results at $t/t_\mathrm{ff} = 1.29, 1.23, 1.21, 1.39, 2.45$ respectively in Fig.~\ref{fig:magnetised_cloud_mu}; in selecting the plotting times we considered two concurrent effects, namely the fact that for larger $\mu$ outflow velocity profiles are more gentle and change is slower, while that for larger $\mu$ collapse is delayed by magnetic forces countering gravity, to nevertheless result in more powerful jets being launched. From the highly super-critical (i.e. weakly magnetised) up to and including the sub-critical (i.e. magnetically dominated) regime, our code proves to be remarkably stable and excels at conserving both linear and angular momentum, yielding highly symmetric results. For $\mu>1$, most of the mass eventually clusters in the centre of the box to form a dense pre-stellar core, while for $\mu < 1$ collapse is magnetically arrested and only yields a core-less thick disk of a mass density that is considerably lower (note also how we plot results at a much later time for $\mu=0.7$, to demonstrate this is not simply a transient phase). 

In our $\mu=20$ run, we observe the emergence of a centrifugally supported disk ($v_r=0$ in the midplane) and a slowly expanding outflow structure, consistent with the magnetic tower described in~\citet{2008A&A...477....9H}; at this modest resolution, these are however, not accompanied by the expected, faster collimated bipolar jets.~\citet{2020A&A...638A.140W} report seeing the latter, but their figures seem to indicate absence of the former, their $\mu=20$ profiles resembling what would rather be expected for higher initial magnetisations; we deem it non-trivial to affirm whether any of the two method behaviours is more desirable, high resolution grid code results clearly showing qualitative differences between strongly and weakly magnetised clouds. As far as we are concerned, we suspect our simulation here lies at a point in $(\mu, n_\mathrm{cloud})$ space where our AR prescription engenders excess dissipation of the particularly weak seeds, constraining their amplification and preventing the anticipated magnetic structures and ensuing phenomenology from fully developing. We note we nevertheless preserve a perfectly axisymmetric structure and do not observe the severe disk fragmentation seen in the results of~\citet{2016MNRAS.455...51H, 2020A&A...638A.140W}, which is not expected in a set-up such as ours according to the extensive investigation of~\citet{2008A&A...477...25H}. 

Our results for $\mu=10, 5, 2$ are in good agreement with those obtained by using SPH at much higher resolution or by employing higher order methods~\citep{2008A&A...477....9H, 2011MNRAS.418.1392P, 2012JCoPh.231..759P, 2016MNRAS.455...51H}, with coherent collimated jets whose launching speed scales with magnetisation, large scale diffuse outflows, accreting pseudo-disks and elongated structures of largely amplified $B_z$ and $\mathbf{J}$ being undeniably present. Most notably, our $\mu=2$ results are not plagued by the spurious emergence of large-scale bubbles (as is the case in the highly magnetised clouds of~\citealt{2017arXiv170607721W, 2020A&A...638A.140W}), nor does our core start drifting, as a consequence of a poor choice of AR prescription~\citep[see][again]{2017arXiv170607721W}. Contrary to the SPMHD scheme of~\citet{2016MNRAS.455...51H}, ours proves capable (through its conservative hyperbolic/parabolic divergence-cleaning prescription) of preventing violent ejection of the core from the centre of the simulation domain and disruption of the disk, by keeping $\nabla \cdot \boldsymbol{B}$ errors consistently low.

Our method can thus reliably be applied to a problem of astrophysical relevance, where a vast range of spatiotemporal scales need to be modeled concurrently. Coupling to SWIFT's gravity solver is seamless, and we improve upon essentially all of the reported shortcomings of `traditional' SPMHD (focusing here on those identified via the magnetised cloud collapse example) through the corrective measures we implement; these are also incidentally tailored to tackle the challenges inherent to modeling galaxy and structure formation, which we shall discuss in the sections that follow.

\subsubsection{Gravitational collapse of a magnetised molecular gas cloud: the impact of Ohmic diffusion}
\label{sec:magnetised_cloud_ohmic_resistivity}

As mentioned in Section~\ref{sec:magnetised_cloud}, there is compelling evidence that gas in molecular clouds is only partly ionised, indicating that magnetic diffusivity is non-zero in such environments; ideal MHD is consequently likely to be an incomplete framework through which to study the early stages of star formation. This is further exacerbated by gravitational collapse simulations, including our own, pointing towards ideal MHD efficiently redistributing angular momentum and preventing the formation of rotationally supported proto-stellar disks in the presence of strong magnetic fields. This numerical result has been termed the `magnetic braking catastrophe'~\citep[see e.g.][]{2003ApJ...599..363A, 2006ApJ...647..374G, 2008ApJ...681.1356M, 2011ApJ...738..180L}, as it is incompatible with the observed co-occurrence of highly magnetised molecular clouds~\citep[see e.g.][]{1999ApJ...520..706C, 2005LNP...664..137H} and Keplerian proto-stellar disks~\citep[][]{2013A&A...560A.103M, 2020ApJ...890..130T} extending to several 10s of AU around young cores.

While turbulence~\citep[e.g.][]{2012ApJ...747...21S, 2012MNRAS.422..347S} or the misalignment between a cloud's rotation axis and the ambient magnetic field~\citep[e.g.][]{2006ApJ...645.1227M, 2009A&A...506L..29H, 2017MNRAS.467.3324L} have been invoked as possible mechanisms through which to reconcile numerical models with observations, the most promising avenue to do so is most likely the consideration of non-ideal MHD. Although only comprehensive, concurrent modeling of all major non-ideal effects (that is Ohmic resistivity, ambipolar diffusion, and the Hall effect) coupled to a full treatment of radiative transfer can give a complete picture~\citep[see e.g.][]{ 2017PASJ...69...95T, 2018MNRAS.481.2450W, 2021MNRAS.507.2354W}, some progress has already been established through less sophisticated models: for instance, \citet{2010A&A...521L..56D, 2011PASJ...63..555M} and~\citet{2019ApJ...876..149M} showed that Ohmic resistivity alone could suppress magnetic braking in the inner, denser regions of collapsing molecular clouds to allow the formation of compact centrifugally supported disks. 

\begin{figure}
 \includegraphics[]{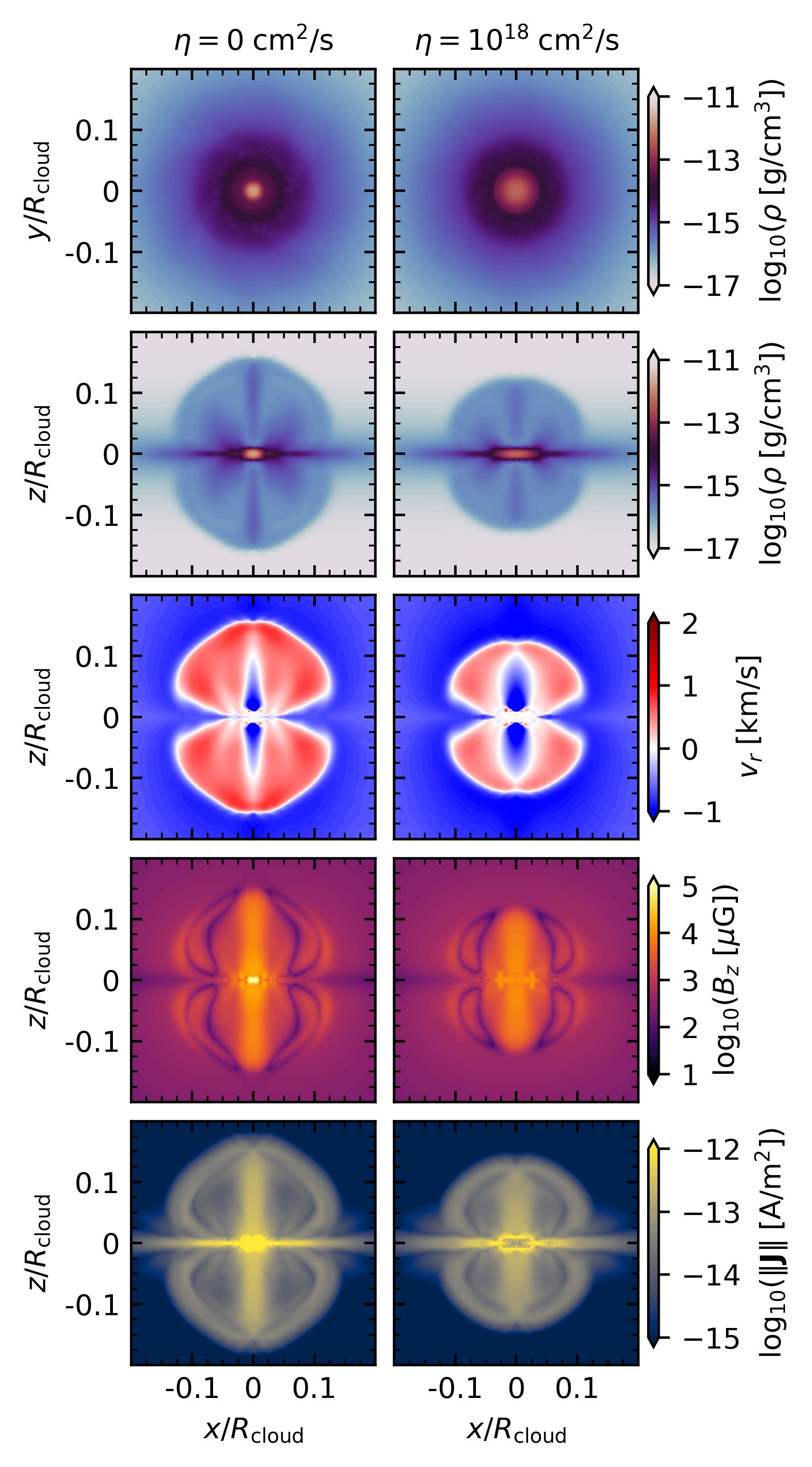} 
 \caption{Similar to Fig.~\ref{fig:magnetised_cloud_convergence} and~\ref{fig:magnetised_cloud_mu} but for re-simulations of the gravitational collapse problem pictured therein at fixed resolution and initial magnetisation (${128}^3$ particles in the molecular cloud region in the initial conditions; $\mu=10$), for which the value of the (constant, spatially uniform) Ohmic diffusivity $\eta$ was varied. We show results at $t/t_\mathrm{ff} = 1.17$ for $\eta=0 \; \mathrm{cm}^2 / \mathrm{s}$ (left column; the ideal MHD case) and for $\eta={10}^{18} \; \mathrm{cm}^2 / \mathrm{s}$ (right column; the non-ideal MHD case). Accounting for Ohmic diffusion leads to the formation of a pre-stellar core that is flatter and less dense. It results in weaker jets being launched, but allows for the emergence of a compact centrifugally supported disk. Diffusive effects partly quench the growth of the magnetic field, particularly within the core.}
 \label{fig:magnetised_cloud_resistivity}
\end{figure}

We here seek to establish whether our simple implementation of Ohmic diffusion is able to reproduce the phenomenology described above, similar experiments having been conducted in studies comparable to ours, such as~\citet{2018MNRAS.476.2476M} or~\citet{2024MNRAS.527.8355Z}. To this end, we repeat the gravitational collapse test of Section~\ref{sec:magnetised_cloud} at a resolution corresponding to $n_\mathrm{cloud} = 128$  and at an initial magnetisation of $\mu=10$; we contrast results from an ideal MHD simulation with those from a run where the magnetic diffusivity was set to (a constant, spatially uniform) $\eta = {10}^{18} \; \mathrm{cm}^2 \; \mathrm{s}^{-1}$. Our choice of $\eta$ follows from the work of~\citet{2002ApJ...573..199N} and~\citet{2007ApJ...670.1198M}, and is expected to be a reasonable estimate of the diffusion coefficient within a dense pre-stellar core. Evidence suggests that the dominant non-ideal effect there is indeed Ohmic diffusion, given the mass densities and magnetic field strengths that typically prevail in such cores (see e.g.~\citealt{2007Ap&SS.311...35W} but also~\citealt{2021MNRAS.501.5873W} for some nuance on which non-ideal effect dominates under which conditions). We compare the outcomes of our two simulations shortly after one free-fall time, at $t / t_\mathrm{ff} = 1.17$, in Fig.~\ref{fig:magnetised_cloud_resistivity}.

We show, as for the ideal MHD case, slices through the simulation box midplane of the mass density $\rho$, the radial/outflow velocity $v_r$, the axial magnetic field $B_z$ and the norm of the magnetic current vector $\boldsymbol{J}$. Results are in excellent qualitative agreement with those obtained by~\citet{2018MNRAS.476.2476M} and~\citet{2024MNRAS.527.8355Z} using AREPO; they considered a set-up very similar to what we describe, employed a finite-volume formulation of non-ideal MHD physically equivalent to our finite-mass one, and used the exact same $\eta$ as we do here. We observe that accounting for Ohmic diffusion alters the profiles of all depicted physical variables: the pre-stellar core is noticeably flattened~\citep[in line with findings presented in the non-ideal MHD studies of][]{2015ApJ...801..117T, 2016A&A...587A..32M}, appearing more oblate in shape rather than nearly spherical, roughly tripling in extent, and reaching densities a factor $\sim 10$ lower as magnetic braking is suppressed and accretion onto the centre made less efficient. Bipolar collimated jets accompanied by diffuse larger scale outflows are seen in both ideal and non-ideal runs~\citep[similarly to e.g.][]{2016MNRAS.457.1037W}, although in the latter case these are launched at a reduced speed and have therefore not propagated as far in the vertical direction~\citep[see also e.g.][]{2016A&A...587A..32M}. Most notably, though, Ohmic diffusion largely impedes mass inflows in the equatorial plane and allows for a compact centrifugally supported disk (i.e. a region of finite support throughout which $v_r=0$) to form~\citep[similarly to e.g.][]{2010A&A...521L..56D, 2011PASJ...63..555M, 2014ApJ...796L..17M, 2019ApJ...876..149M}. The axial magnetic field and magnetic current are both substantially amplified in either simulation, although their growth is partly quenched by magnetic diffusion, most notably within the core~\citep[maximal field strengths are lower by about an order of magnitude, similarly to what e.g.][find when they consider Ohmic diffusion]{2021MNRAS.507.2354W}. The elongated magnetic structure along the rotation/jet axis is less developed and thicker when $\eta \neq 0$, owing to continuous field dissipation.

We interpret the results presented in this section as further evidence that our method couples robustly to SWIFT's gravity solver, and that our implementation of Ohmic diffusion behaves as intended on a complex and highly non-linear problem. Incorporating a more comprehensive treatment of non-ideal effects in our module could be an interesting future development of the work we present here.

\subsection{Coupling MHD to gravity and cosmology}
\label{sec:mhd_and_cosmology_and_gravity}

A major component of SWIFT, which we need to verify our method couples reliably to, in view of running production simulations targeting our main intended scientific applications, is the software framework's cosmology module. It has already been established in~\citet{2025MNRAS.541.1507K} that our method can successfully be used to simulate full magnetohydrodynamical flows in an expanding universe; our scheme was shown to provide accurate numerical predictions for the propagation of cosmological MHD waves, in excellent agreement with analytical expectations, in both the ideal and non-ideal regimes as the assumed cosmological model was varied. We here build upon this and, taking a step up in terms of physical complexity, consider an MHD counterpart to the non-radiative hydrodynamical `Santa Barbara Cluster' (SBC) problem from~\citet{1999ApJ...525..554F}, similarly to~\citet{2011ApJS..195....5M, 2016MNRAS.455...51H}.

\subsubsection{MHD Santa Barbara cluster: Problem statement}
\label{sec:SBC_problem_statement}

In their original work,~\citet{1999ApJ...525..554F} simulated the assembly of a rich $\sim {10}^{15} \; \mathrm{M}_\odot$ galaxy cluster in a cold dark matter universe, using an array of different numerical methods; their goal was to assess whether then state-of-the-art codes concurrently modeling cosmology, gravity and gas dynamics could provide robust, method-agnostic predictions for a representative large-scale structure formation problem. Cosmological simulations were soon thereafter extended to account for magnetic field physics~\citep[e.g.][]{1999A&A...348..351D, 2001A&A...378..777D}, to investigate how theorised high redshift weak magnetic seeds~\citep[see e.g.][]{1994RPPh...57..325K} could lead to the $\sim \mu \mathrm{G}$ cluster fields inferred indirectly from Faraday-rotation measurements~\citep[e.g.][]{1991ApJ...379...80K, 1993ApJ...411..518G} and directly from X-ray~\citep[e.g.][]{1998MNRAS.296L..23B, 1999ApJ...513L..21F} or radio~\citep[e.g.][]{1993ApJ...416..554T, 1995A&A...302..680F} emission in the present-day universe. Subsequent elaborations on the topic~\citep[e.g.][]{2008A&A...482L..13D, 2015MNRAS.453.3999M, 2018MNRAS.480.5113M, 2024A&A...686A.157N, 2025A&A...701A.114T, 2026A&A...705A..41L} indicate that a more comprehensive treatment of galaxy formation physics is likely necessary to have theoretical predictions match a broader range of observed trends; non-radiative cluster simulations nevertheless remain of significant interest, particularly in the context of the study of different amplification mechanisms (see for instance~\citealt{2005ApJ...631L..21B, 2008Sci...320..909R, 2014MNRAS.445.3706V, 2016ApJ...817..127B, 2018MNRAS.474.1672V, 2019MNRAS.486..623D, 2022ApJ...933..131S, 2024ApJ...967..125S}, or~\citealt{2018SSRv..214..122D} for a review). 

Magnetic field amplification in clusters is first driven by gravitational compression, which, under the assumptions of isotropic collapse and magnetic flux conservation (proper to ideal MHD), leads to field strengths $\Vert \boldsymbol{B} \Vert$ scaling with the (increasing) matter density $\rho$ as~\citep[see e.g.][]{2012MNRAS.423.3148S}
\begin{equation}
    \Vert \boldsymbol{B} \Vert \propto \rho^{2/3}
\label{eq:B_rho_isotropic_collapse}
\end{equation}
The case of anisotropic collapse, a more likely evolution scenario for the primordial cosmological density field, was studied by~\citet{2006MNRAS.365.1288K} who applied the Zel’dovich approximation to Gaussian distributed initial density perturbations. They found that the scaling law above then becomes
\begin{equation}
    \Vert \boldsymbol{B} \Vert \propto \rho^{0.86}
\label{eq:B_rho_anisotropic_collapse}
\end{equation}

The second physical process most commonly invoked as the major driving force behind cluster magnetic field amplification is the so called turbulent small-scale dynamo (SSD; see~\citealt{1950RSPSA.201..405B, 1951PhRv...82..863B} for early explorations of the subject, \citealt{1967PhFl...10..859K, 1968JETP...26.1031K} for more rigorous elaborations upon it, or \citealt{2005PhR...417....1B, 2019JPlPh..85d2001R} for comprehensive reviews of the topic). Turbulent velocity fluctuations seeded and continuously driven by structure formation processes such as mergers, accretion and astrophysical feedback~\citep[see e.g.][]{2006MNRAS.366.1437S, 2010A&A...520A..17K, 2006MNRAS.369L..14V, 2009A&A...504...33V, 2011A&A...529A..17V, 2018A&A...611A..15K, 2022MNRAS.515.4229P, 2024MNRAS.528.2308P} induce exponential growth of initially weak magnetic seeds, as turbulent motions stretch, twist, fold and eventually lead to reconnection of magnetic field lines~\citep[see e.g.][]{1972SvPhU..15..159V}; as long as field strengths are dynamically subdominant (i.e. we lie in the so-called kinematic regime) and magnetic tension forces remain too weak to counter this process, flux conservation translates to continuous magnetic field amplification. Dynamo action will eventually be brought to a halt, and the magnetic energy will saturate at a sizable fraction of the kinetic energy, once the field has become strong enough to back-react on the fluid flow through the Lorentz force~\citep[see e.g.][]{1999PhRvL..83.2957S, 2004ApJ...612..276S, 2011PhRvL.107k4504F, 2014ApJ...797L..19F, 2015PhRvE..92b3010S, 2020PhRvF...5d3702S, 2021PhRvF...6j3701S}. After saturation, both turbulent and specific magnetic energies are expected to remain constant, which translates to field strength scaling with matter density as~\citep{2020ApJ...899..115X}
\begin{equation}
    \Vert \boldsymbol{B} \Vert \propto \rho^{1/2}
\label{eq:B_over_rho_saturated_dynamo}
\end{equation}
Capturing the turbulent SSD in cosmological simulations of large clusters has only recently been achieved~\citep[see e.g.][]{2018MNRAS.474.1672V}, given the stringent requirements that resolving the spatial scale at which the amplification mechanism operates imposes~\citep[see e.g.][]{2004ApJ...612..276S}. Its presence can be established through inspection of how the magnetic power spectrum $P_{\boldsymbol{B}} (k)$ scales with wavenumber $k$: \citet{1968JETP...26.1031K} predict that an SSD driven by incompressible subsonic turbulence, i.e. turbulence yielding a kinetic power spectrum that scales as $P_{\boldsymbol{v}} (k) \propto k^{-5/3}$~\citep{1941DoSSR..30..301K}, should result in $P_{\boldsymbol{B}} (k) \propto k^{3/2}$ for $k$ below a threshold $k_\mathrm{thr.}$ set by the characteristic scale of magnetic diffusion. Field amplification and saturation are expected to occur first on small scales, with the peak in $P_{\boldsymbol{B}} (k)$ and $\propto k^{3/2}$ slope being pushed to increasingly smaller $k$ through an inverse cascade as time progresses.

\subsubsection{MHD Santa Barbara cluster: Initialisation}

\begin{figure*}
 \includegraphics[width=\textwidth]{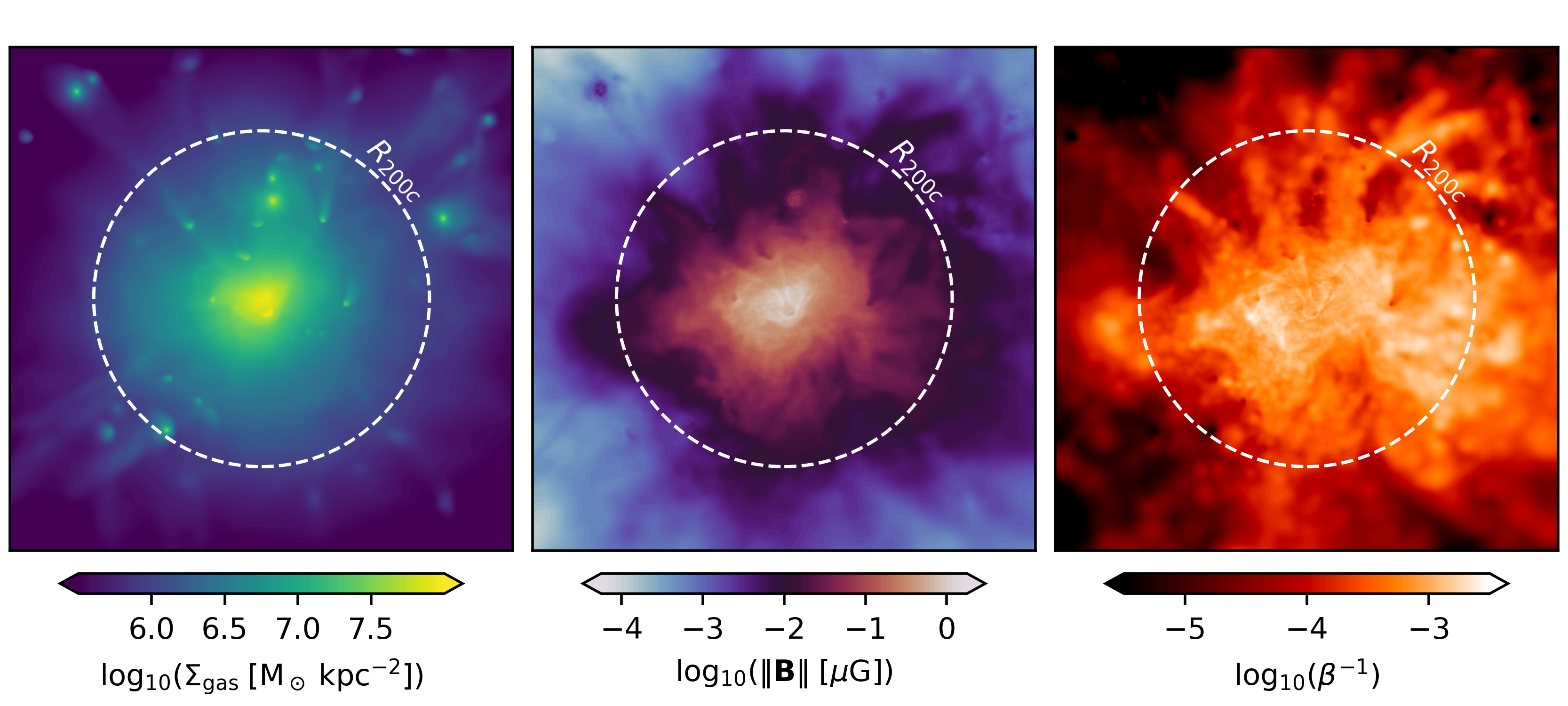} 
 \caption{Different views of the MHD Santa Barbara cluster (SBC) at $z=0$. We show, from left to right, projections of the gas surface density $\Sigma_\mathrm{gas}$, the strength of the magnetic field $\Vert \boldsymbol{B} \Vert$, and the ratio of magnetic to thermal pressure $P_\mathrm{mag} / P_\mathrm{therm} = 1 / \beta$. The maps are centred on the bound particle with the largest value of gravitational potential energy. A white dashed circle of radius $R_{200c}^\mathrm{SBC} = 2.71 \; \mathrm{Mpc}$ delineates the edge of the cluster as identified by the \texttt{VELOCIraptor} structure finder; the average matter density contained within $R_{200c}^\mathrm{SBC}$ is equal to $200$ times the critical density of the universe. The SBC appears highly dynamically active, with clear evidence of continued matter accretion onto its centre; it is thus likely subject to sustained turbulence driving. Magnetic fields are amplified by multiple orders of magnitude (when compared to their seed value of a few $\mathrm{nG}$) to reach strengths of several $\mu \mathrm{G}$ in the cluster core, in good agreement with observations; maxima in their spatial distribution are co-spatial with matter (over)densities. They show an intricate, tangled structure at small scales, and their relative contribution to the energy budget increases within the halo. They nevertheless remain dynamically subdominant with $\beta \gtrsim100$ within $R_{200c}^\mathrm{SBC}$.}
 \label{fig:SBC_projection}
\end{figure*}

To set up our SBC simulations, we start from the initial conditions of~\citet{1999ApJ...525..554F}. They consider a cubic slice of side length $L=64 \; \mathrm{Mpc}$ of an Einstein-de Sitter universe (i.e. one with energy densities $\{ \Omega_m, \Omega_r, \Omega_k, \Omega_\Lambda \} = \{ 1, 0, 0, 0 \}$), where baryons make up $10\%$ of the energy budget, and the reduced Hubble constant is fixed to $h=0.5$. Boundary conditions are taken to be periodic. Assuming an initial fluctuation spectrum consistent with COBE measurements~\citep[see][]{1992MNRAS.258P...1E}, they leverage the algorithm of~\citet{1991ApJ...380L...5H} to generate realisations of the primordial dark matter and baryon matter fields, to which they apply a transfer function proposed by~\citet{1986ApJ...304...15B} to obtain initial conditions at a starting redshift of $z=20$. The simulation box is centred on a local maximum of the field realisations, which constitutes the perturbation that will eventually collapse to form the cluster. ~\citet{1999ApJ...525..554F} provide initial conditions thus created in the form of displacements for $256^3$ points on the vertices of a regular cubic lattice, which is what we use to instantiate $256^3$ dark matter and $256^3$ gas particles of an individual mass of $m_\mathrm{dm} = 9.76 \cdot 10^8 \mathrm{M}_\odot$ and $m_\mathrm{gas} = 1.08 \cdot 10^8 \mathrm{M}_\odot$ respectively (corresponding to a ratio $m_\mathrm{dm} : m_\mathrm{gas} = 9:1$). We note that, in contrast to~\citet{2011ApJS..195....5M} and~\citet{2016MNRAS.455...51H}, we do not make use of the `zoom-in' technique whereby sub-domains of increasingly higher resolution are nested within one another to better resolve the region of interest, but rather choose to sample our simulation volume uniformly. Each gas particle is made to carry the constant magnetic field $\boldsymbol{B} = B_0 \hat{\boldsymbol{e}}_z$, where to facilitate direct comparison we choose the (physical) strength $B_0 = 1.75 \; \mathrm{nG}$, to match the fiducial initial magnetisation used by~\citet{2016MNRAS.455...51H}. We have extrapolated their $z=49$ seed value to our starting $z=20$ under the assumption of high redshift magnetic field evolution being entirely dictated by the cosmological dilution law
\begin{equation}
    \Vert \boldsymbol{B} \Vert \propto \left( 1 + z \right)^2
\label{eq:cosmological_dilution}
\end{equation}
that follows from equation~(\ref{eq:B_rho_isotropic_collapse}) for constant comoving matter density. It is widely established that this is an excellent approximation down to $z \lesssim 10$~\citep[see e.g. Fig. 5 of][]{2015MNRAS.453.3999M}. The topology of the seed field is considered to be largely irrelevant as the anticipated exponential amplification will quickly erase any memory of the initial state; our choice of a homogeneous initial field is nevertheless, incidentally, consistent with plausible magnetogenesis scenarios~\citep[see e.g.][]{2016RPPh...79g6901S}.

We configure SWIFT to concurrently account for cosmology, gravity and MHD, and evolve our system until $z=0$. We subsequently run the \texttt{VELOCIraptor} structure finder~\citep{2011MNRAS.418..320E, 2019PASA...36...21E} on our raw outputs to identify particles tied to the most massive cluster in the simulation; \texttt{VELOCIraptor} operates in phase space as a 6D `friends-of-friends'-like structure finder~\citep{1982ApJ...259..449P}, which we use in its default configuration. Other than halo membership information for individual particles, it also calculates global properties for the retrieved structures; the largest one in our snapshots, which we identify as the SBC, is found to have a total mass of $M_{200c}^\mathrm{SBC} = 1.16 \cdot {10}^{15} \; \mathrm{M}_\odot$ and radius of $R_{200c}^\mathrm{SBC} = 2.71 \; \mathrm{Mpc}$. We here define $M_{200c}^\mathrm{SBC}$ as the aggregate mass within a distance $R_{200c}^\mathrm{SBC}$ from the particle with the largest value of gravitational potential energy (the `centre' of the halo), $R_{200c}^\mathrm{SBC}$ having been chosen to delimit a region with mean matter density $200$ times the critical density of the universe $\rho_c$. The pair $\left( M_{200c}^\mathrm{SBC} , R_{200c}^\mathrm{SBC} \right)$ we obtain is in excellent agreement with that reported in~\citet{1999ApJ...525..554F}, which they calculate as an average over results obtained with 12 different numerical methods applied to the same problem.

\subsubsection{MHD Santa Barbara cluster: Results}

We first present projections of the SBC at $z=0$ in Fig.~\ref{fig:SBC_projection}; we show maps of the gas surface density $\Sigma_\mathrm{gas}$, the magnetic field strength $\Vert \boldsymbol{B} \Vert$, and the ratio of magnetic to thermal pressure (i.e. the inverse of the plasma beta) $P_\mathrm{mag}/P_\mathrm{therm} = 1/\beta$. We observe clear signs of continued accretion of smaller structures onto the central object, with infalling subhalos showing well-developed tails of stripped material; this is indicative of the fundamental processes behind sustained turbulence driving being active throughout our simulation, even if radiative effects are not accounted for. Looking at the matter density values reached, we can get an estimate of the characteristic resolution scale $h_\mathrm{typical}$ inside the main halo. Taking $h_\mathrm{typical}$ to be the median smoothing length of bound particles with $\rho > {10}^3 \rho_c$~\citep[see e.g.][for a rough delimitation of different matter density regimes in units of $\rho_c$ in a similar problem]{2005JCAP...01..009D, 2008SSRv..134..311D}, we can calculate~\citep[similarly to][]{2014MNRAS.445.3706V, 2015MNRAS.453.3999M} a crude estimate of the kinematic Reynolds number $\mathrm{Re}$ within the cluster at $z=0$ as
\begin{equation}
    \mathrm{Re}
    \sim
    \left( \frac{R_{200c}^\mathrm{SBC}}{h_\mathrm{typical}} \right)^{4/3}
    \approx 500
\end{equation}
This estimate can increase to $\mathrm{Re} \gtrsim 3000$ if $h_\mathrm{typical}$ is instead taken to be the maximal smoothing length found inside the cluster region. In either case, we should in principle be in a regime comparable to that of other analogous contemporary works~\citep[e.g.][]{2024ApJ...967..125S}, where a turbulent SSD could plausibly be triggered~\citep[as demonstrated by the idealised simulations of driven turbulence of e.g.][]{2004ApJ...612..276S}. Fig.~\ref{fig:SBC_projection} does indeed depict magnetic fields that have been amplified far beyond their initial $\sim \mathrm{nG}$ value, reaching strengths as large as a few $\mathrm{\mu \mathrm{G}}$ in the cluster centre; this is consistent with recent estimates of the typical magnitude of magnetic fields in observed clusters~\citep[see e.g.][]{2021MNRAS.502.2518S, 2024A&A...691A..23D, 2025A&A...695A.180B}. Higher magnetic field strength is largely co-spatial with larger matter (over)densities, as evidenced by local maxima in magnetisation occurring first and foremost in the core, but also at the location of infalling subhalos. Magnetic fields appear more tangled and show finer structure at smaller scales, in good agreement with the findings of~\citet{2011ApJS..195....5M}. Finally, looking at the $1/\beta$ map, we deduce that although the magnetic fields' relative contribution to the total acceleration increases within the halo, they remain dynamically subdominant down to $z=0$, with $\beta \gtrsim 100$ throughout.

\begin{figure}
 \includegraphics[]{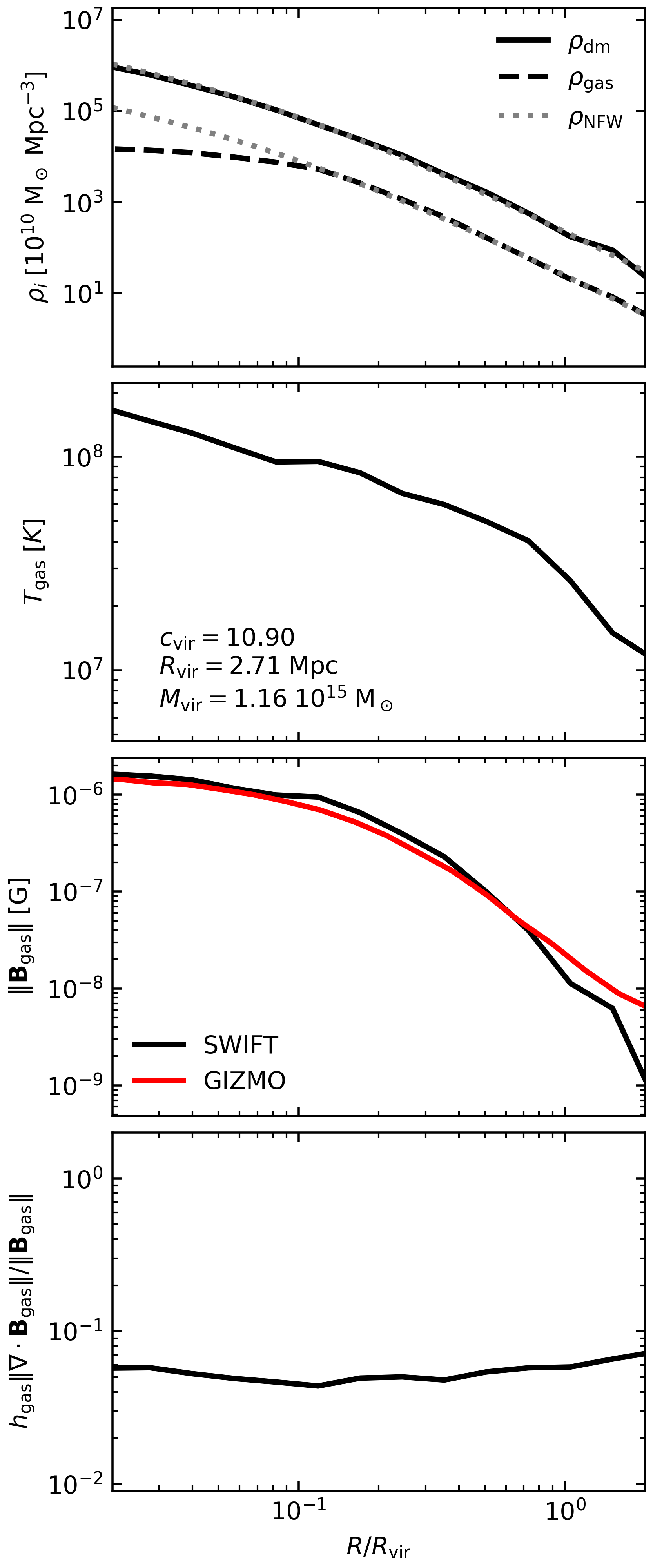} 
 \caption{Radial profiles for the MHD SBC at $z=0$; quantities are binned and subsequently averaged in $25$ logarithmically spaced intervals of distance from the cluster centre. We show, from top to bottom, dark matter and gas density, gas temperature, magnetic field strength, and dimensionless divergence error. In the first panel, we overplot two `NFW' profiles, obtained using the halo properties returned by \texttt{VELOCIraptor} (shown in the legend of the second panel), and each scaled by the fractional contribution of either dark matter or baryons to the universe's energy budget. We compare our magnetic field profile with that~\citet{2016MNRAS.455...51H} obtained using GIZMO MFM.}
 \label{fig:SBC_profiles}
\end{figure}

We further present radial profiles at $z=0$ of physical attributes of interest, binned and subsequently averaged in $25$ logarithmically spaced intervals, in Fig.~\ref{fig:SBC_profiles}. We show, in turn, profiles of the dark matter and baryon densities $\rho_\mathrm{dm}$ and $\rho_\mathrm{gas}$, the gas temperature $T_\mathrm{gas}$, the magnetic field strength $\Vert \boldsymbol{B} \Vert$, and the dimensionless divergence error $h \Vert \nabla \cdot \boldsymbol{B} \Vert / \Vert \boldsymbol{B} \Vert$, as functions of the distance $R$ from the centre of the cluster. We overplot, on top of $\rho_\mathrm{dm}$ and $\rho_\mathrm{gas}$, `Navarro-Frenk-White' (NFW) density profiles $\rho_\mathrm{NFW}$~\citep{1995MNRAS.275..720N}, calculated using the halo properties returned by \texttt{VELOCIraptor} and each scaled by the respective species' relative contribution to the universe's energy budget within our assumed cosmological model. $\rho_\mathrm{dm}$, being entirely set by the gravity solver, is essentially indistinguishable from the corresponding $\rho_\mathrm{NFW}$ over the entire radial range we consider; the shape of our dark matter halo is therefore in excellent agreement with what earlier studies predict~\citep{1996ApJ...462..563N, 1997ApJ...490..493N}. The gas distribution can also be truthfully described by an NFW profile down to radii of $R \sim 0.1 R_{200c}^{SBC}$ as, for a large fraction of the dynamic range probed, gravity remains the dominant force in determining the gas' dynamical evolution. The central region of the cluster sees, however, the flattening of $\rho_\mathrm{gas}$ to form a `low-density core'; this is accompanied by relatively high central temperatures, translating to a large amount of entropy at small radii. Such a configuration is qualitatively very similar to that observed by~\citet{2022MNRAS.511.2367B} on a (purely hydrodynamical) structure formation problem similar to the SBC~\citep{2016MNRAS.457.4063S}; they found that SPHENIX, particularly thanks to its artificial thermal diffusion prescription, is capable of producing `cored' rather than `cuspy' cluster cores, in line with other commonly employed astrophysical codes such as AREPO. Our ability to reproduce this result in the presence of (dynamically subdominant) magnetic fields cements our confidence in our SPMHD scheme being minimally invasive and preserving SPHENIX's successes, coupling seamlessly to the default treatment for cosmology, gravity and gas dynamics in SWIFT. The magnetic field itself decreases monotonically with radius and its profile, similarly to that of gas density, appears to broadly follow a double power law; similarly to~\citet{2011ApJS..195....5M}, we observe a flatter `magnetic core' in the inner regions (of a similar extent to the gas density core), followed by a more rapid decrease in magnetisation at larger distances from the centre. Our measured profile agrees well up to $R \sim R_{200c}^{SBC}$ with that~\citet{2016MNRAS.455...51H} obtained with GIZMO MFM~\citep[itself consistent with that of][]{2011ApJS..195....5M}; beyond the edge of the cluster, we predict a somewhat weaker magnetic field, most likely due to differences in the codes' inherent diffusive properties at low resolution. The divergence error remains at the level of a few percent throughout the entire halo, indicating our solution has most likely not been corrupted by spurious numerical effects; this is particularly notable in light of the major merger that occurs as late as $z \sim 0.5$~\citep[see][]{1999ApJ...525..554F}.

\begin{figure*}
 \includegraphics[width=\textwidth]{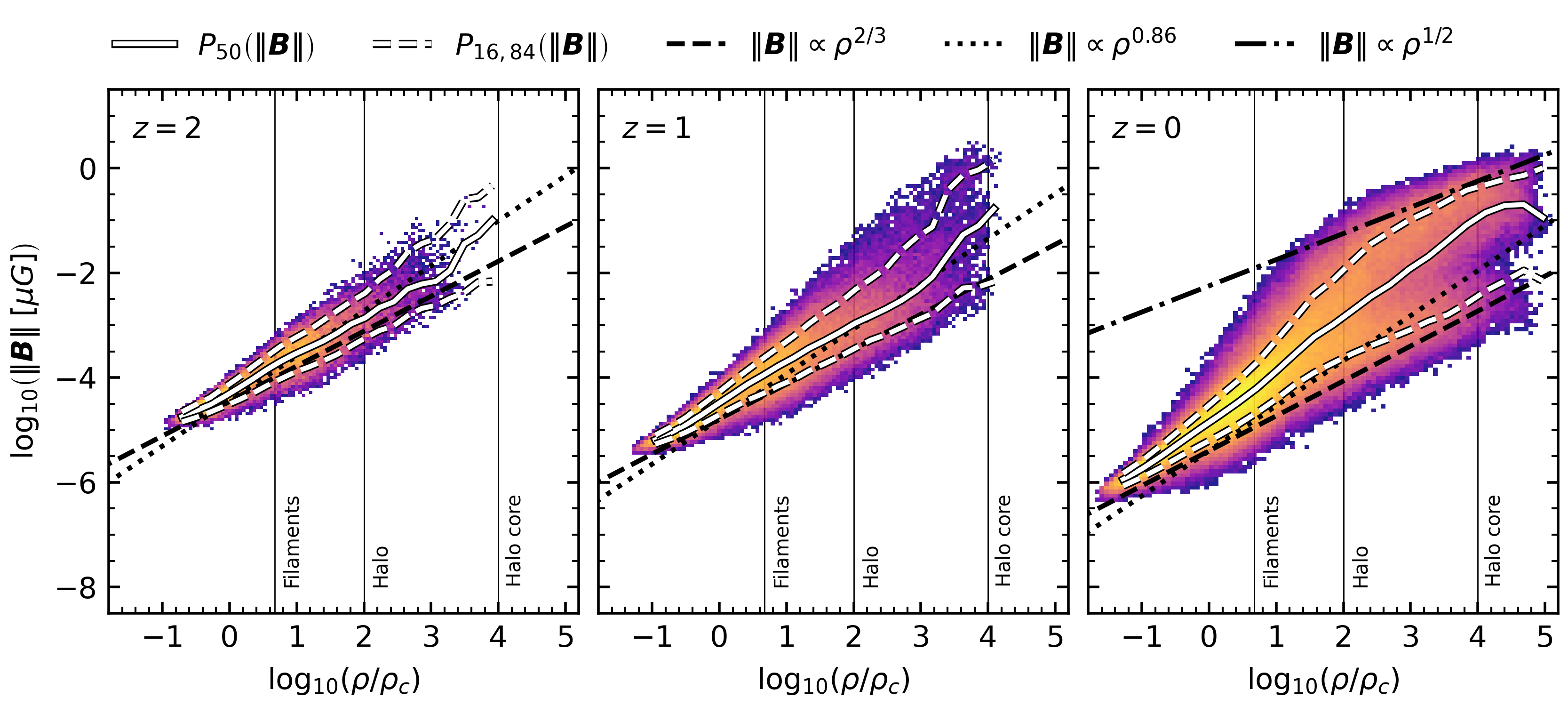} 
 \caption{Evolution of particles tied to the Santa Barbara cluster (SBC) in the matter density $\rho$ (measured in units of the critical density of the universe $\rho_c$) to magnetic field strength $\Vert \boldsymbol{B} \Vert$ phase space, at the three consecutive redshifts $z=2,1,0$ (from left to right). The colour scheme used to paint the $(\rho, \Vert \boldsymbol{B} \Vert)$ distributions ranges from purple to yellow, according to the normalised log-count of particles falling into each 2D bin. We plot the median (solid white line), as well as the ${16}^\mathrm{th}$ and ${84}^\mathrm{th}$ percentiles (dashed white lines) of $\Vert \boldsymbol{B} \Vert$ values binned in logarithmically spaced intervals of $\rho$. We overplot the scaling laws associated with magnetic field amplification owing to isotropic (black dashed lines) and anisotropic (black dotted lines) gravitational collapse, as well the $\rho - \Vert \boldsymbol{B} \Vert$ relation for a saturated small-scale dynamo (SSD; black dotted-dashed line). To guide the eye, thin black solid lines delimit the different $\rho$ regimes according to the cosmic environments to which they correspond at $z=0$, similarly to e.g.~\citet{2005JCAP...01..009D}. $\rho$ and $\Vert \boldsymbol{B} \Vert$ display a clear positive correlation throughout the evolution of the SBC, indicating there is a tight causal link between cosmic mass assembly and magnetic field growth. Field amplification first proceeds through gravitational contraction (see how the $\rho - \Vert \boldsymbol{B} \Vert$ median relation closely follows the black dotted line at $z=2$), to subsequently be overtaken (starting from the cluster core and then cascading down to lower densities) by what appears to be dynamo effects driven by turbulence and shearing motions. The population of particles sitting in the high $\rho$, high $\Vert \boldsymbol{B} \Vert$ part of the diagram at $z=0$ seems to be consistent with a saturated SSD.}
 \label{fig:SBC_phaseSpace}
\end{figure*}

We proceed by presenting plots of the phase space distribution of gas tied to the SBC in Fig.~\ref{fig:SBC_phaseSpace}. We show 2D histograms of particle carried values of $(\rho, \Vert \boldsymbol{B} \Vert)$, from simulation snapshots taken at the three consecutive redshifts $z=2, 1, 0$. We accompany these by lines tracing the median and $16^\mathrm{th}$ and $84^\mathrm{th}$ percentiles of $\Vert \boldsymbol{B} \Vert$ in logarithmically spaced bins of $\rho$. To guide the eye, we further overplot the power laws~(\ref{eq:B_rho_isotropic_collapse}) and~(\ref{eq:B_rho_anisotropic_collapse}), corresponding respectively to magnetic field amplification owing to isotropic and anisotropic gravitational collapse. In the last panel, we further include the scaling relation~(\ref{eq:B_over_rho_saturated_dynamo}), giving the dependence of magnetic field strength on density for a saturated SSD. We note that the (now outdated) cosmological model we have assumed for this test differs significantly from those considered in recent works addressing similar questions; we therefore focus on commenting on general qualitative trends, rather than making one-to-one comparisons at fixed redshift with other studies. There appears to be a clear causal connection between structure formation and magnetic field amplification, $\rho$ and $\Vert \boldsymbol{B} \Vert$ being positively correlated at all redshifts shown; mass assembly proceeding to form denser environments coincides with stronger field strengths being observed. The leftmost panel of Fig.~\ref{fig:SBC_phaseSpace} shows that at $z=2$, when baryonic matter reaches at most $\rho \sim {10}^3 \rho_c$ and $\Vert \boldsymbol{B} \Vert \sim 0.1 \; \mu \mathrm{G}$, the median $\rho - \Vert \boldsymbol{B} \Vert$ relation closely follows equation~(\ref{eq:B_rho_isotropic_collapse}), indicating that magnetic field amplification up to that point is largely driven by gravity. At $z=1$, as matter further contracts, we clearly observe excess amplification beyond that expected from gravitational contraction at the high density end ($\rho \gtrsim {10}^{3.5} \rho_c$). By $z=0$, only relatively under-dense gas follows the adiabatic compression scaling law, which shifts downward under the effect of cosmological dilution (see equation~(\ref{eq:cosmological_dilution})). On the other hand, the median magnetic field for cluster particles with $\rho \gtrsim 10 \rho_c$ shows a departure by at least one order of magnitude from what is expected from flux conservation alone, the trend being more pronounced at larger $\rho$. Although a fraction of high-density particles still lives around the adiabatic contraction line, those that saw their magnetic field most amplified follow closely the $\rho - \Vert \boldsymbol{B} \Vert$ scaling relation expected for a saturated SSD. These observations are in good agreement with what is reported in other SPMHD~\citep[e.g.][]{2005JCAP...01..009D, 2008SSRv..134..311D, 2022ApJ...933..131S, 2024ApJ...967..125S}, high resolution AMR~\citep[e.g.][]{2008A&A...482L..13D} or moving-mesh~\citep[e.g.][]{2015MNRAS.453.3999M, 2025A&A...701A.114T} non-radiative cluster studies, and are clearly suggestive of several magnetic field amplification mechanisms being concurrently at play. This in itself can also be deduced from the increase in scatter around the median $\rho - \Vert \boldsymbol{B} \Vert$ relation as the simulation proceeds~\citep[see][]{2015MNRAS.453.3999M}.

\begin{figure}
 \includegraphics[]{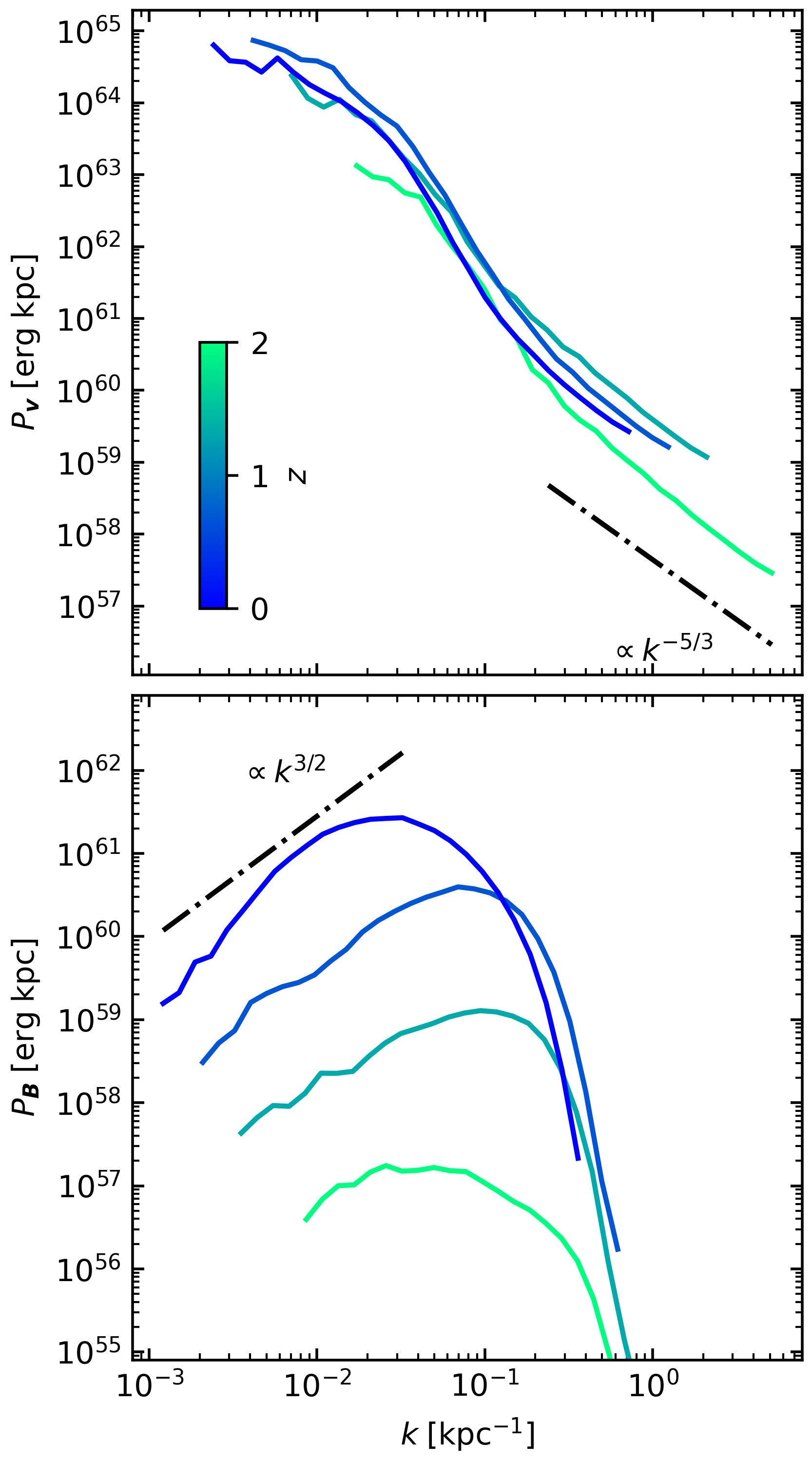} 
 \caption{Kinetic (top) and magnetic (bottom) power spectra $P_{\boldsymbol{v}}$ and $P_{\boldsymbol{B}}$ for the Santa Barbara cluster at the four consecutive redshifts $z=2, 1, 0.5, 0$ (as indicated by line colour). Spectra are are computed by first depositing $\sqrt{\rho/2} \boldsymbol{v}$ and $\sqrt{1/ 2\mu_0} \boldsymbol{B}$ on 3D `zero-padded' Cartesian grids of side length half the size of the cluster $0.5 R_{200c}^\mathrm{SBC}$, centred on the halo particle with largest gravitational potential energy. These are subsequently Fourier transformed and averaged in bins of constant wave vector length $k$. The shape of $P_{\boldsymbol{v}}$ agrees well with the~\citet{1941DoSSR..30..301K} scaling $\propto k^{-5/3}$ at small scales, indicating the continuous presence of subsonic incompressible turbulence in the cluster's core. The shape of $P_{\boldsymbol{B}}$ agrees well with the~\citet{1968JETP...26.1031K} scaling $\propto k^{3/2}$ at large scales, demonstrating the prolonged action of a turbulent small-scale dynamo. The normalisation of $P_{\boldsymbol{B}}$ increases with time, revealing a prolonged accumulation of magnetic energy; the spectrum's peak shifts to increasingly larger scales through an inverse cascade.}
 \label{fig:SBC_power_spectrum}
\end{figure}

To further support the claim that the excess magnetic field amplification we observe (beyond that expected from gravitational compression alone) can indeed be attributed to the action of a turbulent SSD, we complete our investigation by performing a power spectrum analysis of our simulation results~\citep[similarly to e.g.][]{2014MNRAS.445.3706V, 2018MNRAS.474.1672V, 2022ApJ...933..131S, 2024MNRAS.528.2308P, 2025A&A...701A.114T}. Starting from the `vector square roots' of the position-space kinetic and magnetic energy densities
\begin{equation}
    \boldsymbol{\varepsilon}_{\boldsymbol{v}} ( \boldsymbol{x} ) 
    \equiv
        \left(
            \sqrt{\frac{\rho}{2}} \boldsymbol{v}
        \right) ( \boldsymbol{x} )
    \quad \text{and} \quad
    \boldsymbol{\varepsilon}_{\boldsymbol{B}} ( \boldsymbol{x} )
    \equiv
    \sqrt{\frac{1}{2 \mu_0}} \boldsymbol{B} ( \boldsymbol{x} )
\label{eq:position_space_vector_square_roots_energy_density}
\end{equation}
we can compute the 1D kinetic and magnetic power spectra
\begin{equation}
    P_{\boldsymbol{v}} (k) = 
    \underset{\Vert \boldsymbol{k} \Vert = k}{\mathrm{average}}
    \left\{
        4 \pi \boldsymbol{k}^2 
        \widetilde{\boldsymbol{\varepsilon}}_{\boldsymbol{v}} 
        ( \boldsymbol{k} )
        \left[
        \widetilde{\boldsymbol{\varepsilon}}_{\boldsymbol{v}} 
        ( \boldsymbol{k} )
        \right]^*
    \right\}
\label{eq:kinetic_spectrum}
\end{equation}
and
\begin{equation}
    P_{\boldsymbol{B}} (k) = 
    \underset{\Vert \boldsymbol{k} \Vert = k}{\mathrm{average}}
    \left\{
        4 \pi \boldsymbol{k}^2 
        \widetilde{\boldsymbol{\varepsilon}}_{\boldsymbol{B}} 
        ( \boldsymbol{k} )
        \left[
        \widetilde{\boldsymbol{\varepsilon}}_{\boldsymbol{B}} 
        ( \boldsymbol{k} )
        \right]^*
    \right\}
\label{eq:magnetic_spectrum}
\end{equation}
where $\boldsymbol{k}$ is the 3D Fourier wave vector, $\widetilde{\boldsymbol{\varepsilon}}_{\boldsymbol{v}} ( \boldsymbol{k} )$ and $\widetilde{\boldsymbol{\varepsilon}}_{\boldsymbol{B}} ( \boldsymbol{k} )$ the Fourier transformed $\boldsymbol{\varepsilon}_{\boldsymbol{v}} ( \boldsymbol{x} )$ and $\boldsymbol{\varepsilon}_{\boldsymbol{B}} ( \boldsymbol{x} )$, and $*$ denotes complex conjugation, as averages in shells of constant $k \equiv \Vert \boldsymbol{k} \Vert$ in Fourier space. By Parseval's identity, integrating $P_{\boldsymbol{v}} (k)$ and $P_{\boldsymbol{B}} (k)$ over $k$ should respectively return the total kinetic and magnetic energy in the volume considered.

We perform the calculations~(\ref{eq:kinetic_spectrum}) and~(\ref{eq:magnetic_spectrum}) following the procedure laid out in e.g.~\citet{2013ApJ...763...51F, 2017MNRAS.469.3185P, 2020MNRAS.498.3125P, 2024MNRAS.528.2308P, 2022MNRAS.515.4229P, 2025A&A...701A.114T}: we first deposit the energy densities~(\ref{eq:position_space_vector_square_roots_energy_density}) on a ${1024}^3$ regular Cartesian grid centred on the cluster particle with the highest value of gravitational potential energy (as identified by \texttt{VELOCIraptor}). We repeat this for simulation snapshots taken at the four consecutive redshifts $z=2, 1, 0.5, 0$, setting the deposition grid's side length $L_\mathrm{grid}$ to $L_\mathrm{grid} (z) = 0.5 R_{200c}^\mathrm{SBC} (z)$~\citep[see again][]{2025A&A...701A.114T}. To improve power spectrum reconstruction, we `zero-pad' the signal, which we then transform using standard FFT routines provided with the \texttt{scipy} software package. Averaging the Fourier-transformed fields is then performed in discrete bins of $k$. We plot the resulting kinetic and magnetic spectra in Fig.~\ref{fig:SBC_power_spectrum}, together with the scaling relations we expect these to follow, as outlined in Section~\ref{sec:SBC_problem_statement}. At small scales, $P_{\boldsymbol{v}} (k)$ closely follows the~\citet{1941DoSSR..30..301K} scaling $\propto k^{-5/3}$, illustrating the continuous presence of incompressible subsonic turbulence throughout the SBC's history. The amplitude of $P_{\boldsymbol{v}} (k)$ increases by roughly one order of magnitude between $z=2$ and $z=1$ as the cluster feeds on infalling structures, eventually settling at a slightly lower normalisation as the dynamical system becomes more relaxed~\citep{1999ApJ...525..554F}. $P_{\boldsymbol{B}} (k)$ appears at all, but the last redshift probed to closely follow the~\citet{1968JETP...26.1031K} scaling $\propto k^{3/2}$ at large scales, indicating the sustained action of a turbulent SSD. By $z=0$, there is a slight reduction in power at the smallest $k$ captured, something that has been reported in other studies as well~\citep[see e.g. the largest halo considered by][]{2024MNRAS.528.2308P}; it could very well be that our simulations do not meet the resolution requirements to model magnetic field amplification at such large scales~\citep[][claim $m_\mathrm{gas} \sim {10}^6 \mathrm{M}_\odot$ is necessary for fully converged results, a factor ${10}^2$ smaller than what we use here]{2018MNRAS.474.1672V, 2022ApJ...933..131S}. Magnetic energy first saturates on smaller scales, at a fraction of the local kinetic energy, following a steep scaling with $k$ in line with expectations~\citep[e.g.][]{2004ApJ...612..276S, 2015ApJ...810...93P, 2022ApJ...933..131S}. The overall normalisation of $P_{\boldsymbol{B}} (k)$ increases with time, indicating a prolonged accumulation of magnetic energy, as the spectrum's peak shifts to increasingly larger scales through an inverse cascade, again in excellent agreement with previous works.

\subsubsection{MHD Santa Barbara cluster: Method variations }

Fig.~\ref{fig:SBC_phaseSpace} shows that, at $z=0$, the median $\Vert \boldsymbol{B} \Vert$ eventually starts decreasing at the high $\rho$ tail. We found that properties of this feature correlated with resolution; throwing more particles at the problem pushed the `turnover point' to higher densities (noting also that higher resolution allows for larger densities to be reached). Moreover, this turnover was found to be absent in runs where we had disabled our hyperbolic/parabolic divergence-cleaning scheme, for which the median $\Vert \boldsymbol{B} \Vert$ line increased monotonically for all $\rho$. It could very well be that the high degree of dynamical activity in the cluster core is very demanding of the divergence-cleaning algorithm; the latter assists us in obtaining a more sound magnetic field topology, perhaps nonetheless at the price of some excess diffusion. Interestingly, the same feature is present in results from both Eulerian and hybrid codes~\citep{2008A&A...482L..13D, 2015MNRAS.453.3999M}, which employ vastly different strategies to address divergence errors. We leave it for future work to establish whether there exists an intrinsic, resolution-dependent density limit beyond which additional care is necessary to interpret simulation results.

We also ran the SBC problem with a modified version of our code, where our default AR prescription was replaced by that employed by most other contemporary SPMHD solvers~\citep{2018PASA...35...31P, 2020A&A...638A.140W, 2023MNRAS.518.4115G}; the major qualitative difference between the two is that the latter considers the AR signal speed given by equation~(\ref{eq:ar_signal_velocity_others}) which depends on pair-wise velocity differences between particles, whereas the former uses instead a $v_{A}^\mathrm{AR}$ based on individual particles' Alfvén speeds. $v_{\delta \boldsymbol{v}}^\mathrm{AR}$ has been found to limit excess dissipation compared to established alternatives, while simultaneously delivering satisfactory performance on challenging MHD problems~\citep[e.g.][]{2018MNRAS.481.2450W, 2022A&A...659A..91W, 2023A&A...673A..47W}. We nonetheless found that when applied to the SBC it is overly diffusive, pushing the median magnetic field below the adiabatic compression line even at moderate overdensities. We attribute this to the highly nonlinear and disordered velocity field structure proper to cosmological simulations, which inevitably gives rise to exceedingly large $v_{\delta \boldsymbol{v}}^\mathrm{AR}$. The SBC test was thus highly informative in guiding our choice of $v_{A}^\mathrm{AR}$, which showed minimal excess dissipation in both the low (weak seeds) or high (saturated SSD) magnetisation regimes; it had naturally first been established that it also served its purpose on the problems detailed in all preceding sections.

\subsection{Coupling MHD to sub-grid recipes for galaxy formation}
\label{sec:mhd_and_subgrid}

A key component of SWIFT that further cements the software package's wider appeal is the broad range of phenomenological feedback models it comes bundled with; these are designed to capture the repercussions on larger scales of astrophysical processes occurring below a given simulation's resolution limit, and can be `mixed and matched' at will to address the scientific question at hand, thanks to the modular nature of the code. Such complex networks of sub-resolution models have for instance recently been used, in tandem with SWIFT's cosmology, gravity and hydrodynamics modules, to generate realistic synthetic galaxy and cluster populations~\citep{2023MNRAS.526.4978S, 2023MNRAS.526.6103K, 
2026MNRAS.548ag375S, 2026MNRAS.548ag300C}. How such sub-grid recipes for galaxy formation couple to our novel SPMHD scheme therefore comes as a natural point of interest in the context of the present study. The full ramifications of this coupling merit a standalone investigation which should, in turn, inform the astrophysical and cosmological studies that could follow. Here, we shall simply consider an MHD counterpart to the controlled isolated disk galaxy set-up of~\citet{2024MNRAS.532.3299N}; through this somewhat contrived example we mainly aspire~\citep[similarly to e.g.][]{2013MNRAS.432..176P, 2016MNRAS.455...51H, 2016MNRAS.463..477M} to establish the stability properties of our method when combined with a state-of-the-art, complete galaxy formation model, as it comes `out-of-the-box'; having confirmed the broad plausibility of the results we obtain, we leave a more comprehensive exploration thereof (and their implications in terms of both further method development and fundamental science) for future work.

\subsubsection{Magnetic fields in a Milky Way-like isolated disk galaxy: Problem statement}

Isolated disk galaxy set-ups constitute an easily reproducible, largely tractable, yet physically representative configuration that has proven instrumental in the development and calibration of sub-grid feedback models. Simulations of such systems have for instance guided the design of numerical recipes for gas cooling, star formation and evolution, as well as feedback from stellar populations~\citep[see e.g.][]{2003MNRAS.339..289S, 2005MNRAS.361..776S, 2006MNRAS.373.1074S, 2008MNRAS.383.1210S, 2012MNRAS.426..140D, 2016ApJ...826..200S, 2017MNRAS.466...11R, 2018MNRAS.478..302S, 2021MNRAS.506.3882S, 2022MNRAS.514..249C, 2024MNRAS.532.3299N}. Comparable experiments that also account for MHD effects have thus quite naturally been undertaken~\citep[see e.g.][]{2009MNRAS.397..733K, 2009ApJ...696...96W, 2010A&A...523A..72D, 2013MNRAS.432..176P, 2016MNRAS.457.1722R, 2017MNRAS.472.4368R, 2016MNRAS.461.4482D, 2017ApJ...843..113B, 2019MNRAS.483.1008S, 2022MNRAS.515.4229P}, to study the possible origin of galactic magnetic fields and the amplification mechanisms that lead to their observed present-day strengths~\citep[see e.g.][for a recent review]{2023ARA&A..61..561B}. 

The picture the aforementioned studies paint suggests that weak magnetic seeds grow due to the concurrent action of distinct processes. When still dynamically subdominant (the kinematic regime), magnetic fields can be exponentially amplified by several orders of magnitude under the action of a small-scale dynamo~\citep[SSD;][]{1968JETP...26.1031K} driven by turbulence.  Turbulence itself is continuously seeded by e.g. gas motions, astrophysical feedback or fluid instabilities, particularly in the dense interstellar gas constituting galactic disks; an SSD will cause magnetic energy to first saturate on the smallest of scales, to subsequently be transported to increasingly larger scales through an inverse cascade. A second class of processes commonly invoked are large-scale dynamos (LSD), such as the $\alpha-\Omega$ dynamo~\citep[see e.g.][]{1971ApJ...163..255P}, whereby convective/twisting motions and differential shear interfere constructively to promote magnetic field growth. LSDs have been associated with later stages of the amplification process (following an initial exponential phase typically attributed to the SSD), during which magnetic energy buildup slows down as the field assumes an increasingly coherent topology on larger scales, in line with observations~\citep[e.g.][]{1976Natur.264..222S, 2007A&A...470..539B}. Gas compression due to gravitational contraction can further assist magnetic field growth, but its effect will typically be negligible once disk-driven dynamos are in operation. 

Once fields have grown strong enough for the Lorentz force to start back-reacting on the fluid flow, amplification will eventually come to a halt and magnetic energy will saturate once it is in (mildly sub)equipartition with the turbulent and thermal energy components; how and when this happens has been shown to be resolution dependent and still constitutes an open question. There is nevertheless stronger consensus on typical field saturation values, at the $\gtrsim \mu \mathrm{G}$ level, which are in good agreement with observations of both local universe~\citep{2007A&A...470..539B} and high-redshift~\citep{2008Natur.454..302B} galaxies. The perspective we summarise here has been largely corroborated by cosmological simulations~\citep[see e.g.][]{2017MNRAS.469.3185P, 2024MNRAS.528.2308P, 2017MNRAS.472.4368R, 2018MNRAS.479.3343M, 2022MNRAS.513.3326M, 2025A&A...701A.114T}, which in turn have fueled an accruing interest in the physics of the circumgalactic medium~\citep[CGM; see e.g.][]{2019MNRAS.483.1008S, 2020MNRAS.498.3125P, 2021MNRAS.501.4888V} as galactic magnetic fields transported by outflows can pollute it~\citep[see also e.g.][for an observational perspective]{2013Natur.493...66C}.

\begin{figure*}
 \includegraphics[width=\textwidth]{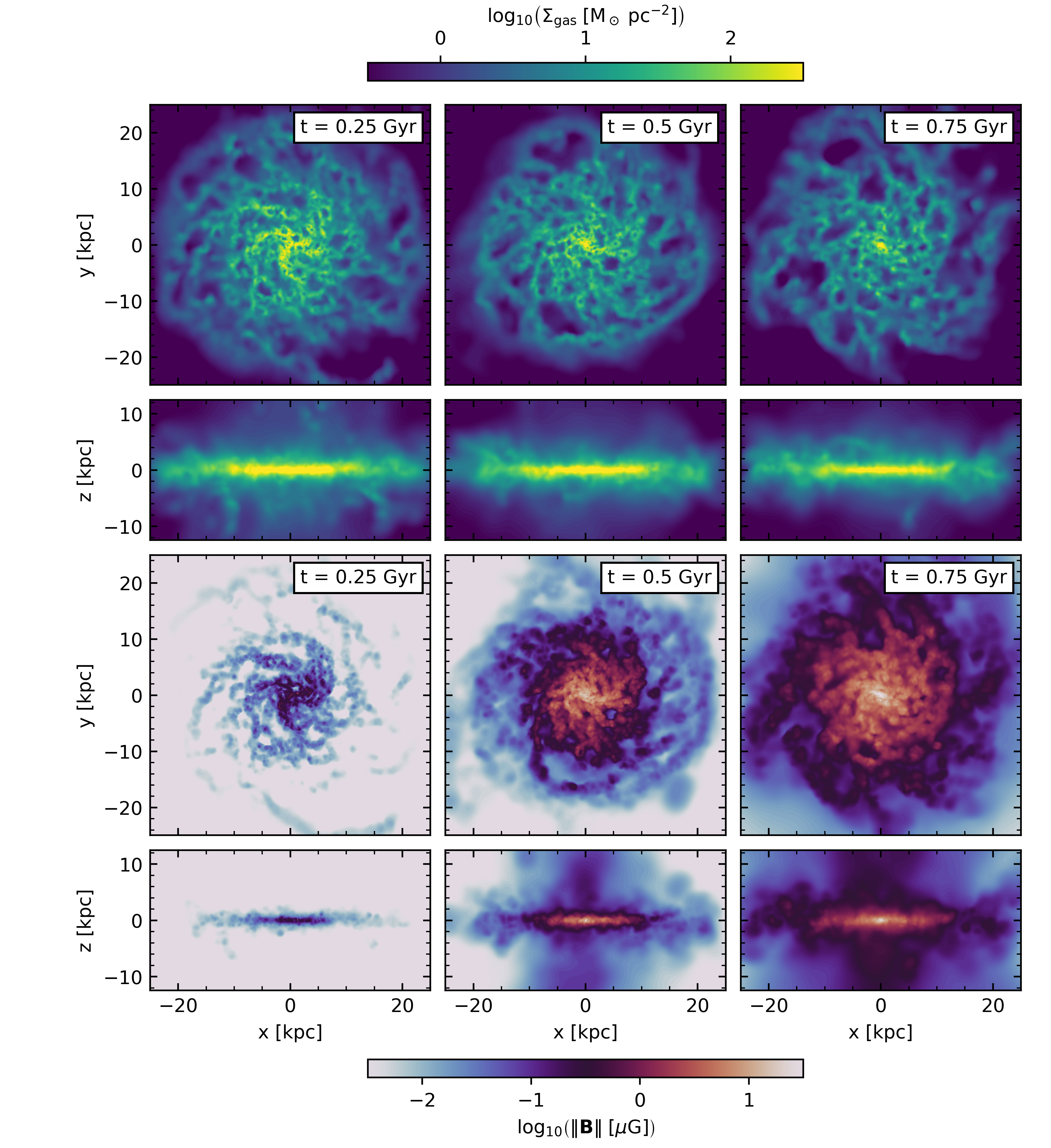} 
 \caption{Evolution of a magnetised Milky Way-like isolated disk galaxy; example reproduced from~\citet{2024MNRAS.532.3299N}, initialised with a uniform trace magnetic field of ${10}^{-3} \; \mu \mathrm{G}$ confined to the disk plane, and run with both gravity and MHD enabled, coupled to the EAGLE~\citep{2015MNRAS.446..521S, 2015MNRAS.450.1937C} sub-grid prescriptions for radiative gas cooling, star formation, stellar feedback and chemical enrichment. We show face-on ($1^\mathrm{st}$ and $3^\mathrm{rd}$ rows) and edge-on ($2^\mathrm{nd}$ and $4^\mathrm{th}$ rows) maps of the gas surface density $\Sigma_\mathrm{gas}$ ($1^\mathrm{st}$ and $2^\mathrm{nd}$ rows)
 and magnetic field strength $\Vert \boldsymbol{B} \Vert$ ($3^\mathrm{rd}$ and $4^\mathrm{th}$ rows) at $t = 250, 500, 750 \; \mathrm{Myr}$ (columns, from left to right); these are integer multiples of the disk plane rotational period $T_\mathrm{rot.} \sim 250 \; \mathrm{Myr}$ at a distance from the galactic centre equal to the characteristic scale-length setting the radial extent of the gaseous and stellar disks. Energetic feedback from stars blows hot, under-dense (super)bubbles, locally disperses the gas, seeds supersonic turbulence and launches large-scale outflows into the circumgalactic medium. Magnetic fields are rapidly amplified to $\gtrsim 10 \; \mu \mathrm{G}$ values, with turbulence and differential shear driving the `inside-out' build-up of a magnetised disk. Fields eventually saturate, first in the galactic centre and subsequently at increasingly larger radial distances, in line with what would be expected for a galactic dynamo (see text). They moreover pollute the circumgalactic medium, as supernova driven winds eject magnetised material from the disk.}
 \label{fig:galaxy_projection}
\end{figure*}

\subsubsection{Magnetic fields in a Milky Way-like isolated disk galaxy: Initialisation}
\label{sec:galaxy_initialisation}

In the name of simplicity, we choose as initial conditions a configuration that corresponds to an already settled galactic disk, comprised of both a gaseous and stellar component, embedded in a static dark matter halo.  We stress that we do not explicitly integrate a CGM component in our set-up~\citep[as opposed to e.g.][]{2019MNRAS.483.1008S}.

We model our dark matter mass distribution using a fixed~\citet{1990ApJ...356..359H} gravitational potential (as opposed to representing it using dark matter particles). Following the parametrisation discussed in~\citet{2024MNRAS.532.3299N}, we set it to have a total mass of $M_{200} = 1.37 \cdot {10}^{12} \; \mathrm{M}_\odot$~\citep[which is compatible with recent estimates of the Milky-Way's virial mass, see e.g.][]{2017MNRAS.465...76M}, a concentration of $c=9$~\citep[according to the $z=0$ concentration-mass relation of][]{2015MNRAS.452.1217C} and spin parameter $\lambda=0.033$~\citep[again typical of ${10}^{12} \; \mathrm{M}_\odot$ dark matter halos, see e.g.][]{2001ApJ...555..240B}. These three parameters, together with an assumed value for the Hubble constant of $H_0 = 70.4 \; \mathrm{km} \; \mathrm{s}^{-1} \; \mathrm{Mpc}^{-1}$, uniquely determine the dark matter component. 

Gas and star particles are instantiated following the procedure outlined in~\citet{2005MNRAS.361..776S} to sample the gaseous and stellar disks, both made to follow exponential surface density profiles whose common scale-length $r_\mathrm{disk}$ is determined through $\lambda$~\citep[see e.g.][for more details]{1998MNRAS.295..319M, 1999MNRAS.307..162S} to be $r_\mathrm{disk} = 4.3 \; \mathrm{kpc}$. The disk components are set to have a combined mass equal to $4 \%$ of $M_{200}$, and the gas fraction is fixed to $30 \%$. We make use of gas particles of an individual mass of $m_\mathrm{gas} = 10^5 \; \mathrm{M}_\odot$, assign them solar metallicities $Z_\mathrm{gas} = 0.0134$ and an initial temperature of $T_\mathrm{gas} = 10^4 \; \mathrm{K}$. We further make each of them carry the uniform magnetic field (which is confined to the disk plane) $\boldsymbol{B} = B_0 \hat{\boldsymbol{e}}_x$, where we set $B_0 = {10}^{-3} \; \mu \mathrm{G}$. This trace magnetic seed is what~\citet{2013MNRAS.432..176P} use to initialise their collapsing gas halo simulations (see also~\citealt{2009MNRAS.397..733K, 2009ApJ...696...96W} who make very similar choices). Although admittedly unrealistic~\citep[see e.g.][for a discussion of how to instantiate topologically sound seed magnetic fields in isolated galaxy simulations]{2010A&A...523A..72D, 2016MNRAS.457.1722R}, our seed field being divergence-less and dynamically weak is largely sufficient for our purposes; it was established that such an initialisation procedure did not bring about unwarranted spurious effects, by verifying the system's early-time evolution was essentially identical in pure hydrodynamical and MHD runs. It was moreover noted that, thanks to the rapid growth from dynamo effects, the memory of the initial field topology is quickly erased, rendering the precise choice of seed field geometry largely unimportant.

Similarly to~\citet{2022MNRAS.514..249C, 2023MNRAS.523.3709C, 2024MNRAS.532.3299N}, so as to prevent the collapse of the gaseous disk and ensuing spurious starburst at the start of the simulation, we assign stellar ages to a sub-set of our stellar particles a distance $10 \; \mathrm{kpc}$ from the galactic centre, to allow them to inject supernova feedback from the get go; this was done by assuming a constant star formation rate of $10 \; \mathrm{M}_\odot \; \mathrm{yr}^{-1}$ during the $100 \; \mathrm{Myr}$ preceding our $t=0$. The vertical structure of the stellar disk is set by assuming it follows an isothermal sheet profile with constant scale-height $z_\mathrm{disk} = 0.1 r_\mathrm{disk}$. The vertical structure of the gaseous disk is computed self-consistently under the joint assumption of hydrostatic equilibrium and the fact that gas is constrained to occupy a portion of space bounded by an `entropy floor'/`effective equation of state'; this is commonly implemented in production galaxy formation simulations that lack the resolution to properly capture the cold phase of the ISM. For our experiment, gravitational softening is fixed to $\varepsilon_\mathrm{soft.} = 0.2 \; \mathrm{kpc}$, and gas particle smoothing lengths $h$ are restricted to be greater than $h_\mathrm{min} = 0.014 \varepsilon_\mathrm{soft.}$, in order to prevent artificial clumping~\citep[adapting the suggestion of][to our different SPH kernel choice]{2024MNRAS.528.2930P}; we note that in practice this value is never reached in our simulation.

\begin{figure}
    \includegraphics[]{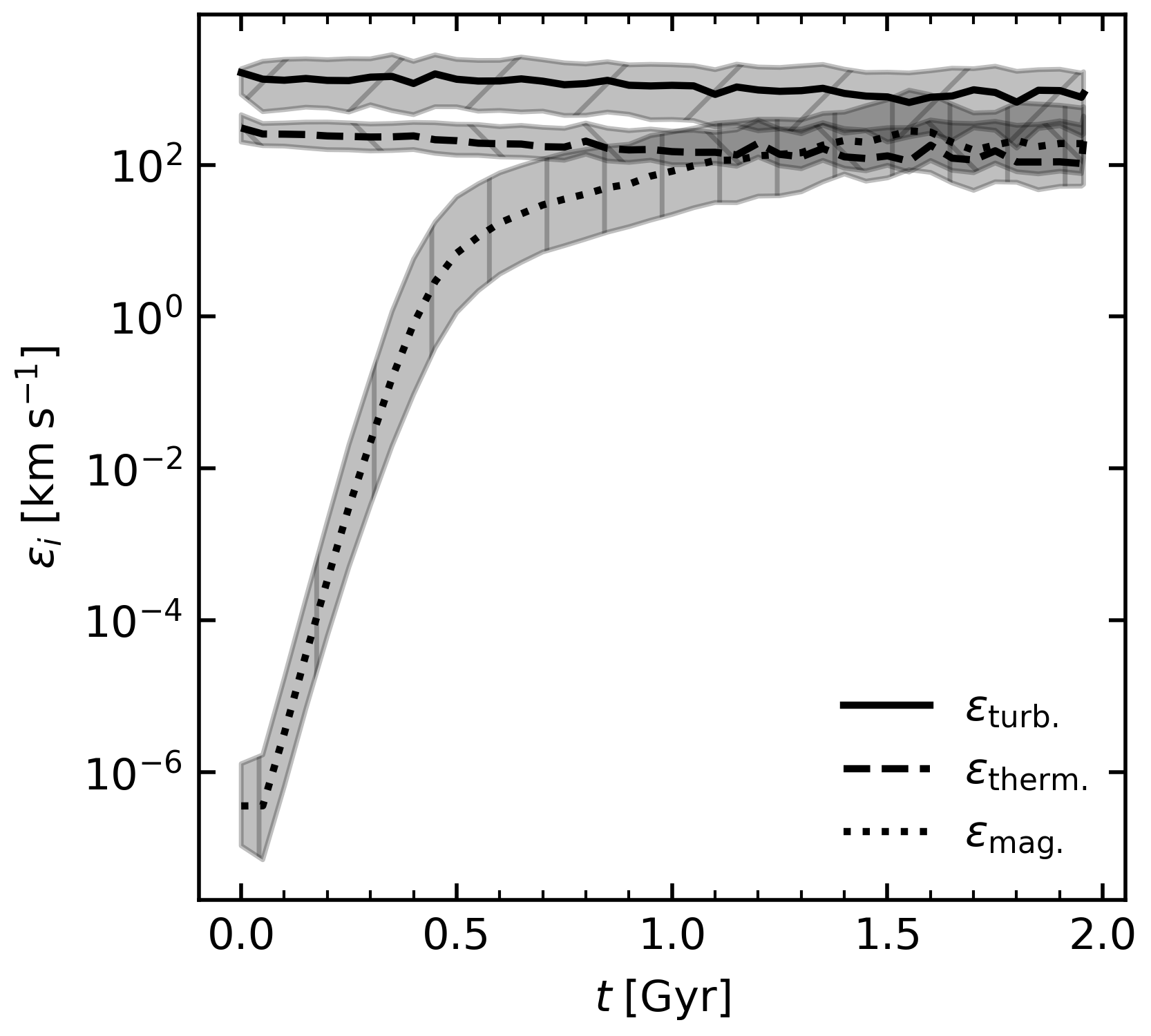}
    \caption{Evolution of different energy components in the galactic disk. We show, as a function of time, the median of specific turbulent kinetic energies $\varepsilon_\mathrm{turb.}$ (solid black line), the median of specific thermal energies $\varepsilon_\mathrm{therm.}$ (dashed black line) and the median of specific magnetic energies $\varepsilon_\mathrm{mag.}$ (dotted black line) for particles found within a cylindrical region $\Omega_\mathrm{disk}$ of radius $10 \; \mathrm{kpc}$ and height $1 \; \mathrm{kpc}$, centred on the midpoint of the simulation volume. We complement these with bands bounded by the ${16}^\mathrm{th}$ and ${84}^\mathrm{th}$ percentiles of the distribution of $\varepsilon_\mathrm{turb.}$, $\varepsilon_\mathrm{therm.}$ and $\varepsilon_\mathrm{mag.}$ in $\Omega_\mathrm{disk}$. Magnetic fields are amplified exponentially over the first $500 \; \mathrm{Myr}$ of the system's evolution, to subsequently grow steadily at a milder rate for the following $\sim 1 \; \mathrm{Gyr}$, eventually saturating completely as $\varepsilon_\mathrm{mag.}$ reaches equipartition with $\varepsilon_\mathrm{therm.}$ at a sizable fraction of $\varepsilon_\mathrm{turb.}$.}
    \label{fig:galaxy_energy_components}
\end{figure}

\subsubsection{Magnetic fields in a Milky Way-like isolated disk galaxy: Sub-grid physics}

We compile our code to run with the full network of sub-grid models used for the EAGLE project~\citep{2015MNRAS.446..521S, 2015MNRAS.450.1937C}, in its default configuration as re-implemented in SWIFT~\citep[see e.g.][for details]{2022MNRAS.516..167B, 2023MNRAS.526.2441B, 2023MNRAS.520.3164A, 2024MNRAS.530.2378S}. This port constitutes only a minor departure from the original implementation, adjusted to our software's code structure and adapted to work with SPHENIX. A similar galaxy formation framework, using a different calibration of free model parameters, was used for the more recent FLAMINGO simulation suite~\citep{2023MNRAS.526.4978S, 2023MNRAS.526.6103K}. 

Radiative gas cooling and heating are modeled on an element-by-element basis following~\citet{2009MNRAS.393...99W}. Individual gas particles track abundances for the 11 chemical species that~\citet{2009MNRAS.393...99W} claim are necessary but sufficient to retrieve cooling/heating rates with percent level accuracy.

To circumvent our inability to capture the physics of the cold phase of the ISM (a byproduct of our finite spatial resolution), we impose the temperature floor $T_\mathrm{min} (\rho)$ of~\citet{2008MNRAS.383.1210S} on gas particles of mass density $\rho$, which translates to a lower bound on their thermal pressure $P$. Once the atomic hydrogen number density $n_\mathrm{H}$ they carry exceeds a value of ${10}^{-4} \; \mathrm{cm}^{-3}$, gas particles are required to have a thermal pressure strictly greater than that predicted by the equation of state (EoS) $P_\mathrm{EoS} \propto \rho^{4/3}$, scaled so that $T_\mathrm{min}$ is normalised to $T_\mathrm{EoS} = 8000 \; \mathrm{K}$ at $n_\mathrm{H, EoS} = 0.1 \; \mathrm{cm}^{-3}$~\citep[which is representative of the warm phase of the ISM, see][]{2015MNRAS.446..521S}. This floor will inevitably limit the development of turbulence, which could in turn potentially hinder or restrict the operation of an active SSD. 

We employ the model of~\citet{2008MNRAS.383.1210S} to stochastically convert gas into collisionless star particles (representing entire stellar populations), at a rate which depends on thermal pressure. Eligibility for star formation is established using the metallicity-dependent atomic hydrogen number density threshold of~\citet{2004ApJ...609..667S}. Such a star formation episode constitutes a challenging situation for our SPMHD method, since it implies an instantaneous, localised removal of magnetic energy (star particles inherit, e.g. their parent gas particle's chemical composition, but not their magnetic field) that is by no means guaranteed to preserve the solenoidal constraint. Once we have established that we are not subject to spurious dissipation of the magnetic field, we are still confronted with a test that is highly demanding of our divergence-cleaning algorithm implementation.

Stellar feedback is simulated following an improved version of the prescription of~\citet{2012MNRAS.426..140D} used in the original EAGLE runs~\citep{2015MNRAS.446..521S}. Of the stars constituting our stellar populations, sampled at birth from the range $[0.01 \; \mathrm{M}_\odot, 100 \; \mathrm{M}_\odot]$ according to a~\citet{2003PASP..115..763C} initial mass function, those with initial masses in the range $[8 \; \mathrm{M}_\odot, 100 \; \mathrm{M}_\odot]$ are assumed to undergo core-collapse supernovae (SNII) at the end of their lives. The impact of SNII on the gas distribution is modeled through stochastic depositions of thermal energy to the nearest gas neighgbour of the exploding star particle~\citep[see][]{2022MNRAS.514..249C}, equating to a temperature increase of $\Delta T = {10}^{7.5} \; \mathrm{K}$ and translating to the typical ${10}^{51} \; \mathrm{erg}$ energy injection per supernova event. Energy injections occur at a time post-star particle formation that is obtained by sampling the stellar lifetime distribution of the respective population. Finally, chemical enrichment and stellar winds from SNII, but also type Ia supernovae (SNIa) and asymptotic giant branch (AGB) and massive stars, are modeled following the prescription of~\citet{2009MNRAS.399..574W} as modified by~\citet{2015MNRAS.446..521S}. Such an implementation of stellar feedback locally makes for abrupt, violent, and discontinuous changes in physical attributes, over intervals much shorter than characteristic timescales of MHD modes, which in turn results in a rapid redistribution of particles. This can instantaneously lead to poor sampling and thus an inaccurate estimate of fields of interest and their derivatives, potentially translating to numerical divergences. It is a study of the occurrence of the latter that prompted the conception of the novel regularisation we propose for the tensile instability correction (TIC); where a naive implementation of existing TIC prescriptions yielded repulsive forces near supernova events that were at times strong enough to dissolve the entire galaxy, our suggestion proved to stabilise runs completely and be remarkably robust.

\begin{figure*}
 \includegraphics[width=\textwidth]{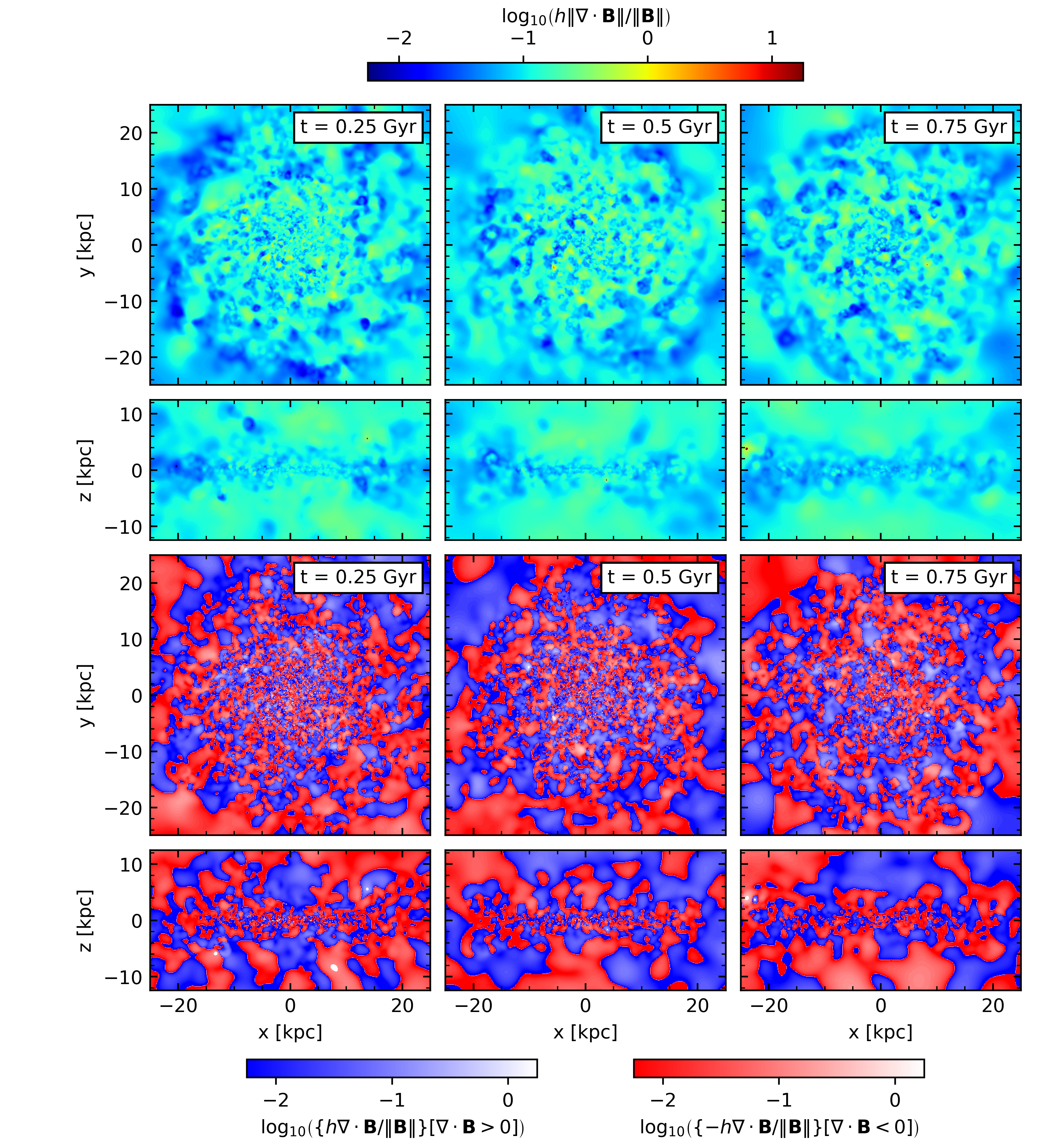} 
 \caption{Divergence errors present in our magnetised Milky Way-like isolated disk galaxy simulation. The plotting configuration and panel layout is identical to Fig.~\ref{fig:galaxy_projection}, the only difference being that instead of gas surface density and magnetic field strength, we here project the standard dimensionless divergence error $\varepsilon_{\nabla \cdot \boldsymbol{B}}$ ($1^\mathrm{st}$ and $2^\mathrm{nd}$ rows) and a variant thereof ($3^\mathrm{rd}$ and $4^\mathrm{th}$ rows); in the latter case we create a composite error map, complementary segments of which are painted using one of two colour maps, according to whether the divergence of the magnetic field there is positive or negative. The relative error stays on average at the $\lesssim 0.1$ level and never exceeds unity, meaning monopole accelerations remain dynamically subdominant throughout the system's evolution. Comparing projections shown here to those presented in Fig.~\ref{fig:galaxy_projection}, we identify a clear spatial correlation between local maxima in $\varepsilon_{\nabla \cdot \boldsymbol{B}}$ and local minima/strong gradients in the gas mass distribution; observed errors are thus likely harmless and attributable to our resolution-limited capabilities at reconstructing physical fields in under-sampled regions. The signed $\varepsilon_{\nabla \cdot \boldsymbol{B}}$ maps in the two bottom rows indicate the divergence fluctuates rapidly on small scales, a likely manifestation of our mixed hyperbolic/parabolic cleaning algorithm efficiently transporting divergence away from where it is sourced in the form of damped waves.}
 \label{fig:galaxy_errors}
\end{figure*}

\subsubsection{Magnetic fields in a Milky Way-like isolated disk galaxy: Results}

We first present in Fig.~\ref{fig:galaxy_projection} projections of the galaxy at three consecutive points in time, taken to be integer multiples of $250 \; \mathrm{Myr}$ (the rotational period at a radial distance $\sim r_\mathrm{disk}$ from the galaxy's rotation axis). We show both face-on and edge-on maps of the gas surface density $\Sigma_\mathrm{gas}$ and norm of the magnetic field vector $\Vert \boldsymbol{B} \Vert$.

The spatial distribution of gas in our simulated galaxy is largely similar (both in terms of morphology and typical $\Sigma_\mathrm{gas} \sim {10}^2 \; \mathrm{M}_\odot \; \mathrm{pc}^{-2}$ values reached) with that shown in e.g.~\citet{2022MNRAS.514..249C, 2023MNRAS.523.3709C, 2024MNRAS.532.3299N}, who considered pure hydrodynamical counterparts to the problem we present here. The thermal energy injections seeded by the stellar feedback prescription we employ disrupt the galactic disk by blowing `super-bubbles' of hot under-dense gas, and are further responsible for launching large-scale outflows (reaching a few $\mathrm{kpc}$ along the $z$ direction) perpendicular to the disk plane, dispersing material in the originally unpopulated CGM. Hints of a large-scale spiral structure are present at all times probed, and while the dynamics is partly dominated by coherent galactic rotation, the flow is undeniably largely turbulent on smaller scales. This is in contrast to other similar investigations that do not explicitly model the effect of supepernovae~\citep[e.g.][]{2009MNRAS.397..733K, 2009ApJ...696...96W, 2013MNRAS.432..176P, 2019MNRAS.483.1008S} and consequently observe smoother field profiles, which in turn is less demanding of the underlying discrete MHD solver. 

Although initially permeated by vanishingly weak magnetic fields, cold dense gas becomes increasingly magnetised, with $\Sigma_\mathrm{gas}$ and $\Vert \boldsymbol{B} \Vert$ being largely spatially correlated~\citep[see however][for how inclusion of an additional diffusion term can reverse this trend]{2019MNRAS.483.1008S}. By $t = 500 \; \mathrm{Myr}$, the $\sim \mathrm{nG}$ seeds have been twisted and amplified to $\gtrsim 10 \; \mu \mathrm{G}$ values in the inner part of the galactic disk, in line with other numerical works such as~\citet{2013MNRAS.432..176P} and the disk galaxy observations they compare their results to~\citep{2007A&A...470..539B}. Again in line with~\citet{2013MNRAS.432..176P}, the build-up of a magnetised disk occurs `inside-out' in our simulation, starting from the galactic centre where star formation, and therefore stellar feedback and the ensuing turbulent velocity flow that can sustain an SSD, are most prominent in the first few $100 \; \mathrm{Myr}$ of the system's evolution~\citep[see][for a discussion of the initial burst in star formation observed in such a system]{2022MNRAS.514..249C, 2023MNRAS.523.3709C}. Amplification is first halted at smaller radii, once magnetic fields there have reached their saturation value, and their subsequent growth occurs primarily at increasingly larger radial distances as time progresses; this is in line with the expectation that growth driven by differential shear (particularly at later times when it takes over as the dominant amplification mechanism, when stellar feedback activity has declined and an SSD cannot operate efficiently anymore) will occur at a rate that scales with the rotational period, itself typically an increasing function of radius. The CGM is polluted by magnetised material evacuated from the disk by supernova-driven winds, as seen also by e.g.~\citet{2010A&A...523A..72D, 2016MNRAS.457.1722R}; it appears most magnetised around the galactic centre, where supernova explosions are most concentrated and gas eligible for ejection has the strongest magnetic fields. We again emphasise that no attempt was made to model a realistic CGM, and results regarding the atmosphere surrounding our galaxy should be interpreted with care. We further note that our system is considered `in vacuum', rather than being embedded in a realistic cosmological environment, meaning processes such as continued mass accretion or mergers, which are likely to play a major role in galaxy evolution, are omitted entirely; we therefore stress that our simulation, as others using a similar set-up, has limited predictive power, particularly at later times.

To look at magnetic field amplification through a different lens, we show in Fig.~\ref{fig:galaxy_energy_components} the time evolution of energy components of interest within the galactic disk. We plot the median, together with the ${16}^\mathrm{th}$ and ${84}^\mathrm{th}$ percentiles (calculated over all particles within a cylindrical region of radius $10 \; \mathrm{kpc}$ and height $1 \; \mathrm{kpc}$, centred on the midpoint of the simulation domain and intended to isolate the gaseous disk) of the distribution of specific thermal and magnetic energies
\begin{equation}
    \varepsilon_\mathrm{therm.} = u
    \quad \text{and} \quad
    \varepsilon_\mathrm{mag.} = \frac{\boldsymbol{B}^2}{2\mu_0\rho}
\end{equation}
as well as of turbulent kinetic energies
\begin{equation}
    \varepsilon_\mathrm{turb.} = \frac{1}{2} \boldsymbol{v}_\mathrm{turb.}^2
\end{equation}
where the turbulent velocity field $\boldsymbol{v}_\mathrm{turb.}$ is defined as the residual after subtraction from the total velocity field $\boldsymbol{v}$ of the global rotation $\boldsymbol{v}_\mathrm{rot.}$ in the azimuthal direction $\hat{\boldsymbol{e}}_\phi$ expected for a~\citet{1990ApJ...356..359H} halo\footnote{We refer the reader to~\citealt{2024MNRAS.532.3299N} for a definition of both $M_\mathrm{Hern.}$ and $r_*$ in terms of the potential parameters we quote in Section~\ref{sec:galaxy_initialisation}}
\begin{equation}
    \boldsymbol{v}_\mathrm{turb.} = \boldsymbol{v} - \boldsymbol{v}_\mathrm{rot.}
    \quad \text{for} \quad
    \boldsymbol{v}_\mathrm{rot}
    =
    \sqrt{\frac{G M_\mathrm{Hern.} \Vert \boldsymbol{r} \Vert}
    {\left( \Vert \boldsymbol{r} \Vert + r_* \right)^2}}
    \hat{\boldsymbol{e}}_\phi.
\end{equation}

Fig.~\ref{fig:galaxy_energy_components} depicts a magnetic field amplification history that can be broken down into three qualitatively distinct stages, and is largely consistent with e.g. the high-resolution AMR results of~\citet{2017MNRAS.472.4368R} for an isolated galaxy or the outcome of the zoom-in cosmological runs of~\citet{2017MNRAS.469.3185P}, carried out using a moving-mesh solver. The initially subdominant magnetic energy component is exponentially amplified over $\sim 500 \; \mathrm{Myr}$ (which is possibly attributable to the action of an SSD in the kinematic phase), before transitioning to a milder linear amplification stage lasting $\sim 1 \; \mathrm{Gyr}$ (possibly attributable to an non-linear SSD or an LSD), prior to coming to a complete halt as $\varepsilon_\mathrm{mag.}$ saturates in equipartition with $\varepsilon_\mathrm{therm.}$ and at a sizable fraction of $\varepsilon_\mathrm{turb.}$; the latter is in excellent agreement with observations of nearby disk galaxies~\citep[see e.g.][]{1996ARA&A..34..155B, 2015A&ARv..24....4B}. Our code proves able to sustain saturated fields late into the system's evolution~\citep[possibly owing to the perpetual action of an LSD, see e.g.][]{2017ApJ...843..113B}, and shows no sign of spurious diffusion as gas particles are converted into stars and their associated magnetic field removed from the simulation box.

We conclude our discussion by considering the divergence errors present in our simulations, which we depict in Fig.~\ref{fig:galaxy_errors} at the three consecutive moments in time at which the galaxy is imaged in Fig.~\ref{fig:galaxy_projection}. We show both face- and edge-on renderings of the standard dimensionless divergence error metric $\varepsilon_{\nabla \cdot \boldsymbol{B}}$ (equation~\ref{eq:dimensionless_divergene_error}), as well as a variation thereof wherein complementary segments of the maps are painted using either a blue- or red-toned colourmap, depending on whether $\nabla \cdot \boldsymbol{B}$ is respectively positive or negative there. We observe that the average error stays at the $\langle \varepsilon_{\nabla \cdot \boldsymbol{B}} \rangle \lesssim 0.1$ level and remains safely below $\mathrm{max} \{ \varepsilon_{\nabla \cdot \boldsymbol{B}} \} < 1$ at all times probed, indicating the dynamics remains unaffected by spurious monopole forces. We note that at later times, as the galaxy settles into a more quiescent state past the initial starburst and ensuing violent feedback events, divergence errors are further reduced. Contrasting Fig.~\ref{fig:galaxy_errors} with Fig.~\ref{fig:galaxy_projection} we can further see that local maxima in the distribution of $\varepsilon_{\nabla \cdot \boldsymbol{B}}$ are largely spatially correlated with local minima or strong gradients in $\Sigma_\mathrm{gas}$, for instance at energy injection sites where gas is rapidly depleted. This hints towards higher measured errors being attributable (at least in part) to inherent limitations in our ability to measure field quantities in sparsely populated and therefore under-resolved regions, rather than being indicative of unphysical artefacts. The last two rows of Fig.~\ref{fig:galaxy_errors} also demonstrate that $\nabla \cdot \boldsymbol{B}$ varies quickly, alternating sign on small scales, wherever a reasonable number of resolution elements are present to sample the gas phase~\citep[][observe the exact same in their results]{2013MNRAS.432..176P}; this oscillatory behaviour can be interpreted as the resultant of our~\citet{2002JCoPh.175..645D}-like divergence-cleaning algorithm continuously and efficiently transporting $\nabla \cdot \boldsymbol{B}$ in the form of damped waves away from where it is sourced in the galactic disk. 

Our results are illustrative of how leveraging upon and refining recent developments in SPMHD research allows the conception of a solution better adapted to simulations of galaxy evolution, when compared to earlier implementations of the method which provided poorer divergence control on similar problems~\citep[e.g.][for whom $\langle \varepsilon_{\nabla \cdot \boldsymbol{B}} \rangle \gtrsim 0.5-1$]{2009MNRAS.397..733K, 2012MNRAS.422.2152B}. We moreover seem to improve upon the results of contemporary SPMHD solvers, who only report matching error levels in comparable numerical experiments when running without self-gravity, star formation and stellar feedback~\citep{2016MNRAS.461.4482D}, or when making use of a factor $\sim 500$ higher mass resolution~\citep{2023A&A...673A..47W}; other modern codes appear to be less capable at preventing $\varepsilon_{\nabla \cdot \boldsymbol{B}}$ from locally reaching values $>1$~\citep{2019MNRAS.483.1008S}. We further seem to slightly outperform the moving-mesh code AREPO~\citep[][who report $\langle \varepsilon_{\nabla \cdot \boldsymbol{B}} \rangle \sim 0.3-0.6$ for a similar test run at a similar mass resolution, but admittedly employ a simpler, yet arguably sufficient for their purposes, divergence-cleaning algorithm]{2013MNRAS.432..176P}, and produce results whose reliability (as evaluated through $\varepsilon_{\nabla \cdot \boldsymbol{B}}$ again) is on par with that of the more computationally demanding AMR code ENZO~\citep{2009ApJ...696...96W} and `mesh-less finite mass'/`mesh-less finite' volume solvers of~\citet{2016MNRAS.455...51H}.

\section{Conclusions}
\label{sec:conclusions}

In this work, we have presented a novel formulation of cosmological SPMHD, chiefly intended for studies of galaxy and cosmic structure formation, which we have implemented in the simulation code SWIFT. Our method was incorporated into the codebase in a manner that allows it to operate on top of any of the mesh-less hydrodynamics solvers the software can be configured with; it has nevertheless been designed and calibrated to work optimally as an extension to the SPHENIX SPH scheme~\citep{2022MNRAS.511.2367B}. It can moreover be seamlessly coupled to any of the numerous complementary physics modules available in SWIFT, including the FMM gravity solver and EAGLE network of sub-grid galaxy formation models. Although our core focus remains Ideal MHD, we have extended our scheme to also account for non-ideal effects by including Ohmic resistivity terms that assume a constant magnetic diffusivity. 

We have sought to maximise performance by integrating into our method recommendations recently put forward in the methods literature, and we provide suggestions for how to modify these in view of achieving code stability when running simulations targeting our science case of predilection. We:
\begin{tasks}
    \task Advocate for a better choice of interpolation stencil, allowing for more accurate reconstruction of physical fields and their spatial gradients~\citep{2012MNRAS.425.1068D}.
    \task Self-consistently derive the set of equations that constitute our model from an action minimisation principle, following the variable smoothing length formalism of~\citet{2013MNRAS.428.2840H} and imposing Maxwell's equations as a constraint, to obtain a conservative density-energy scheme that perfectly preserves mass, linear momentum, energy and entropy at the discrete level.
    \task Mildly modify and augment SPHENIX's state-of-the-art artificial viscosity (AV) prescription~\citep[based on][]{2010MNRAS.408..669C}, so that when coupled to an artificial thermal diffusion (AD) scheme~\citep[based on][]{2008JCoPh.22710040P}, it can effectively respond to MHD shocks. SPHENIX's advantages, including reduction of surface tensions effects, unimpeded phase mixing, and prevention of spurious cooling losses when coupling to sub-grid physics models, are all preserved.
    \task Augment the aforementioned discontinuity-capturing scheme with the artificial resistivity (AR) prescription of~\citet{2013MNRAS.436.2810T}, modified according to the suggestions of~\citet{2016MNRAS.455...51H}, to capture discontinuities in the magnetic field. We argue against the use of the more recent AR scheme proposed by~\citet{2018PASA...35...31P}, which we find to lead to excess magnetic dissipation in gaseous halos in cosmological simulations.
    \task Address any non-vanishing divergence of the magnetic field, which would violate the solenoidal constraint  $\nabla \cdot \boldsymbol{B} = 0$, by including the force source term of~\citet{2001ApJ...561...82B}. This prevents tensile pairing of particles in highly magnetised environments, and enables passively advecting (rather than dispersing) $\nabla \cdot \boldsymbol{B}$. We modulate this non-conservative correction by a novel switch that activates it only when necessary, and prevents spurious particle ejection when coupling to sub-grid models.
    \task Maintain $\nabla \cdot \boldsymbol{B}$ numerically small, using a constrained hyperbolic/parabolic divergence-cleaning scheme inspired from~\citet{2002JCoPh.175..645D}. We combine the suggestions of~\citet{2016JCoPh.322..326T} and~\citet{2016MNRAS.455...51H} to obtain a scheme which removes divergence errors at a critical rate and allows for variable cleaning speeds at no additional computational cost.
    \task Propose novel timestepping criteria to stabilise the time integration of MHD variables in the presence of discontinuous changes in other physical attributes that affect their evolution.
\end{tasks}
The performance of our solver is evaluated using a series of controlled numerical experiments. Importantly, having first used a generalised version of the monopole advection test of~\citet{2002JCoPh.175..645D} to tune our divergence-cleaning scheme and the shock tube of~\citet{1988JCoPh..75..400B} to calibrate our AR prescription, we keep model parameters fixed for all tests presented as part of our method validation procedure. We demonstrate excellent performance on shock tubes involving all possible MHD discontinuities, planar flow problems comprising strong interacting shocks and turbulence, and fluid mixing instability set-ups that have proven notoriously challenging for particle-based methods. Convergence with resolution is shown to be second order in smooth flows, and mildly sub-linear at discontinuities. The joint action of our AV, AD and AR prescriptions allows for an accurate and unique representation of discontinuous jumps in physical attributes, with implemented switches largely limiting unwarranted numerical diffusion. Our formulation of the `tensile instability correction' prevents particle clumping whenever this could have been expected, and the implemented divergence-cleaning algorithm proves highly capable of maintaining divergence errors systematically low, even in highly dynamical environments.

We further successfully apply our method to three challenging astrophysical problems, obtaining results that are both on par with predictions from mesh-based methods and consistent with expectations from theory and observations. We first follow the collapse of a magnetised gas cloud, down to the formation of a dense pre-stellar core and the launching of collimated bipolar jets; where other authors claim that `standard' SPMHD's poor divergence control capabilities and necessary but ad-hoc dissipative corrections prevent the method from delivering stable and physically plausible solutions~\citep{2016MNRAS.455...51H, 2020A&A...638A.140W}, we are able to retrieve the phenomenology reported in high-resolution (unstructured) grid code studies~\citep{2008A&A...477....9H, 2011MNRAS.418.1392P}. We additionally follow the growth of trace magnetic seeds in non-radiative simulations of a massive galaxy cluster, to obtain realistic radial profiles and scaling relations in different mass density regimes. Most importantly, we capture the signature of a turbulent small-scale dynamo, the central mechanism believed to drive the rapid amplification of cluster fields to their observed present-day values. We finally couple, for the first time, our MHD solver to the full EAGLE model of galaxy formation~\citep{2015MNRAS.446..521S, 2015MNRAS.450.1937C}. We study a magnetised Milky Way-like isolated disk, to show an evolution history consistent with previous studies and a saturated magnetic configuration in line with observations of both distant and local universe galaxies. Crucially, we demonstrate code stability when running with a complex network of sub-grid physics models, which engender rapid changes in the particle distribution and various key physical attributes, pushing our numerical method to its limits.

Our study lays the groundwork for numerous follow-up investigations, particularly enabling large-scale magnetohydrodynamical cosmological simulations (thanks to SWIFT's excellent weak- and strong-scaling capabilities), or facilitating the yet largely unexplored coupling of magnetic field physics to sub-resolution models of galaxy formation. More testing is naturally needed to ensure that our method is capable of reproducing key results from the MHD literature, concerning e.g. the magnetorotational instability or driven supersonic turbulence. We nevertheless argue that, beyond that, the present study can serve as an instigator to explore two major research avenues. First, with SPMHD now being a truly mature and stable method, it would be interesting to initiate systematic and controlled code comparison projects of highly non-linear astrophysical systems (such as the cloud collapse example we considered here), to firmly establish the method's capabilities when compared to finite-volume schemes; understanding how these really compare becomes non-trivial when each is considered in isolation. Second, we would encourage the community to look back and improve upon components of SPMHD (such as the tensile instability correction) that have been overlooked and not re-evaluated in years. The holy grail in the field remains the conception of an improved AR prescription, which adequately addresses each distinct type of MHD discontinuity and is truly dissipationless in smooth flows. Nevertheless, we believe that the method is now robust enough and able to capture most of the phenomenology of MHD relevant to galaxy and cosmic structure formation, and can therefore be considered as a viable and complementary approach to existing numerical techniques for simulating magnetic field physics.

\section*{Acknowledgements}

The authors would particularly like to thank Joop Schaye, whose continued support was indispensable to the realisation of this project. They would further wish to extend their gratitude to Josh Borrow, Joey Braspenning, Evgenii Chaikin, Joseph Hennawi, Filip Hu\v{s}ko, Mladen Ivkovic, Roi Kugel, Rob McGibbon, Ruediger Pakmor, Evangelos Petridis, Marco van der Ploeg, Daniel Price, Yves Revaz, Ivan Ridkokasha, Ulrich Steinwandel, Romain Teyssier, Bert Vandenbroucke and many others; the fruitful discussions that were shared were instrumental in improving the quality of this work. They finally feel indebted to all other people in the SWIFT team who contributed with suggestions, comments, and constructive exchanges on discrete methods and code development. This project has received funding from the European
Union’s HORIZON-MSCA-2021-SE-01 Research and Innovation program under the Marie Sklodowska-Curie grant agreement number 101086388 - Project acronym: LACEGAL.

\section*{Data Availability}

SWIFT is fully open-source; extensive documentation describing how to use it and details on how to download it can be found at https://www.swiftsim.com. The MHD module, as well as the initial conditions and code parameter files used to run all examples presented in this work, will be added to the released version of the code upon acceptance of this manuscript. They can be accessed earlier upon request to the authors.



\bibliographystyle{mnras}
\bibliography{spmhdForGalaxyAndCosmicStructureFormation} 



\appendix

\section{Full set of equations of SPMHD in an expanding frame}
\label{appendix:cosmo_SPMHD}
The full set of equations of cosmological SPMHD we consider in SWIFT reads
\begin{strip}
\hrule
    \begin{align}
        \begin{split}
            \frac{\mathrm{d} \boldsymbol{w}_i}{\mathrm{d} t} = 
            \frac{1}{a^{3(\gamma - 1)}} 
            \sum_j m_j 
            \Biggl\{
            & - \left[ 
            \frac{f_{ij}}{{\hat{\rho}}_i^2}
            \boldsymbol{\mathrm{S}}_i \cdot \nabla_i W_{ij} (h_i) +
            \frac{f_{ji}}{{\hat{\rho}}_j^2}
            \boldsymbol{\mathrm{S}}_j \cdot \nabla_i W_{ij} (h_j)
            \right]
            - \zeta_{ij} \Biggl[
            f_{ij} \nabla_i W_{ij} (h_i) + f_{ji} \nabla_i W_{ij} (h_j) 
            \Biggr] \\
            & - \lambda (\beta_i^\mathrm{loc}) \boldsymbol{B}_i
            \Biggl[
            \frac{f_{ij}}{{\hat{\rho}}_i^2}
            \boldsymbol{B}_i \cdot \nabla_i W_{ij} (h_i)
            + \frac{f_{ji}}{{\hat{\rho}}_j^2}
            \boldsymbol{B}_j \cdot \nabla_i W_{ij} (h_j)
            \Biggr]
            \Biggr\}
        \end{split} \\
        \begin{split}
            \frac{\mathrm{d} u_i}{\mathrm{d} t} =
            \frac{1}{a^2} \sum_j m_j \Biggl\{
            & \frac{f_{ij} P_i}{{\hat{\rho}}_i^2}
            \boldsymbol{v}_{ij} \cdot \nabla_i W_{ij} (h_i)
            - \frac{\eta}{\mu_0 \hat{\rho}_i \hat{\rho}_j}
            \Biggl[
            f_{ij} F_{ij} (h_i) + f_{ji} F_{ij} (h_j) \Biggr]
            \frac{\boldsymbol{B}_{ij}^2}{\Vert \boldsymbol{r}_{ij} \Vert} \\
            & 
            - \frac{1}{4 \mu_0}
            \frac{
            \overline{
                \alpha^\mathrm{AR}
            }_{ij}
            \cdot
            \overline{
                v^\mathrm{AR}
            }_{ij}
            }{
            \overline{\rho}_{ij}
            }
            \Biggl[ 
                f_{ij} F_{ij} (h_i) + f_{ji} F_{ij} (h_j)
            \Biggr]
            \boldsymbol{B}^2_{ij}
            \Biggr\} 
        \end{split} \\
        \begin{split}
            \frac{\mathrm{d}}{\mathrm{d} t}
            \left( \frac{\boldsymbol{B}_i}{\hat{\rho}_i} \right) =
            \frac{1}{a^2} \sum_j m_j
            \Biggl\{
            & - \frac{f_{ij} \boldsymbol{w}_{ij}}{\hat{\rho}_i^2}
            \boldsymbol{B}_i \cdot \nabla_i W_{ij} (h_i) 
            + \frac{\eta}{\hat{\rho}_i \hat{\rho}_j}
            \Biggl[
            f_{ij} F_{ij} (h_i) + f_{ji} F_{ij} (h_j) 
            \Biggr]
            \frac{\boldsymbol{B}_{ij}}{\Vert \boldsymbol{r}_{ij} \Vert} \\
            & + \frac{1}{2}
            \frac{
                \overline{
                    \alpha^\mathrm{AR}
                }_{ij}
                \cdot
                \overline{
                    v^\mathrm{AR}
                }_{ij}
            }{
                \overline{\rho}_{ij}
            }
            \Biggl[ 
                f_{ij} F_{ij} (h_i) + f_{ji} F_{ij} (h_j)
            \Biggr]
            \boldsymbol{B}_{ij}
            - \left[
            \frac{f_{ij} \psi_i}{\hat{\rho}_i^2} \nabla_i W_{ij} (h_i) +
            \frac{f_{ji} \psi_j}{\hat{\rho}_j^2} \nabla_i W_{ij} (h_j)
            \right]
            \Biggr\}
            + \Gamma H \frac{\boldsymbol{B}_i}{\hat{\rho}_i}
        \end{split} \\
        \begin{split}
            \frac{\mathrm{d}}{\mathrm{d} t}
            \left( \frac{\psi_i}{c_{h,i}} \right) =
            \frac{1}{a^2} \sum_j m_j
            \Biggl\{
            & \frac{c_{h,i}}{\hat{\rho}_i} f_{ij} \boldsymbol{B}_{ij} \cdot \nabla_i W_{ij} (h_i) 
            + \frac{1}{2 \hat{\rho}_i} f_{ij} \boldsymbol{w}_{ij} \cdot \nabla_i W_{ij} (h_i) \frac{\psi_i}{c_{h,i}}
            \Biggr\}
            - \frac{1}{a^2} \frac{\psi_i}{\tau_{p,i} c_{h,i}}
            + \left( \frac{5}{2} - \frac{\nu}{2} \right) H \frac{\psi_i}{c_{h,i}}
        \end{split}
    \end{align}
\hrule
\end{strip}
where all variables appearing are comoving; we have committed the subscript `c' for clarity. Nevertheless, `primitive' comoving quantities $Q_c$ are as defined in equations~(\ref{eq:comoving_thermo_variables}), (\ref{eq:comoving_B}) and~(\ref{eq:comoving_dedner_variables}) in Section~\ref{sec:cosmological_SPMHD}, and
composite comoving functions $f_c(\{Q_c\})$ have the same dependence on the set of variables $\{Q_c\}$ as their physical counterpart $f(\{Q\})$ on $\{Q\}$.


\bsp	
\label{lastpage}
\end{document}